%% file: PhD_PVRGSilva_f.tex
\documentclass[a4paper,12pt]{report}

\include{meta_f}

\begin{document}

\include{pre_f}   
\include{ch_introduction_f} 
\include{ch_basic_concepts_f} 
\include{ch_data_methods_f} 
\include{ch_ratio_eltot_f} 
\include{ch_rise_sigmatot_f} 
\include{ch_SLT_sigmatot_f} 
\include{ch_conclusions_f} 


\makeatletter
\renewcommand\@biblabel[1]{#1.}
\makeatother

\cleardoublepage
\bibliographystyle{naturemag.bst} 
\bibliography{PhD_references_f.bib}


\appendix

\include{app_ddr_f}
\include{app_mellin_f}

\include{app_publications_f}

\end{document}

%% file: meta_f.tex
\usepackage[english,brazilian]{babel}

\usepackage[utf8]{inputenc}

\usepackage{microtype}
\microtypesetup{activate={true,nocompatibility}}
\microtypesetup{factor=1100, stretch=10, shrink=10}
\microtypesetup{tracking=true, spacing=true, kerning=true}
\SetTracking{encoding={T1}, shape=sc}{0}
\microtypesetup{final, babel}

\usepackage[namelimits]{amsmath}
\usepackage{amssymb,amsfonts,amsthm,textcomp}
\usepackage[a4paper,right=2.5cm,left=2.85cm,bottom=2.25cm,top=3cm]{geometry}           
\usepackage{setspace}

\usepackage{graphicx}
\usepackage{epstopdf}

\usepackage[normalem]{ulem} 
\usepackage[usenames,dvipsnames,svgnames]{xcolor}

\usepackage{caption}

\usepackage{subfig}
\usepackage{etoolbox} 
\robustify\subref
\usepackage{multicol}

\usepackage{array}
\usepackage{multirow}

\usepackage{booktabs}  

\usepackage{longtable}

\usepackage{lscape}

\usepackage[notintoc,portuguese]{nomencl}
\makenomenclature

\usepackage{footnote}
\makesavenoteenv{tabular}

\usepackage{lastpage}

\usepackage[nottoc,notlof,notlot]{tocbibind}

\usepackage{icomma}

\usepackage[tight]{units}

\usepackage[Lenny]{fncychap}

\usepackage{fancyhdr}
\fancypagestyle{plain}
{\fancyhf{}
 \fancyhead[R]{\thepage}
}
\newcommand{\nomedoaluno}{Paulo Victor Recchia Gomes da Silva}
\newcommand{\titulo}{Phenomenological Analyses on Hadronic Cross-Sections at High and Asymptotic Energies}

\usepackage{hyperref}
\hypersetup{colorlinks=true, linkcolor=black, citecolor=black, filecolor=black, pagecolor=black, urlcolor=black,
            pdfauthor={\nomedoaluno},
            pdftitle={\titulo},
            pdfsubject={Assunto do Projeto},
            pdfkeywords={palavra-chave, palavra-chave, palavra-chave},
            pdfproducer={LaTeX},
            pdfcreator={pdfTeX}}

\usepackage{url}
\usepackage{indentfirst}
\usepackage{comment}

\usepackage{pgfplots}
\usepackage{tikz}
\usetikzlibrary{decorations.markings}

\usepackage{rotating}

\usepackage{cite}

\usepackage{threeparttable}

\usepackage{bbm}      
\usepackage{dsfont}   

\newcommand{\sigmatot}{\sigma_\text{tot}}
\newcommand{\sigmael}{\sigma_\text{el}}
\newcommand{\sigmain}{\sigma_\text{inel}}
\newcommand{\sigmadiff}{\sigma_{\text{diff}}}

\newcommand{\etal}{\textit{et al. }}

\DeclareMathOperator{\Real}{Re}
\DeclareMathOperator{\Imag}{Im}

\DeclareMathOperator{\Trace}{Tr}

\DeclareMathOperator{\corr}{corr}

\newtheorem*{corollary}{Corollary} 

\newcommand{\reggeon}{\mathds{R}}
\newcommand{\pomeron}{\mathds{P}}
\newcommand{\AP}{A_\mathds{P}}   
\newcommand{\pp}{pp}
\newcommand{\ppbar}{\bar{p}p}

\newcommand{\LT}{\text{LT}}
\newcommand{\SLT}{\text{SLT}}
\newcommand{\LTth}{\text{LT}_\text{th}}
\newcommand{\SLTth}{\text{SLT}_\text{th}}
\newcommand{\LTthk}{\text{LT}_{\text{th},\kappa}}
\newcommand{\SLTthk}{\text{SLT}_{\text{th},\kappa}}

\newcommand{\QLT}{\text{QLT}}
\newcommand{\QSLT}{\text{QSLT}}
\newcommand{\QLTth}{\text{QLT}_\text{th}}
\newcommand{\QSLTth}{\text{QSLT}_\text{th}}

\newcommand{\QSLTthk}{\text{QSLT}_{\text{th},\kappa}}

\usepackage{notoccite}   
\usepackage{pdfpages}

 \usepackage[backgroundcolor=gray!30!white]{todonotes} 

%% file: pre_f.tex
%

\pagestyle{empty}
\pagenumbering{arabic}



 
 \begin{center}
  
  \begin{minipage}{0.2\textwidth}
   \includegraphics[scale=0.15]{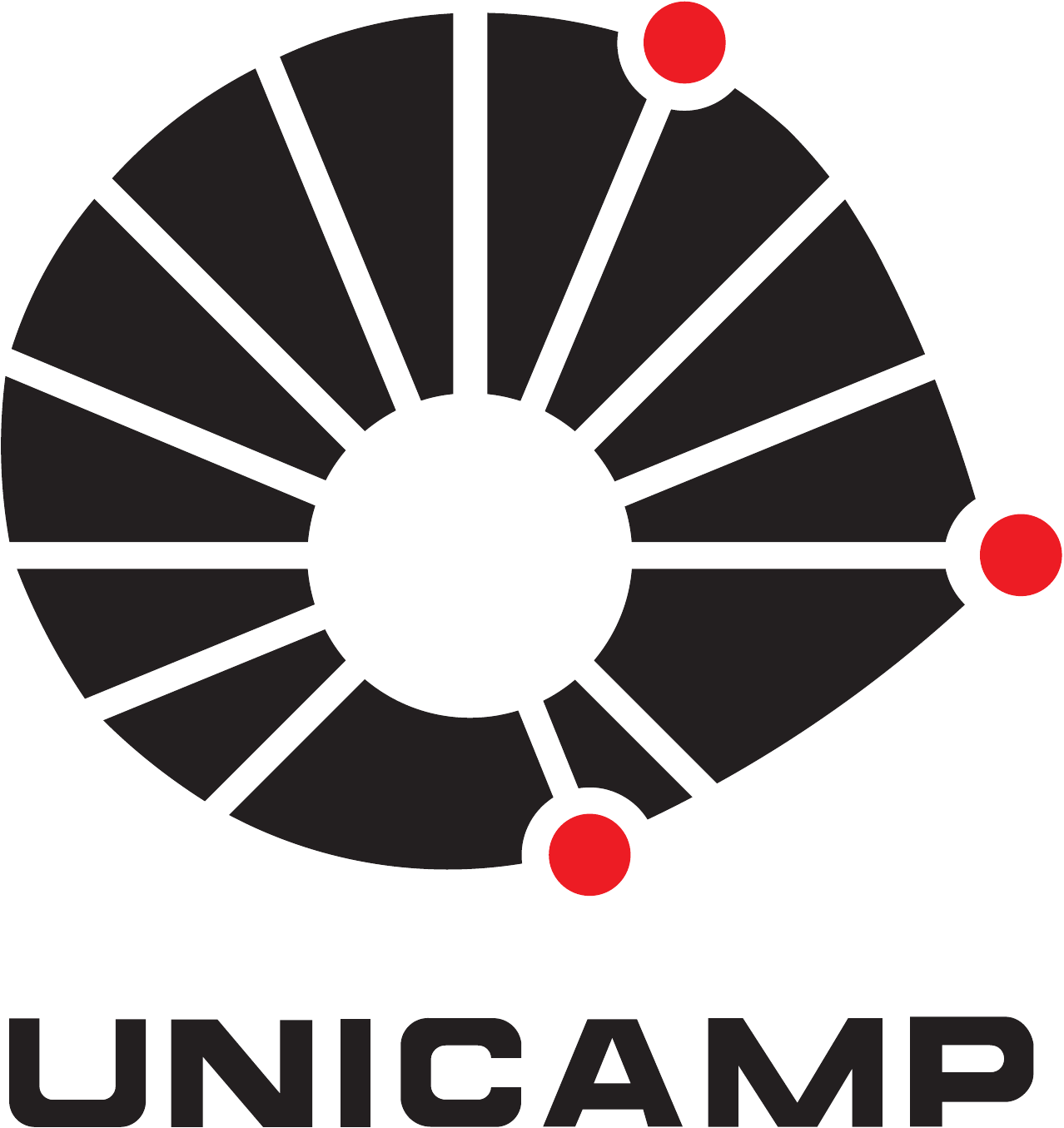}
  \end{minipage}
  \hfill
  \begin{minipage}{0.7\textwidth}
   \hspace*{-1.75cm}{\large UNIVERSIDADE ESTADUAL DE CAMPINAS \\ Instituto de Física ``Gleb Wataghin''}
  \end{minipage}

  \par
  \vspace{200pt}

  {\large PAULO VICTOR RECCHIA GOMES DA SILVA}
  
  \par
  \vspace{100pt}

  {\large PHENOMENOLOGICAL ANALYSES ON HADRONIC CROSS-SECTIONS AT HIGH AND ASYMPTOTIC ENERGIES}

  \par
  \vspace{100pt}
  
  {\large ANÁLISES FENOMENOLÓGICAS DE SEÇÕES DE CHOQUE HADRÔNICAS EM ENERGIAS ALTAS E ASSINTÓTICAS}

  
  \vfill
{\large CAMPINAS}\\
{\large \the\year}
  
 \end{center}
 
 \newpage
 



%
 \begin{center}
 
 \vspace*{1.5cm}
 
 {\large PAULO VICTOR RECCHIA GOMES DA SILVA}

 \par
 \vspace{60pt}
 {\large PHENOMENOLOGICAL ANALYSES ON HADRONIC CROSS-SECTIONS AT HIGH AND ASYMPTOTIC ENERGIES}
 
 \vspace{60pt}

 {\large ANÁLISES FENOMENOLÓGICAS DE SEÇÕES DE CHOQUE HADRÔNICAS EM ENERGIAS ALTAS E ASSINTÓTICAS}
 
 \end{center}
 \par
 \vspace{55pt}
 \hspace*{200pt}\parbox{6.6cm}{{Tese apresentada ao 
 Instituto de Física ``Gleb Wataghin'' da Universidade Estadual
de Campinas como parte dos requisitos
exigidos para a obtenção do título de
Doutor em Ciências.}}

 \vspace{25pt}
 \hspace*{200pt}\parbox{6.6cm}{{Thesis presented to the
Institute of Physics ``Gleb Wataghin'' of the University of
Campinas in partial fulfillment of the
requirements for the degree of Doctor of Science.}}

 \par
\vspace*{1.5cm}
 \parbox{7.6cm}{{Orientador: Marcio José Menon}}

 \par
 \vspace{25pt}
  
%

 \parbox{6.25cm}{{Este exemplar corresponde à versão final da tese
defendida pelo aluno Paulo Victor Recchia Gomes da Silva,
e orientada pelo Prof.~Dr. Marcio José Menon.}}

 \par
 \vfill
 \begin{center}
 {CAMPINAS}\\
 {\the\year}
 \end{center}

 \newpage



 \includepdf{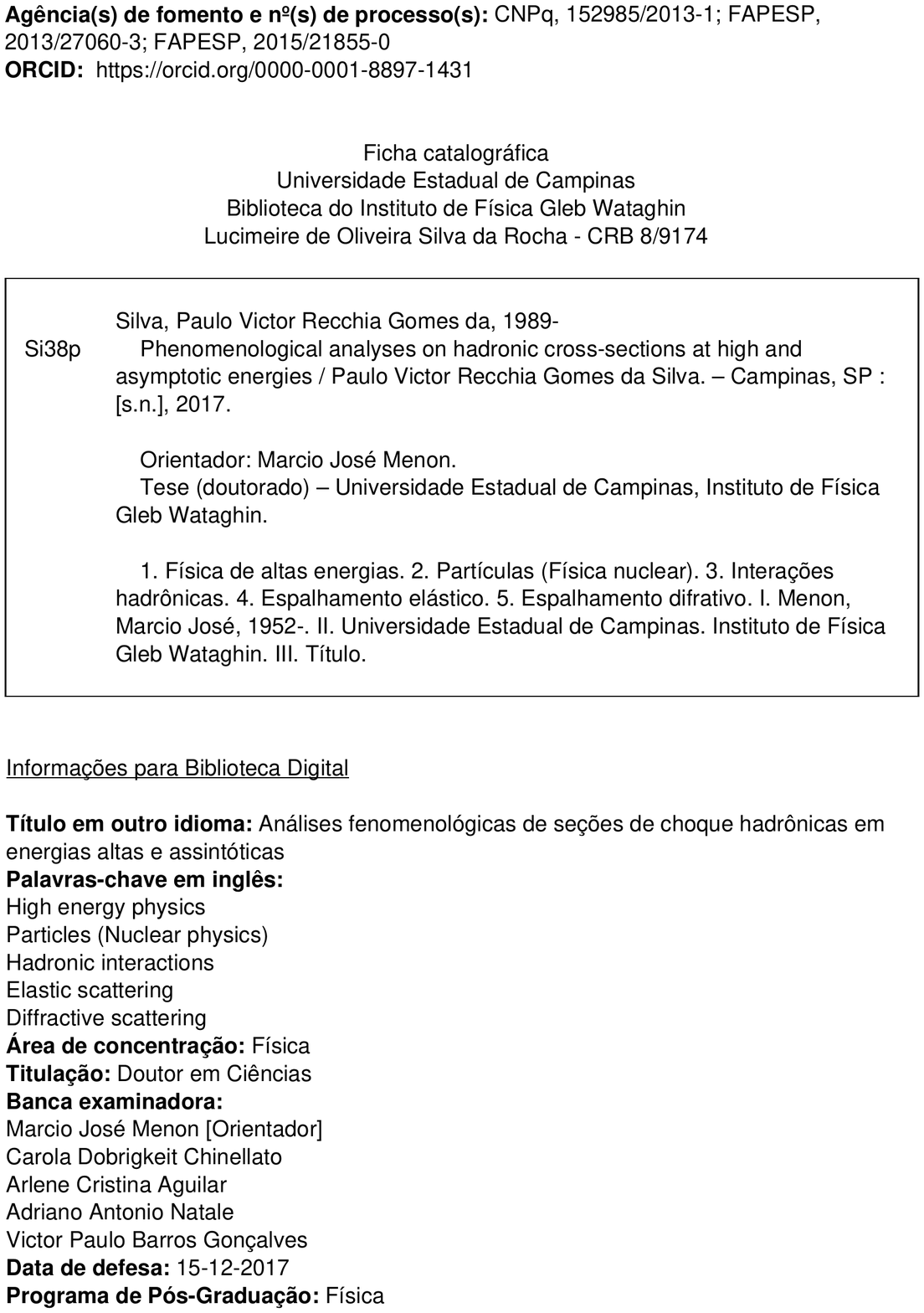}

\newpage



\includepdf{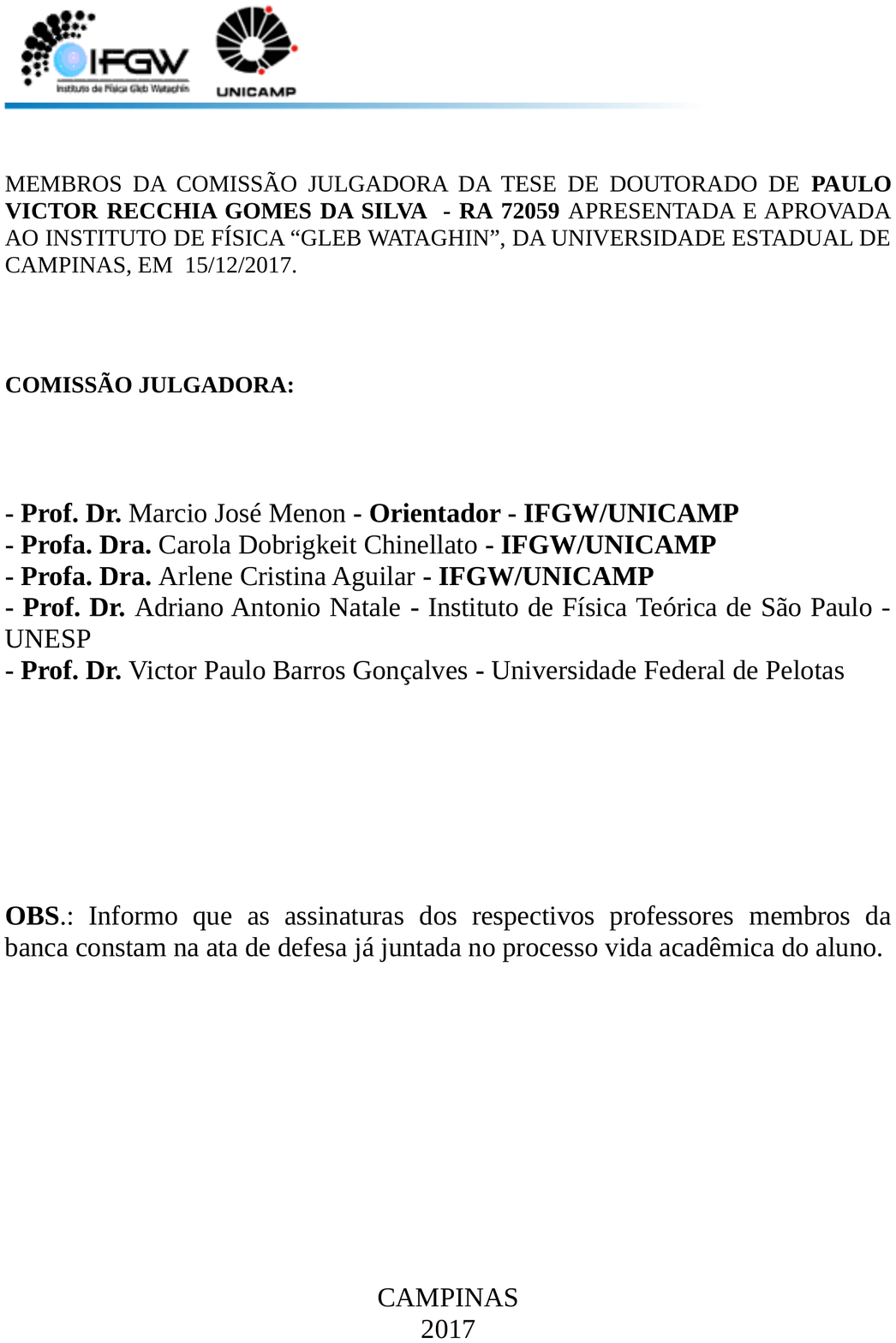}




\cleardoublepage



\vspace*{0.75\textheight}
 \begin{flushright}
   \emph{To my parents, José and Lurdinha,\\ and my wife, Silvia.}
 \end{flushright}

\cleardoublepage


\vspace*{15cm}

\begin{flushright}
\noindent
{\it The tides all come and go witnessed by no waking eye\\
The willows mark the wind\\
And all we know for sure amidst this fading light\\
We'll not go home again\\
Come and see}
\vspace*{10pt}

\noindent``The Island''\\
Written by Colin Meloy\\
Performed by The Decemberists\\
(in \textit{The Crane Wife}, 2006)

\end{flushright}


\cleardoublepage



\singlespacing


 \begin{center}
   \emph{\begin{LARGE}\textbf{Acknowledgments}\end{LARGE}}\label{abstract}
 \vspace{2pt}
 \end{center}
\vspace*{0.5cm}

To end a Ph.D. is like to end a very long trip. A trip where a lot of people cross our paths. 
People that we are thankful and that have helped us in several manners, both in direct and indirect ways.

My special thanks to Silvia Cucatti, my beloved wife, which has been by my side 
since I started my Master degree. I do not have the words to say how much you help me, 
how much life is easy and pleasant at your side and how much I'm thankful. I love you!

I thank my advisor, Prof. Márcio Menon, for everything. Thank you for the friendship,
for the good moments,  for the opportunity to work with you. I'll never forget what
I have learned with you.

The family is the ground where things start to grow. I'll be eternally thankful to
my parents, my sister, and relatives for all the support during my life.

Good friends are those who stay with us in easy and difficult times. 
And I can say that I have good friends. I'm thankful to Pedro Pasquini,
Gabriela Stenico, Bruno Daniel, Lucas Santos, Mary Díaz, Bruno Miguez, 
Mônica Nunes, Felipe Penha, Luiz Stuani and Renan Picoreti. You have 
turned the walk more easy to be done. Special thanks to Prof. Daniel 
Fagundes, for the friendship and for all the help during my research. 

I'm grateful to Prof. Enrico Meggiolaro for the reception, collaboration
and for all the help during my stay in Pisa, Italy. I'm also thankful to 
Matteo Giordano for the collaboration. I thank Claudio Bonati, Francesco
Bigazzi,  Michele Mesiti, Giacomo Ceccarelli, Domenico Logoteta and Thomas
Elze for the company and the good moments during my stay at the Physics 
Department of the University of Pisa. I also thank the department for the 
reception.

I'm thankful to the Professors of DRCC, especially Profs. Arlene Cristina Aguilar, 
David Dobrigkeit Chinellato and José Augusto Chinellato, for the discussions 
and opinions during the Ph.D.

I'm also thankful to the examination board, Profs. Carola Dobrigkeit Chinellato, 
Arlene Cristina Aguilar, Adriano Antônio Natale and 
Victor Paulo Barros Gonçalves, for the discussions and comments about the present text.

I'm thankful to Prof. László Jenkovszky and Prof. Giulia Pancheri 
for our work collaboration.

I would like to thank all the colleagues and Professors that I have met 
during the 10 years that I have stayed at UNICAMP. I have learned a lot 
of things during this time. I also would like to thank the secretariat 
for all the help.

I'm thankful to CNPq (contract 152985/2013-1) and FAPESP (contracts 2013/27060-3 and 2015/21855-0) 
for the financial support.

\cleardoublepage


\singlespacing 


 \begin{center}
   \emph{\begin{LARGE}\textbf{Resumo}\end{LARGE}}\label{resumo}
 \vspace{2pt}
 \end{center}
  
\selectlanguage{brazilian}
\noindent

A Cromodinâmica Quântica (QCD) constitui a teoria quântica de campos
da interação forte.
Apesar do seu sucesso na descrição de diversos processos
envolvendo hádrons, o espalhamento elástico
ainda é um desafio para a teoria. Este processo é caracterizado por
pequeno momento transferido e, nessa escala,
não é possível utilizar a abordagem perturbativa. Embora
resultados não-perturbativos tenham sido obtidos recentemente,
ainda não possuímos uma descrição completa (para qualquer energia e
momento transferido) no âmbito da QCD para grandezas físicas
associadas ao espalhamento elástico,
por exemplo, a seção de choque total ($\sigmatot$), o parâmetro
$\rho$, a seção de choque elástica
($\sigmael$) e a seção de choque diferencial.

As descrições dos dados experimentais associados baseiam-se em abordagens
empíricas e fe\-no\-me\-no\-ló\-gi\-cas tais como
o Formalismo de Regge-Gribov e modelos inspirados em QCD. Desde o
início das operações do LHC,
nós tivemos a oportunidade de estudar essas grandezas nas energias
mais altas disponíveis em experimentos
de aceleradores, especificamente 7 e 8 TeV  para o espalhamento
próton-próton ($pp$).

Nesta tese, o interesse principal é a dependência com a energia de
três grandezas, especificamente a razão $X=\sigmael/\sigmatot$,
$\sigmatot$ e o parâmetro $\rho$, ou seja, o comportamento destas
grandezas em energias altas e assintóticas, bem como a influência
de contribuições sub-dominantes para $\sigmatot$. Estes tópicos estão
divididos em três estudos independentes, mas complementares, envolvendo
aspectos empíricos, fenomenológicos e teóricos.

No primeiro tópico, desenvolvemos uma análise empírica da razão
$\sigmael/\sigmatot$, a qual está
relacionada com a função perfil com parâmetro de impacto nulo
(opacidade hadrônica central). Através
de parametrizações apropriadas, com um número pequeno de parâmetros
livres, obtivemos boas descrições dos dados experimentais
dos espalhamentos $pp$ e antipróton-próton ($\ppbar$). A partir dos
ajustes aos dados utilizando quatro variantes, concluímos que
o cenário assintótico de disco negro não é a única solução e, além
disso, os resultados favorecem um cenário de disco cinza.

No segundo tópico, estudamos o crescimento de $\sigmatot$ 
em função da
energia através de pa\-ra\-me\-tri\-za\-ções
baseadas no formalismo de Regge-Gribov e testamos dois termos
dominantes, um logaritmo ao quadrado e um logaritmo elevado
a um número real $\gamma$, onde $\gamma$ é um parâmetro livre de
ajuste. Adicionalmente, discutimos dois métodos analíticos
para conectar as partes real e imaginária da amplitude de espalhamento
elástico, especificamente Relações de Dispersão Derivativas (RDD)
e Unicidade Assintótica (UA), os quais resultam em diferentes
parametrizações para $\sigmatot$ e parâmetro $\rho$. Por sua vez, essas
diferenças são também discutidas. Os resultados favorecem a método RDD
tanto no contexto formal, quanto no contexto prático. A recente tensão
entre os dados das Colaborações TOTEM e ATLAS em 7 e 8 TeV também é
discutida e considerada nas reduções de dados.

No terceiro e último tópico, dois termos subdominantes de $\sigmatot$
obtidos em uma abordagem não-perturbativa da QCD para o espalhamento
elástico
são considerados em ajustes aos dados de $pp$ e $\ppbar$, bem como em
ajustes aos dados de outras reações bárion-bárion e méson-bárion.
Nesta análise, com um parâmetro extra e com informações teóricas
adicionais, também obtemos um cenário assintótico de disco cinza.

\par
\vspace{0.5em}
\noindent\textbf{Palavras-chave:} Física de altas energias; 
Fenomenologia de partículas; Física ha\-drô\-ni\-ca; Es\-pa\-lha\-men\-to elástico e difrativo.


\cleardoublepage

\selectlanguage{english}


 \begin{center}
   \emph{\begin{LARGE}\textbf{Abstract}\end{LARGE}}\label{abstract}
 \vspace{2pt}
 \end{center}
  
\noindent

Quantum Chromodynamics (QCD) constitutes the quantum field theory 
of the strong interaction. Despite the success of this theory 
in the description of several processes involving hadrons, the elastic
scattering is still a theoretical challenge. 
This process is characterized by a small transferred momentum
and, in this range, the perturbative techniques are not applicable. 
Although nonperturbative results have been
obtained in recent years, we still do not have a full 
description within QCD of the quantities
related to the elastic scattering, valid for all the energies and transferred momentum,
for example, the total cross section ($\sigmatot$), 
the $\rho$ parameter, the elastic cross section ($\sigmael$) and the 
differential cross section.

The attempts to describe experimental data rely on empirical and 
phenomenological approaches
such as Regge-Gribov Formalism and QCD inspired models. Since the 
start of Run 1 at the LHC,
we have the opportunity to study the quantities above in the largest 
energies available in accelerator
experiments, namely 7 and 8 TeV for proton-proton ($pp$) scattering.

In this thesis, the main interest is in the energy dependence of three quantities, 
the ratio $X=\sigmael/\sigmatot$,
the $\sigmatot$ and the $\rho$ parameter, namely the behaviour at high and asymptotic
energies, as well as the influence of sub-leading
contributions to $\sigmatot$.
These topics are divided into three different, but complementary studies,
involving empirical, phenomenological and theoretical aspects.

In the first topic, we develop an empirical analysis on 
the ratio $\sigmael/\sigmatot$, a quantity related to the profile function 
at impact parameter zero (the hadronic central opacity).
By means of suitable parameterizations, with a small number of free parameters,
we have obtained good descriptions of the experimental data on $pp$ and antiproton-proton ($\ppbar$)
data. 
From the fits with four variants, we conclude that the asymptotic 
black-disk scenario is not a unique solution and, moreover,
the results favour a grey-disk scenario.

In the second topic, we study the rise of $\sigmatot$ with the energy
through parameterizations based on the Regge-Gribov
formalism and we consider two options for the leading terms: a log-square and a log-raised-to-$\gamma$,
with $\gamma$ a free fit parameter. In addition, we discuss two analytic 
methods to connect the real
and imaginary parts of the elastic scattering amplitude, namely 
Derivative Dispersion Relations (DDR)
and Asymptotic Uniqueness (AU), which lead to different 
parameterizations for $\sigmatot$
and the $\rho$ parameter; these differences are critically discussed.
The results favour the DDR method in both formal and practical contexts.
The recent tension between the TOTEM and ATLAS data at 7 and 8 TeV
is discussed and considered in the data reductions.

In the third topic, two sub-leading terms for $\sigmatot$, obtained in a 
nonperturbative QCD approach to
the elastic scattering, are considered in fits to $pp$ and $\ppbar$ 
data and also in fits to data
from meson-baryon and other baryon-baryon scattering. In this 
analysis, with an extra parameter,
and with theoretical inputs, we also obtain an asymptotic grey-disk 
scenario for the colliding particles.

\par
\vspace{1em}




\microtypesetup{protrusion=false}

%
%
%
%


\cleardoublepage

\onehalfspacing


\fancypagestyle{plain}
{
    \fancyhead{}
    \fancyfoot[]{}
}

\listoffigures

\cleardoublepage


\listoftables

\cleardoublepage

\tableofcontents

\cleardoublepage

\fancypagestyle{plain}
{\fancyhf{}
 \fancyhead[R]{\thepage}
}

\microtypesetup{protrusion=true}

%% file: ch_introduction_f.tex
%
%
%
%
%

\cleardoublepage

\pagestyle{fancy}

\chapter{Introduction}\label{chapt:intro}

The study of High-Energy Physics aims to describe the fundamental constituents 
of matter and the interactions among them.
The Standard Model constitutes today the most accepted approach to describe the great amount of
phenomena observed in experiments. This model is the result of the combination of the Electroweak Theory
and Quantum Chromodynamics (QCD), aimed to describe the electromagnetic and weak interactions (in a unified way) 
and strong interactions, respectively.

Among the processes that still lack description inside the theories of the Standard Model, the elastic scattering
of hadrons (for instance proton-proton scattering, as performed in the Large Hadron Collider, LHC)
does not have a full description based on QCD. Although being a simple process,
since no particle is produced in the final state (i.e., the particles remain the same in the final state, only with the kinematic
configuration changed), elastic scattering is characterized by small scattering angles, therefore small transferred momenta ($|t|$).

In this regime, we cannot apply the perturbative techniques of QCD, since the strong coupling constant assumes large values~\cite{Halzen_Martin_book:1984}
and, consequently, the perturbative expansion does not converge. On the other hand, we do not have yet a full nonperturbative QCD description,
based on first principles of the theory, able to describe the amount of experimental data presently available on
elastic scattering. However, of interest here, recent developments have been made by Giordano and Meggiolaro~\cite{Giordano_Meggiolaro:2014} in calculating
the forward elastic scattering amplitude in the \textit{asymptotic limit} (limit of infinite energies) based on a nonperturbative approach.

As a consequence, for finite values of the square of energy in the center of mass ($s$), 
the energy dependence of the total cross section $\sigmatot(s)$ (related to the total number of particles scattered)
does not have a full description within QCD, since this quantity is connected to the imaginary part of the elastic scattering amplitude through the optical
theorem. Elastic scattering is then usually treated by means of empirical studies and phenomenological approaches such as
the Regge-Gribov formalism and QCD inspired models.

Perturbative techniques can be applied only at large transferred momentum squared ($\gtrsim 3\text{ GeV}^2$). 
In fact, Donnachie and Landshoff \cite{Donnachie:1979} have calculated the amplitude
considering the exchange of three gluons between the partons. The result gives a $|t|^{-8}$ behavior to the differential cross section. 
This behavior is indeed observed in experimental data \cite{TOTEM:2011a}.
Although successful, this result is not gauge invariant, since not all possible gluon exchanges were considered in the calculation.

From the experimental side, the TOTEM Collaboration at the LHC has as its main goals to measure diffractive scattering, which is related to 
scattering at small transferred momenta (small angles). The quantities measured are, for instance, the total cross section, 
the differential elastic cross section, the integrated elastic cross section, and the single and double diffractive cross sections.

\begin{figure}[htb]
 \centering
 \includegraphics[scale=0.5]{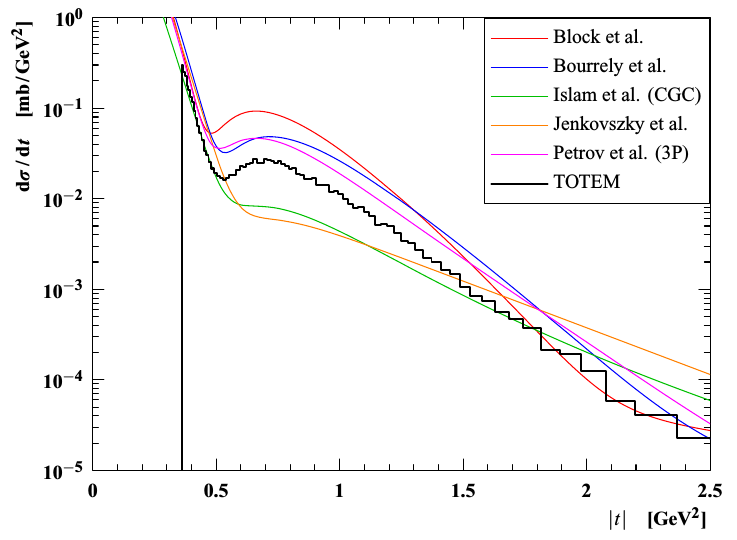}
 \caption{\label{fig:totem_data}Comparison between differential cross section data at 7 TeV measured by the TOTEM Collaboration and 
 those predicted by several models. 
 Figure taken from Ref.~\cite{TOTEM:2011a}.}
\end{figure}

The first measurement of the differential cross section at 7 TeV as a function of the transferred momentum 
is displayed in Figure~\ref{fig:totem_data},
where the experimental data (solid black line) is compared to predictions of several 
models~\cite{Block:2011b,Bourrely:2003,Islam:2009,Jenkovszky:2011,Petrov:2003} tunned with pre-LHC experimental data.
It is clear that all predictions miss the main features presented by the data: the position of the dip (the minimum around $0.5~\text{GeV}^2$) 
and the behaviour of the data at larger $|t|$, where the data show a smooth decrease, while some models predict oscillations.

Measurements of $\sigmatot$ at 7 TeV were performed by TOTEM Collab. considering different methods which all give
compatible results \cite{TOTEM:2011b,TOTEM:2013a,TOTEM:2013c}. At 8 TeV,
$\sigmatot$ was also determined \cite{TOTEM:2013d}. 
Moreover, at this energy, data in the Coulomb-Nuclear interference region was obtained allowing the determination of the $\rho$
parameter \cite{TOTEM:2016}, which is associated with the forward phase of the elastic scattering amplitude. 

Preliminary differential cross section data at 13 TeV by TOTEM present the same features of the 7 TeV data~\cite[see slide 8]{TOTEM_Kaspar:2017}.
Again, most models (now tunned with LHC 7 TeV data) present divergence with the experimental data, 
specially in what concerns the large $|t|$ behaviour:
the data has a smooth decrease with $|t|$ while most models show oscillations.

The ATLAS Collaboration, with the ALFA detector, also measured the elastic differential cross section, 
the total and the elastic cross sections at 7 and 8 TeV~\cite{ATLAS:2014,ATLAS:2016}. As will be discussed in more detail through the text,
there is a tension between the total cross section measured at 7 TeV and mainly at 8 TeV by TOTEM and ATLAS Collaborations.

Nevertheless, the amount of experimental information available since the start of operation of the LHC has definitely allowed
several interesting studies on elastic scattering. The new data that will become available at 13 TeV will certainly shed light in this subject.

In this thesis, we present phenomenological and empirical analyses on the
elastic hadron scattering, related to the energy dependence of three main physical
quantities: the total cross section ($\sigmatot$) and the $\rho$ parameter (forward
amplitude) and the ratio between the elastic and total cross section ($\sigmael/\sigmatot$).
Although the main focus concerns proton-proton ($pp$) and antiproton-proton ($\bar{p}p$)
scattering, one of the studies related to $\sigmatot$ includes also
meson-baryon and other baryon-baryon scattering. The energy interval covers the region from 5 GeV up to 8 TeV.

Three different, but complementary, analyses are developed. A global
description and some of the main results are summarized in what follows.

The first analysis concerns an \textit{empirical} study on the energy
dependence of the ratio $\sigmael/\sigmatot$ (a quantity related to the evolution of
the hadronic opacity), as well as the asymptotic scenarios associated ($s \rightarrow
\infty$). The analysis on $pp$ and $\bar{p}p$ data, through suitable empirical
parameterizations and data reductions, does not favor the black disk limit (1/2) but a
semi-transparent (grey) scenario at asymptotic energies (below 1/2). Possible physical
interpretations of the empirical parameterizations are discussed.

In the second analysis, based on the Regge-Gribov phenomenology,
analyticity, crossing and uniqueness concepts, we investigate the forward amplitude ($\sigmatot$
and $\rho$), related to $pp$ and $\bar{p}p$ scattering. The leading component for
$\sigmatot(s)$ is expressed by $\ln^{\gamma}s$ and two variants are considered: either
$\gamma = 2$ fixed (standard case) or $\gamma$ as a free fit parameter. Simultaneously,
two analytic methods to connect $\sigmatot(s)$ and $\rho(s)$ are investigated:
Derivative Dispersion Relations (DDR) and Asymptotic Uniqueness (AU), which is based on the
Phragm\'en-Lindel\"off theorems. The data reductions using DDR favor $\gamma$ values exceeding 2
and in case of AU values below 2. The results indicate that, on the formal and
practical contexts, the DDR method is more adequate for the energy interval investigated than the AU approach.

In the third analysis, recent nonperturbative QCD results obtained by
Giordano and Meggiolaro in Ref.~\cite{Giordano_Meggiolaro:2014} are discussed, with focus on two subleading
components predicted for $\sigmatot(s)$. With two Reggeons, a critical Pomeron and $\ln^2s$
as leading component, the two subleading contributions, given by $\ln s \ln\ln s$ and
$\ln s$, have been tested through several fits to $pp$, $\bar{p}p$, meson-baryon and
other baryon-baryon data. Although the data reductions do not allow to identify which of the
two components is dominant at present energies (with the available data), several characteristics
related to the QCD spectrum are inferred. Among them, it is shown that the $2^{++}$
glueball state leads to an asymptotic prediction for the ratio $\sigmael/\sigmatot$
consistent with a semi-transparent scenario (in accordance with our first analysis).

The text is organized as follows. We begin presenting a short review on
some basic concepts in Chapter~\ref{chapt:basic_concepts}. In Chapter~\ref{chapt:data_fits}, we discuss the methodology employed
through this work, in all the analyses developed, presenting also some comments on the LHC data
at 7 and 8 TeV. The three aforementioned analyses are presented in Chapters~\ref{chapt:ratio_eltot}, \ref{chapt:rise_sigmatot}
and  \ref{chapt:SLT_sigmatot}, respectively. Each one of these three Chapters ends with a summary and
corresponding conclusions. Ours final conclusions and final remarks are the contents of
Chapter~\ref{chapt:conclusions}. In Appendix~\ref{app:DDR} we show how the Derivative Dispersion Relations can be obtained
from the Integral Dispersion Relations. In Appendix~\ref{app:mellin}, we present a result of interest to this work using 
the Mellin transform, concerning the relation between asymptotic behavior of a function and the associated singularity.
A list of publications related to this thesis is presented in Appendix~\ref{app:publications}.

%% file: ch_basic_concepts_f.tex
%
%
%
%
%

\cleardoublepage



\chapter{Basic Concepts}
\label{chapt:basic_concepts}

In this chapter, we present basic concepts and definitions associated to the elastic scattering
that will be useful in the following chapters.

\section{Elastic Scattering}
\label{sec:basic_elastic_scattering}

In High-Energy Physics, \textit{elastic scattering} is a particular case of \textit{diffractive processes}, 
which can be define as (quoting Barone and Predazzi~\cite{Barone_Predazzi_book:2002})

\begin{quotation}
 \textit{``A reaction in which no quantum numbers are exchanged between the colliding particles is, at high energies, a diffractive reaction.''}
\end{quotation}

\noindent Therefore, the elastic scattering is \textit{the diffractive process in which the initial and final particles are the same:}
\begin{equation}
 1 + 2\rightarrow 1'+2',
\end{equation}

\noindent where $'$ indicates that the particles in the final states are in 
a different kinematic configuration. On the other hand, it is important to stress that
the final particles have the same quantum numbers of the initial ones.

Besides the elastic scattering, we have two more particular types of \textit{soft diffractive process}\footnote{We have one more type of diffractive process, 
called central diffraction, that does not constitute a soft process.}:

\begin{enumerate}
 \item Single diffraction: when one of the initial particles remains the same (with only its kinematic configurations altered) 
 and the other gives origin to a bunch of particles (or resonances) with the resulting quantum numbers equal to the initial particle,
 \begin{equation}
  1 + 2 \rightarrow 1' + X_2;
 \end{equation}

 \item Double diffraction: the same as the above case, but now the two particles gives origin to a bunch of particles (or resonances),
 \begin{equation}
  1 + 2 \rightarrow X_1 + X_2.
 \end{equation}
 
\end{enumerate}

\begin{figure}[htb!]
\centering
\begin{tikzpicture}[scale=1,>=stealth]

\foreach \y/\z in {-1.5/1.5,-5.5/-2.5,-9.5/-6.5}{ \draw (-5,\y) rectangle (2,\z);\node() at (-5.3,\z){$\phi$};}


\foreach \x in {-1.9,-5.9,-9.9}{ \node() at (-1.5,\x){0}; \node() at (0,\x){5}; \node() at (1.5,\x){10};
\node() at (-3,\x){-5}; \node() at (-4.5,\x){-10}; \node() at (2.25,\x){$y$};
}


\foreach \y in {0,-4,-8}{ \node() at (-5.3,\y){0}; \draw (-5.1,\y) -- (-5,\y);}


\foreach \x in {-5,-4.5,...,2} \draw (\x,-1.6) -- (\x,-1.5);

\foreach \x in {-5,-4.5,...,2} \draw (\x,-5.6) -- (\x,-5.5);

\foreach \x in {-5,-4.5,...,2} \draw (\x,-9.6) -- (\x,-9.5);



\foreach \x /\y in {-4.7/1,1.7/-1} \draw[fill=red,draw=red] (\x,\y) circle (1mm);


\draw[fill=red,draw=red] (-4.7,-3) circle (1mm);

\foreach \x in {0.5,1,...,1.9} {\foreach \y in {-3,-4,...,-5} \draw[fill=blue,draw=blue] (\x,\y) circle(1mm);}

\foreach \x/\y in {0.73/-3.4,1.4/-3.2,1.5/-3.6,1.8/-3.5,0.3/-3.7,0.2/-3.4,1.1/-3.7,0.75/-3.7} \draw[fill=blue,draw=blue] (\x,\y) circle(1mm);

\foreach \x/\y in {0.6/-4.5,0.3/-4.7,1.25/-4.2,1.8/-4.5,1.3/-4.6,1.2/-5.1,1.56/-4.3,0.9/-4.7,0.9/-4.2,0.2/-4.3} \draw[fill=blue,draw=blue] (\x,\y) circle(1mm);


\foreach \x in {0.5,1,...,1.9} {\foreach \y in {-7,-8,...,-9} \draw[fill=blue,draw=blue] (\x,\y) circle(1mm);}

\foreach \x/\y in {0.6/-7.3,1.3/-7.7,1.25/-7.2,1.8/-7.5,0.8/-7.6,0.25/-7.1,1/-7.3,0.3/-7.6} \draw[fill=blue,draw=blue] (\x,\y) circle(1mm);

\foreach \x/\y in {0.75/-8.4,1.4/-8.2,1.3/-8.6,1.8/-8.5,0.7/-8.7,1.7/-8.8,0.7/-8.2,0.3/-8.2,0.4/-8.8,0.35/-8.5} \draw[fill=blue,draw=blue] (\x,\y) circle(1mm);

\foreach \x in {-4.5,-4,...,-3.5} {\foreach \y in {-7,-8,...,-9} \draw[fill=blue,draw=blue] (\x,\y) circle(1mm);}

\foreach \x/\y in {-3.3/-8.3,-3.6/-8.2,-3.35/-8.8,-3.9/-8.5,-4.3/-8.7,-4.8/-8.8,-4.7/-8.3,-4.8/-7.9} \draw[fill=blue,draw=blue] (\x,\y) circle(1mm);

\foreach \x/\y in {-3/-7.6,-3.3/-7.7,-3.25/-7.2,-3.8/-7.5,-4.3/-7.6,-4.25/-8.3,-4.8/-7.5,-4.5/-7.3} \draw[fill=blue,draw=blue] (\x,\y) circle(1mm);


\begin{scope}[thick,decoration={markings,mark=at position 0.4 with {\arrow{>}}}]
\draw[thick,postaction={decorate}] (-9.5,-1) -- (-8,0);
\draw[thick,postaction={decorate}] (-9.5,1) -- (-8,0);

\draw[thick,postaction={decorate}] (-9.5,-5) -- (-8,-4);
\draw[thick,postaction={decorate}] (-9.5,-3) -- (-8,-4);

\draw[thick,postaction={decorate}] (-9.5,-9) -- (-8,-8);
\draw[thick,postaction={decorate}] (-9.5,-7) -- (-8,-8);
\end{scope}

\begin{scope}[thick,decoration={markings,mark=at position 0.75 with {\arrow{>}}}]

\draw[thick,postaction={decorate},red] (-8,0) -- (-6.5,-1);
\draw[thick,postaction={decorate},red] (-8,0) -- (-6.5,1);

\draw[thick,postaction={decorate},blue] (-8,-4) -- (-6.4,-4.8);
\draw[thick,postaction={decorate},blue] (-8,-4) -- (-6.5,-5);
\draw[thick,postaction={decorate},blue] (-8,-4) -- (-6.6,-5.2);

\draw[thick,postaction={decorate},red] (-8,-4) -- (-6.5,-3);

\draw[thick,postaction={decorate},blue] (-8,-8) -- (-6.6,-9.2);
\draw[thick,postaction={decorate},blue] (-8,-8) -- (-6.5,-9);
\draw[thick,postaction={decorate},blue] (-8,-8) -- (-6.4,-8.8);

\draw[thick,postaction={decorate},blue] (-8,-8) -- (-6.4,-7.2);
\draw[thick,postaction={decorate},blue] (-8,-8) -- (-6.5,-7);
\draw[thick,postaction={decorate},blue] (-8,-8) -- (-6.6,-6.8);
\end{scope}

\filldraw[fill=gray!50] (-8,0) circle (0.5cm);
\filldraw[fill=gray!50] (-8,-4) circle (0.5cm);
\filldraw[fill=gray!50] (-8,-8) circle (0.5cm);


\node() at (-8,1.5) {Elastic Scattering};

\node() at (-8,-2.5) {Single Diffraction};

\node() at (-8,-6.3) {Double Diffraction};

\foreach \y in {1,-3,-7}\node() at (-9.75,\y){$1$};

\foreach \y in {-1,-5,-9}\node() at (-9.75,\y){$2$};

\foreach \y in {1,-3}\node() at (-6.2,\y){$\textcolor{red}{1'}$};

\node() at (-6.2,-1){$\textcolor{red}{2'}$};

\node() at (-6.1,-6.9){$\textcolor{blue}{X_1}$};

\foreach \y in {-5.15,-9.15}\node() at (-6.1,\y){$\textcolor{blue}{X_2}$};


\draw[dashed,thick] (-3.7,-1.5) -- (-3.7,1.5);
\draw[dashed,thick] (0.8,-1.5) -- (0.8,1.5);

\draw[dashed,thick] (-3.7,-5.5) -- (-3.7,-2.5);
\draw[dashed,thick] (-0.3,-5.5) -- (-0.3,-2.5);

\draw[dashed,thick] (-2.7,-9.5) -- (-2.7,-6.5);
\draw[dashed,thick] (-0.3,-9.5) -- (-0.3,-6.5);

\end{tikzpicture}
\caption{\label{fig:diff_rapidity_gap}Rapidity ($y$) gap present in soft diffractive process:
elastic scattering, simple and double diffraction ($\phi$ represents the azimuthal angle).}
\end{figure}
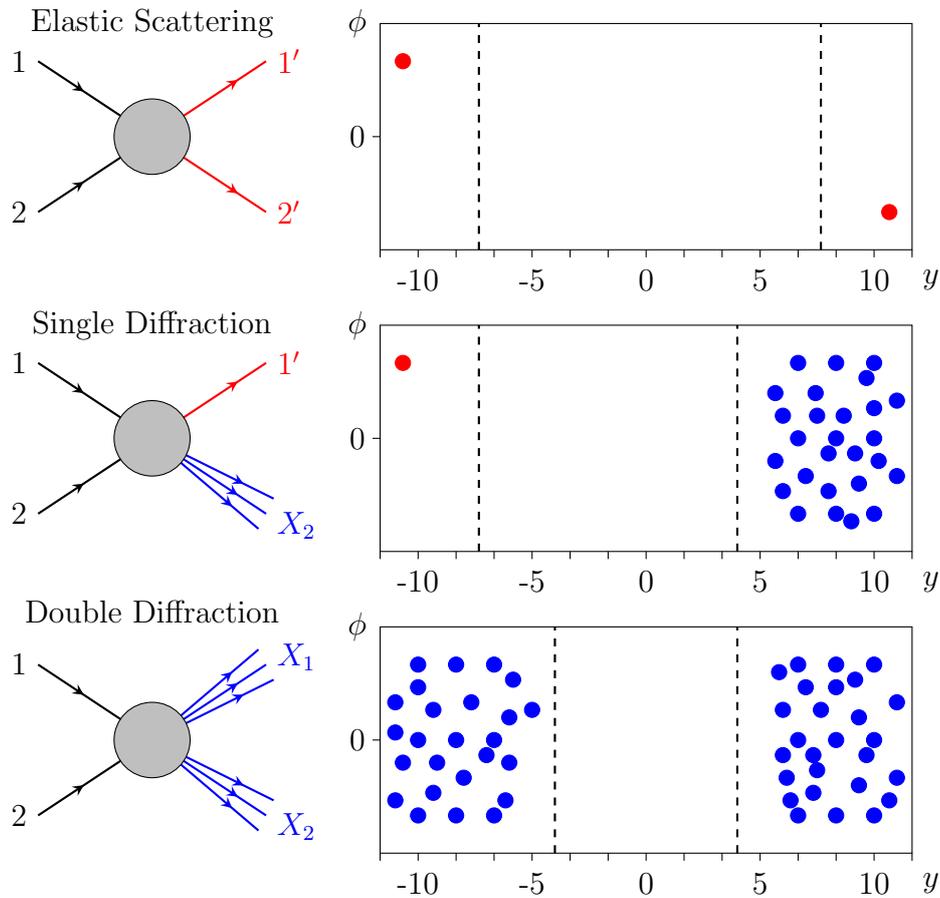

From the experimental point of view, the diffractive processes are characterized by 
presenting a rapidity\footnote{Rapidity is given by $y=\dfrac{1}{2}\ln \left(\dfrac{E+p_z}{E-p_z}\right)$
where $E$ is the energy of the particle and $p_z$ is the component of the momentum
of the particle in the beam direction. In the limit of large momenta (compared to the particle mass)
we have $y\approx \eta$, where $\eta = -\ln[\tan(\theta/2)]$ is the pseudo-rapidity and $\theta$ is the scattering angle.} 
gap in the final state, i.e. a large angular separation between 
the final states in the plot of the azimuthal $\phi$ angle versus rapidity $y$,
as illustrated in Fig.~\ref{fig:diff_rapidity_gap}.

\section{The Mandelstam Variables and the Center of Mass System}
\label{sec:basic_var_mandelstam_cm_system}

Let us consider the following process
\begin{equation}
 1 + 2 \rightarrow 3 + 4,
 \label{eq:esp_canals}
\end{equation}

\noindent where each particle has a mass $m_i$ and a four-momentum $p_i = (E_i,\mathbf{p}_i)$, where $E_i$ is the energy 
and $\mathbf{p}_i$ is the momentum. Therefore
\begin{equation}
 p_i^2 = m_i^2 \quad (i=1,\dots,4).
\end{equation}

The Mandelstam variables are three relativistic invariants defined by~\cite{Barone_Predazzi_book:2002}
\begin{align}
 s & = (p_1 + p_2)^2,\label{eq:s-channel}\\
 t & = (p_1 - p_3)^2,\label{eq:t-channel}\\
 u & = (p_1 - p_4)^2.\label{eq:u-channel}
\end{align}

With energy-momentum conservation, one can show that
\begin{equation}
 s + t + u = \sum^4_{i=1} m_i^2.
 \label{eq:soma-stu}
\end{equation}

Therefore, only two Mandelstam variables are independent. 
Usually, we chose $s$ and $t$ to be independent, 
as we will see in the next section.


Let us consider the center of mass (c.m.) system, where
\begin{equation}
 \mathbf{p}_1 + \mathbf{p}_2 = 0,
\end{equation}

\noindent therefore,
\begin{equation}
 \mathbf{p}_1 = -\mathbf{p}_2 \equiv \mathbf{p}.
\end{equation}

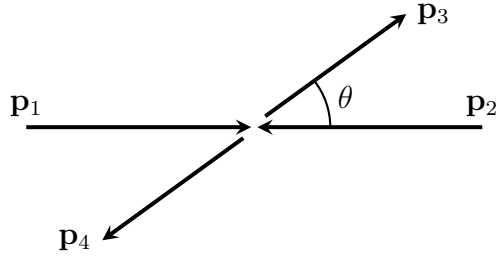
\begin{figure}[h!]
 \centering
\begin{tikzpicture}[scale=2,>=stealth]


\node(O) at (0,0){};

\node[above,scale=1](p1) at (-1.5,0){$\mathbf{p}_1$};
\node[above,scale=1](p2) at (1.5,0){$\mathbf{p}_2$};
\node[right,scale=1](p3) at (1,0.75){$\mathbf{p}_3$};
\node[left,scale=1](p4) at (-1,-0.75){$\mathbf{p}_4$};

\draw[->,line width=1.5pt] (p1.south) -- (-0.02,0);
\draw[->,line width=1.5pt] (p2.south) -- (0.02,0);

\draw[->,line width=1.5pt] (O.north east) -- (p3.west);
\draw[->,line width=1.5pt] (O.south west) -- (p4.east);

\draw[thick] (0.5,0) arc(0:38:0.5cm);
\node[scale=1](angulo) at(0.6,0.2){$\theta$};

\end{tikzpicture}
 \caption{\label{fig:espalhamento_cm}Scattering in the center of mass system (c.m.).}
\end{figure}

Of interest in this work, we shall consider particles with equal masses $m$. In this case,
\begin{align}
 s & =  4(|\mathbf{p}|^2 + m^2),\label{eq:s-equal-mass}\\
 t & = -2|\mathbf{p}|^2(1-\cos\theta)\label{eq:t-equal-mass},\\
 u & = -2|\mathbf{p}|^2(1+\cos\theta)\label{eq:u-equal-mass},
\end{align}

\noindent where $\theta$ is the scattering angle in the c.m. system defined in Figure~\ref{fig:espalhamento_cm}.

Therefore, the variable $s$ is equal to the square of the energy 
in the c.m. and $t$ corresponds to the momentum transferred squared
\begin{equation}
t=-q^2.\nonumber
\end{equation}

\noindent
The process of Eq.~\eqref{eq:esp_canals} is called $s$-channel.

We recall that in relativistic field theory, an incoming particle $1$ with
momentum $p_1$ can be seen as an outgoing antiparticle $\bar{1}$ with momentum $-p_1$. 
Therefore, if we change the particles $2$ and $3$ of side in the $s$-channel reaction, Eq.~\eqref{eq:esp_canals}, 
the reaction becomes
\begin{equation}
1+\bar{3} \to \bar{2}+4
\label{eq:t-channel-reac}
\end{equation}

\noindent and we must also change the signs of the momenta in Eqs.~\eqref{eq:s-channel}-\eqref{eq:u-channel}.
Consequently, $t$ now is the square of the energy in c.m. The process of Eq.~\eqref{eq:t-channel-reac} is called $t$-channel. 

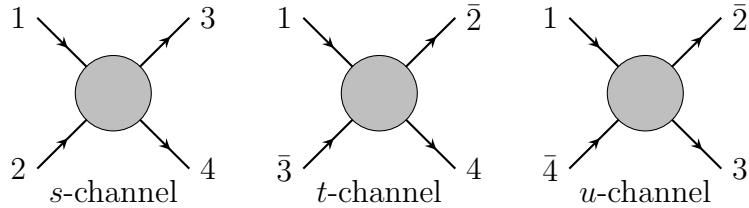
\begin{figure}[h!]
 \centering
 \begin{tikzpicture}[scale=1,>=stealth]

\begin{scope}[thick,decoration={markings,mark=at position 0.4 with {\arrow{>}}}]
\draw[thick,postaction={decorate}] (-4.5,-1) -- (-3.5,0);
\draw[thick,postaction={decorate}] (-4.5,1) -- (-3.5,0);

\draw[thick,postaction={decorate}] (-1,-1) -- (0,0);
\draw[thick,postaction={decorate}] (-1,1) -- (0,0);

\draw[thick,postaction={decorate}] (2.5,-1) -- (3.5,0);
\draw[thick,postaction={decorate}] (2.5,1) -- (3.5,0);
\end{scope}

\begin{scope}[thick,decoration={markings,mark=at position 0.75 with {\arrow{>}}}]

\draw[thick,postaction={decorate}] (-3.5,0) -- (-2.5,-1);
\draw[thick,postaction={decorate}] (-3.5,0) -- (-2.5,1);

\draw[thick,postaction={decorate}] (0,0) -- (1,-1);
\draw[thick,postaction={decorate}] (0,0) -- (1,1);

\draw[thick,postaction={decorate}] (3.5,0) -- (4.5,-1);
\draw[thick,postaction={decorate}] (3.5,0) -- (4.5,1);
\end{scope}

\filldraw[fill=gray!50] (-3.5,0) circle (0.5cm);
\filldraw[fill=gray!50] (0,0) circle (0.5cm);
\filldraw[fill=gray!50] (3.5,0) circle (0.5cm);


\node [left] (s1) at (-4.5,1) {$1$};
\node [left] (s2) at (-4.5,-1) {$2$};
\node [right] (s3) at (-2.5,1) {$3$};
\node [right] (s4) at (-2.5,-1) {$4$};
\node [below](chs) at (-3.5,-1) {$s$-channel};

\node [left] (t1) at (-1,1) {$1$};
\node [left] (t3bar) at (-1,-1) {$\bar{3}$};
\node [right] (t2bar) at (1,1) {$\bar{2}$};
\node [right] (t4) at (1,-1) {$4$};
\node [below](cht) at (0,-1) {$t$-channel};

\node [left] (u1) at (2.5,1) {$1$};
\node [left] (u4bar) at (2.5,-1) {$\bar{4}$};
\node [right] (u3) at (4.5,-1) {$3$};
\node [right] (u2bar) at (4.5,1) {$\bar{2}$};
\node [below](chu) at (3.5,-1) {$u$-channel};
\end{tikzpicture}
\caption{\label{fig:channels_stu}$s$-, $t$- and $u$-channels.}
\end{figure}

Analogously, we have the so-called $u$-channel: $1+\bar{4} \to \bar{2}+3$. 
These three reactions are summarysed in Fig.~\ref{fig:channels_stu}.

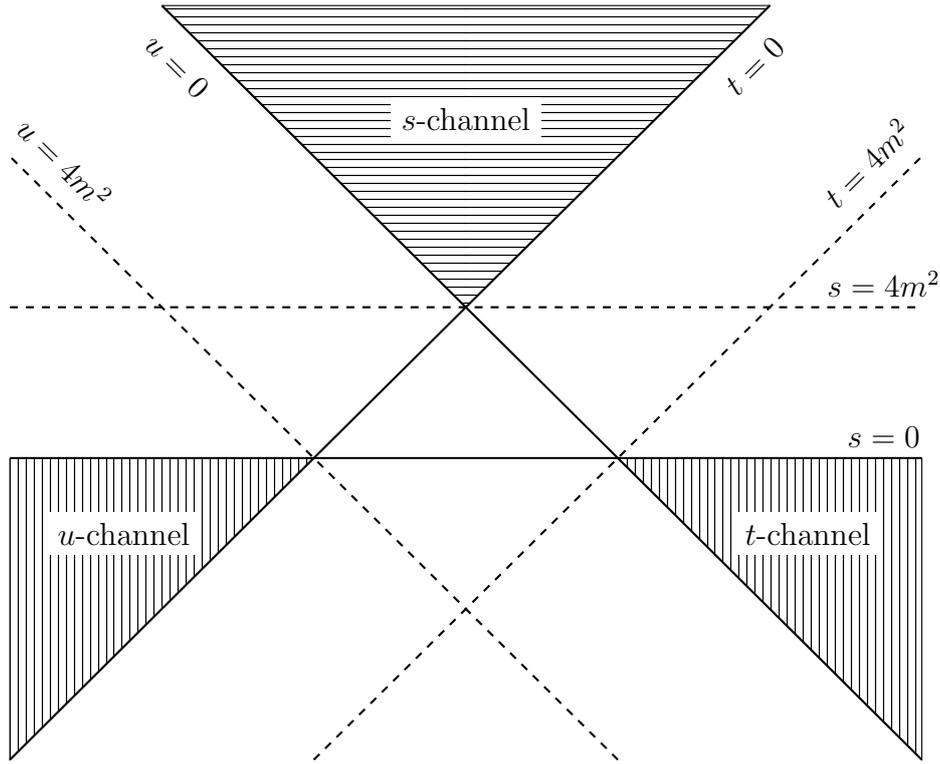
\begin{figure}[h!]
 \centering
\begin{tikzpicture}[scale=1]
\usetikzlibrary{patterns}

\draw[thick] (-6,0) -- (6,0);
\draw[thick,dashed] (-6,2) -- (6,2);

\draw[thick] (-6,-4) -- (4,6);
\draw[thick,dashed] (-2,-4) -- (6,4);

\draw[thick] (6,-4) -- (-4,6);
\draw[thick,dashed] (2,-4) -- (-6,4);

\filldraw[pattern =horizontal lines] (-4,6) -- (4,6) -- (0,2) -- (-4,6);
\filldraw[pattern =vertical lines] (6,0) -- (2,0) -- (6,-4) -- (6,0);
\filldraw[pattern =vertical lines] (-6,0) -- (-2,0) -- (-6,-4) -- (-6,0);

\node [above] (s0) at (5.5,0) {$s=0$};
\node [above] (s) at (5.5,2) {$s=4m^2$};

\node [above,rotate=45] (t0) at (4,5) {$t=0$};
\node [above,rotate=45] (t) at (5.5,3.7) {$t=4m^2$};

\node [above,rotate=-45] (u0) at (-4,5) {$u=0$};
\node [above,rotate=-45] (u) at (-5.5,3.7) {$u=4m^2$};

\node (canals) at (0,4.5) [fill=white] {$s$-channel};

\node (canalt) at (4.5,-1) [fill=white] {$t$-channel};

\node (canalu) at (-4.5,-1) [fill=white] {$u$-channel};
\end{tikzpicture}
\caption{\label{fig:stu_physical_region}Physical region (filled areas) of $s$-, $t$- and $u$-channels for particles with equal masses.}
\end{figure}

From the physical conditions $|\mathbf{p}| \geq 0$ and $-1\leq \cos\theta \leq 1$, 
we have the following physical regions (see Fig.~\ref{fig:stu_physical_region}) 
for the Mandelstam variables for each of the channels discussed above, 
considering particles with equal masses,

\begin{align}
 & s \geq 4m^2; \quad t \leq 0; \quad u \leq 0 \quad (s\text{-channel}), \label{eq:s-channel-physical}\\
 & t \geq 4m^2; \quad u \leq 0; \quad s \leq 0 \quad (t\text{-channel}), \label{eq:t-channel-physical}\\
 & u \geq 4m^2; \quad s \leq 0; \quad t \leq 0 \quad (u\text{-channel}). \label{eq:u-channel-physical}
\end{align}

\section{Elastic Scattering Amplitude and Physical Quantities}
\label{sec:basic_concepts_amplitude}

\subsection{The Elastic Amplitude}
\label{subsec:basic_concepts_amplitude_theory}

The elastic scattering amplitude, $F$, is a complex-valued function and it is one of the elements of the scattering matrix $S$,
which is required to be \textit{relativistically invariant}. As a consequence, the elastic scattering amplitude must be written in terms of
relativistic invariants, chosen to be the Mandelstam variables, specifically $s$ and $t$. Therefore $F=F(s,t)$.

Beyond the relativistic invariance, we have three postulates~\cite{Barone_Predazzi_book:2002}, listed below.

\begin{itemize}
\item \textit{Unitarity}: The $S$ matrix is required to be a \textit{unitary} operator, 
i.e. $SS^\dagger = 1$, where $\dagger$ is the Hermitian conjugate. 
This postulate is directly related to the conservation of probability.

\item \textit{Analyticity}: It states that the amplitudes are
analytic functions of its variables when these are analytically continued to complex values. Moreover, the \textit{physical}
amplitudes are the limits of these functions when the variables reduce to real values. It is assumed that the only singularities 
of these analytic functions are those dictated by unitarity, namely, the poles and branch points associated with the exchange of
particles \cite{Barone_Predazzi_book:2002}.

\item \textit{Crossing symmetry}: It states that the same amplitude describes the three different channels
discussed in the previous section, namely $s$-, $t$- and $u$-channels,
and the amplitudes for each process are obtained by means of analytical continuations from one kinematic domain to the other.
\end{itemize}

These postulates are the basis of the study of the elastic scattering amplitude and, consequently, the basis of several theoretical/formal results,
for instance, the Froissart-Lukaszuk-Martin bound and 
Dispersion Relations, just to keep within the topics of this thesis.

\subsection{Physical Quantities}
\label{subsec:basic_concepts_physical_quantities}

Neglecting spin effects, the elastic hadron scattering \textit{at high energies} is characterized by seven physical quantities,
expressed in terms of the scattering amplitude and denoted as follows~\cite{Barone_Predazzi_book:2002}.

 \subsubsection{Elastic differential cross section}
 
 The elastic differential cross section is related to the absolute value of the elastic amplitude, namely
\begin{equation}
 \frac{d\sigma}{dt}(s,t) = \frac{1}{16\pi s^2}|F(s,t)|^2.
 \label{eq:dsigdt-amplitude}
\end{equation}

\subsubsection{Elastic (integrated) cross section}

The integral in $t$ of $d\sigma/dt$ gives the total elastic cross section as a function of the energy
\begin{equation}
 \sigmael(s)= \int_{-\infty}^0 \frac{d\sigma}{dt}dt.
 \label{eq:sigmael-def}
\end{equation}

\subsubsection{Total cross section}

The optical theorem connects the imaginary part of the elastic scattering amplitude in the forward direction
with the total cross section ($\sigmatot$), 
\begin{equation}
 \boxed{\sigmatot(s) = \frac{\Imag F(s,t=0)}{s}.}
\label{eq:optical-theorem}
\end{equation}

\subsubsection{Inelastic (integrated) cross section}

The total cross section, in turn, is associated with the total number of particles scattered
in elastic and inelastic events:
\begin{equation}
 \sigmatot = \sigmael + \sigmain.
 \label{eq:unitarity-cross-section}
\end{equation}

The equation above is related to the unitarity principle, and from it, we can determine the inelastic
cross section
\begin{equation}
 \sigmain = \sigmatot - \sigmael.
 \label{eq:sigmainel-def}
\end{equation}

\subsubsection{Optical point and $\rho$ parameter}

The optical point is the differential cross section at $t=0$.
Using the optical theorem, Eq.~\eqref{eq:optical-theorem}, and Eq.~\eqref{eq:dsigdt-amplitude}, it is expressed by
\begin{eqnarray}
 \left.\frac{d\sigma}{dt}\right|_{t=0} & = & \frac{1}{16\pi s^2}|\Real F(s,0) + i\Imag F(s,0)|^2\nonumber\\
  & = &  \frac{[\Imag F(s,0)]^2}{16\pi s^2}\left[\left(\frac{\Real F(s,0)}{\Imag F(s,0)}\right)^2 + 1\right]\nonumber.
\end{eqnarray}

In the above formula, the ratio between the real and imaginary parts of 
the amplitude is related to the \textit{phase of the forward amplitude}
and is named $\rho$ parameter
\begin{equation}
 \boxed{\rho(s) = \frac{\Real F(s,0)}{\Imag F(s,0)}.}
 \label{eq:par-rho-def}
\end{equation}

With this definition the optical point reads
\begin{equation}
 \left.\frac{d\sigma}{dt}\right|_{t=0} = \frac{\sigmatot^2(1+\rho^2)}{16\pi}.
 \label{eq:optical-point}
\end{equation}

\subsubsection{Slope parameter}

The logarithm of the differential cross section data in terms of $t$ (linear scale) is characterized by showing a sharp forward peak, called \textit{diffraction peak}.
Empirically, it can be parametrized (for small values of transferred momenta) by
\begin{equation}
 \frac{d\sigma}{dt} = \left.\frac{d\sigma}{dt}\right|_{t=0} e^{-B|t|}
 \label{eq:dsigdt-exp}
\end{equation}

\noindent where $B=B(s)$ is the (constant in $t$) forward slope.


\subsection*{Comments on notation}

Given the classification of diffractive processes discussed in Section~\ref{sec:basic_elastic_scattering}, 
it is worth making some comments on the notation used to refer to diffractive process in the text.

In general, we will refer to \textit{elastic} scattering as those processes at which the internal structure 
of the particles remain unchanged (no production of particles in the final state) and \textit{inelastic} scattering
as those cases in which particles are produced in the final state. Within this view, the single and double diffractive process
are types of inelastic events. Specifically, we can write the inelastic cross section as
\begin{equation}
 \sigmain = \sigmadiff + \sigma_\text{non-diff},
\end{equation}

\noindent i.e. a sum of two components: the diffractive one (diff), 
characterized by a rapidity gap in the final state, and the non-diffractive one (non-diff) in which no rapidity gap is observed.
In turn, the diffractive cross section can be written as the sum of the central and \textit{soft} diffraction cross sections
\begin{equation}
\sigmadiff = \sigmadiff^\text{central} + \sigmadiff^\text{soft}, 
\label{eq:sigmadiff-def}
\end{equation}

\noindent and the soft (inelastic) cross section includes single (sd) and double (dd) diffraction
\begin{equation}
\sigmadiff^\text{soft} = \sigma_\text{sd} + \sigma_\text{dd}. 
\label{eq:sigmadiff-def-soft}
\end{equation}

At last, we can try to organize those types of events taking into account the presence of 
a hard scale in QCD (i.e., a scale that allows the applicability of the perturbative techniques).
Non-diffractive inelastic events are grouped in what we call \textit{hard}
process and it is possible to calculate the associated amplitude
with perturbation theory. Elastic, single and double diffractive 
events are grouped in the \textit{soft} process. This classification depends on the momentum
exchanged between the interacting particles and the transition from one regime to the other
(soft to hard, for instance) is not clear. This transition
may be referred as a \textit{semi-hard} process and is model dependent.
Within QCD language, elastic and diffractive scattering are associated with colourless exchanges.

\section{Profile Function and the Eikonal Representation}
\label{sec:basic_profile_eikonal}

The elastic scattering amplitude can also be represented in the impact parameter space ($b$) by
the profile function $\Gamma(s,b)$. The relation between $F(s,t)$ and $\Gamma(s,b)$ is given by
the Fourier-Bessel transform (assuming azimuthal symmetry)
\begin{equation}
 F(s,t) =4\pi i s\int_0^\infty  bdb J_0(b\sqrt{-t})\Gamma(s,b), 
 \label{eq:amplitude-profile-FBtransform}
\end{equation}

\noindent where $J_0$ is the Bessel function of first kind.

In terms of the profile function, the total, elastic and inelastic cross section read~\cite{Barone_Predazzi_book:2002}
\begin{gather}
\sigmatot(s)  = 4 \pi \int_0^{\infty} b\, db\, \Real \Gamma(s,b),
\qquad
\sigmael(s)  = 2 \pi \int_0^{\infty} b\, db\, |\Gamma(s,b)|^2 \label{eq:tot-el-cross-section-profile} \\
\intertext{and}
\sigmain(s)  = 2 \pi \int_0^{\infty} b\, db\, G_\text{inel}(s,b),
\label{eq:inel-cross-section-profile}
\end{gather}

\noindent where
\begin{equation}
G_\text{inel}(s,b) = 2 \Real \Gamma(s,b) - |\Gamma(s,b)|^2
\label{eq:g-inel-unitarity}
\end{equation}

\noindent is the inelastic overlap function. The above relation also constitutes an unitarity relation.

We also introduce the \textit{eikonal representation}, in which the profile function is written as
\begin{equation}
 \Gamma(s,b) \equiv 1-e^{i\chi(s,b)},
 \label{eq:eikonal-def}
\end{equation}

\noindent where $\chi(s,b)$ is a complex valued function called eikonal function. In terms of this function,
Eq.~\eqref{eq:g-inel-unitarity} reads
\begin{equation}
 G_\text{inel}(s,b) = 1-e^{-2\Imag \chi(s,b)}.
\end{equation}

From unitarity, $\Imag \chi(s,b) \geq 0$, we have $G_\text{inel}(s,b) \leq 1$, which allows a probability interpretation
to the inelastic overlap function: $G_\text{inel}$ gives the probability of an inelastic event to happen at given $s$ and $b$.
From that, we may also associate the imaginary part of $\chi(s,b)$ with absorption in the scattering process. For this reason,
and using an optical analogy, we introduce the \textit{opacity} function
\begin{equation}
 \Omega(s,b)\equiv \Imag\chi(s,b).
 \label{eq:opacity-def}
\end{equation}

Now, neglecting the real part of the elastic amplitude\footnote{As it will be shown 
in Chapter~\ref{chapt:rise_sigmatot}, in the present energies the real part is not zero yet,
but it is small enough to be neglected in a first approximation.}, we see that $\Gamma(s,b)$ is a real valued function,
and
\begin{equation}
 \Gamma(s,b) = 1 -e^{-\Omega(s,b)}.
\end{equation}

Expanding the exponential, we have in first order
\begin{equation}
 \Gamma(s,b) \approx \Omega(s,b).
\end{equation}

Therefore, the profile function is connected with the opacity function.

\section{Simple Models for the Profile Function}
\label{sec:basic_profile_models}

Let us now discuss two simple and useful models for the profile function: 
the Grey-Disk and the Gaussian profile. 

Consider first the grey-disk (GD) model. Let $R(s)$ be the radius of the disk. 
We write the profile function as
\begin{equation}
 \Gamma_\text{GD}(s,b) = \left\{
 \begin{array}{ll}
  \Gamma_0(s), & 0 \leq b \leq R(s)\\
  0, & b > R(s),
 \end{array}\right.
 \label{eq:profile-gd}
\end{equation}

\noindent where $\Gamma_0(s) = \Gamma(s,b=0)$ is the central value of profile function. 
In turn, the Gaussian (G) profile is given by
\begin{equation}
 \Gamma_\text{G}(s,b) = \Gamma_0(s)e^{-b^2/R^2}.
 \label{eq:profile-gaussian}
\end{equation}

The elastic scattering amplitudes associated to them, calculated using Eq.~\eqref{eq:amplitude-profile-FBtransform}, read
\begin{align}
 F_\text{GD}(s,t) & = 4\pi i sR(s)\Gamma_0(s) \frac{J_1(R\sqrt{-t})}{\sqrt{-t}},\\[10pt]
 F_\text{G} (s,t) & = 2\pi i s R^2(s) \Gamma_0(s) e^{-|t|R^2}.
\end{align}

Calculating $\sigmatot$ and $\sigmael$ by substituting Eqs.~\eqref{eq:profile-gd} and \eqref{eq:profile-gaussian}
in Eq.~\eqref{eq:tot-el-cross-section-profile}, we see that the ratio $\sigmael/\sigmatot$ is proportional 
to the profile function at the center $\Gamma_0(s)$ 
(and to the central opacity, in first order as discussed in the previous section)
\begin{equation}
 \frac{\sigmael}{\sigmatot}(s)=\left\{
  \begin{array}{ll}
   \Gamma_0(s)/2, & \text{Grey-disk}\\
   \Gamma_0(s)/4, & \text{Gaussian}.
  \end{array}
 \right.
 \label{eq:ratio-eltot-grey-gaussian}
\end{equation}

In the grey-disk model, the limit of the opacity going to infinity ($\Omega\to\infty$) is called \textit{black-disk}.
In this case $\Gamma_0(s)\to 1$ and $\sigmael/\sigmatot = 1/2$.

It is important to stress that the connection of the ratio $\sigmael/\sigmatot$ with the opacity or
profile function is \textit{only at zero impact parameter}. When considering a grey-disk model, 
we assume a constant value (in $b$) for the profile, Eq.~\eqref{eq:profile-gd}. However this is a 
na\"ive model and may not be the real case. For example, we may have a torus-like configuration, 
as discussed by Dremin~\cite{Dremin:2014,Dremin:2015a,Dremin:2015b}, in which at the center we have a 
more transparent/grey behaviour with some maximum value (maybe black) at some $b\neq0$.

\section{Relation between ratios $\sigmael/\sigmatot$ and $\sigmatot/B$}
\label{sec:basic_eltot_totB}

For further reference, let us recall the connection between the ratios involving $\sigmatot$, $\sigmael$ and $B$.

From Eqs.~\eqref{eq:sigmael-def} and \eqref{eq:dsigdt-exp}, we obtain
\begin{equation}
 \sigmael(s) = \frac{(1+\rho^2)}{B(s)}\frac{\sigmatot^2}{16\pi}.
 \label{eq:sigmael-integrated-exp}
\end{equation}

Neglecting the value of the real part ($1+\rho^2 \approx 1$),
we obtain an approximate relation between the ratios
\begin{equation}
 \frac{\sigmatot(s)}{B(s)}\approx 16\pi\frac{\sigmael(s)}{\sigmatot(s)}.
 \label{eq:ratios-X-Y}
\end{equation}

\section{Froissart-Lukaszuk-Martin Bound}
\label{sec:basic_FLM_bound}

The Froissart-Lukaszuk-Martin (FLM) bound is one of the most important results in hadronic scattering. 
First derived by Froissart using partial wave expansion for the amplitude~\cite{Froissart:1961} and later 
re-derived by Martin twice (one in studies of analyticity of the amplitude in the enlarged Lehmman 
ellipse~\cite{Martin:1963} and other in the context of Axiomatic Field Theory~\cite{Martin:1965}),
it states that the hadronic total cross section can grow asymptotically at most as the square of the
logarithm of the center of mass energy, namely
\begin{equation}
 \sigmatot(s) \leq B_\text{FLM} \ln^2 (s/s_0) \quad (s\to\infty),
 \label{eq:FLM-bound}
\end{equation}

\noindent where $s_0$ is an undetermined energy scale and the coefficient $B_\text{FLM}$ 
is also bounded, being its maximum value obtained by Martin and Lukaszuk~\cite{Lukaszuk_Martin:1967}
\begin{equation}
 B_\text{FLM} \leq \frac{\pi}{m_\pi^2} \approx 60 \text{ mb},
 \label{eq:FLM-bound-coeff}
\end{equation}

\noindent with $m_\pi$ denoting the mass of pion $\pi^0$.

All these derivations consider the finite range of the strong interaction, the unitarity principle and 
the assumption usually made that the elastic amplitude has a polynomial bound in the asymptotic limit \cite{Eden:1971}
\begin{equation}
 F(s,t) \sim |s|^N \quad (s\to\infty),
\end{equation}

\noindent where $N$ is some constant. This assumption is important, for instance, 
when dealing with dispersion relations (these relations will be discussed in more details in Chapter~\ref{chapt:rise_sigmatot}).

Recently, Azimov~\cite{Azimov:2011,Azimov:2012a,Azimov:2012b} discussed the implications
of violating the FLM bound. In Ref.~\cite{Azimov:2011}, a re-derivation of the FLM bound with the smallest possible number of assumptions is presented.
The main assumptions, according to Ref.~\cite{Azimov:2011}, are the unitarity principle and the absence of massless intermediary states. Azimov argues that
the violation of the Froissart bound is not necessarily related to the violation of the unitarity principle but with the violation
of the polynomial bound in the non-physical region.

Independent of this, we recall that the FLM bound gives us two bounds: an analytical one and a numerical one, both up to some $s_0$ scale. 
The analytical bound is in respect to the maximum rate of increase of the total cross section and is represented by the $\ln^2 s$ itself.
The numerical one is the combination of the analytical bound and the maximum value allowed 
for the coefficient $B_\text{FLM}$ given by Eq.~\eqref{eq:FLM-bound-coeff}. For instance, for $s_0=1\text{ GeV}$, the maximum total cross section at 7 TeV is
$\sim 10^3$ mb, while the experimental data is around 95 mb \cite{TOTEM:2011b}.
Therefore, even if we have a rise faster than $\ln^2 s$, we do not necessarily exceed the numerical limit imposed by the FLM bound.

At last, for further reference, we mention that Martin has recently derived a similar bound to the inelastic cross section~\cite{Martin:2009}
\begin{equation}
 \sigmain(s) \leq \frac{1}{4}\frac{\pi}{m_\pi^2} \ln^2(s/s_0) \quad (s\to\infty)
 \label{eq:bound-inel-Martin}
\end{equation}

\noindent with $s_0$ undetermined.

%% file: ch_data_methods_f.tex
%
%
\cleardoublepage



\chapter{Experimental Data and Fit Procedures}
\label{chapt:data_fits}

In this chapter, we present the experimental data used in the analyses (fits) discussed 
in Chapters~\ref{chapt:ratio_eltot}, \ref{chapt:rise_sigmatot}
and \ref{chapt:SLT_sigmatot}. 
We also discuss general aspects of the methodology employed in the fits performed. 
Specific details, as initial values for the free parameters in non-linear data reductions, will be discussed separately in each case.

\section{Experimental Data}
\label{sec:data_fits-data}

In this section, we treat general information about the datasets used in the analyses discussed in the thesis.

\subsection{Cross sections and $\rho$ Parameter from $pp$ and $\ppbar$ Scattering}

The total cross section, elastic cross section and $\rho$ parameter are determined from the differential 
elastic scattering data $d\sigma/dt$. The total cross sections are obtained from the extrapolation to $t=0$ 
of the fit to the diffraction peak (region of small $t$) using, usually, Eq.~\eqref{eq:dsigdt-exp} 
and the elastic cross section $\sigmael$ is obtained from the integral of $d\sigma/dt$ over an specific range of $|t|$.

On the other hand, in order to measure the $\rho$ parameter, it is necessary to reach the Coulomb-Nuclear interference region,
typically with $|t| < 0.01$ GeV$^2$.
The experimental data from $pp$ and $\ppbar$ scattering of $\sigmatot$ and $\rho$ parameter used in the analyses here discussed 
are available in the Particle Data Group website \cite{PDG_data_website}.
We will consider data at $\sqrt{s} \geq 5$ GeV, as used in the analyses by COMPETE Collab.~\cite{COMPETE:2002b} and PDG~\cite{PDG:2016}. 
We will also consider statistic and systematic uncertainties added in quadrature.
We do not apply any selective criteria to the datasets and we consider points at the same energy as being independent.

Concerning $\sigmatot$ data in the LHC energies, we display in Table~\ref{tab:data_sigtot_LHC} all values considered here
with their uncertainties and the collaboration name with the reference. 
We also introduce a label that will be used for further references.

\begin{table}[htb]
 \centering
 \caption{\label{tab:data_sigtot_LHC}Experimental data of $\sigmatot$ from $pp$ scattering obtained by 
 TOTEM and ATLAS Collaborations at the LHC. 
 Central values, statistic ($\Delta\sigmatot^\text{stat}$), systematic ($\Delta\sigmatot^\text{syst}$) and total uncertainties ($\Delta\sigmatot$) are displayed with the references. 
 In the first column, we define a label that will be used in the text to refer to the data.}
 \small
 \begin{tabular}{ccccccc}\hline\hline
  Label & $\sqrt{s}$ (TeV)& $\sigmatot$ (mb) & $\Delta\sigmatot^\text{stat}$ (mb) & $\Delta\sigmatot^\text{syst}$ (mb) & $\Delta\sigmatot$ (mb) & Collaboration            \\\hline
  T1    & 7               & 98.3             & 0.2                                & 2.8                                & 2.8                    & TOTEM \cite{TOTEM:2011b} \\
  T2    & 7               & 98.6             & -                                  & 2.2                                & 2.2                    & TOTEM \cite{TOTEM:2013a} \\
  T3    & 7               & 99.1             & -                                  & 4.3                                & 4.3                    & TOTEM \cite{TOTEM:2013c} \\
  T4    & 7               & 98.0             & -                                  & 2.5                                & 2.5                    & TOTEM \cite{TOTEM:2013c} \\
  T5    & 8               & 101.7            & -                                  & 2.9                                & 2.9                    & TOTEM \cite{TOTEM:2013d} \\
  T6    & 8               & 101.5            & -                                  & 2.1                                & 2.1                    & TOTEM \cite{TOTEM:2015}  \\
  T7    & 8               & 101.9            & -                                  & 2.1                                & 2.1                    & TOTEM \cite{TOTEM:2015}  \\
  T8    & 8               & 102.9            & -                                  & 2.3                                & 2.3                    & TOTEM \cite{TOTEM:2016}  \\
  T9    & 8               & 103.0            & -                                  & 2.3                                & 2.3                    & TOTEM \cite{TOTEM:2016}  \\\hline
  A1    & 7               & 95.35            & 0.38                               & 1.304                              & 1.36                   & ATLAS \cite{ATLAS:2014}  \\
  A2    & 8               & 96.07            & 0.18                               & 0.85 $\pm$ 0.31                    & 0.92                   & ATLAS \cite{ATLAS:2016}  \\\hline\hline
 \end{tabular}
 \normalsize
\end{table}

In the last years, the TOTEM Collaboration has published several values for $\sigmatot$ at 7 and 8 TeV using different methods.
The measurements T1 \cite{TOTEM:2011b} and T2 \cite{TOTEM:2013a} at 7 TeV were obtained through a luminosity-dependent 
method using as input for $\rho$ value the prediction of COMPETE highest-rank result \cite{COMPETE:2002b}. 
To obtain the value T3 \cite{TOTEM:2013c}, TOTEM Collab. has considered a $\rho$-independent method. 
Measurements T4 \cite{TOTEM:2013c} and T5 \cite{TOTEM:2013d} at 7 and 8 TeV, respectively, follow from a luminosity 
independent method, using again the extrapolation for $\rho$ from COMPETE.
The values T6 and T7 at 8 TeV are reported in Ref.~\cite{TOTEM:2015}, where they have observed a deviation
from the pure exponential behaviour of the diffraction peak 
(see in Ref.~\cite{Fagundes_etal:2015c} a simple analysis that accounts for this deviation). 
The values T6 and T7 are obtained considering two different forms of the argument 
of the exponential: a polynomial of second and third degree, respectively. In both cases, 
the value of $\rho$ considered was again the extrapolation of the COMPETE result.
In Ref.~\cite{TOTEM:2016}, TOTEM Collaboration reported the first measurement of the $\rho$ parameter at the LHC by measuring 
the differential cross section in the Coulomb-Nuclear interference region. At 8 TeV, this parameter reads
\begin{equation}
 \rho_\text{TOTEM}\, (8\text{ TeV}) = 0.12 \pm 0.03,
 \label{eq:totem-rho-8TeV}
\end{equation}

\noindent and the total cross sections obtained in the analysis considering two different profiles for the proton are T8 and T9.
Although the value of $\sigmatot$ is affected by the choice of the profile function, the value of $\rho$ does not change.
This measurement is compatible with the prediction of the COMPETE preferred model, however
it is interesting to note that the central value of Eq.~\eqref{eq:totem-rho-8TeV} is below this prediction.

The ATLAS Collaboration also reported the total cross section measured by the ALFA detector at 7 and 8 TeV. 
They considered the luminosity-dependent method and used for the $\rho$ value the extrapolation of COMPETE at 
7 TeV and of PDG 2014 \cite{PDG:2014} result at 8 TeV. 
These values are labeled in Table~\ref{tab:data_sigtot_LHC} by A1 and A2, respectively.

Cosmic-ray data were not included in the fits. It is not expected that their inclusion would change the results since 
they have large uncertainties. Apart from that, we show some of them in the figures of $\sigmatot$ to illustrate the trend of the data.
We have considered the experimental information reported by the ARGO-YBJ Collaboration \cite{ARGO_YBJ:2009}, 
the Pierre Auger Observatory \cite{PierreAuger:2012}, and the Telescope Array \cite{TA:2015}.

The datasets of $\sigmatot$ and $\rho$ for $pp$ and $\ppbar$ scattering are considered in Chapter \ref{chapt:rise_sigmatot}.
In Chapter \ref{chapt:SLT_sigmatot}, we also used $\sigmatot$ data from $pp$ and $\ppbar$ scattering.

\subsection{$\sigmatot$ from other reactions}

In Chapter \ref{chapt:SLT_sigmatot}, besides $pp$ and $\ppbar$ data,
we also consider $\sigmatot$ data from reactions involving mesons and other baryons. 
These datasets are available at the PDG website \cite{PDG_data_website}. 
In Table~\ref{tab:data_sigtot_otherreac}, we display the reactions considered, the energy range of each dataset,
and the number of points.
Once more, we consider statistic and systematic uncertainties added in quadrature, 
without selective criteria.

\begin{table}[htb]
\centering
\caption{\label{tab:data_sigtot_otherreac} Information of datasets from reactions involving mesons and baryons. 
We show the minimum energy, maximum energy, and number of points for each dataset.} 
\begin{tabular}{c c c c}\hline\hline
 Reaction    & $\sqrt{s_\text{min}}$ (GeV) & $\sqrt{s_\text{max}}$ (GeV) & Nº points\\\hline
 $pn$        & 5.30 & 26.40 & 34\\
 $\bar{p}n$  & 5.18 & 22.98 & 33\\\hline
 $\pi^+p$    & 5.21 & 25.28 & 50\\
 $\pi^-p$    & 5.03 & 34.67 & 95\\
 $K^+p$      & 5.13 & 24.14 & 40\\
 $K^-p$      & 5.11 & 24.14 & 63\\
 $K^+n$      & 5.24 & 24.16 & 28\\
 $K^-n$      & 5.11 & 24.16 & 36\\\hline\hline
\end{tabular}
\end{table} 

\subsection{Ratio $\sigmael/\sigmatot$ from $pp$ and $\ppbar$ Scattering}
\label{subsec:data_ratioX}

The empirical analysis discussed in Chapter \ref{chapt:ratio_eltot} consists in fits performed directly to $\sigmael/\sigmatot$ data.
Our dataset, with minimum energy of 5 GeV, comprises values of this ratio from $pp$ and $\ppbar$ scattering with
29 and 13 points respectively (42 in total). In this case, our selection criterion was to 
calculate the ratio using cross sections that were determined in the same analysis, i.e. $\sigmatot$ and $\sigmael$ 
determined from the same $d\sigma/dt$ data. The cross sections used in these calculations are 
available at the PDG website \cite{PDG_data_website} for 5 GeV $< \sqrt{s} \leq$ 1.8 TeV.
We show in Table~\ref{tab:data_eltot_LHC} the experimental information at LHC energies
\cite{TOTEM:2011b,TOTEM:2013a,TOTEM:2013c,TOTEM:2013d,ATLAS:2014}.
Once more, we have considered each point as being independent,
and we added the statistic and systematic uncertainties in quadrature.

\begin{table}[htb]
\centering
\caption{\label{tab:data_eltot_LHC}Experimental data for ratio $\sigmael/\sigmatot$ at the LHC energies ($pp$ scattering). 
We use the same label of Table~\ref{tab:data_sigtot_LHC}.}   
\begin{tabular}{c c c c c c}\hline\hline
Label & $\sqrt{s}$ (TeV) & $\sigmatot$ (mb) & $\sigmael$ (mb) & $\sigmael/\sigmatot$ & Collaboration \\\hline
T1 & 7 & 98.3  $\pm$ 2.8 & 24.8 $\pm$ 1.2 & 0.252 $\pm$ 0.014 & TOTEM \cite{TOTEM:2011b} \\
T2 & 7 & 98.6  $\pm$ 2.2 & 25.4 $\pm$ 1.1 & 0.258 $\pm$ 0.013 & TOTEM \cite{TOTEM:2013a} \\
T3 & 7 & 99.1  $\pm$ 4.1 & 25.4 $\pm$ 1.1 & 0.256 $\pm$ 0.015 & TOTEM \cite{TOTEM:2013c} \\
T4 & 7 & 98.0  $\pm$ 2.5 & 25.1 $\pm$ 1.1 & 0.256 $\pm$ 0.013 & TOTEM \cite{TOTEM:2013c} \\
T5 & 8 & 101.7 $\pm$ 2.9 & 27.1 $\pm$ 1.4 & 0.266 $\pm$ 0.016 & TOTEM \cite{TOTEM:2013d} \\\hline
A1 & 7 & 95.4  $\pm$ 1.4 & 24.0 $\pm$ 0.6 & 0.252 $\pm$ 0.004 & ATLAS \cite{ATLAS:2014}  \\\hline\hline
\end{tabular}
\end{table}

It is also possible to estimate the ratio $\sigmael/\sigmatot$ at 57 TeV,
evaluated from the experimental information on $\sigmatot$ and $\sigmain$
obtained by the Pierre Auger Collaboration \cite{PierreAuger:2012}:
\begin{eqnarray}
\frac{\sigmael}{\sigmatot}(\sqrt{s} = 57\, \mathrm{TeV}) = 0.31 ^{+0.17}_{-0.19},
\nonumber
\end{eqnarray}

\noindent where statistical, systematic and Glauber uncertainties \cite{PierreAuger:2012}
have been added in quadrature. We stress that this point did not
take part of our data reductions. It will be included in the figure as illustration.

\section{Fit Procedures}
\label{sec:data_fits_procedures}

All fits were done using the objects of the \texttt{TMinuit} class available in the ROOT Framework (CERN) \cite{ROOT_website} 
considering a confidence level of 68\% ($1\sigma$) \cite{Bevington_Robinson_book:1992}.
The parameters are determined from the minimization of the $\chi^2$ function, defined as
\begin{equation}
 \chi^2 = \sum_i \left[\frac{y_i - f(x_i,\mathbf{a})}{\Delta y_i}\right]^2,
\end{equation}

\noindent where $y \pm \Delta y$ denote the experimental data with its uncertainty, 
$x$ is the independent variable from which $y$ depends on, $f(x,\mathbf{a})$ is the
parametrization, and the vector $\mathbf{a}$ represents the free parameters. 
The sum runs over all data considered (index $i$). 
For example, $x$ can be the energy, and $y$, the total cross section.

We consider as a measure of goodness of fit the reduced $\chi^2$ given by the ratio $\chi^2/\nu$, 
where $\nu$ is the degree of freedom (the difference between the number 
of data points considered in the fit and the number of free parameters). 
We also considered the Estimated Distance to Minimum (\texttt{EDM}) \cite{MINUIT_Manual:1994,James:1975}, 
which is an internal parameter of \texttt{MINUIT}. 
We say that the fit has converged when \texttt{EDM} $\leq 10^{-4}$ (considering a confidence level of $1\sigma$).

To be selected as the final result, the fit must have the smallest $\chi^2$ value, 
the smallest \texttt{EDM}, and a positive-definite error matrix.
We have tried to avoid the results that have a non-positive definite error matrix 
since the error estimation in these cases may have problems
and, in general, we cannot trust the estimated uncertainties of the parameters \cite{MINUIT_Manual:1994,James:1975}.

Together with the error matrix, the code also provides the correlation matrix, 
whose elements are the correlation between the fit parameters. 
In some cases, we will use these values in the discussion of our results.

We also show in some cases an uncertainty region in the figures and predictions for some quantities. 
In these cases, the errors were calculated using standard error propagation \cite{Bevington_Robinson_book:1992} from the variances and covariances
of the parameters determined in the fit (error matrix). Therefore, they represent a confidence level of $1\sigma$.

The integrated $\chi^2$ probability is also shown in some cases. This probability is given by \cite{Bevington_Robinson_book:1992}
\begin{equation}
 P(\chi^2,\nu) = \int_{\chi^2}^\infty \frac{(x^2)^{1/2(\nu-2)}e^{-x^2/2}}{2^{\nu/2}\Gamma(\nu/2)}
\end{equation}

\noindent where $\Gamma(x)$ is the Gamma function. For a good fit with large $\nu$ (say, beyond $10^{2}$), 
we expect $\chi^2/\nu \sim 1.0$, corresponding to $P(\chi^2,\nu)\sim 0.5$.

It is important to note that the inclusion of systematic uncertainties in the fit 
puts some limits in the interpretation of the $\chi^2/\nu$ and $P(\chi^2,\nu$) values.
The $\chi^2$ assumes that the experimental value is Gaussian distributed, i.e., 
the uncertainty is directly correlated to the confidence level of the measurement. 
On the other hand, the systematic uncertainty follows a uniform distribution, 
meaning that the probability of the true value is constant inside the error bars. 
Therefore, we will use these statistical parameters as a measure that the fit is
reasonable and that we have reached a minimum in 
the parameter space and not to select a model or result based on small differences in these values. 
Of course, large deviations from the expected $\chi^2/\nu$ value (close to zero or 2, 
for instance) is an indication that we do not have a good result.

At last, there is an important aspect of the fits considered here that is their non-linear dependence on the free parameters. 
In that case, it is important to give to the code the initial values for the parameters. 
Since here we are dealing with different parametrizations, 
we will explain the methodologies used for the initial value in each analysis separately.

More discussion on experimental data, fits procedures and critical remarks can be
found in Refs.~\cite{Fagundes_Menon_Silva:2013a,Menon_Silva:2013a,Menon_Silva:2013b,Fagundes_Menon_Silva:2017b},
in special Appendix A in Ref.~\cite{Fagundes_Menon_Silva:2017b}.

%% file: ch_ratio_eltot_f.tex
%
%
%
%
\cleardoublepage



\chapter[Empirical Studies on $\sigmael/\sigmatot$]{Empirical Studies on the Ratio Between Elastic and Total Cross Sections}
\label{chapt:ratio_eltot}

\section{Introduction}
\label{sec:ratioX_intro}

In this chapter, we discuss an empirical study of the ratio between the elastic
(integrated) cross section ($\sigmael$) and the total cross section ($\sigmatot$),
the first of three topics that will be covered through the text.
The chapter is based on the research presented in
Refs.~\cite{Fagundes_Menon_Silva:2015,Fagundes_Menon_Silva:2016a,Fagundes_Menon_Silva:2016b}.

\textit{Empirical} approaches consist in the development of \textit{model-independent} 
descriptions of experimental data,  in which we look for quantitative results 
that may work as an effective bridge for further developments of QCD in the 
soft-scattering sector and even selecting phenomenological pictures.

The ratio
\begin{equation}
X = \frac{\sigmael}{\sigmatot},
\label{eq:x-def}
\end{equation}

\noindent as a function of the c.m. energy $\sqrt{s}$, plays an important role for several reasons.

First, in the impact parameter representation (see Section~\ref{sec:basic_profile_models}), $X$ is connected to the
opacity of the colliding hadrons and, in the cases of the grey/black disk or Gaussian profiles,
it is proportional to the central opacity ($b = 0$).

We recall that the $s$-channel unitarity states that $\sigmain = \sigmatot - \sigmael$. With this equation,
one can determine the inelastic cross section from the total and elastic cross sections and this approach constitutes
the less-biased way to obtain $\sigmain$ since the direct measurement involves model-dependent extrapolations.
The same is true for the ratio $\sigmain/\sigmatot = 1-X$, which can be associated with the inelasticity of the 
collision \cite{Bellandi:1991}.

Moreover, empirical information on $X(s)$ 
and $1 - X(s)$ at the highest $s$ and as $s \rightarrow \infty$ 
can provide crucial information on the asymptotic properties of the hadronic interactions,
i.e. the information on how $\sigmatot$ and $\sigmael$ reach simultaneously their respective unitarity bounds 
(which is one of the main prospects in the LHC forward-physics program \cite{ForwardLHC:2016}). 
Empirical information on the asymptotic limit is also important in the construction and selection of 
phenomenological models, including those based or inspired in nonperturbative QCD. 

Finally, the ratio $X$ can be connected, through an approximated relation (see Section~\ref{sec:basic_eltot_totB}),
to the ratio between the total cross-section and the elastic slope parameter ($B$) 
\begin{equation}
Y(s) = \frac{1}{16 \pi}\frac{\sigmatot}{B}(s) \approx 16\pi X(s).
\label{eq:y-def}
\end{equation}

This ratio
plays an important role
in the study of extensive air-showers (EAS) \cite{Ulrich_etal:2009},
specially in the estimation of the proton-proton cross-section~\cite{Engel:1998,Engel:2000}
from the proton-air production cross section at energies above 50 TeV.
The interpretation of the EAS development depends on
extrapolations from theoretical formalisms that have been tested only in the
accelerator energy region, resulting in rather large theoretical uncertainties,
mainly in the estimation of the total cross-section.
Therefore, model-independent information on $Y(s)$, from $X(s)$, may be important in these extrapolations.
Moreover, the ratio $Y$ gives also information on the connection between $\sigmatot$ and $B$
at the highest and asymptotic energies.
    
\begin{figure}[htb]
\centering
\includegraphics[scale=0.5]{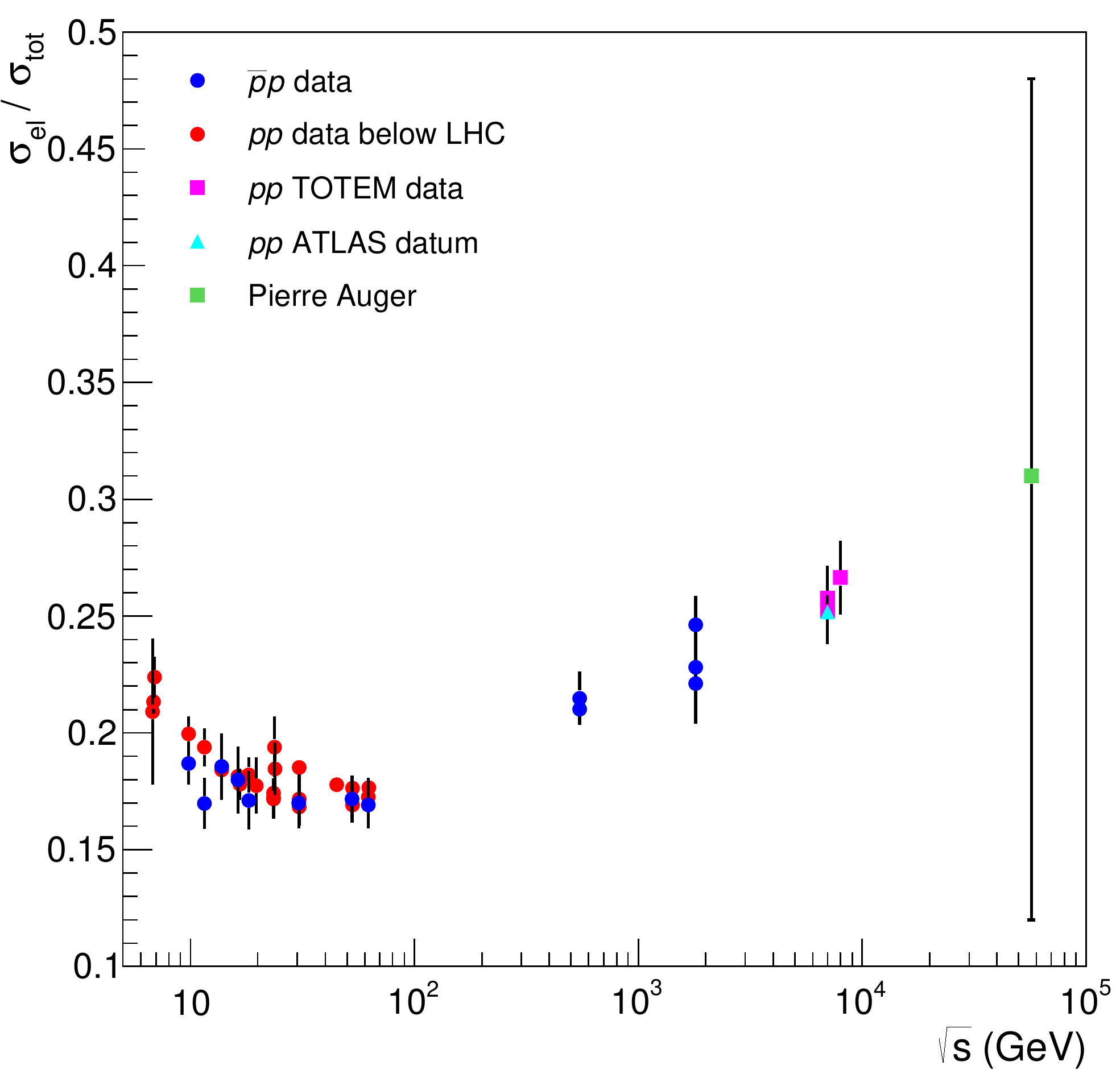}
\caption{\label{fig:data_X}Experimental data of the ratio $X=\sigmael/\sigmatot$ used in this analysis, from $pp$ and $\bar{p}p$
elastic scattering in the energy region 5 GeV $< \sqrt{s} \leq$ 8 TeV (accelerator data). 
At $\sqrt{s}$ = 57 TeV we show, as illustration, the ratio obtained from the estimations
of $\sigmatot$ and $\sigmain$ in a cosmic-ray experiment (Pierre Auger Observatory).}
\end{figure}

Our goal here is to obtain an empirical description of the ratio $X$ and discuss 
the implications of the results along the aforementioned lines, with main focus on 
asymptotic scenarios ($s \rightarrow \infty$). We will consider data on $X$ from $pp$ and $\ppbar$
scattering above 5 GeV (see Fig.~\ref{fig:data_X}), including all the TOTEM data at 7 and 8 TeV and also the ATLAS datum at 7 TeV.
References and further details of the dataset are given in Section~\ref{subsec:data_ratioX}.

We introduce four analytical parametrizations for $X(s)$, looking for good descriptions
with an economic number of free parameter.
We investigate all the three possible asymptotic scenarios: either the standard black disk limit or scenarios above or bellow that limit. 
The results for $X(s)$ are extended to the inelastic channel (with the $s$-channel unitarity),
and we discuss the connection to dissociative processes (single, double diffraction), including the Pumplin bound. 
We also treat the connection between $X(s)$ and the ratio $Y(s)$, Eq.~\eqref{eq:y-def}, together with discussions
on the applicability of the results in studies of extensive air showers in cosmic-ray experiments.

The chapter is organised as follows.
In Section~\ref{sec:ratioX_asymp_values}, ee start presenting the asymptotic scenarios for $X$ considered in our analysis.
Next, in Section~\ref{sec:ratioX_parametrization}, we discuss our empirical parametrization and the variants considered.
In Section~\ref{sec:ratioX_fit_results} we present the fit results, which are compared and discussed in Section~\ref{sec:ratioX_fit_discussion_results}.
Predictions and extensions to other quantities are presented in Section~\ref{sec:ratioX_fit_results_extension}. 
In Section~\ref{sec:ratioX_comments_interpretations}, we discuss some possible physical aspects related to the analytical structure of the empirical parametrizations,
presenting also comments on a semi-transparent asymptotic scenario. A summary and our final conclusions are the contents of  
Section~\ref{sec:ratioX_fit_conclusion_final}.

\section{Asymptotic Scenarios}
\label{sec:ratioX_asymp_values}

In order to build our analytical parametrizations, we will start with the asymptotic scenarios
considered in this analysis.
Let $A$ denote the asymptotic value of the ratio $X$:
\begin{equation}
\lim_{s \rightarrow \infty} X(s) \equiv A.
\label{eq:A-def}
\end{equation}

Below we discuss and display some numerical values for $A$,
which will be considered as typical of each scenario to be investigated in this work.

\subsection{The Black Disk}
\label{subsec:ratioX_blackdisk}

The black-disk limit represents a standard phenomenological
expectation. It is typical of eikonalized formalisms, as the traditional models by
Chou and Yang~\cite{Chou:1969}, Bourrely, Soffer and Wu~\cite{Bourrely:1984,Bourrely:1988,Bourrely:2011}, the hybrid approach by
Block and Halzen~\cite{Block:2012} and a number of models that have been continuously refined and developed 
(for example, \cite{Khoze:2015a,Gotsman:2015a,Gotsman:2015b,Block:2015,Nemes:2015,Selyugin:2014,Grau_etal:2012,Wibig:2011,Fagundes:2011}).

As already mentioned in Section~\ref{sec:basic_profile_models}, this limit reads
\begin{equation}
\boxed{A = \frac{1}{2}.}
\nonumber
\end{equation}

\subsection{Above the Black Disk}
\label{subsec:ratioX_above_blackdisk}

Scenarios above the black disk can be inferred from theoretical results
and appear in some phenomenological approaches, as discussed below.

\begin{enumerate}

\item In the formal context, the $s$-channel unitarity,
\begin{equation}
\frac{\sigmael}{\sigmatot} + \frac{\sigmain}{\sigmatot} = 1,
\label{eq:unitarity}
\end{equation}

\noindent imposes an obvious maximum bound for $X(s)$, namely
\begin{equation}
\boxed{A = 1.}
\nonumber
\end{equation}

\item Two well known bounds have been
established for the total cross-section~\cite{Froissart:1961,Lukaszuk_Martin:1967} and the
inelastic cross-section~\cite{Martin:2009,Wu:2011} (see also Section~\ref{sec:basic_FLM_bound}),
\begin{equation}
\sigmatot(s) < \frac{\pi}{m_{\pi}^2} \ln^2(s/s_0)
\quad \mathrm{and} \quad
\sigmain(s) < \frac{\pi}{4m_{\pi}^2} \ln^2(s/s_0).
\nonumber
\end{equation}

In case of simultaneous saturation of both bounds
as $s \rightarrow \infty$, it is mathematically possible that 
$\sigmain/\sigmatot \rightarrow 1/4$, which from unitarity, Eq.~\eqref{eq:unitarity}, implies in
\begin{equation}
\boxed{A = \frac{3}{4}.}
\nonumber
\end{equation}

We note that this limit does not correspond to the usual interpretation of
the aforementioned asymptotic bounds. In fact, two fractions $1/2$ are usually associated 
with each bound, in place of $1/4$ in the inelastic case (although without formal derivation~\cite{Martin:2009}),
favouring the black-disk scenario~\cite{Dremin:2013,Cartiglia:2013,Martin:2009,Block:2011}.
Even if a simultaneous saturation of both bounds might be questionable \cite{Troshin:2012}, 
the number 0.75 can be considered as an instrumental choice for data reductions, lying between the black disk 
and the maximum value allowed by unitarity.

\item In the phenomenological context, the $U$-matrix unitarization scheme by Troshin and Tyurin
predicts an asymptotic limit beyond the black disk~\cite{Troshin_Tyurin:1993,Troshin_Tyurin:2007}. In this approach,
the reflective scattering mode takes place at small values of the impact parameter~\cite{Alkin:2014}.

\end{enumerate}

\subsection{Below the Black Disk}
\label{subsec:ratioX_below_blackdisk}

Together with scenarios above the black-disk limit, there are also some results that
suggest and indicate asymptotic limits below the black disk.
 
\begin{enumerate}

\item In the publications of the TOTEM Collaboration, the authors quote the prediction
for the total cross-section obtained by the COMPETE Collaboration (in 2002), 
with parametrization RRPL2, energy cutoff at 5 GeV and given by \cite{COMPETE:2002b}
\begin{equation}
\sigmatot^\text{COMPETE}(s) = 42.6\, s^{-0.46} - 33.4\, s^{-0.545} + 35.5 + 0.307 \ln^{2}(s/29.1),
\label{eq:res-compete}
\end{equation}

\noindent where all coefficients are in mb and $s$ is in GeV$^2$.
TOTEM Collaboration also present a fit to the $\sigmael$ data above 10 GeV \cite{TOTEM:2013d}
\begin{equation}
\sigmael^\text{TOTEM}(s) = 11.7 - 1.59 \ln(s) + 0.134 \ln^{2}(s).
\label{eq:res-el-totem}
\end{equation}

Using the above parametrizations, the ratio $X$ for $s \rightarrow \infty$ reads
\begin{equation}
\boxed{A =  0.436,}
\nonumber
\end{equation}

\noindent which suggests a scenario below the black disk.

\item Fagundes, Menon and Silva \cite{Fagundes_Menon_Silva:2013a} and Menon and Silva \cite{Menon_Silva:2013a,Menon_Silva:2013b}
have developed several analyses of the experimental data on
$\sigmatot$, the $\rho$ parameter and $\sigmael$, including the TOTEM
Collaboration results at 7 and 8 TeV. For our purposes, we recall that the parametrization
for the total cross section is expressed by
\begin{equation}
\sigmatot(s) = \mathrm{Regge}\ \mathrm{terms}\ + \alpha + \beta \ln^{\gamma}(s/s_h)
\nonumber
\end{equation}

\noindent where $\alpha$, $\beta$ and $\gamma$ are free parameters.
Fits to data on $\sigmatot$ and $\rho$ (using derivative dispersion relations) from
$pp$ and $\bar{p}p$ scattering above 5 GeV, have led to statistically consistent solutions
either with $\gamma = 2$ (fixed) or  $\gamma > 2$ (free fit parameter).
In \textit{both cases}, extension of the parametrization to $\sigmael$ data (same $\gamma$ value) 
allowed to extract the ratio $X(s)$ and its asymptotic value $A$. In all cases we have obtained
$A < 1/2$, with lowest central value around $1/3$ (see a summary of the results in \cite{Menon_Silva:2013b},
figure 10).
For future use, as a typical input in data reductions, we shall consider the lowest value obtained
in these analyses, which, within the uncertainties, reads
\begin{equation}
\boxed{A = 0.3.}
\nonumber
\end{equation}

\end{enumerate}

\section{Model-Independent Parametrization}
\label{sec:ratioX_parametrization}

We present in this Section a detailed
discussion on the choices and steps that led us to the construction of an empirical
ansatz for $X(s)$.

\subsection{Empirical and Analytical Arguments}
\label{subsec:ratioX_emp_an_arguments}

\subsubsection{General Aspects}

The available data for the ratio $X$ are shown in Fig.~\ref{fig:data_X}. It is easy to see that
as the energy increases above 5 GeV, the $X$ data decrease
up to the CERN-ISR region ($\approx$ 20 - 60 GeV), where they remain approximately constant and then
begin to increase smoothly. From a strictly empirical point of view, this rise in the linear-log
plot scale may suggest a parabolic parametrization in terms of $\ln s$, that is, a growth with positive curvature.
However, as discussed in Section~\ref{sec:ratioX_asymp_values}, 
the Unitarity Principle demands the obvious bound $1$ for $X(s)$ as
$s \rightarrow \infty$ and finite values
are also dictated by the Froissart-Martin bounds and all phenomenological 
and empirical analysis, independently of the asymptotic scenario considered.
Therefore, except in case of existence of an unexpected singular behaviour at
some finite value of the energy, the above facts indicate a constant finite
asymptotic limit for the ratio $X(s)$, i.e. a smooth \textit{saturation effect}
as $s \rightarrow \infty$.
That in turn, demands a change in the sign of the curvature at some finite value
of the energy, so that $X(s)$ goes asymptotically to a constant limit with negative curvature.

The above arguments suggest an analytical parametrization
related to a sigmoid function (``S-shaped" curve), in order to impose the change of 
the curvature sign and an asymptotic limit (a constant).  
However, the behavior of the data at low energies and the necessity to obtain the correct
curvature to describe the experimental data, indicate that deviations from the pure sigmoid
function must be taken into account or, at least, investigated.
That led us to express the parametrization for $X$ as a composite function of the sigmoid $S$ and
some function $f$ of the energy:
\begin{equation}
X(s) = A\,S(f), \qquad f = f(s),
\label{eq:X-par-gen}
\end{equation}

\noindent where, from Eq.~\eqref{eq:A-def},
\begin{equation}
\lim_{s \rightarrow \infty} S(f(s))= 1.
\nonumber
\end{equation}

Therefore,
the point would be to test different forms 
for $S(f)$ and $f(s)$ through
fits to the experimental data, looking for statistically consistent descriptions
with an \textit{economical number} of free parameters.
Obviously, the problem does not have a unique solution and it is not possible
to test all the analytical possibilities. 

However, the use of different classes of sigmoid functions combined with different classes
of phenomenologically-based elementary functions is a way to take into account the 
intrinsic uncertainty in the choice of the complete parametrization.
With that in mind,
we shall select two forms for $S(f)$, combined (each one) with two forms for $f(s)$,
as explained in what follows.

\subsubsection{Sigmoid Functions $S(f)$}

Several classes of sigmoid functions have applications in different scientific contexts
and that includes the logistic, hyperbolic tangent, error function, algebraic ratios 
and many others (see for example \cite{Menon_etal_sig:1994,Kucharavy_deGuio:2011}; 
we shall return to its applications in Section~\ref{sec:ratioX_comments_interpretations}). 
Here we consider two classes of sigmoid functions. One of them, already used in all our
previous analyses \cite{Fagundes:2012,Fagundes:2013,Fagundes_Menon_Silva:2015,Fagundes_Menon_Silva:2016b},
is the \textit{Hyperbolic Tangent}, denoted by
\begin{equation}
S^{HT}(f) = \tanh f = \frac{1 - \exp\{{-2f}\}}{1 + \exp\{{-2f}\}}.
\label{eq:sigmoid-tanh}
\end{equation}

In addition,  we now consider a \textit{Logistic} function, with notation
\begin{equation}
S^L(f)  = \frac{1}{1 + \exp\{-f\}},
\label{eq:sigmoid-def}
\end{equation}

\noindent which can be connected with $S^{HT}(f)$ by translation and scaling transformation. 

\subsubsection{Elementary Functions $f(s)$}

We shall express $f(s)$ in terms of elementary
functions of the \textit{standard soft variable}
\begin{equation}
\ln{(s/s_0)},
\label{eq:var-v-def}
\end{equation}

\noindent where $s_0$ is a fixed energy scale to be discussed later.
Different functions and different conditions have already been
investigated in previous analyses \cite{Fagundes:2012,Fagundes:2013,Fagundes_Menon_Silva:2015,Fagundes_Menon_Silva:2016b}.
Here, we first express $f$ as a sum of two terms: a linear function of the
standard variable $\ln(s/s_0)$ and a function $g(s)$ which
can account for possible deviations from linearity:
\begin{equation}
f(s) = \alpha + \beta\, \ln(s/s_0) + \gamma\, g(s),
\label{eq:function-f-def}
\end{equation}

\noindent where $\alpha$, $\beta$ and $\gamma$ are real free fit parameters.

Different tests with distinct datasets
(only $pp$ or including $\bar{p}p$) and different energy cutoffs
led us to choose two typical forms in soft scattering for $g(s)$:
a \textit{Power-Law},
\begin{equation}
g_{PL}(s) = \ln^{\delta}(s/s_0),
\label{eq:func-g-PL}
\end{equation}

\noindent where $\delta$ is an (additional) free fit parameter, and a \textit{Logarithmic-Law},
\begin{equation}
g_{LL}(s) = \ln \ln(s/s_0),
\label{eq:func-g-LL}
\end{equation}

\noindent without an additional parameter.

For each sigmoid function (logistic or hyperbolic tangent) we consider
two \textit{variants} associated and denoted by the functions $g(s)$. Throughout the chapter, we
shall use the notation defined below.

\begin{itemize}
\item[-] Logistic ($L$) with variant Power-Law ($PL$)
\begin{equation}
X_{PL}^{L}(s) =  \frac{A}{1 + \exp\{-[\alpha + \beta \ln (s/s_0) + \gamma \ln^{\delta} (s/s_0)]\}};
\label{eq:X-par-Log-PL}
\end{equation}

\item[-] Logistic ($L$) with variant Logarithmic-Law ($LL$)
\begin{equation}
X_{LL}^{L}(s) =  \frac{A}{1 + \exp\{-[\alpha + \beta \ln (s/s_0) + \gamma \ln \ln (s/s_0)]\}};
\label{eq:X-par-Log-LL}
\end{equation}

\item[-] Hyperbolic Tangent ($HT$) with variant Power-Law ($PL$)
\begin{equation}
X_{PL}^{HT}(s) = A\,\tanh\{\alpha + \beta \ln (s/s_0) + \gamma \ln^{\delta} (s/s_0)\};
\label{eq:X-par-tanh-PL}
\end{equation}

\item[-] Hyperbolic Tangent ($HT$) with variant Logarithmic-Law ($LL$)
\begin{equation}
X_{LL}^{HT}(s) = A\,\tanh\{\alpha + \beta \ln (s/s_0) + \gamma \ln \ln (s/s_0)\};
\label{eq:X-par-tanh-LL}
\end{equation}

\end{itemize}

\noindent with the condition
\begin{equation}
s > s_0.
\nonumber
\end{equation}

At last, and still as a matter of a short notation, we shall refer to the sigmoid functions as $logistic$ or $tanh$
and to the variants as $PL$ (Power-Law) or $LL$ (Logarithmic-Law).

\subsection{Review on Previous Results}
\label{subsec:ratioX_previous_results}

In this Section, we review some previous results
obtained in fits with the tanh, special
cases of the variant $PL$ and energy scales at 1 and 25 GeV$^2$.

In the 2012 analysis by Fagundes and Menon \cite{Fagundes:2012,Fagundes:2013}, the dataset was restricted to $pp$
scattering above 10 GeV and included only the first TOTEM datum at 7 TeV. In order to infer
uncertainty regions in the extrapolation to higher energies, two extreme asymptotic
limits have been tested by either fixing $A~=~1/2$ (black-disk limit) or $A~=~1$ (maximum
unitarity). The experimental data have been well described through variant $PL$ with fixed $\delta = 2$, 
fixed $s_0 = 1$ GeV$^2$,  and only three fit parameters: $\alpha$, $\beta$ and $\gamma$
(denoted $\gamma_1$, $\gamma_2$ and $\gamma_3$ in Ref.~\cite{Fagundes:2012}).
Through the approximate relation $Y \approx 16\pi X$ (discussed in Section~\ref{sec:basic_eltot_totB}), it was possible
to extend the extrapolation of the uncertainty regions to the ratio $\sigmatot/B$, 
which, in the context of the Glauber model \cite{Fagundes:2012}, plays an important role in the determination
of the $pp$ total cross-section from proton-air production cross-section in cosmic-ray experiments.

This empirical analysis was pushed forward in Refs.~\cite{Fagundes_Menon_Silva:2015,Fagundes_Menon_Silva:2016b}, where
the energy cutoff has been extended down to 5 GeV
and the dataset included all the TOTEM measurements at 7 and 8 TeV \cite{Fagundes_Menon_Silva:2015} and the 
ATLAS datum at 7 TeV \cite{Fagundes_Menon_Silva:2016b}. Preliminary fits to only $pp$ data with variant $PL$, 
different $A$ values and energy scale fixed in $25$ GeV$^2$ (the energy cutoff) led to
solutions with the parameter $\delta$ consistent with 
0.5, within the uncertainties. Next, the $\delta$ parameter was fixed in 0.5 
and new fits were developed with the inclusion of
$\bar{p}p$ data, considering, again, fits with either $A$ fixed or $A$ free.
All data reductions presented consistent
descriptions of the experimental data analyzed and in the case of the fit with $A$ free, all fits
converged to
a unique solution with asymptotic limit below the black disk, namely 
$A = 0.332 \pm 0.049$ \cite{Fagundes_Menon_Silva:2016b}.

\subsection{The Energy Scale}
\label{subsec:ratioX_energy_scale}

As discussed in the above section, different energy scales have been considered in previous works:
$s_0=1~\text{ GeV}^2$ (which consists in an usual choice in phenomenology) in Ref.~\cite{Fagundes:2012,Fagundes:2013}
and $s_0=25~\text{ GeV}^2$ (corresponding to the energy cutoff) in Refs.~\cite{Fagundes_Menon_Silva:2015,Fagundes_Menon_Silva:2016b}.
In the present work, we will consider a third possibility that has a more physical meaning, namely the energy threshold for the scattering states
\begin{equation}
s_0 = 4 m_p^2,
\nonumber
\end{equation}

\noindent where $m_p$ is the proton mass.
We will show (Section~\ref{subsec:ratioX_results_conclusion}) 
that the selected results do not depend on the aforementioned choices.

Evidently, a more ``general'' result could be obtained by letting $s_0$ be a free parameter. However,
this would introduce additional non-linearity in the fits that, in turn, would led to additional correlations
between the free parameters which are not easy to control. For this reason, we decided to consider
here only the case of a fixed energy scale.

\subsection{Constrained and Unconstrained Fits}
\label{subsec:ratioX_cons_unconst_fits}

For each combination of sigmoid and elementary function 
[see Eqs.~\eqref{eq:X-par-Log-PL}, \eqref{eq:X-par-Log-LL}, \eqref{eq:X-par-tanh-PL} and \eqref{eq:X-par-tanh-LL}], 
two types of fits will be considered: those with $A$ as a fixed parameter and those whith $A$ as a free parameter.
With that in mind, we introduce the following notation that will be used in the discussion of the fit results.

\begin{itemize}
 \item \textbf{Constrained fit}: consists in the fit in which the parameter $A$ is fixed in one of the values discussed
 in Section~\ref{sec:ratioX_asymp_values}, therefore imposing an asymptotic scenario. In the case of the variant $PL$ there are
 four free parameters and in the case $LL$ only three free parameters.
 
 \item \textbf{Unconstrained fit}: the case in which $A$ is a free parameter. This case has five free parameters in the variant $PL$
 and four free parameters in the $LL$ case.
\end{itemize}

In any case, these choices of sigmoids and elementary functions represent an economic number of free parameters.
For instance, compare with the individual fits to $\sigmatot$ and $\sigmael$, 
as those in Eqs.~\eqref{eq:res-compete} and \eqref{eq:res-el-totem},
which demand ten or more parameters for the ratio $X$.

\section{Fit Procedures and Results}
\label{sec:ratioX_fit_results}

\subsection{Initial Values}
\label{subsec:ratioX_initial_values}

The methodology used in the fits in order to select the best fit result, and also how the uncertainties of 
the experimental data were taken into account, were discussed in Chapter~\ref{chapt:data_fits}.
Nevertheless, we recall that we use as a test of goodness of fit the reduced chi-squared ($\chi^2/\nu$) and the integrated probability $P(\chi^2,\nu)$.

In this section, our focus will be on the initial values used for the free parameters in the fits.
Given the non-linearity of the parametrizations, different initial 
values have been tested in order to check the stability of the result. 
In this respect, we developed the procedure described below.

For each sigmoid (logistic or tanh) and each variant ($PL$ or $LL$),
Eqs.~\eqref{eq:X-par-Log-PL} to \eqref{eq:X-par-tanh-LL}, 
we first develop the constrained fits
($A$ fixed). In this case we consider the five numerical values displayed in Section~\ref{sec:ratioX_asymp_values}
as representative of the three scenarios investigated:

\begin{itemize}
 \item above the black disk: $\left\{\begin{array}{ll}
                               A=1 & \text{(maximum unitarity)}\\
                               A=0.75 & \text{(a possible ``formal" result)}
                              \end{array}\right.$
 
 \item the black disk: $A = 0.5$;

 \item below the black disk: $\left\{\begin{array}{ll}
                               A=0.436 & \text{(from TOTEM/COMPETE, see Sect.~\ref{subsec:ratioX_below_blackdisk})}\\
                               A=0.3 & \text{(lowest value of Refs.\cite{Fagundes_Menon_Silva:2013a,Menon_Silva:2013a,Menon_Silva:2013b}, see Sect.~\ref{subsec:ratioX_below_blackdisk})}
                              \end{array}\right.$
\end{itemize}

For each fixed $A$, different initial values have been tested for the other parameters 
(four with $PL$ and three with $LL$), until convergence and consistent statistical results have been reached.

In a second step, using the values of the parameters from the constrained fit result
as initial values, we have developed the unconstrained fits, with each $A$ 
now as a free parameter too.

As already stated, by fixing $A$  we \textit{impose} an asymptotic limit and by
letting $A$ as a free fit parameter, we  \textit{select} an asymptotic
scenario. 

In the next section, we display the fit results beginning with those obtained with the logistic function
and followed by those with the hyperbolic tangent.
All these results will be discussed in Section~\ref{sec:ratioX_fit_discussion_results}.

\subsection{Fit Results with the Logistic}
\label{subsec:ratioX_results_Log}

We first present the constrained and unconstrained fit results with variant $PL$, followed by those with variant $LL$.

\begin{figure}[htb!]
\centering
\subfloat{\label{fig:res_Log_PL_panel_a}\includegraphics[scale=0.39]{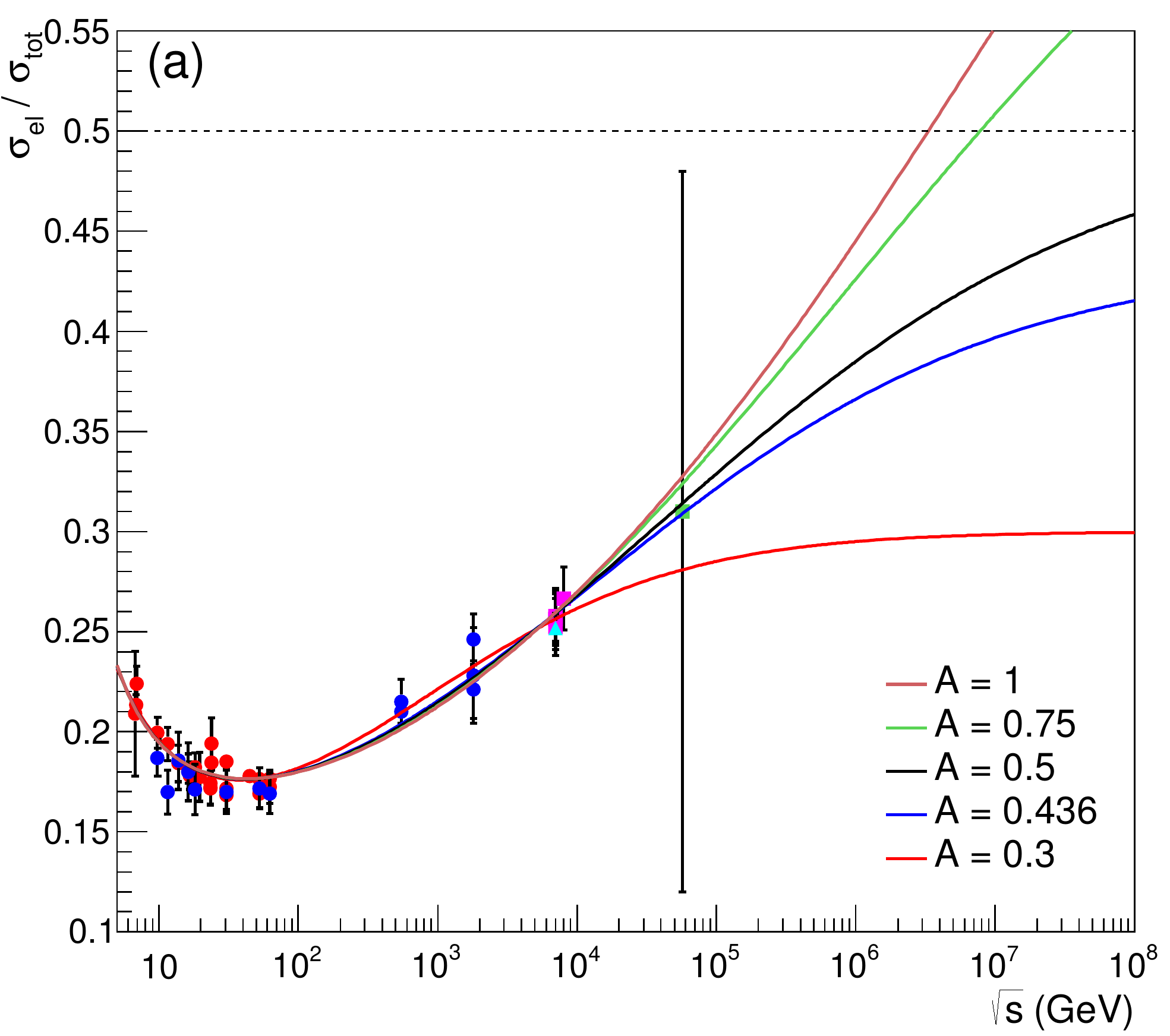}}\hfill
\subfloat{\label{fig:res_Log_PL_panel_b}\includegraphics[scale=0.39]{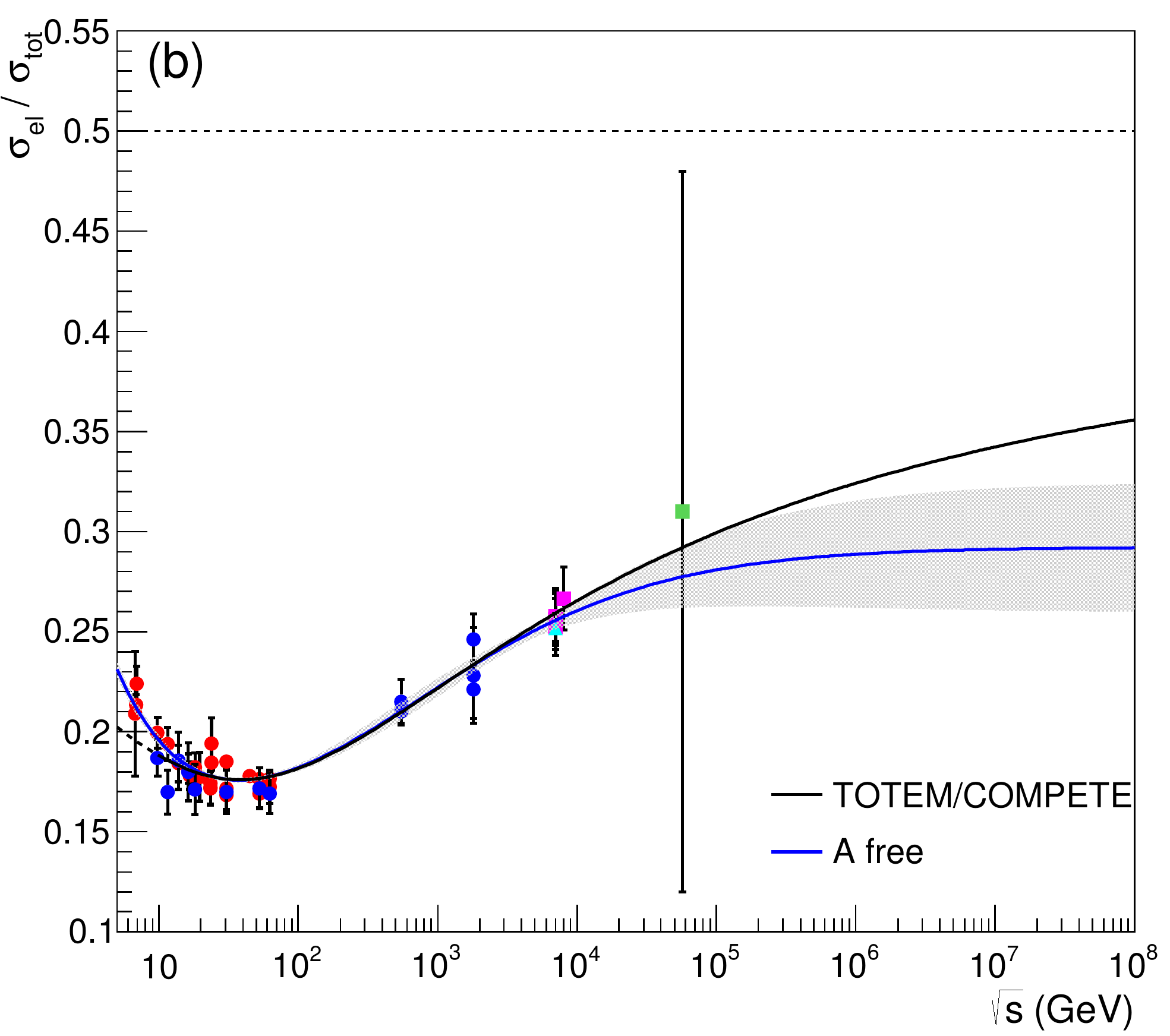}}
\caption{\label{fig:res_Log_PL}Fit results with the logistic function, variant $PL$ [Eq.~\eqref{eq:X-par-Log-PL}]: \subref{fig:res_Log_PL_panel_a}
constrained case ($A$ fixed, Table \ref{tab:res_Log_PL_Afixed}); 
\subref{fig:res_Log_PL_panel_b} unconstrained case ($A$ free, Table \ref{tab:res_Log_PL_Afree})
with the corresponding uncertainty region and the result from the TOTEM and COMPETE parametrizations, 
Eqs.~\eqref{eq:res-compete} and \eqref{eq:res-el-totem}. Legend for data given in Figure \ref{fig:data_X}.}
\end{figure}

The results and statistical information of each constrained fit with variant $PL$ are displayed in Table~\ref{tab:res_Log_PL_Afixed}.
The comparison of the corresponding curves with the experimental data are show
in Fig.~\ref{fig:res_Log_PL}\subref{fig:res_Log_PL_panel_a}, where, for clarity, the plotted curves correspond only to the
central values of the free parameters (i.e. without the uncertainty regions).

With these results for the parameters as initial values, including
each $A$ value, the unconstrained fits have been developed. The results
are displayed in Table~\ref{tab:res_Log_PL_Afree} and show that all data reductions
have approximately the same goodness of fit 
and the \textit{same asymptotic central value}, namely $A = 0.292$.
Although, in each case, the values of the parameters $\alpha$, $\beta$, $\gamma$
and $\delta$ may differ, once plotted together all curves corresponding
to the central values of the parameters overlap, see Fig.~\ref{fig:res_Log_PL}\subref{fig:res_Log_PL_panel_b}.
In this figure, we display the corresponding uncertainty region 
evaluated through error propagation from the fit parameters (Table~\ref{tab:res_Log_PL_Afree})
within one standard deviation.
For comparison, we also show the ratio $X(s)$ obtained
through the TOTEM and COMPETE parametrizations, Eqs.~\eqref{eq:res-compete} and \eqref{eq:res-el-totem}.

\begin{table}[htb!]
\centering
\caption{\label{tab:res_Log_PL_Afixed}Fit results with the logistic function, 
variant $PL$ [Eq.~\eqref{eq:X-par-Log-PL}] and constrained case 
($A$ fixed), $\nu = 38$ (Fig.~\ref{fig:res_Log_PL}\subref{fig:res_Log_PL_panel_a}).} 
\begin{tabular}{c c c c c c c}\hline\hline
$A$ fixed & $\alpha$ &        $\beta$       &       $\gamma$       & $\delta$ & $\chi^2/\nu$ & $P(\chi^2,\nu)$\\\hline
0.3   & 125.5(1.5)  & 0.328(22)  & -123.7(1.5)  & 1.56(12)$\times 10^{-2}$  & 0.811 & 0.790 \\
0.436 & 169.12(10)  & 0.1828(72) & -168.59(10)  & 6.62(30)$\times 10^{-3}$  & 0.882 & 0.677 \\
0.5   & 211.37(96)  & 0.1627(63) & -211.156(95) & 4.74(21)$\times 10^{-3}$  & 0.899 & 0.647 \\
0.75  & 69.517(80)  & 0.1315(51) & -70.027(79)  & 1.148(52)$\times 10^{-2}$ & 0.932 & 0.589 \\
1.0   & 82.898(75)  & 0.1195(46) & -83.825(73)  & 8.78(40)$\times 10^{-3}$  & 0.944 & 0.568 \\\hline\hline
\end{tabular}
\end{table}

\begin{table}[htb!]
\centering
\caption{\label{tab:res_Log_PL_Afree}Fit results with the logistic function, variant $PL$ [Eq.~\eqref{eq:X-par-Log-PL}]
and unconstrained case ($A$ free), $\nu=37$ (Fig.~\ref{fig:res_Log_PL}\subref{fig:res_Log_PL_panel_b}).}
\small
\begin{tabular}{c c c c c c c c}\hline\hline
$A$ initial & $A$ free &     $\alpha$     &     $\beta$      &     $\gamma$      & $\delta$ & $\chi^2/\nu$ & $P(\chi^2,\nu)$\\\hline
0.3     & 0.292(33) & 125.6(1.5) & 0.35(11) & -123.6(1.5) & 1.66(51)$\times 10^{-2}$ & 0.831 & 0.756 \\
0.436   & 0.292(33) & 169.9(1.3) & 0.35(11) & -167.9(1.3) & 1.22(38)$\times 10^{-2}$ & 0.831 & 0.756 \\
0.5     & 0.292(33) & 212.2(1.2) & 0.35(11) & -210.2(1.2) & 9.8(3.0)$\times 10^{-3}$ & 0.831 & 0.757 \\
0.75    & 0.292(32) & 70.5(1.9)  & 0.36(12) & -68.5(1.9)  & 2.98(91)$\times 10^{-2}$ & 0.832 & 0.755 \\
1.0     & 0.292(32) & 70.5(1.9)  & 0.36(12) & -68.5(1.9)  & 2.98(91)$\times 10^{-2}$ & 0.832 & 0.755 \\\hline\hline
\end{tabular}
\normalsize
\end{table}

\begin{figure}[htb!]
\centering
\subfloat{\label{fig:res_Log_LL_panel_a}\includegraphics[scale=0.39]{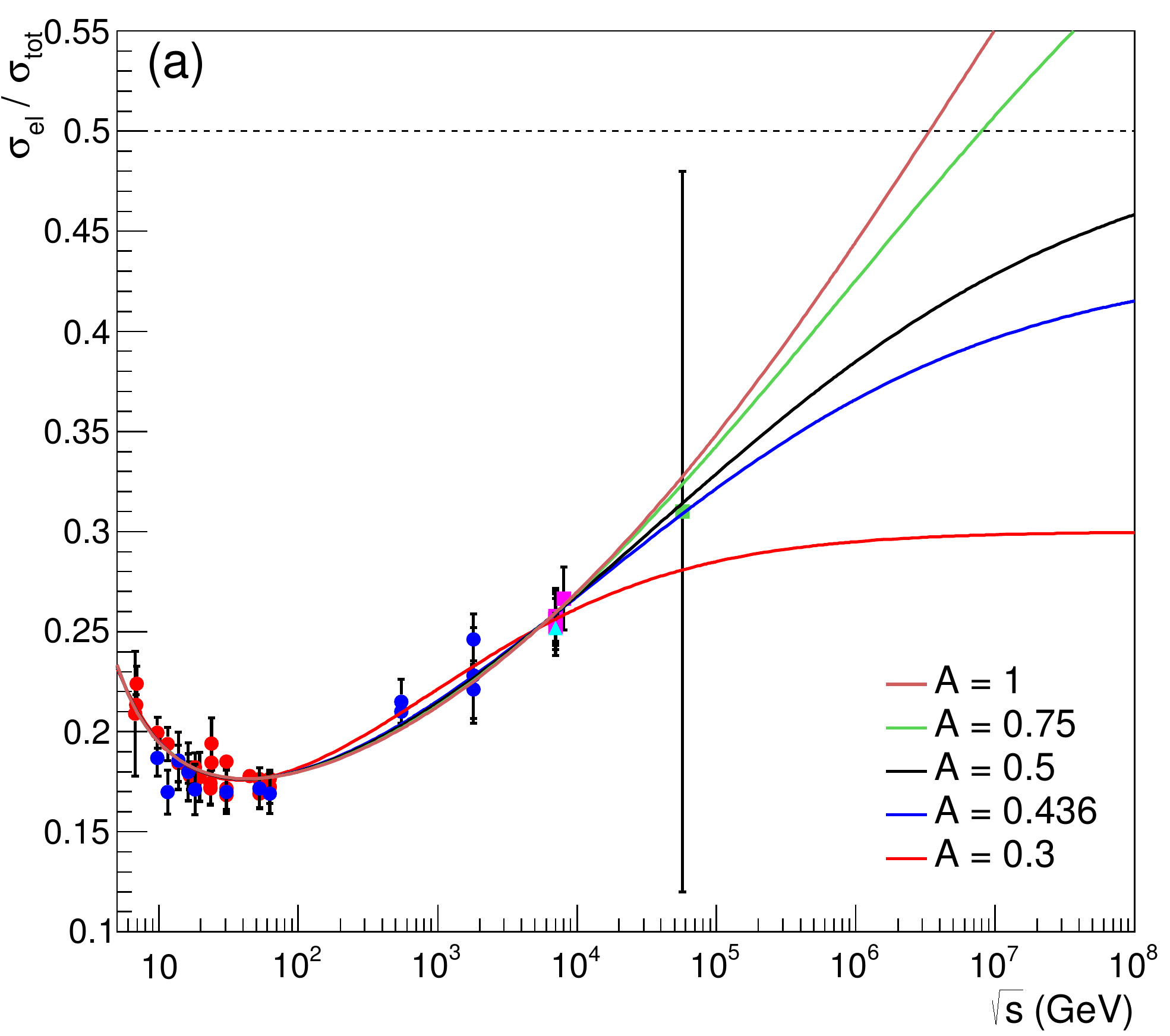}}\hfill
\subfloat{\label{fig:res_Log_LL_panel_b}\includegraphics[scale=0.39]{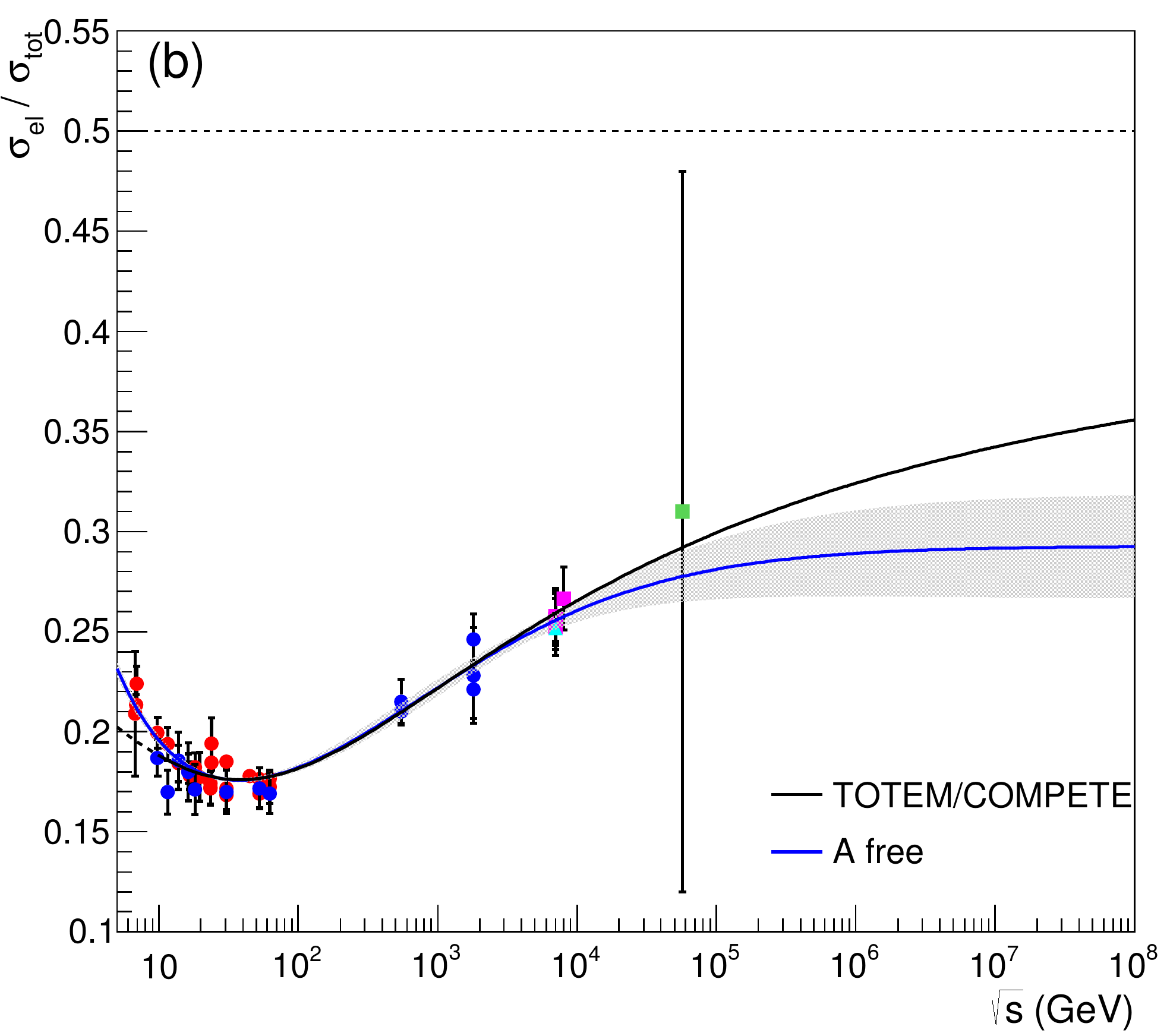}}
\caption{\label{fig:res_Log_LL}Fit results with the logistic function, variant $LL$ [Eq.~\eqref{eq:X-par-Log-LL}]: 
\subref{fig:res_Log_LL_panel_a} constrained case ($A$ fixed, Table \ref{tab:res_Log_LL_Afixed}); 
\subref{fig:res_Log_LL_panel_b} unconstrained case ($A$ free, Table \ref{tab:res_Log_LL_Afree}) with the corresponding
uncertainty region and the result from the TOTEM and COMPETE parametrizations, Eqs.~\eqref{eq:res-compete} 
and \eqref{eq:res-el-totem}. Legend for data given in Figure \ref{fig:data_X}.}
\end{figure}

The same procedure has been developed with the variant $LL$. In this case we have only
three parameters ($A$ fixed) or four ($A$ free). The constrained fit results
are displayed in Table \ref{tab:res_Log_LL_Afixed} and Figure~\ref{fig:res_Log_LL}\subref{fig:res_Log_LL_panel_a} and the unconstrained
case in Table \ref{tab:res_Log_LL_Afree}.
Once more, all fit results converged to the same solution
in statistical grounds and in the values of all the parameters. The corresponding
curve including the uncertainty region is shown in Figure~\ref{fig:res_Log_LL}\subref{fig:res_Log_LL_panel_b}.

\begin{table}[htb!]
\centering
\caption{\label{tab:res_Log_LL_Afixed}Fit results with the logistic function, variant $LL$ [Eq.~\eqref{eq:X-par-Log-LL}] and constrained case 
($A$ fixed), $\nu = 39$ (Fig.~\ref{fig:res_Log_LL}\subref{fig:res_Log_LL_panel_a}).} 
\begin{tabular}{c c c c c c}\hline\hline
$A$ fixed &       $\alpha$      &        $\beta$       &       $\gamma$       &  $\chi^2/\nu$ & $P(\chi^2,\nu)$\\\hline
0.3   & 1.90(13)   & 0.324(22)  & -1.95(14)  &  0.789 & 0.824 \\
0.436 & 0.534(79)  & 0.182(10)  & -1.122(78) &  0.858 & 0.720 \\
0.5   & 0.221(71)  & 0.1620(91) & -1.005(69) &  0.875 & 0.692 \\
0.75  & -0.503(58) & 0.1302(71) & -0.813(56) &  0.907 & 0.638 \\
1.0   & -0.922(53) & 0.1186(65) & -0.742(51) &  0.918 & 0.617 \\\hline\hline
\end{tabular}
\end{table}

\begin{table}[htb!]
\centering
\caption{\label{tab:res_Log_LL_Afree}Fit results with the logistic function, variant $LL$ [Eq.~\eqref{eq:X-par-Log-LL}] and unconstrained case ($A$ free),
$\nu=38$ (Fig.~\ref{fig:res_Log_LL}\subref{fig:res_Log_LL_panel_b}).}
\begin{tabular}{c c c c c c c}\hline\hline
$A$ initial & $A$ free  &     $\alpha$     &     $\beta$      &     $\gamma$      & $\chi^2/\nu$ & $P(\chi^2,\nu)$\\\hline
0.3     & 0.293(26) & 2.05(59) & 0.346(87) & -2.07(50) & 0.808 & 0.794 \\
0.436   & 0.293(26) & 2.05(59) & 0.346(88) & -2.07(50) & 0.808 & 0.794 \\
0.5     & 0.293(26) & 2.05(59) & 0.346(88) & -2.07(50) & 0.808 & 0.794 \\
0.75    & 0.293(26) & 2.05(59) & 0.346(88) & -2.07(50) & 0.808 & 0.794 \\
1.0     & 0.293(26) & 2.05(59) & 0.346(88) & -2.07(50) & 0.808 & 0.794 \\\hline\hline
\end{tabular}
\end{table}

\subsection{Fit Results with the Hyperbolic Tangent}
\label{subsec:ratioX_results_Tanh}

\begin{figure}[htb!]
\centering
\subfloat{\label{fig:res_Tanh_PL_panel_a}\includegraphics[scale=0.39]{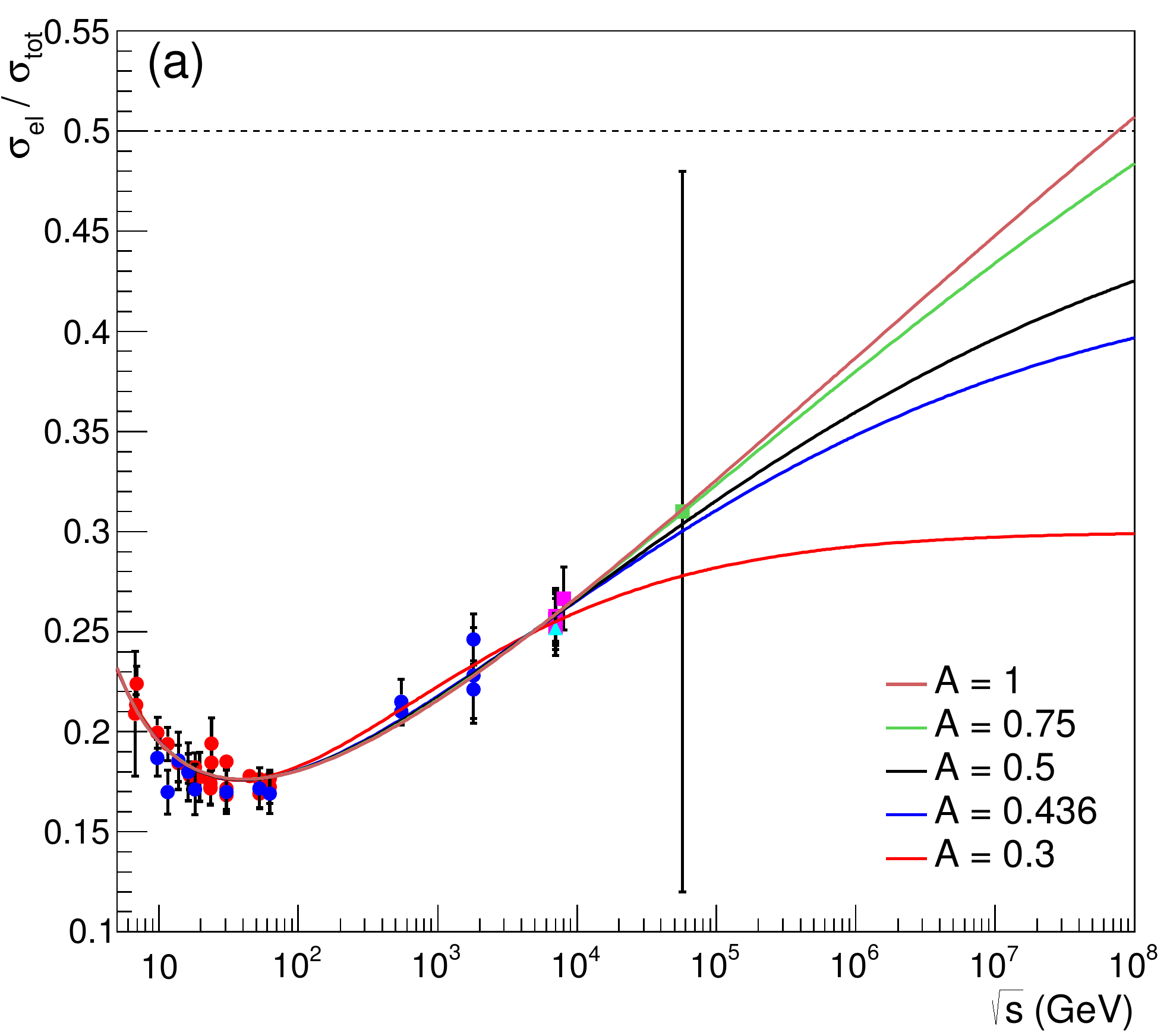}}\hfill
\subfloat{\label{fig:res_Tanh_PL_panel_b}\includegraphics[scale=0.39]{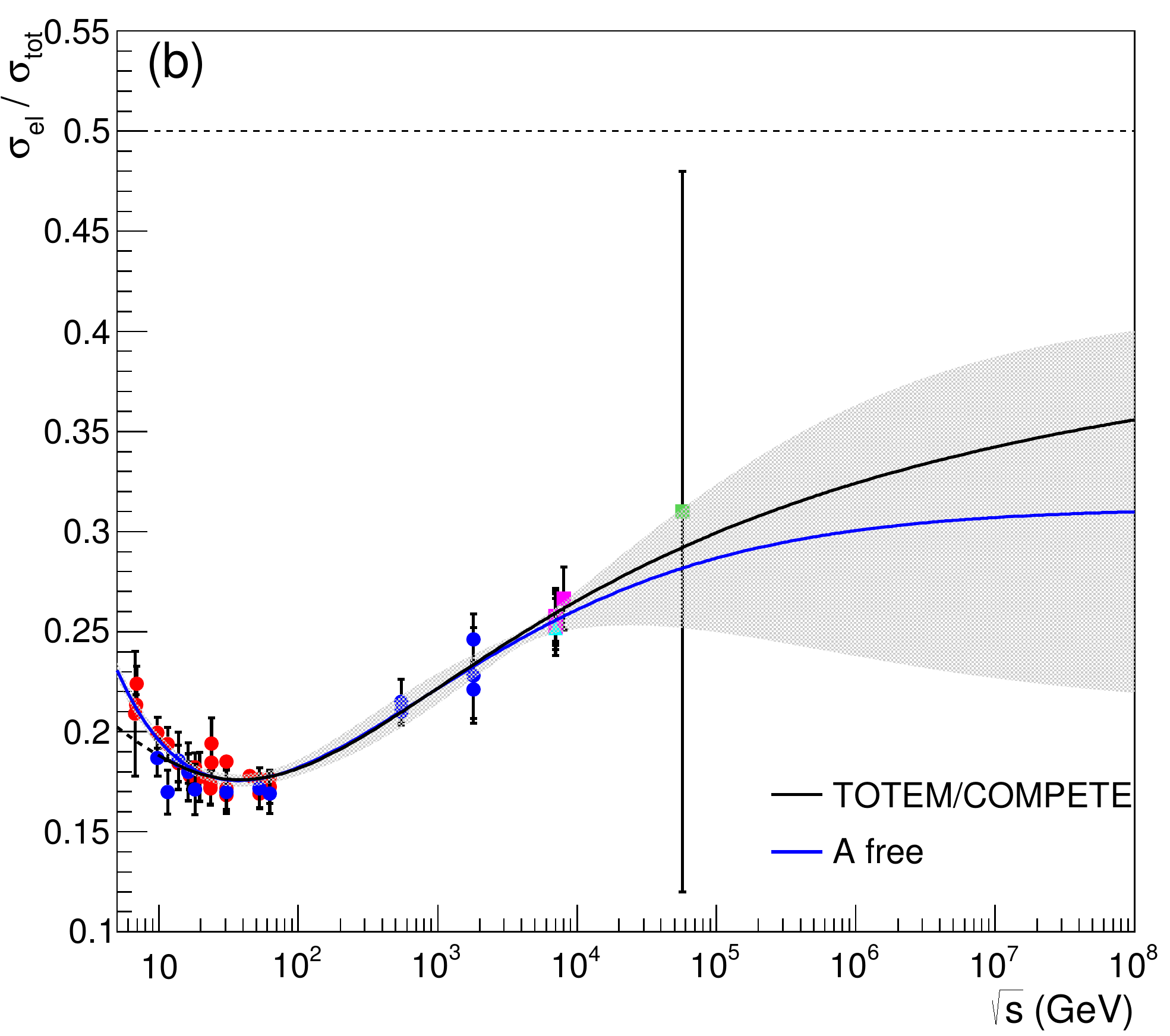}}
\caption{\label{fig:res_Tanh_PL}Fit results with the tanh, variant $PL$ [Eq.~\eqref{eq:X-par-tanh-PL}]: 
\subref{fig:res_Tanh_PL_panel_a} constrained case ($A$ fixed, Table \ref{tab:res_Tanh_PL_Afixed}); 
\subref{fig:res_Tanh_PL_panel_b} unconstrained case ($A$ free, Table \ref{tab:res_Tanh_PL_Afree})
with the corresponding uncertainty region and the result from the TOTEM and COMPETE parametrizations,
Eqs.~\eqref{eq:res-compete} and \eqref{eq:res-el-totem}. Legend for data given in Figure \ref{fig:data_X}.}
\end{figure}

\begin{figure}[htb!]
\centering
\subfloat{\label{fig:res_Tanh_LL_panel_a}\includegraphics[scale=0.39]{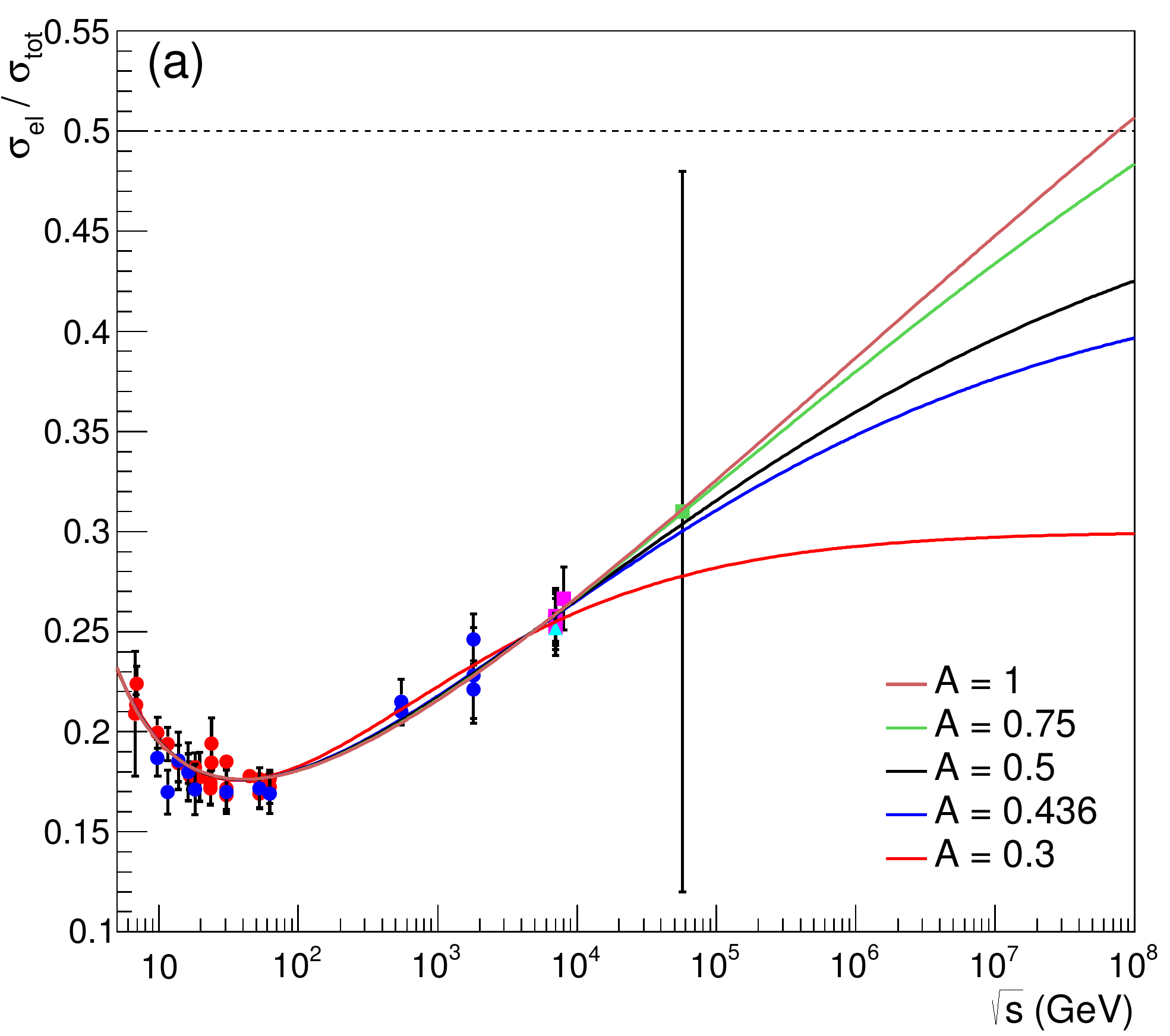}}\hfill
\subfloat{\label{fig:res_Tanh_LL_panel_b}\includegraphics[scale=0.39]{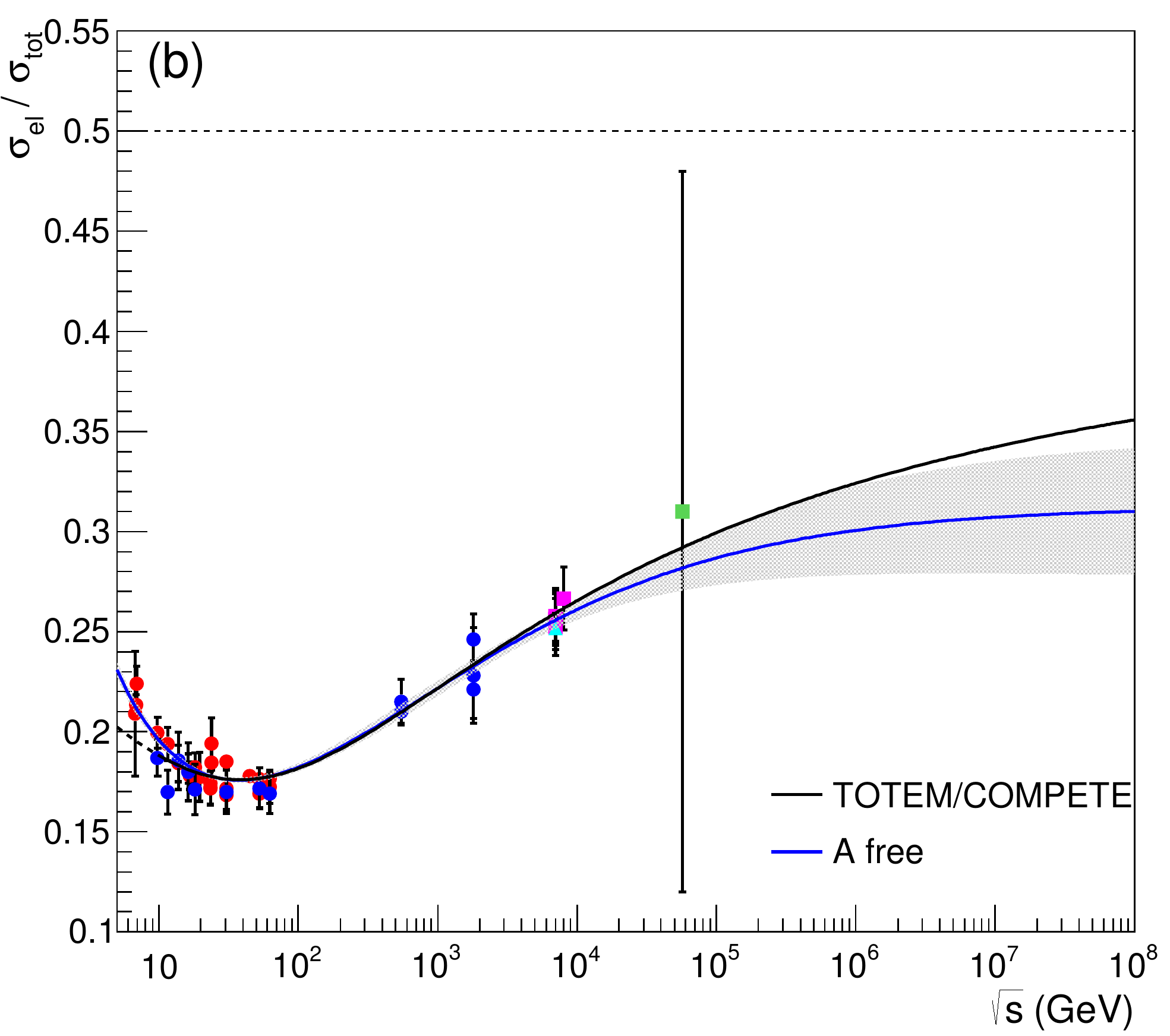}}
\caption{\label{fig:res_Tanh_LL}Fit results with the tanh, variant $LL$ [Eq.~\eqref{eq:X-par-tanh-LL}]: 
\subref{fig:res_Tanh_LL_panel_a} constrained case ($A$ fixed, Table \ref{tab:res_Tanh_LL_Afixed}); 
\subref{fig:res_Tanh_LL_panel_b} unconstrained case ($A$ free, Table \ref{tab:res_Tanh_LL_Afree}) with the corresponding
uncertainty region and the result from the TOTEM and COMPETE parametrizations, Eqs.~\eqref{eq:res-compete} and \eqref{eq:res-el-totem}.
Legend for data given in Figure \ref{fig:data_X}.}
\end{figure}

The same procedures discussed for the logistic function
have been applied in the case of the tanh.
%
%
The fit results with the variant $PL$ are displayed in Table~\ref{tab:res_Tanh_PL_Afixed} (constrained
case), Table~\ref{tab:res_Tanh_PL_Afree} (unconstrained case) and Fig.~\ref{fig:res_Tanh_PL}.
In turn, the results with the variant $LL$ are displayed in Table \ref{tab:res_Tanh_LL_Afixed} (constrained
case), Table~\ref{tab:res_Tanh_LL_Afree} (unconstrained case), and Fig.~\ref{fig:res_Tanh_LL}.

\begin{table}[htb!]
\centering
\caption{\label{tab:res_Tanh_PL_Afixed}Fit results with the tanh, variant $PL$ [Eq.~\eqref{eq:X-par-tanh-PL}] and 
constrained case ($A$ fixed), $\nu = 38$ (Fig.~\ref{fig:res_Tanh_PL}\subref{fig:res_Tanh_PL_panel_a}).} 
\small
\begin{tabular}{c c c c c c c}\hline\hline
$A$ fixed &  $\alpha$      &        $\beta$       &       $\gamma$       & $\delta$ & $\chi^2/\nu$ & $P(\chi^2,\nu)$\\\hline
0.3   & 125.23(24)   & 0.13243(93) & -123.95(24) & 6.31(48)$\times 10^{-3}$  &   0.818          & 0.780  \\
0.436 & 169.21(13)   & 0.0588(35)  & -168.50(13) & 2.23(15)$\times 10^{-3}$  &   0.844          & 0.740 \\
0.5   & 211.44(10)   & 0.0477(28)  & -210.83(10) & 1.380(96)$\times 10^{-3}$ &   0.853          & 0.724 \\
0.75  & 68.06(14)    & 0.0283(16)  & -67.67(14)  & 2.56(18)$\times 10^{-3}$  &   0.870          & 0.697 \\
1.0   & 83.50(10)    & 0.0204(16)  & -83.22(10)  & 1.51(10)$\times 10^{-3}$  &   0.875          & 0.687 \\\hline\hline
\end{tabular}
\normalsize
\end{table}

\begin{table}[htb!]
\centering
\caption{\label{tab:res_Tanh_PL_Afree}Fit results with the tanh, variant $PL$ [Eq.~\eqref{eq:X-par-tanh-PL}] and 
unconstrained case ($A$ free), $\nu=37$ (Fig. \ref{fig:res_Tanh_PL}\subref{fig:res_Tanh_PL_panel_b}).} 
\small
\begin{tabular}{c c c c c c c c}\hline\hline
$A$ initial & $A$ free &     $\alpha$    &     $\beta$      &     $\gamma$      & $\delta$ & $\chi^2/\nu$ & $P(\chi^2,\nu)$\\\hline
0.3     & 0.31(10)   & 125.18(43) & 0.12(11)  & -123.99(43) & 5.7(4.9)$\times 10^{-3}$   & 0.838    & 0.746 \\
0.436   & 0.31(10)   & 169.45(43) & 0.12(11)  & -168.25(42) & 4.2(3.8)$\times 10^{-3}$   & 0.838    & 0.746 \\ 
0.5     & 0.312(49)  & 211.73(18) & 0.118(52) & -210.54(17) & 3.3(1.4)$\times 10^{-3}$   & 0.837    & 0.746 \\
0.75    & 0.31(10)   & 68.42(48)  & 0.12(11)  & -67.23(48)  & 1.04(0.93)$\times 10^{-2}$ & 0.838    & 0.745 \\
1.0     & 0.31(10)   & 83.92(46)  & 0.12(11)  & -82.73(45)  & 8.5(7.4)$\times 10^{-3}$   & 0.838    & 0.746 \\\hline\hline
\end{tabular}
\normalsize
\end{table}

%
%

\begin{table}[htb!]
\centering
\caption{\label{tab:res_Tanh_LL_Afixed}Fit results with the tanh, variant $LL$ [Eq.~\eqref{eq:X-par-tanh-LL}] and constrained case ($A$ fixed),
$\nu = 39$ (Fig. \ref{fig:res_Tanh_LL}\subref{fig:res_Tanh_LL_panel_a}).} 
\begin{tabular}{c c c c c c}\hline\hline
$A$ fixed  &   $\alpha$      &        $\beta$     &       $\gamma$       & $\chi^2/\nu$ & $P(\chi^2,\nu)$\\
\hline
0.3   & 1.289(54)   & 0.1317(90)  & -0.786(58)   &   0.796          & 0.813  \\
0.436 & 0.718(25)   & 0.0587(35)  & -0.358(25)   &   0.822          & 0.777 \\
0.5   & 0.605(20)   & 0.0476(28)  & -0.291(20)   &   0.831          & 0.763 \\
0.75  & 0.381(12)   & 0.0283(16)  & -0.174(12)   &   0.847          & 0.737 \\
1.0   & 0.2811(89)  & 0.0204(11)  & -0.1259(87)  &   0.853          & 0.729 \\\hline\hline
\end{tabular}
\end{table}

\begin{table}[htb!]
\centering
\caption{\label{tab:res_Tanh_LL_Afree}Fit results with the tanh, variant $LL$ [Eq.~\eqref{eq:X-par-tanh-LL}] and unconstrained
case ($A$ free), $\nu=38$ (Fig. \ref{fig:res_Tanh_LL}\subref{fig:res_Tanh_LL_panel_b}).} 
\begin{tabular}{c c c c c c c}\hline\hline
$A$ initial & $A$ free   &     $\alpha$    &     $\beta$      &     $\gamma$    & $\chi^2/\nu$ & $P(\chi^2,\nu)$\\ 
\hline
0.3     & 0.312(35)   & 1.19(25) & 0.117(37)  & -0.70(21) & 0.815    & 0.783 \\
0.436   & 0.312(48)   & 1.19(35) & 0.117(51)  & -0.70(29) & 0.815    & 0.783 \\
0.5     & 0.312(35)   & 1.19(25) & 0.117(36)  & -0.70(21) & 0.815    & 0.783 \\
0.75    & 0.312(35)   & 1.19(25) & 0.117(37)  & -0.70(21) & 0.815    & 0.783 \\
1.0     & 0.312(35)   & 1.19(25) & 0.117(37)  & -0.70(21) & 0.815    & 0.783 \\\hline\hline
\end{tabular}

\end{table}



\section{Discussion and Conclusions on the Fit Results}
\label{sec:ratioX_fit_discussion_results}

In this section we discuss all the fit results presented in the previous sections
in a comparative way with respect to: 
constrained/unconstrained fits (Section~\ref{subsec:ratioX_discussion_const_unconst_fits}), 
logistic/tanh (Section~\ref{subsec:ratioX_discussion_sigmoid}) 
and variants $PL$/$LL$ (Section~\ref{subsec:ratioX_discussion_PL_LL}). 
At last, we present our final 
conclusions on the fit results (Section~\ref{subsec:ratioX_results_conclusion}).

\subsection{Constrained and Unconstrained Fits}
\label{subsec:ratioX_discussion_const_unconst_fits}

In the case of \textit{constrained} fits ($A$ fixed),
the values of the parameters, statistical information and curves are presented in
Table~\ref{tab:res_Log_PL_Afixed}, Fig.~\ref{fig:res_Log_PL}\subref{fig:res_Log_PL_panel_a} (logistic - $PL$),
Table~\ref{tab:res_Log_LL_Afixed}, Fig.~\ref{fig:res_Log_LL}\subref{fig:res_Log_LL_panel_a} (logistic - $LL$), 
Table~\ref{tab:res_Tanh_PL_Afixed}, Fig.~\ref{fig:res_Tanh_PL}\subref{fig:res_Tanh_PL_panel_a} (tanh - $PL$) and
Table~\ref{tab:res_Tanh_LL_Afixed}, Fig.~\ref{fig:res_Tanh_LL}\subref{fig:res_Tanh_LL_panel_a} (tanh - $LL$).
For clarity, we decided to not show the uncertainty regions of the constrained fits, therefore 
the curves in the figures correspond to the central values 
of the parameters.
However, error propagation from the fit parameters
lead to typical uncertainty 
regions as those displayed in part (b) of each figure.

From the Tables, for $\nu$ = 38 or 39, the values of the $\chi^2/\nu$ lie in a 
typical interval 0.79 - 0.94 and those of the $P(\chi^2, \nu)$ in the corresponding
interval 0.82 - 0.57, indicating, therefore, statistically consistent fit results
in all investigated cases. We also note that the corresponding curves are consistent
with all the experimental data analysed (mainly within the uncertainty regions,
not shown in the figures).

In the case of \textit{unconstrained} fits ($A$ free),
the value of the parameters, statistical information and curves are presented in
Table~\ref{tab:res_Log_PL_Afree}, Fig.~\ref{fig:res_Log_PL}\subref{fig:res_Log_PL_panel_b} (logistic - $PL$), 
Table~\ref{tab:res_Log_LL_Afree}, Fig.~\ref{fig:res_Log_LL}\subref{fig:res_Log_LL_panel_b} (logistic - $LL$), 
Table~\ref{tab:res_Tanh_PL_Afree}, Fig.~\ref{fig:res_Tanh_PL}\subref{fig:res_Tanh_PL_panel_b} (tanh - $PL$), 
Table~\ref{tab:res_Tanh_LL_Afree}, Fig.~\ref{fig:res_Tanh_LL}\subref{fig:res_Tanh_LL_panel_b} (tanh - $LL$).

Here we note that
all fit results converged to practically the
same solution, especially in what concerns the asymptotic limit. These values 
are summarized in Table \ref{tab:res_asymp_A}. Small differences in the value 
of the other fit parameters are discussed in what follows.

\begin{table}[htb!]
\centering
\caption{\label{tab:res_asymp_A}Summary of the asymptotic limits obtained in the unconstrained fits
($A$ free) with the logistic, tanh, variants $PL$, $LL$ and
the corresponding $\nu$ and $\chi^2/\nu$. The values of $\nu$ and $\chi^2/\nu$ are also shown
in the case of the black disk ($A~=~0.5$).} 
\begin{tabular}{c c c c c}
\hline\hline
sigmoid: &\multicolumn{2}{c}{logistic}&\multicolumn{2}{c}{tanh} \\ \cmidrule(lr){2-3} \cmidrule(lr){4-5} 
variant: &    $PL$       &   $LL$     & $PL$      &   $LL$         \\
\hline
$A$ free        &   0.292(33)   & 0.293(26)  & 0.31(10)  & 0.312(48)       \\
$\nu$           &   37          & 38         & 37        & 38              \\
$\chi^2/\nu$    &   0.832       & 0.808      & 0.838     & 0.815           \\ 
\hline                 
$A$ fixed       &  0.5          & 0.5        & 0.5       & 0.5\\
$\nu$           &  38           & 39         & 38        & 39              \\
$\chi^2/\nu$    &  0.899        & 0.875      & 0.853     & 0.831           \\ 
\hline\hline
\end{tabular}
\end{table}

\subsection{Sigmoid Functions: Logistic and Hyperbolic Tangent}
\label{subsec:ratioX_discussion_sigmoid}

First, let us compare the \textit{constrained} results in part (a) of
Figs.~\ref{fig:res_Log_PL} and \ref{fig:res_Log_LL} (logistic) with those in part (a) of Figs.~\ref{fig:res_Tanh_PL} and \ref{fig:res_Tanh_LL}
(tanh). We note that, above the experimental data,  the rise of $X(s)$ 
with the logistic function is faster than with the tanh one, and the differences
increase as $A$ increases (from 0.3 up to 1.0). For example, in the case
of $A$ = 1 (fixed), at $\sqrt{s}$ = 10$^7$ GeV, $X \approx$ 0.55 with the logistic
and $X \approx$ 0.45 with the tanh in both variants, $PL$ and $LL$.

On the other hand, in the \textit{unconstrained} cases (part (b) of the same figures),
these differences are negligible, even in the asymptotic region. The small differences
in the central values of $A$ (namely $\approx$ 0.29 for the logistic and $\approx$
0.31 for the tanh) are also negligible within the uncertainties
in these parameters (Table~\ref{tab:res_asymp_A}).

\subsection{Variants $PL$ and $LL$} 
\label{subsec:ratioX_discussion_PL_LL}

In the \textit{constrained} fits, the values of the free parameters differ
substantially with variants $PL$ and $LL$, in both cases:
logistic (Tables~\ref{tab:res_Log_PL_Afixed} and \ref{tab:res_Log_LL_Afixed}) and tanh
(Tables~\ref{tab:res_Tanh_PL_Afixed} and \ref{tab:res_Tanh_LL_Afixed}). Despite these differences, once plotted
together, all curves overlap,
as can be seen in parts (a) of Fig. \ref{fig:res_Log_PL} ($PL$) and Fig. \ref{fig:res_Log_LL} ($LL$) for
the logistic and parts (a) of Fig. \ref{fig:res_Tanh_PL} ($PL$) and Fig. \ref{fig:res_Tanh_LL} ($LL$) for the tanh.

In the \textit{unconstrained} case and variant $PL$ (Table \ref{tab:res_Log_PL_Afree} for the logistic
and Table \ref{tab:res_Tanh_PL_Afree} for the tanh) all results lead to practically the same
asymptotic values, namely $A$ = 0.292 (logistic) and $A$ = 0.31 (tanh),
with some differences in the central values of the other parameters.
On the other hand, with variant $LL$ the central values of the parameters are
all equal up to three figures with the two sigmoid functions considered (see Table~\ref{tab:res_Log_LL_Afree} for the logistic
and Table~\ref{tab:res_Tanh_LL_Afree} for tanh). In each case, the goodness of fit presents this same feature.

\subsection{Conclusions on the Fit Results}
\label{subsec:ratioX_results_conclusion}


Concerning \textit{constrained} fits, given the relative large uncertainties
in the experimental data and the small differences in the values of
$\chi^2/\nu$ and  $P(\chi^2, \nu)$ for $\nu$ = 38 - 39, 
we understand that all the fit results are statistically consistent 
with the dataset and equally probable on statistical grounds,
even in the extreme cases of $A$ = 0.3 and $A$ = 1. In other words,
despite the large differences in the extrapolated results (at the highest and asymptotic 
energy regions), the constrained fits do not allow us to select an asymptotic
scenario.
That leads to an important consequence: \textit{although consistent with the
experimental data, the black disk does not represent an unique or definite
solution.}

With regard to \textit{unconstrained} fits,
independently of the sigmoid function (logistic or tanh) and variant ($PL$ or $LL$),
the data reductions lead to unique solutions within the uncertainties,
indicating a scenario below the black disk, with central value of $A$
around 0.29 (logistic) and around 0.31 (tanh),
as summarized in Table~\ref{tab:res_asymp_A}.
Given the convergent character of these
solutions, we consider \textit{the unconstrained fits as the preferred results of
this analysis}.
In other words, we understand that \textit{the data reductions favor a semi-transparent
(or grey) scenario}.

As discussed in Section~\ref{sec:ratioX_parametrization}, we have chosen for this analysis two classes of sigmoid functions
and with two elementary functions based on their applicability in describing soft processes.
In the end, we have four empirical parametrizations
formed by the combination of the sigmoid and elementary functions. 
As can be seen in Table~\ref{tab:res_asymp_A}, we have small differences among the four asymptotic results (value of $A$ parameter). These differences 
may be associated with
a kind of ``uncertainty" in the choices of $S(f)$ and $f(s)$. In this sense,
once constituting independent results,
we can infer a \textit{global} ($g$) asymptotic value by considering
the arithmetic mean and adding
the uncertainties in quadrature (Table~\ref{tab:res_asymp_A}):
\begin{equation}
A_g = 0.30 \pm 0.12.
\label{eq:A-mean-g}
\end{equation}

It is important to notice that this result is consistent, within the uncertainties,
with previous analyses using the
tanh, $\delta$ = 0.5 (fixed) and  energy scale at 25 GeV$^2$ (the energy
cutoff), namely  $A= 0.36(8)$ in \cite{Fagundes_Menon_Silva:2015} and $A = 0.332(49)$ in \cite{Fagundes_Menon_Silva:2016b}.
This suggests that \textit{the convergent result does not depend on the energy scale}.
Moreover, within the uncertainties, the above value is in plenty agreement
with the limits obtained through
individual fits to $\sigmatot$ and $\sigmael$ data, using different
variants and procedures (see Fig. 10 in Ref.~\cite{Menon_Silva:2013b}).

Concerning the variants $PL$ and $LL$, given the goodness of fits,
the same central values of the parameters up to three figures and the smaller
number of free parameters, we select the \textit{variant $LL$ as our representative result}.
From Table~\ref{tab:res_asymp_A}, 
we can also infer another mean value now \textit{restricted} ($r$) to variant $LL$
and the two sigmoid functions:
\begin{equation}
A_r = 0.303 \pm 0.055.
\label{eq:A-mean-r}
\end{equation}

\noindent Within the uncertainties, this result is also in plenty agreement with the aforementioned analyses.
In what follows we shall focus our discussion on predictions and extension to other
quantities, to this particular variant.

\subsection{Selected Results and Scenarios}
\label{subsec:ratioX_selected_results}

In the previous section, we have selected the $LL$ variant as our representative result and the grey-disk scenario
as the favored one by the unconstrained fits ($A$ free). On the other hand, the black-disk case is also consistent with the
experimental data analyzed (constrained case, $A=0.5$ fixed). Given the importance of this scenario in the phenomenological context,
we will also consider it in the discussions below.

\begin{figure}[htb!]
\centering
\subfloat{\label{fig:res_selected_LL_panel_a}\includegraphics[scale=0.39]{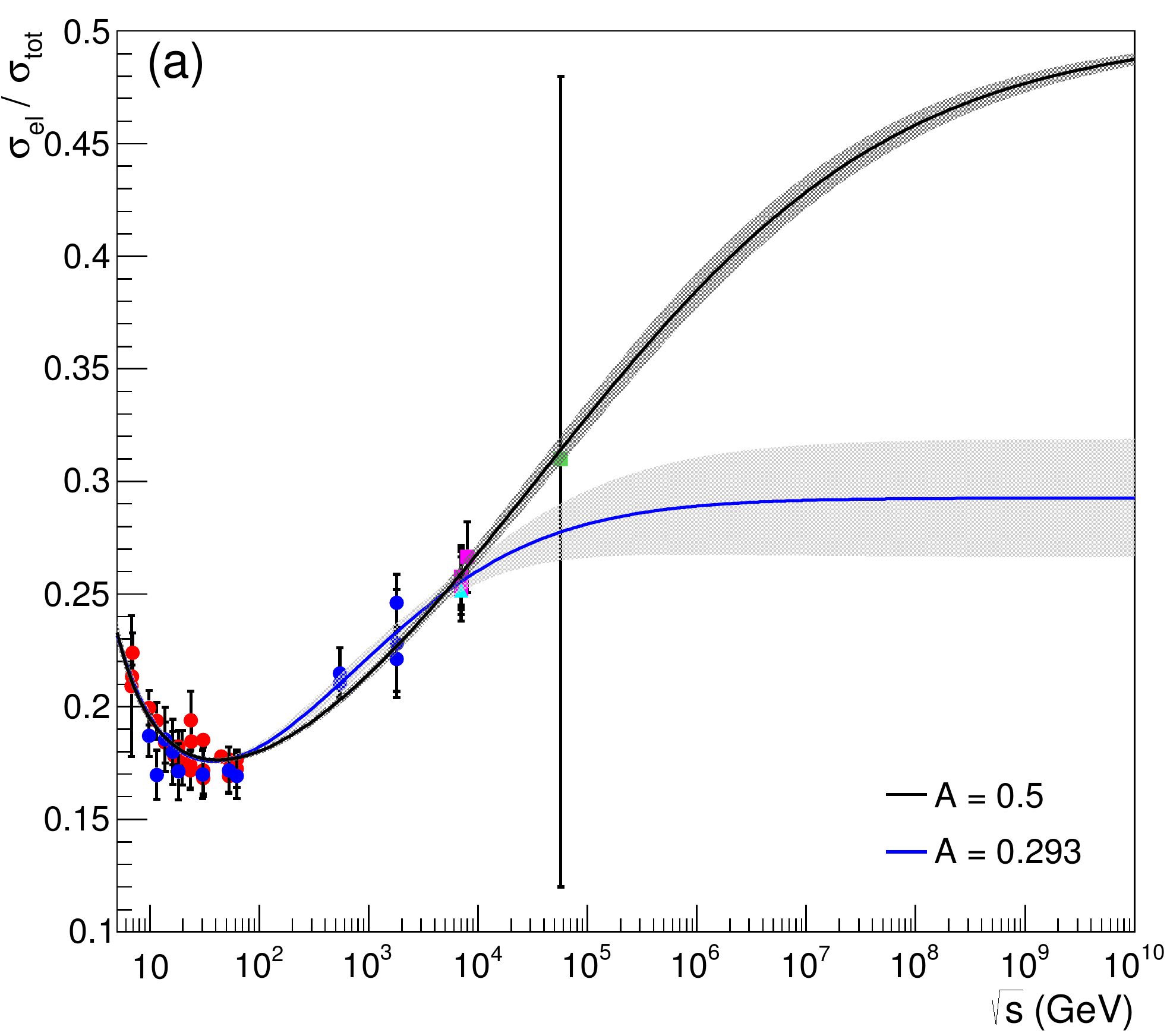}}\hfill
\subfloat{\label{fig:res_selected_LL_panel_b}\includegraphics[scale=0.39]{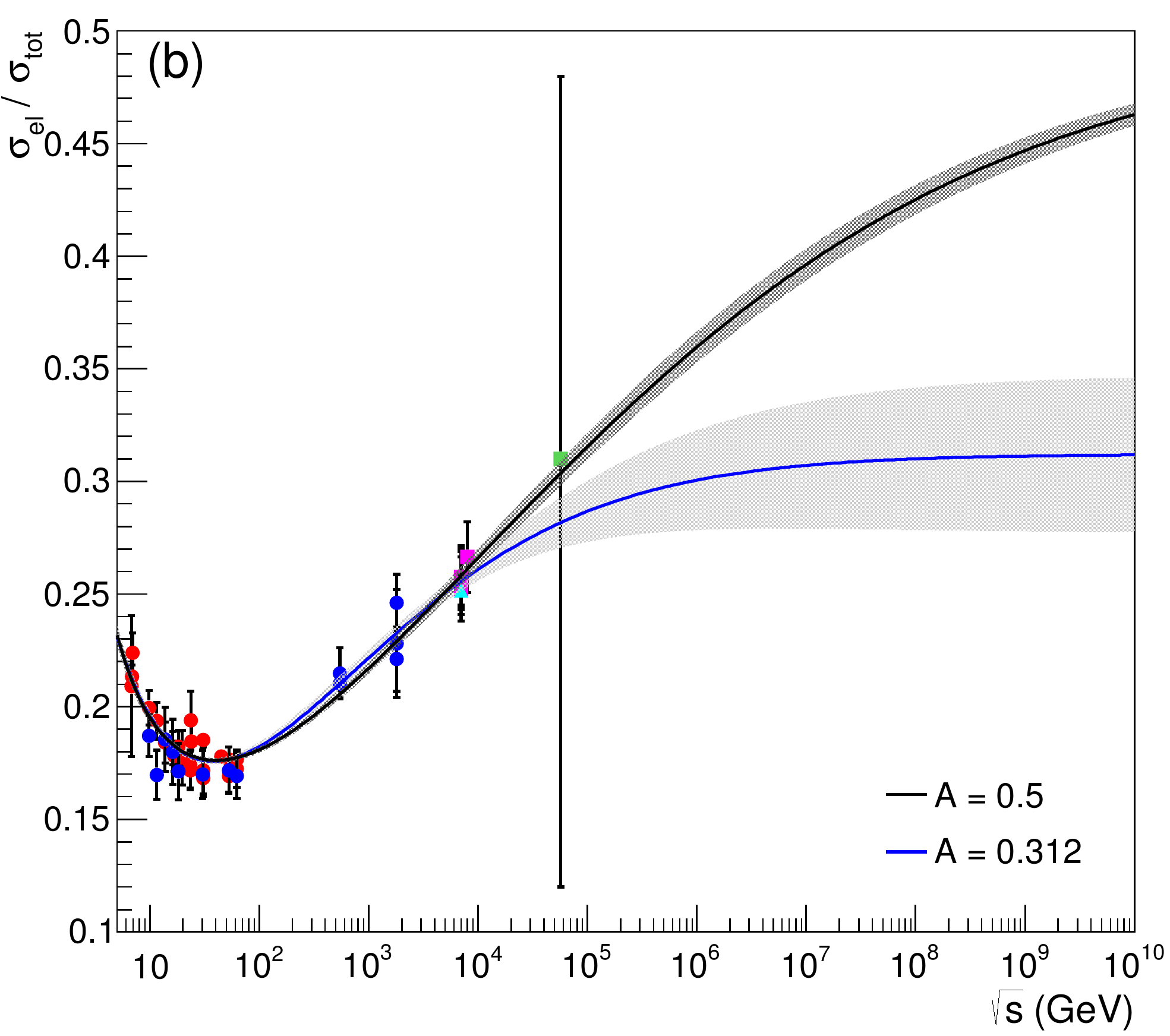}}
\caption{\label{fig:res_selected_LL}Selected fit results for $X(s)$ with variant $LL$ [Eq.~\eqref{eq:X-par-Log-LL}], $A$ free (grey),
$A$ fixed (black) and sigmoid: \subref{fig:res_selected_LL_panel_a} logistic; \subref{fig:res_selected_LL_panel_b} tanh.}
\end{figure}

The comparison between grey and black-disk results together with 
the uncertainty region for the $LL$ variant is shown in Fig.~\ref{fig:res_selected_LL}.

Numerical predictions at some energies of interest for $X(s)$ 
are shown in the third column of Table~\ref{tab:pred_ratios_Log_LL} for the logistic ($A$ = 0.293
and $A~=~0.5$) and Table~\ref{tab:pred_ratios_Tanh_LL} for tanh ($A$ = 0.312 and $A~=~0.5$).

\begin{table}[htb!]
\centering
\caption{\label{tab:pred_ratios_Log_LL}Predictions for different ratios at the LHC energy region
and beyond, with the \textit{logistic}, variant $LL$, $A$ = 0.293 (grey)
and $A~=~0.5$ (black).} 
\begin{tabular}{llllll} 
\hline\hline
$A$ &$\sqrt{s}$ (TeV)&$\sigmael/\sigmatot$&$\sigmain/\sigmatot$&$\sigmadiff/\sigmatot$&
$\sigmatot/B$ \\
\hline
      & 2.76   &  0.2408(37) & 0.7592(37) & 0.2592(37) & 12.11(19) \\
      &   8    &  0.2575(49) & 0.7425(49) & 0.2425(49) & 12.94(25) \\
 0.293&   13   &  0.2637(65) & 0.7363(65) & 0.2363(65) & 13.26(33) \\  
      &   57   &  0.278(13)  & 0.722(13)  & 0.222(13)  & 13.95(64) \\
      &   95   &  0.281(15)  & 0.719(15)  & 0.219(15)  & 14.11(74) \\
\hline
      & 2.76 & 0.2366(32) & 0.7634(32) & 0.2634(32) & 11.89(16) \\
      &   8    & 0.2625(44) & 0.7375(44) & 0.2375(44) & 13.20(22) \\
 0.5  &   13   & 0.2750(49) & 0.7250(49) & 0.2250(49) & 13.82(25) \\
      &   57   & 0.3141(64) & 0.6859(64) & 0.1859(64) & 15.79(32) \\
      &   95   & 0.3276(69) & 0.6724(69) & 0.1724(69) & 16.46(34) \\  
\hline\hline
\end{tabular}
\end{table}

\begin{table}[htb!]
\centering
\caption{\label{tab:pred_ratios_Tanh_LL}Predictions for different ratios at the LHC energy region
and beyond, with the \textit{tanh}, variant $LL$, $A$ = 0.312 (grey) and $A~=~0.5$ (black).} 
\begin{tabular}{llllll}
\hline\hline
$A$ &$\sqrt{s}$ (TeV)&$\sigmael/\sigmatot$&$\sigmain/\sigmatot$&$\sigmadiff/\sigmatot$&
$\sigmatot/B$ \\
\hline
     &2.76 & 0.2404(36) & 0.7596(36) & 0.2596(36) & 12.08(18) \\
     &   8    & 0.2577(46) & 0.7423(46) & 0.2423(46) & 12.95(23) \\
0.312&   13   & 0.2647(58) & 0.7353(58) & 0.2353(58) & 13.30(29) \\
     &   57   & 0.282(11)  & 0.718(11)  & 0.218(11)  & 14.16(57) \\
     &   95   & 0.286(13)  & 0.714(13)  & 0.214(13)  & 14.39(67) \\
\hline
     &2.76 & 0.2381(33) & 0.7619(33) & 0.2619(33) & 11.97(17) \\
     &   8    & 0.2612(42) & 0.7388(42) & 0.2388(42) & 13.13(21) \\
0.5  &   13   & 0.2719(46) & 0.7281(46) & 0.2281(46) & 13.66(23) \\
     &   57   & 0.3038(57) & 0.6962(57) & 0.1962(57) & 15.27(28) \\
     &   95   & 0.3145(60) & 0.6855(60) & 0.1855(60) & 15.81(30) \\
\hline\hline
\end{tabular}
\end{table}

It is interesting to note that, from Fig.~\ref{fig:res_selected_LL}, 
in the case of $A$ free,
the extrapolations are consistent with our asymptotic value 0.303 $\pm$ 0.055 for energies above
10$^3$ TeV. 
This suggests that the asymptotic region
might already be reached around 10$^3$ TeV. On the other hand, for the black disk case, 
the asymptotia is predicted to energies far beyond 10$^{10}$ TeV.

\begin{figure}[htb]
\centering
\includegraphics[scale=0.5]{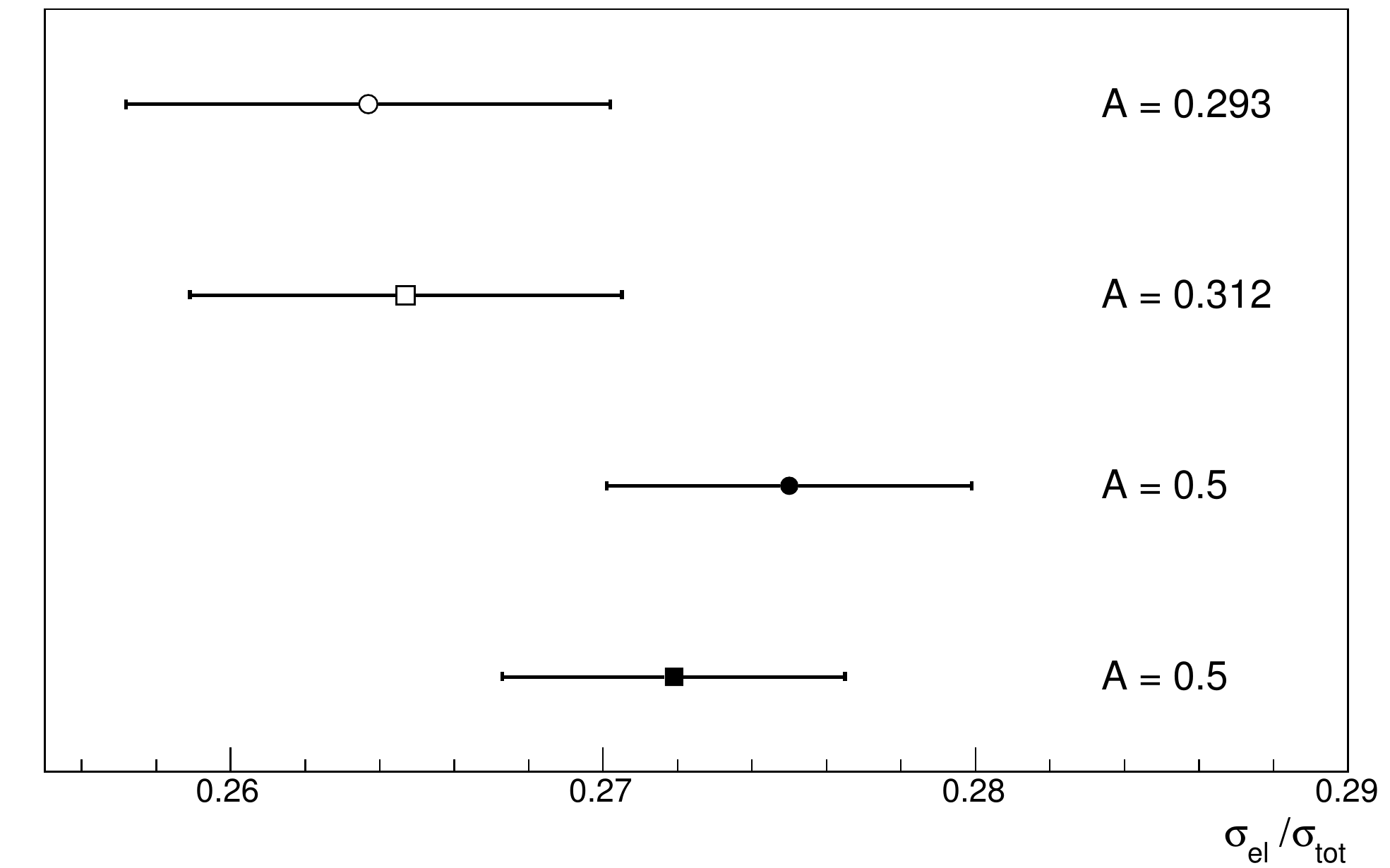}
\caption{\label{fig:pred_X_LHC_13TeV}Predictions for the ratio $X$ at 13 TeV, with variant $LL$, grey ($A$ free) and black
($A~=~0.5$) scenarios and sigmoids logistic (circles) and tanh (squares).}
\end{figure}

Of interest to Run 2 of LHC, we display in Fig.~\ref{fig:pred_X_LHC_13TeV} our predictions to $X$ at 13 TeV. 
Although the error bars will not allow the selection between the two scenarios when the data become available,
the experimental value at this energy may indicate some preference for one of them. In any case, this new data will be useful 
for a better determination of the curvature of the parametrization and, consequently, to a better estimation of the uncertainties in the free parameters.

\section{Extension to Other Quantities}
\label{sec:ratioX_fit_results_extension}

With the selected results, let us present extensions
and predictions to some other quantities of interest,
that did not take part in the data reductions.

\subsection{Inelastic Channel: Ratios and Diffractive Dissociation}
\label{subsec:ratioX_extension_inel}

\begin{figure}[htb]
\centering
\subfloat{\label{fig:pred_ineltot_LL_panel_a}\includegraphics[scale=0.39]{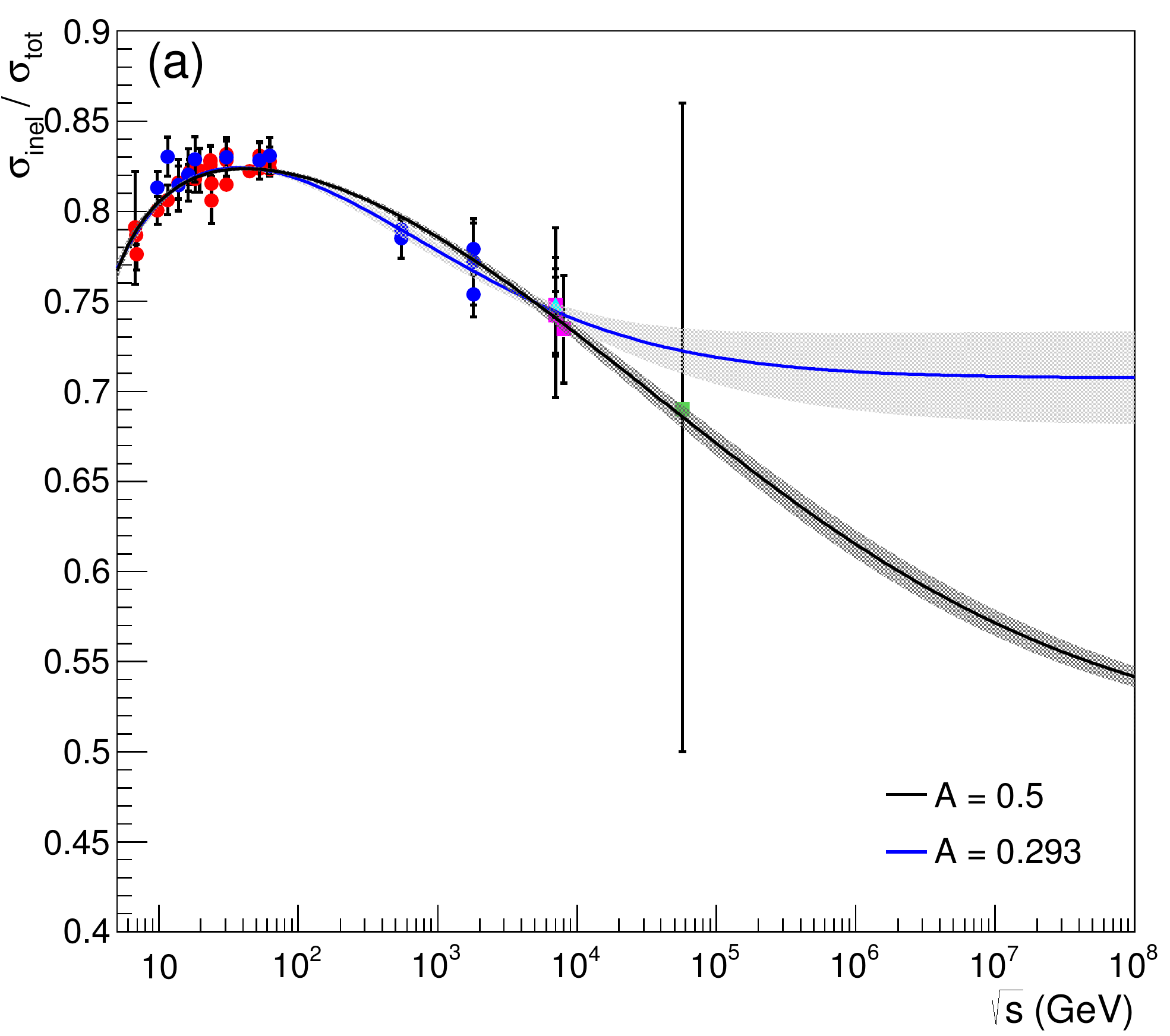}}\hfill
\subfloat{\label{fig:pred_ineltot_LL_panel_b}\includegraphics[scale=0.39]{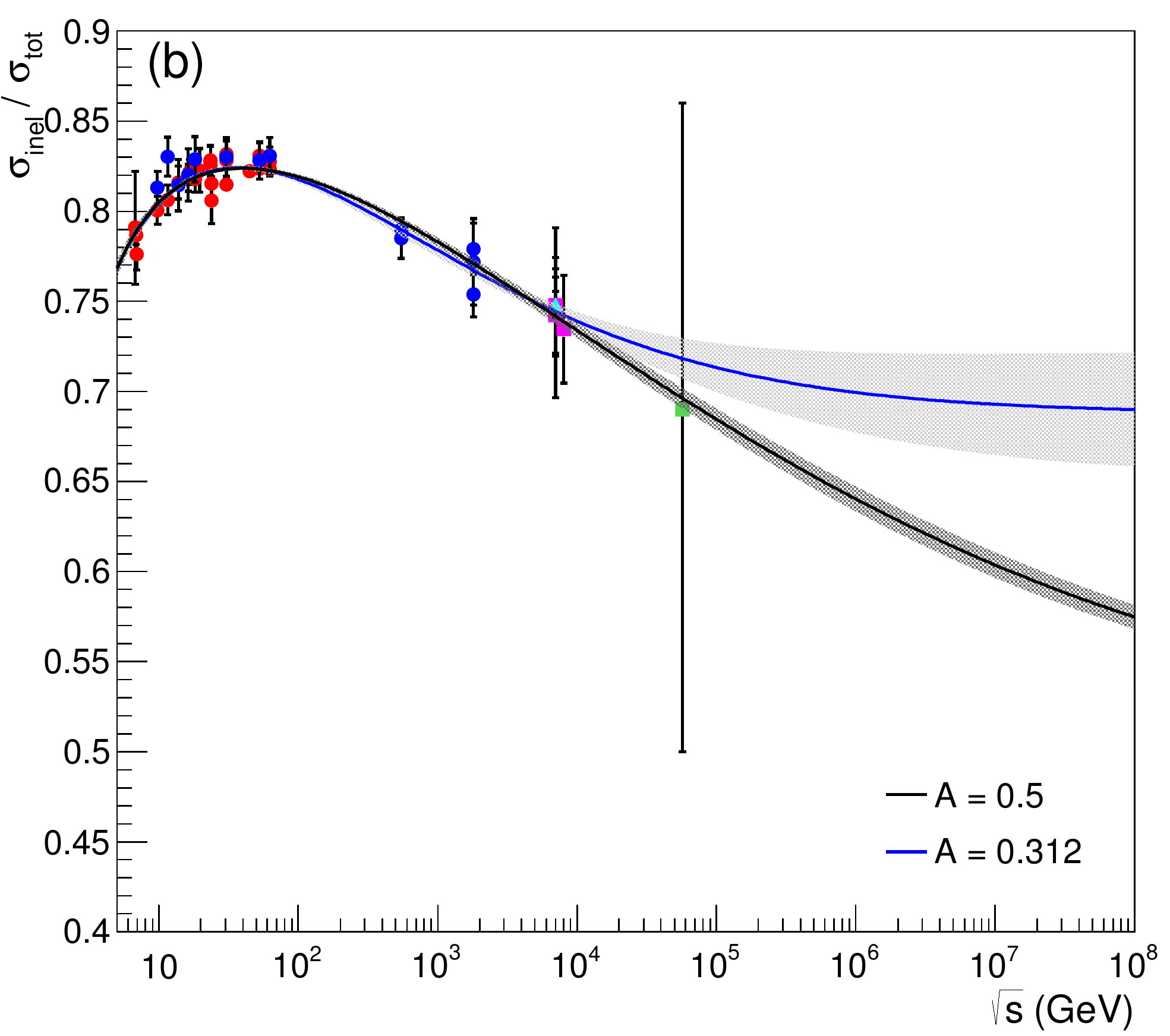}}
\caption{\label{fig:pred_ineltot_LL}Predictions, via unitarity, for the ratio between the inelastic and total cross 
sections, with variant $LL$ [Eq.~\eqref{eq:X-par-Log-LL}], $A$ free (grey),
$A$ fixed (black) and sigmoids: \subref{fig:pred_ineltot_LL_panel_a} logistic; \subref{fig:pred_ineltot_LL_panel_b} tanh.}
\end{figure}

The ratio between the inelastic and total cross-section is directly
obtained via unitarity: 1 - $X(s)$. The results 
are displayed in Fig.~\ref{fig:pred_ineltot_LL},
together with the uncertainty regions and the experimental data available.
As expected, all cases present consistent descriptions of the experimental data.
Numerical predictions at the energies of interest for the ratio $\sigmain/\sigmatot$
are shown in the fourth column of Table~\ref{tab:pred_ratios_Log_LL} with logistic ($A$ = 0.293
and $A~=~0.5$) and Table~\ref{tab:pred_ratios_Tanh_LL} with tanh ($A$ = 0.312 and $A~=~0.5$).

In what concerns the asymptotic limit,
\begin{equation}
\lim_{s \rightarrow \infty} \frac{\sigmain}{\sigmatot} = 1 - A,
\nonumber
\end{equation}

\noindent contrasting with the black disk (1/2), our global ($g$) and restricted ($r$) estimations,
Eqs.~\eqref{eq:A-mean-g} and \eqref{eq:A-mean-r}, predict
\begin{equation}
1 - A_g = 0.70 \pm 0.12
\qquad
\mathrm{and}
\qquad
1 - A_r = 0.697 \pm 0.055.
\nonumber
\end{equation}

Beyond elastic scattering, the soft inelastic diffractive  processes
(single and double dissociation) play a fundamental role in the
investigation of hadronic interactions. An important formal result
on the diffraction dissociation cross-section concerns the Pumplin 
upper bound \cite{Pumplin:1973,Pumplin:1982}:
\begin{equation}
 \frac{\sigmael}{\sigmatot} + \frac{\sigmadiff}{\sigmatot} \leq \frac{1}{2},
 \label{eq:pumplin}
\end{equation}

\noindent where $\sigmadiff$ stands for the soft inelastic diffractive cross section
(the sum of the single and double dissociation cross sections, see Eq.~\eqref{eq:sigmadiff-def}).
In this context, the black disk limit (1/2) may be associated with a combination
of the soft processes (elastic + diffractive), giving room to a semi-transparent scenario.

In this respect, Lipari and Lusignoli have discussed the possibility that the Pumplin bound 
may be already saturated in the LHC energies \cite{Lipari:2013}. 
The argument is based on
a combination of the measurements by the TOTEM and ALICE Collaborations at 7 TeV,
which indicates
\begin{equation}
\frac{\sigmadiff}{\sigmatot} \approx 0.24_{-0.06}^{+0.05},
\qquad
\frac{\sigmael + \sigmadiff}{\sigmatot} = 0.495_{-0.06}^{+0.05},
\qquad
\frac{\sigmadiff}{\sigmael} = 0.952_{-0.24}^{+0.20},
\nonumber
\end{equation}

\noindent suggesting therefore that the Pumplin bound is close to saturation.

In case of saturation, namely the equality
in Eq.~\eqref{eq:pumplin}, it is possible to estimate the ratio $\sigmadiff/\sigmatot$
at the LHC energies and beyond. The numerical predictions
for this ratio
are shown in the fifth column of Table \ref{tab:pred_ratios_Log_LL} (logistic, $A$ = 0.293
and $A~=~0.5$) and Table \ref{tab:pred_ratios_Tanh_LL} (tanh, $A$ = 0.312 and $A~=~0.5$).
Obviously, the asymptotic value of this ratio is zero in the case of a black disk scenario strictly
associated with the elastic channel.

Moreover, using the Pumplin bound and Unitarity, we can also infer an upper bound for 
the ratio $\sigmadiff/\sigmain$, namely
\begin{equation}
R(s) \equiv \frac{\sigmadiff}{\sigmain} \leq \frac{1/2 - X(s)}{1-X(s)}.
\nonumber
\end{equation}

\begin{figure}[htb!]
\centering
\includegraphics[scale=0.5]{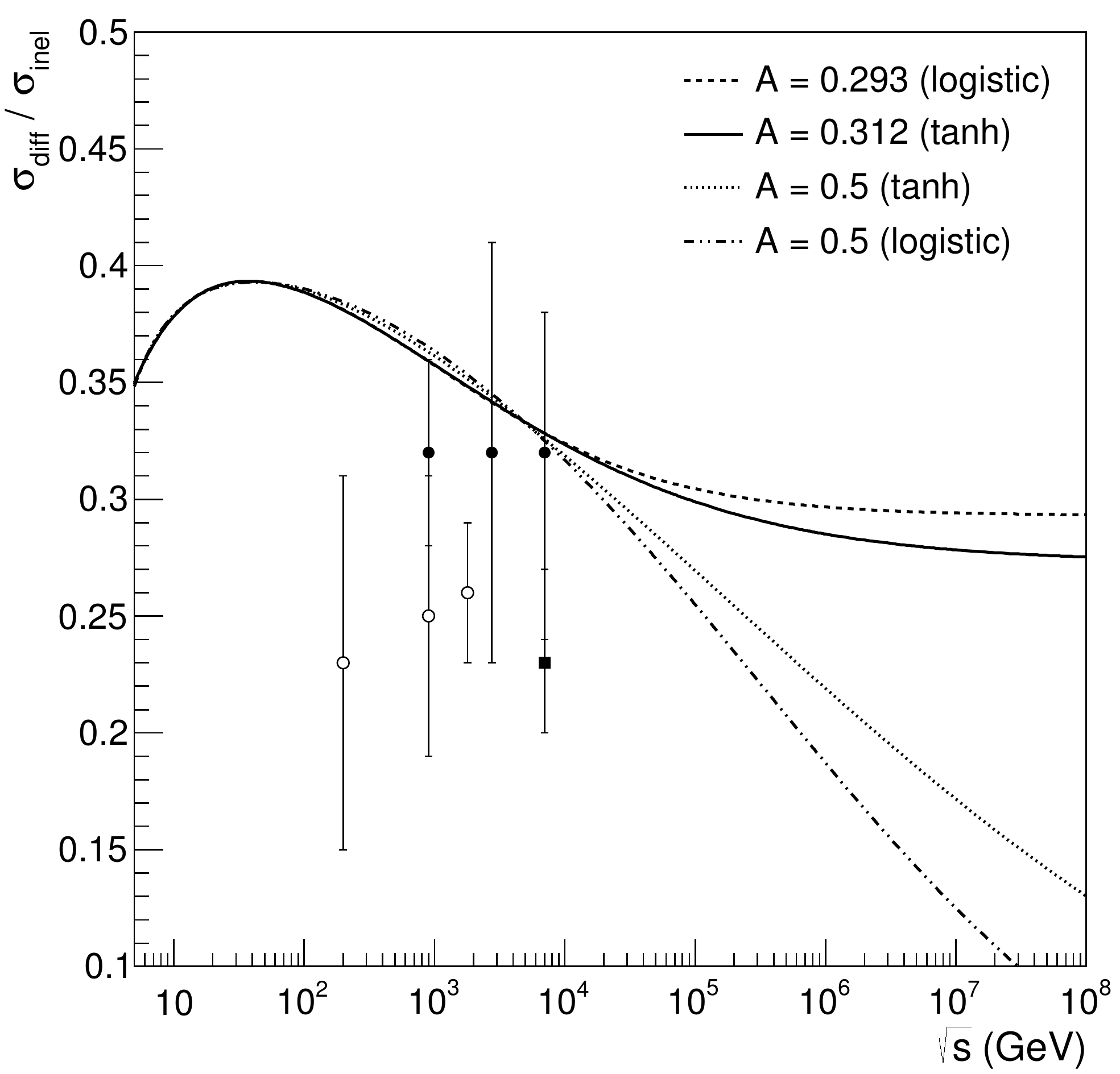}
\caption{\label{fig:pred_upper_bound_diffinel}Upper bounds for the ratio between the dissociative 
(single plus double) and inelastic cross sections 
from the selected results for the grey and black scenarios (logistic and tanh). Experimental information from $pp$ scattering 
at 0.9, 2.76 and 7 TeV (black marks) \cite{ALICE:2013,CMS:2013,CMS:2015} 
and $\bar{p}p$ scattering at 0.2, 0.9 and 1.8 TeV (white marks) \cite{UA5:1986a,UA5:1986b,CDF:2001,CDF:1994b,CDF:1994a}.}
\end{figure}

The curves corresponding to this bound in all cases treated in this Section are shown in 
Fig.~\ref{fig:pred_upper_bound_diffinel}, together with experimental data from 
$pp$ \cite{ALICE:2013,CMS:2013,CMS:2015} and $\ppbar$ \cite{UA5:1986a,UA5:1986b,CDF:2001,CDF:1994b,CDF:1994a} scattering.
Once more, contrasting with the asymptotic null limit in a black disk scenario, our
global and reduced estimations ($A_g$ and $A_r$),  lead to the predictions:
\begin{equation}
R_g = 0.29 \pm 0.12
\qquad
\mathrm{and}
\qquad
R_r = 0.283 \pm 0.055
\qquad
\mathrm{for}
\qquad
s \rightarrow \infty.
\nonumber
\end{equation}

\subsection{Ratio Y Associated with Total Cross-Section and Elastic Slope}
\label{subsec:ratioX_extension_ratioY}

In cosmic-ray studies, the determination of the $pp$ total cross-section from the proton-air
production cross-section is based on the Glauber formalism \cite{Ulrich_etal:2009,Engel:1998,Engel:2000}.
In this context, the nucleon-nucleon impact parameter amplitude
(profile function) constitutes an important ingredient for the connection between hadron-hadron and
hadron-nucleus scattering.
The profile function is typically parametrized by 
\begin{equation}
a_j(s,\vec{b}_j) = \frac{[1 + \rho(s)]}{4\pi}\, \frac{\sigmatot(s)}{B(s)}\, 
e^{-\vec{b}_j^{2}/[2B(s)]},
\end{equation}
\noindent where $\rho$, $\sigmatot$ and $B$
demand inputs from models to complete the connection. However, models have been tested
only in the accelerator energy region and in general are characterized by  different
physical pictures and different predictions at higher energies. As a consequence, the 
extrapolations result in large theoretical
uncertainties, as clearly illustrated by Ulrich \etal \cite{Ulrich_etal:2009}. 
From the above equation, any extrapolation is strongly dependent on the ratio
$\sigmatot/B$,
namely the unknown correlation between $\sigmatot$ and $B$ in terms of energy.

One way to overcome this difficulty is to estimate the ratio
\begin{equation}
Y(s) = \frac{\sigmatot}{B}(s)
\end{equation}
\noindent through its approximate relation with the ratio $X(s)$, treated in Section~\ref{sec:basic_eltot_totB},
\begin{equation}
Y(s) \approx 16 \pi X(s).
\end{equation}

\begin{figure}[htb]
\centering
\subfloat{\label{fig:pred_totB_LL_panel_a}\includegraphics[scale=0.39]{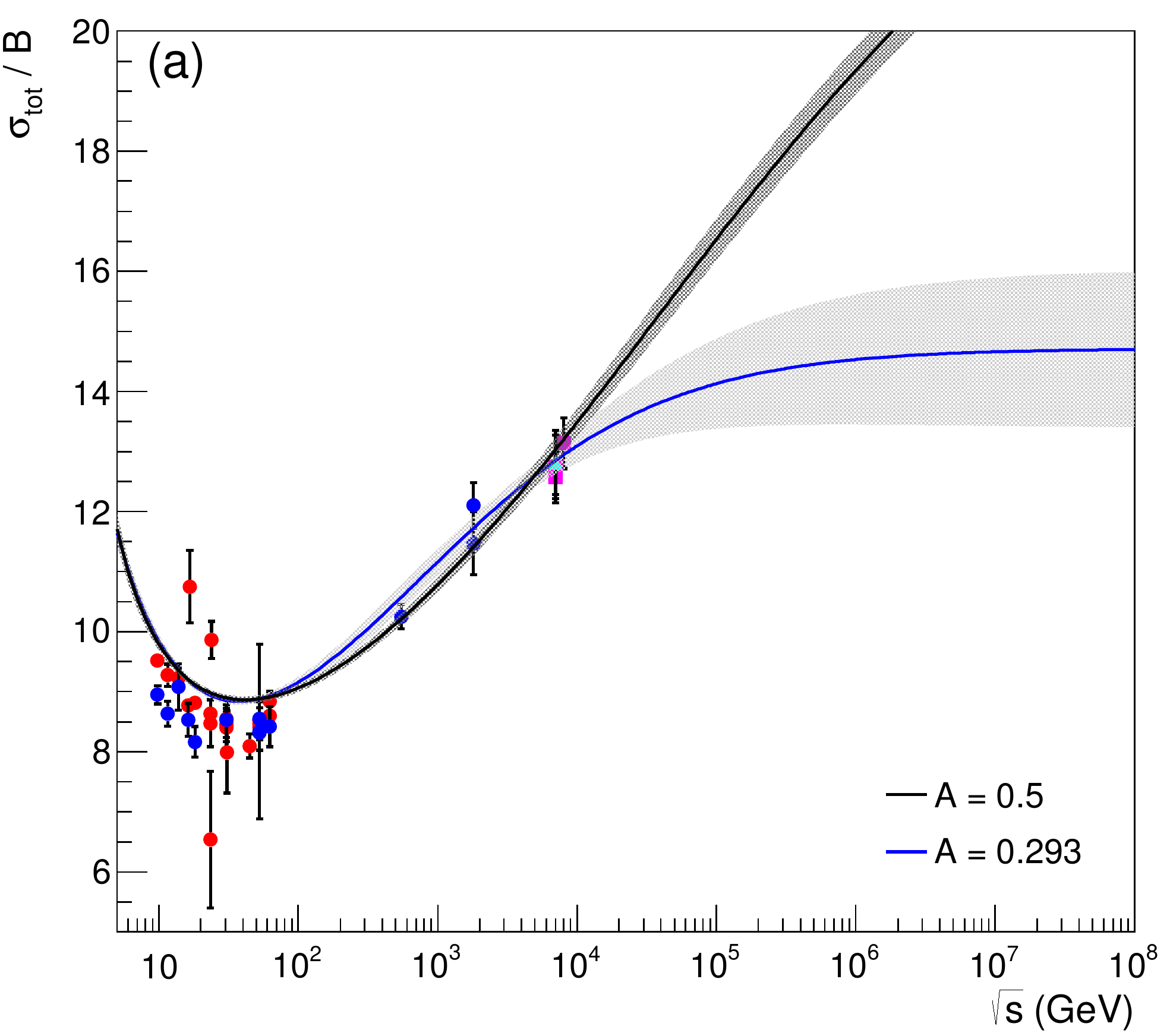}}\hfill
\subfloat{\label{fig:pred_totB_LL_panel_b}\includegraphics[scale=0.39]{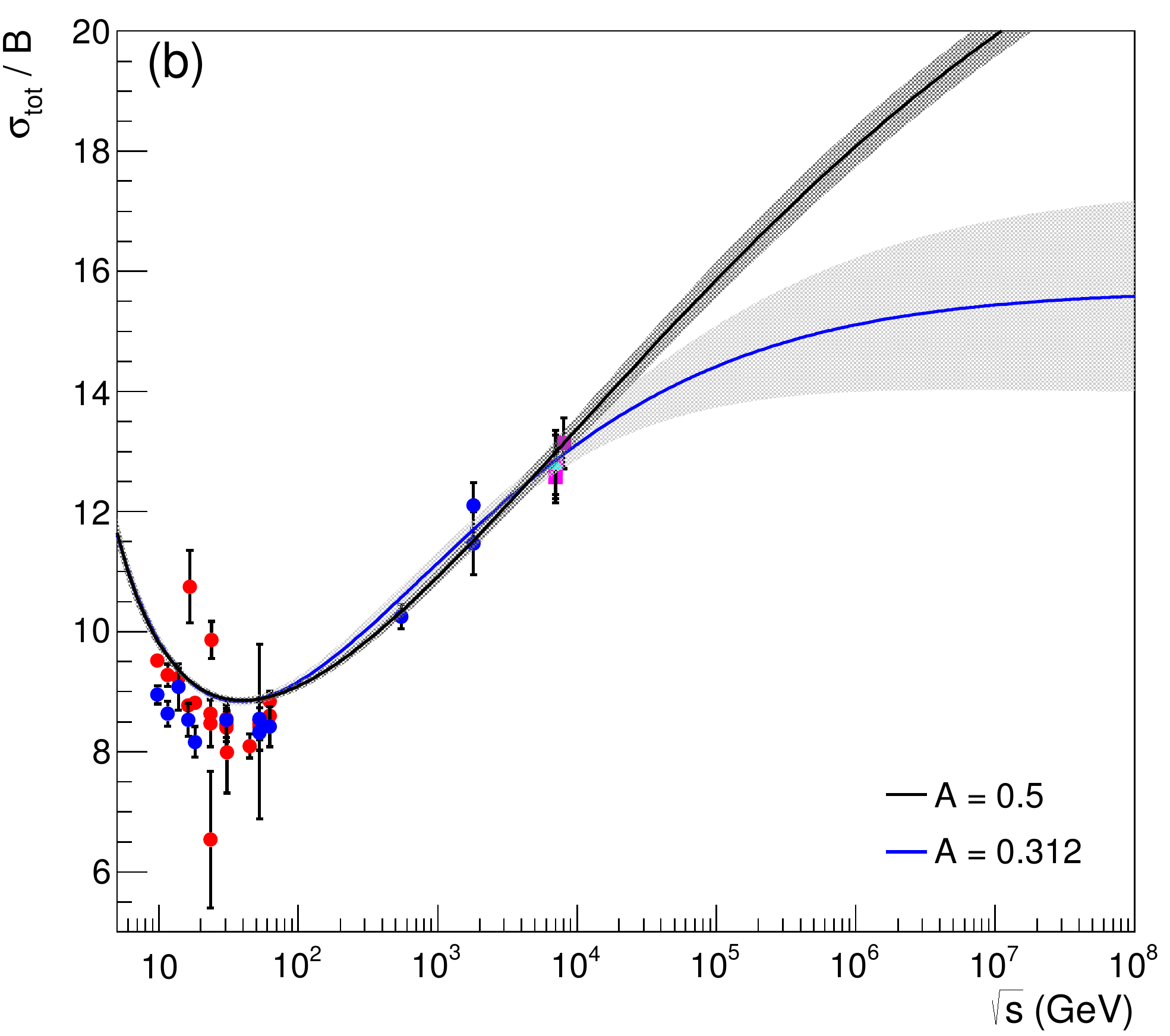}}
\caption{\label{fig:pred_totB_LL}Predictions for the ratio between the total cross 
sections and the elastic slope, with variant $LL$, [Eq.~\eqref{eq:X-par-Log-LL}], $A$ free (grey),
$A$ fixed (black) and sigmoid: \subref{fig:pred_totB_LL_panel_a} logistic; \subref{fig:pred_totB_LL_panel_b} tanh.}
\end{figure}

The behavior of $Y(s)$ extracted in this way and in all cases treated in this Section
are shown in Fig.~\ref{fig:pred_totB_LL}; the corresponding numerical predictions
at the energies of interest
are displayed in the sixth column of Table \ref{tab:pred_ratios_Log_LL} (logistic, $A$ = 0.293
and $A~=~0.5$) and Table \ref{tab:pred_ratios_Tanh_LL} (tanh, $A$ = 0.312 and $A~=~0.5$).

\section{On Physical Interpretations}
\label{sec:ratioX_comments_interpretations}

In this section, we discuss some possible physical concepts related to the results obtained with the sigmoid ansatz.
Our discussion is focused on the choices for the sigmoid and elementary functions
rather than a specific asymptotic scenario.

We treat two aspects: (1) the possible connection of $X(s)$ with 
the contribution of effective partonic interactions through the opacity
concept; (2) the relation of this process with a saturation effect
associated to a change of curvature in $X(s)$.

Sigmoid functions \cite{Menon_etal_sig:1994,Kucharavy_deGuio:2011} have great applicability in several topics of technology and exact sciences,
but not restrict to these areas. They are also applied in biological, humanities and social sciences problems, for instance 
language change, diffusion of an innovation and many others~\cite{Sunday_etal:2012,Tsoularis:2001,Lopez_etal:2010}. 
Of interest in high-energy physics, application in unitarization schemes (eikonal/$U$-matrix have been considered by Cudell, Predazzi, Selyugin \cite{Selyugin:2008,Cudell:2009} and also application related to polarized gluon density has been recently discussed by Bourrely \cite{Bourrely:2015}).

In general, these sigmoid functions are associated with the Pearl-Verhulst logistic
processes, in which the growth of a population is bounded and proportional to its
size, as well as, to the difference between the size and its bound, resulting in the 
logistic differential equation \cite{Sunday_etal:2012,Tsoularis:2001}
\begin{equation}
\frac{dN}{dt} = r N \left[1 - \frac{N}{M}\right],
\end{equation}

\noindent where $N = N(t)$ is the population at a time $t$, $M$ is its maximum value 
and $r > 0$ is the intrinsic rate of growth.
The equation has variable coefficients if $r=r(t)$ and/or $M=M(t)$ \cite{Lopez_etal:2010}.

In our case, the logistic ansatz for $S(f)$, Eq.~\eqref{eq:sigmoid-def}, is a trivial solution of the
differential equation
\begin{equation}
\frac{dS}{df} = S \left[1 - S\right].
\end{equation}

In terms of the ratio $X$, Eq.~\eqref{eq:X-par-gen}, and variable $v\equiv \ln s/s_0$
this differential equation reads
\begin{equation}
\frac{dX}{dv} = \frac{df}{dv} X \left[1 - \frac{X}{A}\right],
\label{eq:eq-dif-X-v} 
\end{equation}

\noindent corresponding to a logistic equation with variable
coefficient $df/dv$.

Therefore, the evolution of the ratio $X$ with energy may be associated with a Pearl-Verhulst logistic process.
Now, in case of the elastic scattering and in what concerns a corresponding ``population growth'', it seems reasonable
to think in terms of an effective number of partonic interactions taking part in the collision process, as the energy increases.
In this respect we have the comments that follows.

In the phenomenological context, 
the QCD inspired models are, generally, based on the concepts
of semi-hard QCD or mini-jet models~\cite{Cline:1973,Afek:1980,Gaisser:1985,LHeureux:1985,Pancheri:1985,Durand:1987,
Capella:1986,DiasdeDeus:1987,Block_Gregores:1999}.
These models are constructed with the separation of soft and semi-hard contributions to the scattering amplitude.
The soft ones are treated on phenomenological grounds and the semi-hard ones are determined with the parton model,
probabilistic arguments, and perturbative QCD calculations.

In terms of the opacity function ($\Omega$) introduced in Section~\ref{sec:basic_profile_eikonal}, 
the probability that an inelastic event takes place at $b$ and $s$
is given by
\begin{equation}
G_\text{inel}(s,b) = 1 - e^{-2\Omega}.
\end{equation}

Now, denoting the probability that there are \textit{no} soft (semi-hard) inelastic interaction by
$\bar{P}_\text{S}$ ($\bar{P}_\text{SH}$), we can associate
\begin{equation}
G_\text{inel}(s,b) = 1 - \bar{P}_\text{S} \bar{P}_\text{SH} = 1 - e^{-2\Omega_\text{S}} e^{-2\Omega_\text{SH}},
\label{eq:G-inelastic-S-SH}
\end{equation}

\noindent so that the total opacity reads
\begin{equation}
\Omega(s,b) = \Omega_\text{S}(s,b) + \Omega_\text{SH}(s,b).
\end{equation}

Now let us focus on the semi-hard opacity, which is constructed under probabilistic arguments
and the QCD parton model. 
Let $n(s,b)$ be the average number of parton-parton collisions at $s$ and $b$, which
is associated with the probability of semi-hard inelastic scattering. Assuming that the parton-parton
collisions are independent and that the probability that $n$ parton-parton collisions occur follows the Poisson distribution,
the probability that hadrons do not undergo semi-hard
scattering can be expressed by
\begin{equation}
 \bar{P}_\text{SH} = e^{- n(s,b)}. \nonumber
\end{equation}

\noindent
Comparing with $P_\text{SH}=e^{-2\Omega_\text{SH}}$,
we conclude that the semi-hard opacity reads
\begin{equation}
\Omega_\text{SH}(s,b) = \frac{1}{2} n(s,b).
\end{equation}

In the theoretical context, $n(s,b)$ is expressed in terms of parton-parton cross section
and hadronic matter distribution, which is related to form factors in the $q^2$-space
(see, for example, \cite{Fagundes:2015} for a recent version).

The key point here is the fact that the average number
of partonic interactions is clearly connected with the opacity and
the profile functions and, consequently, with the ratio $X(s)$
(at least in what concerns the central opacity in grey disk and Gaussian profiles,
as discussed in Chapter~\ref{chapt:basic_concepts}).
Therefore, this result corroborates the logistic interpretation of Eq.~\eqref{eq:eq-dif-X-v}
as associated with an effective number of partonic interactions taking part in the collision process.

Going one step further, we note that in this context,
the change of curvature presented by the sigmoid functions
suggests a change in the rate of effective partonic interactions,
namely a fast rise at low energies is followed
by a saturation effect starting at the inflection point, which represents
a change in the dynamics of the interaction.
In order to look for a quantitative connection, we have the comments that follows.

On the one hand, these inflection points (energies $\sqrt{s_\text{ip}}$ at which the change of curvature takes place)
in terms of the asymptotic ratio $A$ for all cases investigated
are displayed in Table~\ref{tab:res_inflection_pt}.
We see that despite the different values of $A$, the position of
the inflection point lies in a rather restrict interval:
$\sqrt{s_\text{ip}}~\approx~80~-~100~\text{GeV}$. 

\begin{table}[htb]
\centering
\caption{\label{tab:res_inflection_pt}Inflection points: roots of the second derivative of $X(\sqrt{s})$ in GeV
for each fit developed.} 
\begin{tabular}{c c c c c}
\hline\hline
 &\multicolumn{2}{c}{logistic}&\multicolumn{2}{c}{tanh} \\ \cmidrule(lr){2-3} \cmidrule(lr){4-5} 
   $A$ &    $PL$  &   $LL$  & $PL$  &   $LL$        \\
\hline
free   &   81.6   &  81.2   &  80.1     & 80.0  \\
0.3    &   82.9   &  82.4   &   78.7    & 78.5  \\
0.436  &   92.2   &  92.0   &   86.7    & 86.6  \\ 
0.5    &   93.7   &  93.5   &   87.8    & 87.8  \\
0.75   &   96.5   &  96.1   &   89.7    & 89.6  \\
1.0    &   97.4   &  97.1   &   90.3    & 90.2  \\ 
\hline\hline
\end{tabular}
\end{table}

\begin{figure}[h!]
 \centering
 \includegraphics[scale=0.25]{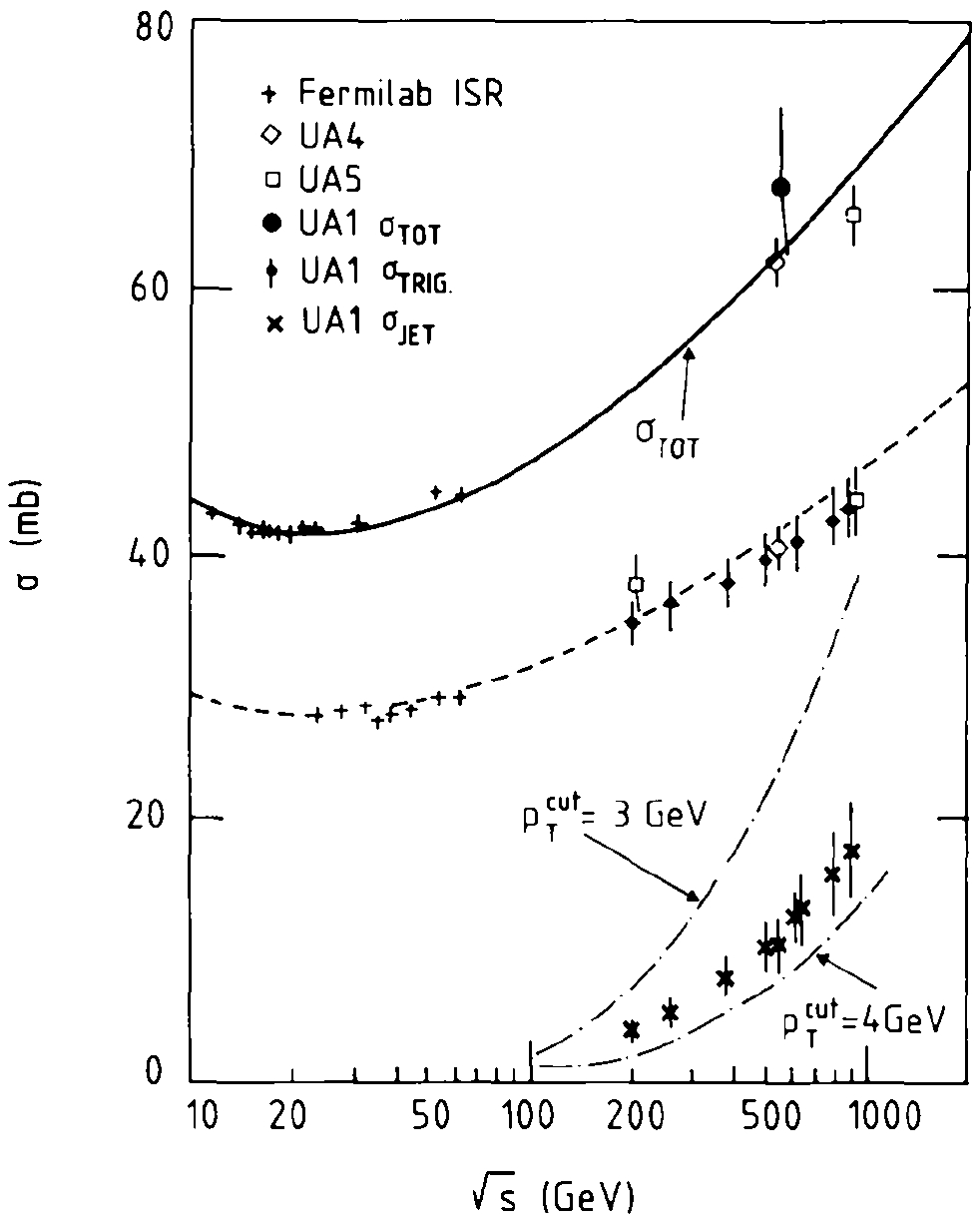}
 \caption{\label{fig:minijet_cross_section}Minijet cross-section ($\sigma_\text{jet}$) from $\ppbar$ scattering measured by UA1 Collaboration.
 Figure taken from Ref.~\cite{UA1_Albajar:1988}.}
\end{figure}

On the other hand, concerning this energy region, 
the UA1 Collaboration have reported the measurement of low transverse energy clusters (mini-jets) 
in $\bar{p}p$ collisions at the CERN Collider for $\sqrt{s}$: 200 - 900 GeV
\cite{UA1_Albajar:1988}. Extrapolation of the observed mini-jet cross-section
to lower energy (see Fig.~\ref{fig:minijet_cross_section}) suggests that the 
region 80 - 100 GeV is consistent with the beginning of the mini-jet production.

Now, the rise of the mini-jet cross-section has been associated with
the observed faster rise of the inelastic and, consequently the total cross-section 
\cite{UA1_Albajar:1988} (see also references there in). Therefore, once $X = \sigmael/\sigmatot$, 
it seems reasonable to associate this behavior with a change of curvature in $X$ 
and the beginning of a saturation effect.

Although suggestive, we stress that the above arguments are certainly limited 
as effective physical interpretations of our analysis and results.
However, we hope they may be useful for further investigations.

\section{On a Semi-Transparent Asymptotic Scenario}
\label{sec:ratioX_comments_semitransparent}

From the discussion in Section~\ref{sec:ratioX_fit_discussion_results}, our analysis favors
the global result $A_g = 0.30 \pm 0.12$ which means a scenario below the black disk. 

This type of semi-transparent or grey scenario has already been suggested in the past. For instance,
in the mid-seventies, Fia\l{}kowski and Miettinen suggested a semi-transparent scenario in the context 
of a multi-channel approach \cite{Fialkowski:1975}. Based on the Pumplin bound, 
Sukhatme and Henyey \cite{Sukhatme_Henyey:1976} have also conjectured this scenario.

More recently, different authors have discussed this possibility, as pointed out below.

\begin{enumerate}
\item Lipari and Lusignoli \cite{Lipari:2009} and Achilli \etal \cite{Achilli:2011} have discussed
the observed overestimation of $\sigmael$ (or underestimation of $\sigmain$)
in the context of one channel eikonal models. That led Grau \etal \cite{Grau_etal:2012} to re-interpret
the Pumplin upper bound as an effective asymptotic limit,
\begin{equation}
\frac{\sigmael}{\sigmatot} + \frac{\sigmadiff}{\sigmatot} \rightarrow \frac{1}{2} \qquad \mathrm{as} \qquad s \rightarrow \infty.
\nonumber
\end{equation}

Based on the behavior of the experimental data of $X(s)$ and a combination 
of the results at 57 TeV by the Auger Collaboration together with the
predictions by Block and Halzen, Grau \etal
have conjectured a rational limit $1/3$ as a possible asymptotic value for the ratio $\sigmael/\sigmatot$
\cite{Grau_etal:2012}.

\item In the QCD-motivated model by Kohara, Ferreira and Kodama 
\cite{Kohara_etal:2014a,Kohara_etal:2014b,Kohara_etal:2015a,Kohara_etal:2015b}, the scattering amplitude is constructed
under both perturbative and non-perturbative QCD arguments (related to extensions
of the Stochastic Vacuum Model). The model leads to consistent descriptions
of the experimental data on $\bar{p}p$ and $pp$ elastic scattering (forward quantities
and differential cross sections) at energies above 20 GeV, 
including the LHC and with extensions to the cosmic-ray energy region. 
The predictions for the asymptotic ratio $X$ lie below 1/2 and are close to 1/3.

\item As already mentioned, the combination of the parametrizations by the COMPETE ($\sigmatot$) and TOTEM 
($\sigmael$) Collaborations indicates the asymptotic value $A = 0.436$.
Moreover, combination of the ATLAS parametrization ($\sigmael$) \cite{ATLAS:2014} with
that by COMPETE reads $A = 0.456$. Both, therefore, below the black disk.

\item We stress that through a completely different approach, several individual and simultaneous fits to 
$\sigmatot$ and $\rho$ data, extended to fit the $\sigmael$
data \cite{Fagundes_Menon_Silva:2013a,Menon_Silva:2013a,Menon_Silva:2013b}, have led to asymptotic ratios in plenty agreement
with our inferred global limit $A_g = 0.30 \pm 0.12$.

\item At last, we notice that more recent phenomenological analysis (2017) have also pointed to a grey scenario,
as discussed by Roy~\cite{Roy:2017}, Dremin~\cite{Dremin:2017} and Kohara~\cite{Kohara:2017}.
\end{enumerate}

All the facts above corroborate the results and conclusions 
presented here, indicating the semi-transparent limit as a possible
asymptotic scenario for hadronic interactions.

\section{Summary and Conclusions}
\label{sec:ratioX_fit_conclusion_final}

We have presented a study on the asymptotic scenario of hadronic scattering through
an empirical analysis of the energy dependence of the ratio $X=\sigmael/\sigmatot$.
Our study is based on the choice of four analytical parametrizations written as a
composite function of sigmoid ($S$) and elementary functions of the energy ($f(s)$). The sigmoid functions considered
are the logistic and the hyperbolic tangent and the two elementary functions are written in the
variable $\ln s$ and consist in a linear term plus a power law ($PL$) or a logarithmic law ($LL$).
Another important ingredient is the asymptotic value $A$ of the ratio $X$, so that our parametrization is written
as $X(s)= A S(f(s))$. Two types of fits to $pp$ and $\ppbar$ data were considered: one where $A$ is fixed at
some determined values (\textit{constrained} fits) and another one with $A$ as a free parameter (\textit{unconstrained} fits).

In the constrained fits, we covered all the range allowed for $A$ by unitarity, with representative values 
chosen from other empirical, theoretical and phenomenological studies. The lowest value was $0.3$ 
(based on results of Refs.~\cite{Fagundes_Menon_Silva:2013a,Menon_Silva:2013a,Menon_Silva:2013b}) and the 
highest was $1$ (the maximum value allowed by unitarity), passing through the black-disk ($0.5$) scenario. 
All fits converged to statistically consistent solutions with good
description of the experimental data analysed (including data at LHC energies)
and equivalent results on statistical grounds.
As a consequence, this result indicates that the black-disk does not constitute a unique or exclusive solution.

On the other hand, in the unconstrained variant, all fits converged to values compatible with a semi-transparent scenario, i.e. $A<0.5$. 
A global average value of $A$ (considering all parametrizations) is $A_g = 0.30\pm 0.12$. 
Based on the number of free parameters, we chose the $LL$ variants as the preferable results. 
In this case, the average value of $A$ reads $A_r = 0.303 \pm 0.055$ (restricted to $LL$ variant).

In what concerns our empirical parametrizations, it seems important
to stress their efficiency in describing the experimental data analyzed,
independently of the fixed physical value of $A$. 
An important feature of our parametrizations is the small number of free parameters. For example,
the constrained black-disk fit with the variant $LL$ demands only three
fit parameters and led to consistent description of all data.
We stress, once more, the contrast with ten or more parameters typical
of individual fits to $\sigmatot$ and $\sigmael$ data.
All these features indicate the good quality of our analytical choices for $S(f)$ and $f(s)$, on empirical
and phenomenological grounds. Moreover, the simple relations among $X$ and other ratios, specially $\sigmatot/B$, 
allow us to predict and extrapolate quantities in a large range of energy.

We recall that sigmoid functions are characterized by having a change of curvature at some point, 
which could indicate a change of dynamics. 
We have determined the energy range at which this change of curvature occurs and, 
interesting enough, we found energies close to the region where the mini-jet cross section starts to rise.
Specifically, all inflection points determined are in the energy range 80 - 100 GeV and the value increases as the value of $A$ increases.

Given the empirical efficiency of these \textit{analytical representations} 
for $X(s)$, we have attempted to look for possible physical connections
with the underline theory/phenomenology of soft strong interactions.
Presently, we can only devise some suggestive ideas
relating a ``population growth", represented by the logistic
differential equation, with the number of effective parton interactions
taking part in the collision, which can be connected to mini-jet and semi-hard QCD models. 
Despite suggestive, this conjecture is rather limited on physical grounds 
and must be investigated in more detail.

At last, it is interesting to note that, when comparing the grey disk scenario from the unconstrained 
fit ($A$ free) and the black-disk scenario ($A=0.5$ fixed),
we see that they differ in the energy at which the asymptotic region may be reached. 
For the grey-disk case, asymptotia may already be reached around 10$^3$ TeV, and in case
of the black disk, only far beyond 10$^{10}$ TeV.

%% file: ch_rise_sigmatot_f.tex
%
%
%
%
\cleardoublepage



\chapter[Phenomenological Studies on the Rise of the $\sigmatot$ and the LHC Data]{Phenomenological Studies on the Rise of the Total Cross-Section and the LHC Data}
\label{chapt:rise_sigmatot}

\section{Introduction}
\label{sec:rise_intro}

This chapter is devoted to a phenomenological analysis on the problem of the rise of hadronic cross section. 
It is based on the research presented in Refs.~\cite{Fagundes_Menon_Silva:2017a,Fagundes_Menon_Silva:2017b}.
As discussed in Chapter~\ref{chapt:intro},
we do not have a full description for the energy dependence of $\sigmatot$ from first principles of QCD, since this quantity
is related to the elastic scattering amplitude in the forward direction through the optical theorem~\cite{Barone_Predazzi_book:2002}
\begin{equation}
\sigmatot(s) = \frac{\Imag F(s,t=0)}{s}
\label{eq:rise-optical-theo}
\end{equation}

\noindent and in the \textit{soft regime} (limit of high energies and small momentum transfer) we cannot apply perturbative techniques to calculate
the amplitude. On the other hand, nonperturbative results are only available in the asymptotic limit~\cite{Giordano_Meggiolaro:2014},
which is the topic of Chapter~\ref{chapt:SLT_sigmatot}.

In the last decades, among several phenomenological models (see, for instance, the reviews~\cite{Pancheri_Srivastava:2017,Dremin:2013,Kaspar:2011,Fiore:2009}), 
the Regge-Gribov formalism~\cite{Collins_book:1977,Barone_Predazzi_book:2002,Donnachie_etal_book:2002}, a pre-QCD theory to treat hadronic scattering, has shown to be useful in describing the main 
features presented by the experimental data: the decrease of $\sigmatot$ until approximately 20 GeV and the rise
as the energy increases. Within the Regge-Gribov notation, the first regime is described by the exchange of Reggeons, and the latter one by
the exchange of the \textit{ad hoc} object called Pomeron. 
Although useful, the connection between these objects and the QCD
description in the soft regime is not clear~\cite{Donnachie_etal_book:2002}.

Usually, the models are based on general results from axiomatic field theory
and are derived by using general concepts like analyticity and unitarity.
The Froissart-Lukaszuk-Martin (FLM) bound~\cite{Froissart:1961,Martin:1963,Martin:1965,Lukaszuk_Martin:1967}, 
already presented in Section~\ref{sec:basic_FLM_bound}, is an important example of these results. It reads
\begin{equation}
\sigmatot(s) < c \ln^2(s/s_0) \quad (s\to\infty)
\label{eq:rise-FLM-bound}
\end{equation}

\noindent where $s_0$ is an unspecified energy scale and the pre-factor on the right-hand side is bounded by
\begin{equation}
c \leq  \frac{\pi}{m_{\pi}^2} \approx\ \mathrm{60\ mb},
\label{eq:rise-FLM-bound-coeff}
\end{equation}

\noindent where $m_{\pi}$ is the pion mass. We call to the attention that this bound is derived in the limit of asymptotic energies ($s\to\infty$).

Still in the context of Regge-Gribov, the COMPETE Collaboration has performed a complete and comprehensive study~\cite{COMPETE:2002a,COMPETE:2002b}
of possible parametrizations of the total cross section, testing several properties like factorization, universality of the leading term, among others. 
By means of fits to experimental data of several reactions involving hadrons and with a rank procedure,
the highest-rank result has as its leading term the $\ln^2 s$ dependence, therefore in accordance with the FLM bound.

A similar analysis with the COMPETE highest-rank result is still performed by the COMPAS Group (IHEP, Protvino) 
and published in the \textit{Review of Particle Physics} (RPP)
edited by the Particle Data Group (PDG) every two-years~\cite{PDG:2010,PDG:2012,PDG:2014,PDG:2016}.

In this chapter, we are interested in performing a \textit{forward-amplitude analyses}. This approach consists in analytic
parametrizations for the total cross section, connected with the ratio $\rho$ between
the real and imaginary parts of the forward amplitude,
\begin{equation}
\rho(s) = \frac{\Real F(s,t=0)}{\Imag F(s,t=0)},
\label{eq:rise-rho-def}
\end{equation}

\noindent by means of analytic or numerical methods
and simultaneous fits to the experimental data available on these two quantities. For examples, 
see Refs.~\cite{Amaldi:1977,Augier:1993,Bueno_Velasco:1996,Cudell:1997,COMPETE:2002a,
COMPETE:2002b,Luna_Menon:2003,Luna_Menon_Montanha2004,Igi:2002,Igi:2005,Block_Halzen:2004,Block_Halzen:2005}.

Specifically, we will consider two types of leading terms: the $\ln^2 s$ and the parametrization proposed
by Amaldi \etal~\cite{Amaldi:1977}, namely $\ln^\gamma s$, where $\gamma$ is a real free fit parameter.
Fits to experimental data performed by several authors with different datasets (including recent studies with LHC data)
point to a possible rise faster than $\ln^2 s$, i.e. $\gamma >2$~\cite{Amaldi:1977,Augier:1993,Bueno_Velasco:1996,
Fagundes_Menon_Silva:2012a,Fagundes_Menon_Silva:2013a,Menon_Silva:2013a,Menon_Silva:2013b}. 
As already commented in Section~\ref{sec:basic_FLM_bound}, Azimov~\cite{Azimov:2011,Azimov:2012a,Azimov:2012b} has argued that there is a possibility of a faster rise than $\ln^2s$,
without violating unitarity.
Moreover, as recalled above, the FLM upper bound is an asymptotic result ($s\to\infty$) and presently we investigate data up to $\sim 10$~TeV.

We also compare two possible analytic methods to connect the real and imaginary parts of 
the forward elastic amplitude (and consequently $\sigmatot$ and $\rho$):
the Derivative Dispersion Relations (DDR) and the Asymptotic Uniqueness (AU),
which is based on the Phragmén-Lindelöff theorems.
The main point is to confront the DDR and AU methods with focus on the leading $\ln^\gamma s$ contribution ($\gamma$ real),
as well as to stress the importance of this component in the empirical and phenomenological contexts.

At last, we mention that all data on $\sigmatot$ and $\rho$ obtained
at the LHC by TOTEM and ATLAS Collaborations at 7 and 8~TeV are included in the dataset.
Moreover, we take account of the tension observed in the measurements of 
$\sigmatot$ at 8 TeV by TOTEM and ATLAS~\cite{ATLAS:2016}.

The chapter is  organized as follows. In Section~\ref{sec:rise_ReggeGribov} we present a
short review of results from the Regge-Gribov formalism that are of interest here, including also some COMPETE and PDG results,
as well as results obtained with the Amaldi \etal parametrization.
In Section~\ref{sec:rise_RealImagParts} we discuss the two methods employed in the analytical
connection between the real and imaginary parts of the amplitude and we derive the formulas from which we will construct our parametrizations.
The parametrizations (also called here as Models) and their notation are defined in Section~\ref{sec:rise_analytic_models}.
Some critical comments on experimental data presently available are presented in Section~\ref{sec:rise_expdata} and the fit results in Section~\ref{sec:rise_results}.
In Section~\ref{sec:rise_discussion_comments} we discuss and compare all the results obtained. Finally, a summary and our conclusions are presented
in Section~\ref{sec:rise_conclusions}.

\section{Regge-Gribov Formalism}
\label{sec:rise_ReggeGribov}

The Regge-Gribov formalism is based on the analytical continuation of
the partial wave expansion of the elastic scattering amplitude to \textit{complex} angular momentum.
This process results in the so-called Watson-Sommerfeld transform.
The main results of this approach are usually used as parametrization to describe the energy dependence of the total cross section.

This approach was first formulated by T. Regge at the end of 1950s, therefore before the birth of QCD, 
for non-relativistic Quantum Mechanics. Later, in the 1960s, this approach was generalized by V. Gribov, Chew and Frautschi and
other authors to the relativistic case.

In this Section, we present the main results that are of interest in this work.
We refer to classical texts, like Collins' book~\cite{Collins_book:1977},
Svensson's lectures \cite{Svensson:1967} and also more recent reviews by Barone and Predazzi~\cite{Barone_Predazzi_book:2002} 
and Donnachie \etal~\cite{Donnachie_etal_book:2002} for further details.
We also discuss results that were obtained by other authors using parametrizations inspired by Regge-Gribov results.

\subsection{Main Results}
\label{subsec:rise_Regge_main_results}

While performing the analytical continuation to complex angular momentum $\ell$ values, we assume the presence
of simple poles in the plane, which we will denote by $\alpha(t)$. In this framework, the elastic scattering
amplitude in the asymptotic limit is expressed by~\cite{Barone_Predazzi_book:2002,Collins_book:1977}
\begin{equation}
F(s,t) =  \sum_i \beta_i(t) \xi_i(t) s^{\alpha_i(t)},
\label{eq:amp-Regge}
\end{equation}

\noindent where the sum runs over all poles in the $\ell$-plane, $\beta_i(t)$ is called the residue function (strength) 
and $\xi_i(t)$ the signature factor, given by
\begin{equation}
\xi_i(t) = - \frac{1 + \zeta_i e^{-i \pi\alpha_i(t)}}{\sin \pi \alpha_i(t)}, 
\label{eq:signature-factor-Regge}
\end{equation}

\noindent with $\zeta = + 1$ or  $\zeta = - 1$ for analytic continuations 
through even or odd integer values of the angular momenta, respectively. 

The poles $\alpha(t)$ are also called \textit{Regge trajectories} or \textit{Reggeons} and correspond to objects that are
exchanged in the $t$-channel between the interacting particles. Contrary to electromagnetic interactions
where an elementary particle is exchanged, the Reggeon corresponds to a family of particles that may be exchanged, which have in common
the quantum numbers of the interaction. Chew and Frautschi realized~\cite{Chew:1961,Chew:1962} that by grouping mesons or baryons that have in common 
the same quantum numbers (except spin) into families and ploting their spin as a function of their masses, a regularity can be found. 
This plot, called Chew-Frautschi diagram, gives us the $t$-dependence of the Regge trajectories, and they have approximately a linear dependence
\begin{equation}
 \alpha(t) = \alpha(0) - \alpha't,
 \label{eq:Regge-trajectory}
\end{equation}

\noindent where $\alpha(0)$ and $\alpha'$ are the intercept and the slope of the Reggeon, respectively.

\begin{figure}[htb]
 \centering
 \includegraphics[scale=1]{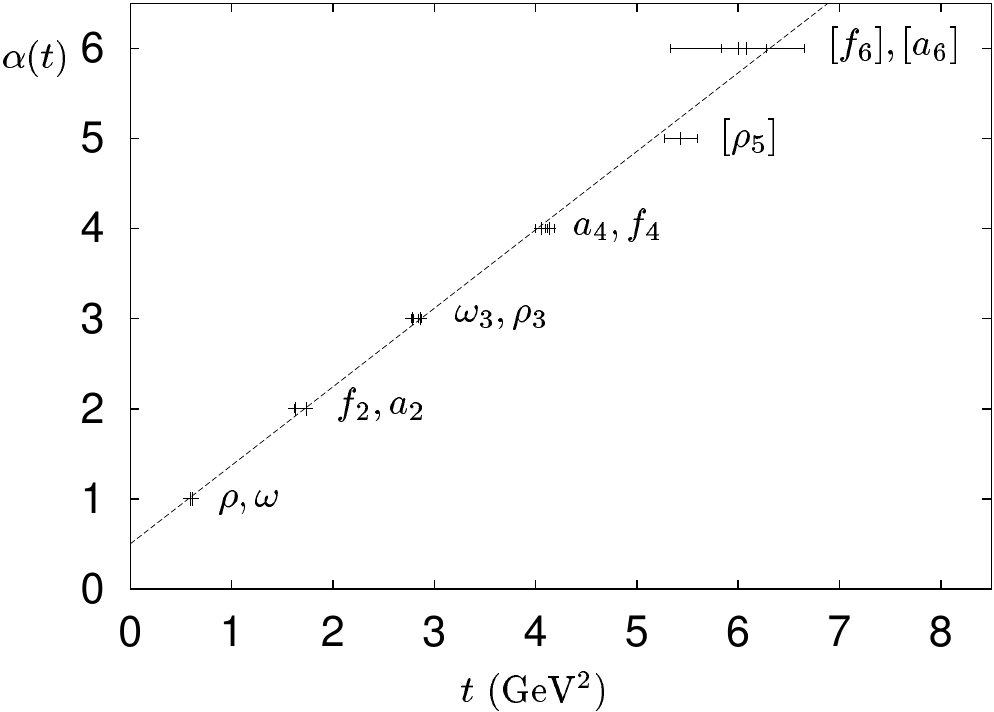}
 \caption{\label{fig:chew-frautschi}Chew-Frautschi plot for mesonic families $a_2/f_2$ and $\rho/\omega$ and the corresponding trajectory $\alpha(t)$. Figure taken from Ref.~\cite{Donnachie_etal_book:2002}.}
\end{figure}

In Fig.~\ref{fig:chew-frautschi}\footnote{Reproduced with permission from \textit{Pomeron Physics and QCD} by Sandy Donnachie,
Guenter Dosch, Peter Landshoff and Otto Nachtmann (Cambridge University Press, 2002).}, 
we show the Chew-Frautschi plot for the mesonic families $a_2/f_2$ ($+$) and $\rho/\omega$ ($-$).

In what concerns the study of forward quantities like $\sigmatot$, this linear dependence is enough to describe the data. 
The non-linearity of the Regge trajectory is associated with structures in the differential 
cross section at small $|t|$~\cite{Fagundes_etal:2015c}. 
From the experimental point of view, this structure has negligible effects on the total cross section~\cite{TOTEM:2015}.

One can show \cite{Barone_Predazzi_book:2002} that in the $t$-channel, where $t>0$, the simple pole contribution to the amplitude, namely
\begin{equation*}
 \frac{1}{\ell-\alpha(t)},
\end{equation*}

\noindent has the form of the Breit-Wigner amplitude for resonances when $t$ equals the squared 
mass of a particle lying on the Regge trajectory.

Next, we present the full contribution of a Reggeon to the elastic scattering amplitude and discuss some particular cases.


\subsection{Simple Pole Contributions to the Amplitude}
\label{subsec:rise_Regge_simple_pole}

In this section, we discuss in more detail the results from the Regge-Gribov formalism and their connections with
experimental data. Besides the Reggeon concept, we discuss the \textit{Pomeron}, a special Reggeon that was introduced
to describe the rise of the total cross section.

\subsubsection{Reggeon Contribution}

Let us consider the contribution of a single Reggeon exchange in Eq.~\eqref{eq:amp-Regge}~\cite{Collins_book:1977,Barone_Predazzi_book:2002}
\begin{equation}
F(s,t) =  \beta(t) \xi(t) s^{\alpha(t)}.
\nonumber
\end{equation}

Recalling that the Reggeon can have an even ($+$) or odd ($-$) signature and
writing explicitly the signature factor, Eq.~\eqref{eq:signature-factor-Regge}, for each case, we have
\begin{align}
\xi_{+}(t) & = - \cot \frac{\pi}{2} \alpha(t) + i,\label{eq:Regge-even-sig-factor}\\
\xi_{-}(t) & = - \tan \frac{\pi}{2} \alpha(t) - i,\label{eq:Regge-odd-sig-factor}
\end{align}

\noindent so that the \textit{forward} ($t = 0$) complex even and odd \textit{amplitudes} read
\begin{align} 
F_{+}(s,0) &= \beta_{+}(0) \left[i - \cot \frac{\pi}{2} \alpha_{+}(0)\right] s^{\alpha_{+}(0)}, \label{eq:Reggeon-even-amplitude} \\[10pt] 
F_{-}(s,0) &=- \beta_{-}(0)\left[i + \tan \frac{\pi}{2} \alpha_{-}(0)\right] s^{\alpha_{-}(0)}. \label{eq:Reggeon-odd-amplitude}
\end{align}

These are crossing \textit{symmetric} and \textit{antisymmetric} functions of the energy,
\begin{equation}
F_{+}(- s) = + F^{*}_{+} (s), 
\qquad
F_{-}(- s) = - F^{*}_{-} (s), 
\end{equation}

\noindent where $*$ denotes complex conjugation. The physical amplitudes for $pp$ and $\ppbar$ scattering
are given by~\cite{Eden_book:1967},
\begin{equation}
F_{pp} = F_{+} + F_{-}, \qquad F_{\bar{p}p} = F_{+} - F_{-}.
\label{eq:Regge-pp-ppbar-even-odd}
\end{equation}

Using the optical theorem~\eqref{eq:rise-optical-theo} and denoting the strengths and intercepts by
\begin{equation}
\beta_{+}(0) = a_1, \qquad \alpha_{+}(0) - 1 = - b_1, \qquad 
\beta_{-}(0) = a_2, \qquad \alpha_{+}(0) - 1 = - b_2,
\nonumber
\end{equation}

\noindent the Reggeons ($\reggeon$) contributions to the total cross section and $\rho$ parameter [Eq.~\eqref{eq:rise-rho-def}] are given by
\begin{align}
\sigma^{\reggeon}(s) & = a_1 s^{-b_1} + \tau a_2 s^{-b_2}, \label{eq:Reggeon-sigtot}\\[10pt]
\rho^{\reggeon}(s) &= \frac{1}{\sigma^\reggeon(s)} \left\{- a_1 \tan \left(\frac{\pi b_1}{2}\right) s^{- b_1} + 
\tau a_2 \cot \left(\frac{\pi b_2}{2} \right) s^{-b_2}\right\} \label{eq:Reggeon-rho}
\end{align}

\noindent where $\tau = -1$ for $pp$ and $\tau = +1$ for $\bar{p}p$.

For $pp$ and $\ppbar$ scattering, the trajectories that contribute most to the amplitude correspond
to the non-degenerated $a_2/f_2$ and $\omega/\rho$ mesonic families with even and odd signature, respectively.
Additionally, from the Chew-Frautschi plot for the $a_2/f_2$ and $\omega/\rho$ trajectories (see Fig.~\ref{fig:chew-frautschi}),
$b_1 > 0$ and $b_2 >0$ (approximately $1/2$).
Therefore, these cross sections are decreasing functions of the energy, 
in agreement with the experimental data for energies $\lesssim 20$ GeV (Fig.~\ref{fig:sigmatot_rho_data_pp_ppbar}). 

For further reference we note that the above results imply
\begin{equation}
\Imag F_+(s) = a_1s^{-b_1} > 0 \quad \text{and} \quad
\Imag F_-(s) = - a_2s^{-b_2} < 0.
\label{eq:Regge-Imag-even-odd-amp}
\end{equation}

\noindent Associated with the $\tau$ values, we then have $\sigma_{\bar{p}p} > \sigma_{pp}$ in the energy region
where these cross sections are not equal, also in agreement with the experimental data. 

\subsubsection{Pomeron Contribution}

The value of the intercept $\alpha(0)$ is bounded by Unitarity: $\alpha (0) \leq 1$~\cite{Barone_Predazzi_book:2002}.
The contribution ot the total cross section of this maximum-allowed value ($\alpha(0)=1$) reads
\begin{equation}
 \sigmatot^\text{max}(s) \sim a s^{\alpha(0)-1} = a.
\end{equation}

Therefore, it contributes as a \textit{constant} to $\sigmatot$ and constitutes a special
trajectory called \textit{critical Pomeron} (or constant Pomeranchuk limit).

At the time (late 1960s), it was expected that the total cross section would assume a constant value for high-enough energies \cite{Barger:1968}.
Therefore, the idea of the critical Pomeron was the perfect explanation: at low energies we have the dominance of the Reggeon exchanges and for energies
above 20 GeV the critical Pomeron becomes dominant and the total cross section assumes a constant value.
Moreove, according to the Pomenranchuck theorem~\cite{Barone_Predazzi_book:2002},
the asymptotic value of particle-particle and antiparticle-particle total cross sections should be the same indicating that
the Pomeron must be associated to the symmetric amplitude, having therefore an even signature. Since in the elastic scattering the particles do not
have their quantum numbers changed, the Pomeron carries the quantum numbers of the vacuum.

This scenario changed in the 1970s when experimental results by 
the IHEP-CERN Collaboration at Serpukhov and at the CERN-ISR indicated the
\textit{rise} of $\sigmatot$ above $\sim$ 20 GeV. To describe this rise within the Regge approach, 
an \textit{ad hoc} trajectory has been introduced, with intercept slightly greater than one, so that we have
a \textit{contribution that increases with the energy}. This even trajectory
(in order to account for an asymptotic equality between $pp$ and $\bar{p}p$ scattering)
has been associated with a simple pole in the amplitude, corresponding
to a power law in $s$, with \textit{positive} exponent slightly greater than 0. This trajectory is called
\textit{supercritical Pomeron} or \textit{soft Pomeron}.

Denoting the soft Pomeron ($\pomeron$) intercept $\alpha_\pomeron(0) = 1 + \epsilon$
(with $\epsilon \gtrsim 0$) and $\delta=\beta_\pomeron(0)$ the strength, 
we obtain for a simple pole ($S$) Pomeron:
\begin{align}
\frac{\Imag F_{S}^{\pomeron}(s)}{s} &= \delta s^{\epsilon}, \ \epsilon > 0, \label{eq:Pomeron-soft-simple-imag} \\[10pt]
\frac{\Real F_{S}^{\pomeron}(s)}{s} &= \delta \tan \left(\frac{\pi \epsilon}{2}\right) s^{\epsilon}.
\label{eq:Pomeron-soft-simple-real}
\end{align}

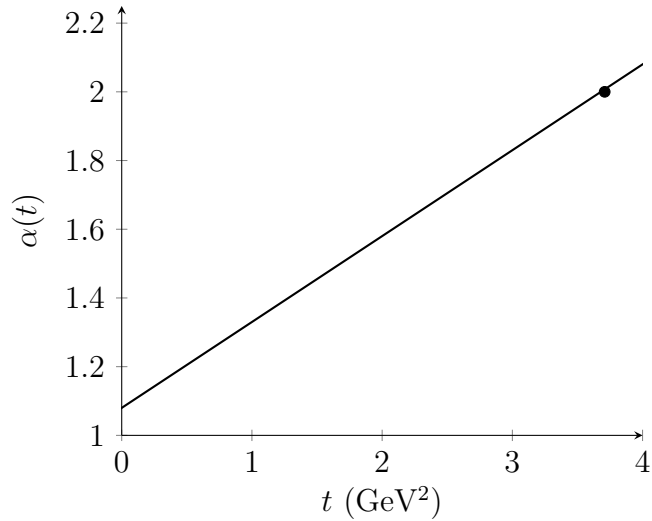
\begin{figure}[htb!]
 \centering
 \begin{tikzpicture}
 \begin{axis}[axis lines = left, xlabel={$t$ (GeV$^2$)},ylabel={$\alpha(t)$},xmin=0, xmax=4, ymin=1, ymax=2.25]
  \addplot[color=black,thick]{1.08 + 0.25*x};
  \addplot[mark=*]
  coordinates {(3.709, 2)};
 \end{axis}
\end{tikzpicture}
 \caption{\label{fig:pomeron_trajectory}Soft Pomeron trajectory with parameters $\epsilon\sim 0.08$ and $\alpha_\pomeron'\sim 0.25$ GeV$^2$
 and the $2^{++}$ glueball candidate \cite{Abatzis:1994}.}
\end{figure}

Contrary to the Reggeon trajectories, the soft Pomeron trajectory does not have any particle on it. 
Its intercept and slope are estimated from fits to $\sigmatot$ and $d\sigma/dt$ data~\cite{Barone_Predazzi_book:2002,Donnachie_etal_book:2002}:
$\epsilon \sim 0.08$ and $\alpha_\pomeron' \sim 0.25$~GeV$^{-2}$.
The candidates to be in the Pomeron trajectory, which must have the quantum numbers of the vacuum, are the \textit{glueballs}.
With the above parameters, there is a glueball candidate ($2^{++}$) \cite{Abatzis:1994} that may be on the trajectory,
as illustrated in Fig.~\ref{fig:pomeron_trajectory}.

Although above the limit imposed by unitarity for the intercept, the soft Pomeron only violates the Froissart-Martin bound
at asymptotic energies. At present energies (LHC),
this $s^\epsilon$ term is below the numerical limit given by Eqs.~\eqref{eq:rise-FLM-bound}
and \eqref{eq:rise-FLM-bound-coeff}.
Moreover, the analytic model with two non-degenerated Regge trajectories and the soft Pomeron
\begin{align}
\sigmatot(s) & = a_1 \left[\frac{s}{s_0}\right]^{-b_1} + \tau a_2 \left[\frac{s}{s_0}\right]^{-b_2} +
\delta \left[\frac{s}{s_0}\right]^{\epsilon}, \label{eq:sigtot-2Reggeons-SoftPomeron}\\[10pt]
\rho(s) & = \frac{1}{\sigmatot(s)} \left\{- a_1 \tan \left(\frac{\pi b_1}{2}\right) \left[\frac{s}{s_0}\right]^{- b_1}\!\!\! + 
\tau a_2 \cot \left(\frac{\pi b_2}{2} \right) \left[\frac{s}{s_0}\right]^{-b_2} \!\!\! + 
\delta \tan \left(\frac{\pi \epsilon}{2}\right) \left[\frac{s}{s_0}\right]^{\epsilon}\right\},\label{eq:rho-2Reggeons-SoftPomeron}
\end{align}

\noindent with $\tau = -1 (+1)$ for $pp$ ($\bar{p}p$) and
where we have included the energy scale $s_0$, present consistent descriptions of
the LHC data, as recently discussed by Donnachie and Landshoff~\cite{Donnachie:2013}
and also Menon and Silva~\cite{Menon_Silva:2013b}.

A fundamental aspect of this simple pole Pomeron contribution
is the fact that the power law with positive exponent ($\epsilon \sim 0.08 - 0.09$, for example, \cite{Donnachie:2013,Menon_Silva:2013b})
implies in a strict rise of the associated cross section with energy.
This behavior is directly related to
the standard or original concept of the super-critical (or soft) Pomeron,
i.e. a \textit{rising cross section}. However, this is not the case in 
some amplitude analyses, in which this behavior is not reproduced in the whole
range of investigated energy (as will be discussed in Section~\ref{subsec:rise_Regge_COMPETE_PDG}).

As mentioned in the introduction, a QCD description of the soft Pomeron is not yet available. However, in the
context of the perturbative QCD, the simplest picture of the Pomeron is the exchange of two gluons~\cite{Low:1975,Nussinov:1975}.
Within this approach, nonperturbative effects in the gluon propagator are usually included 
to obtain finite results for forward quantities, see for instance Refs.~\cite{Cudell:1991,Halzen:1992,Canfora:2017}.
Moreover, in the context of the BFKL equation, the (hard) Pomeron corresponds to the exchange
of a ladder of reggeized gluons between the partons~\cite{Barone_Predazzi_book:2002}.

\subsection{Contributions of Higher-order Poles to the Amplitude}
\label{subsec:rise_Regge_higher_order_poles}

In the Regge context, we have worked with the presence of simple poles in the complex $\ell$-plane,
which gives us the power-law contribution $s^{\alpha}$ to the amplitude.
Although not entirely justified within the Regge-Gribov formalism, it is mathematical possible
to have poles of higher order in the complex plane~\cite{Eden_Landshoff:1964}. 
As discussed in~\cite{Donnachie_etal_book:2002}, these higher-order
poles are associated with the $n$-th derivative with respect to $\alpha$, where $n+1$ is the order of the pole.
Specifically, the $n$-th derivative of a simple pole reads
 \begin{equation}
 \frac{d^n}{d \alpha^n}\left[\frac{1}{\ell - \alpha}\right] = 
 (-1)^{n-1} \frac{n!}{[\ell - \alpha]^{n+1}},
 \quad
 n = 1, 2,\dots,
 \label{eq:higher-order-pole-l-plane}
 \end{equation}
 
 \noindent and for the power law,
\begin{equation}
\frac{d^n}{d \alpha^n} s^{\alpha} = s^{\alpha} \ln^n s, \quad n = 1, 2,\dots.
\label{eq:higher-order-pole-power-law}
\end{equation}

Therefore, associated with a pole of order $N=n+1$ ($t$-channel), the contribution to the amplitude
in the $s$-channel
is $s^{\alpha} \ln^{N-1}(s)$. Considering a Pomeron ($\pomeron$) with intercept $\alpha_\pomeron(0) = 1$, 
the contribution to the total cross section is 
 \begin{equation}
\sigma^{\pomeron}(s) = \frac{\Imag F(s)}{s} \propto \ln^{N-1} s.
\nonumber
\end{equation}

Now, taking into account the Froissart-Lukaszuk-Martin bound, Eq.~\eqref{eq:rise-FLM-bound}, the possible leading contributions 
are $\ln s$ (double pole) or $\ln^2 s$ (triple pole).

We recall that in these cases, and yet in the context of Regge-Gribov, the real part of the amplitude can
be evaluated through a representation of the Watson-Sommerfeld integral, introduced by
Gribov and Migdal at the end of 1960s~\cite{Gribov_Migdal:1968a, Gribov_Migdal:1968b}.
In the forward direction, we have (see Eq. 44.b in Ref.~\cite{Gribov_Migdal:1968b})
\begin{equation}
\frac{\Real F(s,0)}{s} = \frac{\pi}{2} \frac{d}{d\ln s} \left[\frac{\Imag F(s,0)}{s} \right].
\end{equation}

As will be discussed in Section~\ref{subsec:rise_DDR}, this relation corresponds to the first order series-expansion
of a derivative dispersion relation for even amplitudes. For the double pole ($D$) and triple pole
($T$) we have, respectively
\begin{align}
\frac{\Imag F^\pomeron_{D}(s)}{s} = \beta \ln s \qquad & \rightarrow \qquad \frac{\Real F^\pomeron_{D}(s)}{s} = \frac{\pi}{2}\beta,\label{eq:Regge-Gribov-double-pole-1storder}\\[10pt]
\frac{\Imag F^\pomeron_{T}(s)}{s} = \beta \ln^2 s \qquad & \rightarrow \qquad \frac{\Real F^\pomeron_{T}(s)}{s} = \pi \beta\ln s. \label{eq:Regge-Gribov-triple-pole-1storder}
\end{align}

It is important to note that the logarithmic laws demand an energy scale $s_0$ for consistency,
implying in a null value at $s = s_0$. More importantly, if the energy
increases in the region $s < s_0$, the double-pole contribution increases through \textit{negative values} and the
triple pole contribution \textit{decreases} through positive values until $s=s_0$.
Obviously, both a negative contribution and a decreasing contribution,
as the energy increases are not consistent with the standard 
super-critical (soft) Pomeron concept.

\subsection{COMPETE and PDG Analyses}
\label{subsec:rise_Regge_COMPETE_PDG}

In this Section, we summarize two important analyses on the forward data performed by 
the COMPETE Collaboration~\cite{COMPETE:2002a,COMPETE:2002b} and by the COMPAS Group (IHEP, Protvino)
published in the Review of Particle Physics (RPP) by the Particle Data Group (PDG)~\cite{PDG:2012,PDG:2014,PDG:2016}.

These analyses consider parametrizations inspired by the Regge-Gribov formalism. In special, the results obtained by
COMPETE Collaboration have become a sort of a guide to describe the energy evolution of $\sigmatot$.

\subsubsection{COMPETE Collaboration}
 
The analysis made by the COMPETE Collaboration consisted in a great effort to
find the most suitable parametrization to describe the data of $\sigmatot$ and 
$\rho$ for several reactions available at the time (2002). They considered several 
parametrizations based on Regge-Gribov Theory (for instance, soft Pomeron, double and triple-pole Pomeron)
and tested several properties, for instance, the hypothesis of factorization of the residues 
of the Regge poles and universality of leading term, among others. 
To find the best parametrization, they created a ranking 
procedure (see Ref.~\cite{COMPETE:2002a} for details), resulting as the highest-rank
the parametrization that corresponds to two Reggeon exchanges, a critical Pomeron and a triple-pole Pomeron with $\alpha_\pomeron(0)=1$,
namely (using their notation)
\begin{equation}
\sigmatot^{a_\mp b}(s) = Y_1^{ab} \left(\frac{s}{s_1}\right)^{\alpha_1(0)-1} \pm Y_2^{ab}\left(\frac{s}{s_1}\right)^{\alpha_2(0)-1} 
+ Z^{ab} + B\ln^2\left(\frac{s}{s_0}\right),
\label{eq:sigtot-par-COMPETE}
\end{equation}

\noindent where we are considering the scattering between particles $a$ and $b$ and $s_1 = 1$ GeV$^{2}$.

The fits to $\sigmatot$ data were constrained to the real part of the forward elastic 
amplitude through the $\rho$ data (when available) using Derivative Dispersion Relations~\cite{Bronzan_Kane_Sukhatme:1974,Avila_Menon:2004}
(that will be discussed with more details in Section~\ref{subsec:rise_DDR}).
 
Their dataset (with cutoff energy $\sqrt{s_\text{min}}= 5$ GeV) includes the following 
reactions: $\sigmatot$ of $pp$, $\ppbar$, $\pi^\pm p$, $K^\pm p$, $\Sigma^-p$, $\gamma p$, 
and $\gamma\gamma$ scatterings and $\rho$ data for $pp$, $\ppbar$, $\pi^\pm p$, and 
$K^\pm p$ scatterings. They also included cosmic-ray data for $pp$.
 
As already mentioned in the previous section, the use of a logarithmic function demands the inclusion of
an energy scale. 
In COMPETE analysis, the energy scale $s_0$ is a \textit{free parameter} and universal, 
i.e., does not depend on the scattering particles. 
As we will see, different analyses consider different energy scales. Therefore, it is important to keep in
mind that the value of the energy scale inside the logarithmic term can affect the value of the coefficient  
$B$ obtained in the fit due the correlation among the parameters. 
For further reference, the result obtained by COMPETE for their $B$ and $s_0$ parameter are
\begin{equation}
 B_\text{COMPETE} = 0.3152 \pm 0.0095 \text{ mb}\quad \text{and} \quad s_0 = 34.0 \pm 5.4 \text{ GeV}^2.
 \label{eq:result-compete}
\end{equation}

We call the attention to the fact that the energy scale obtained by COMPETE is greater than the energy cutoff considered,
$s_\text{min} = 25 \text{ GeV}^2$. As a consequence, the leading term $\ln^2(s/s_0)$ decreases as the energy increases
in the range $s_\text{min} \leq s \leq s_0$, contradicting the standard soft Pomeron concept, as already mentioned in
Section~\ref{subsec:rise_Regge_simple_pole}
and discussed in more detail by Menon and Silva in Ref.~\cite{Menon_Silva:2013a}, Section 4.2.
 
\subsubsection{COMPAS Group (PDG)}
 
The fits to $\sigmatot$ and $\rho$ data presented in the Review of Particle Physics (RPP)
by the Particle Data Group (PDG) \cite{PDG:2014} have been performed with the same highest rank parametrization
selected by COMPETE, except for some changes in the energy scale that have been done in the last years.

Until the 2010 edition~\cite{PDG:2010}, they have used exactly the same highest-rank result obtained by COMPETE. 
From the 2012 edition on, they introduced two modifications in the parametrization.
The first consists in the substitution of  the energy scale $s_1$ and $s_0$
appearing, respectively, in the Reggeons terms and in the leading $\ln^2 s$ term by
the new scale $s_M$ given by
\begin{equation}
s_M= (m_a + m_b +M)^2,
\label{eq:pdg2012-energy-scale}
\end{equation}

\noindent where $m_a$ and $m_b$ are the masses of the particles and $M$ is a free universal parameter,
i.e. independent of the particles. Therefore, the energy scale is now reaction-dependent. 
The second involves the pre-factor $H$ of the $\ln^2 s$ term. Although still universal, in 2012 edition, the authors
have written it in terms of the new parameter $M$
\begin{equation}
 H_\text{PDG2012} = \pi \frac{(\hbar c)^2}{M^2}.
 \label{eq:pdg2012-B-parameter}
\end{equation}

\noindent We recall that in the 2010 edition, the $H$ parameter was independent.
This structure was maintained in the subsequent editions, 2014~\cite{PDG:2014} and 2016~\cite{PDG:2016}.

Concerning the dataset, they enlarged the one used by COMPETE including data from $pn$, $\bar{p}n$,
$pd$, $\bar{p}d$, $\pi^\pm d$, $K^\pm d$ and $\gamma d$, where $d$ stands for the deuteron.
Over the years, they have also updated the dataset with new experimental 
information for $pp$ scattering: cosmic-ray data at the energy range of RHIC 
($\sim$ 200 GeV) obtained by the ARGO-YBJ Collab. \cite{ARGO_YBJ:2009} and the datum at 57 TeV by the Pierre Auger
Observatory \cite{PierreAuger:2012}. Of course, they also added the data obtained at the LHC (TOTEM and ATLAS).
These datasets are available as \texttt{ASCII} files on the PDG website \cite{PDG_data_website}.

In particular, in the 2014 edition~\cite{PDG:2014}, they have explored the possibility to vary 
$\sqrt{s_\text{min}}$, selecting three values: 5, 6 and 7 GeV. For 5 GeV (the same cutoff energy used by COMPETE),
the result does not describe the TOTEM data at 7 TeV, besides presenting $s_0>s_\text{min}$ (decreasing Pomeron contribution).
They argue that the best result obtained is for $\sqrt{s_\text{min}}=7$ GeV, describing the aforementioned data at 7 TeV.
 
Their results for $M$, $H$ and for $s_M^{pp}$ ($pp$ and $\ppbar$ scattering) in these representative cases are
\begin{equation}
\sqrt{s_\text{min}}=5 \text{ GeV}:\left\{
\begin{array}{l}
  M = 2.127\pm 0.015 \text{ GeV},\\
  H_\text{PDG2014} = 0.2704 \pm 0.0038\text{ mb},\\
  s_M^{pp} = 16.03 \pm 0.12 \text{ GeV}^2,                                   
\end{array}\right.
 \label{eq:result-pdg2014-5GeV}
\end{equation} 
\begin{equation}
\sqrt{s_\text{min}}=7 \text{ GeV}:\left\{
\begin{array}{l}
  M = 2.076\pm 0.016 \text{ GeV},\\
  H_\text{PDG2014} = 0.2838 \pm 0.0045\text{ mb},\\
  s_M^{pp} = 15.62 \pm 0.13 \text{ GeV}^2.                                  
\end{array}\right.
 \label{eq:result-pdg2014-7GeV}
\end{equation} 


In the last edition (2016), they have considered only one energy cutoff, $\sqrt{s_\text{min}} = 5$~GeV.
The parameters of interest here are
\begin{equation}
 \begin{array}{l}
  M = 2.1206\pm 0.0094 \text{ GeV},\\
  H_\text{PDG2016} = 0.2720 \pm 0.0024\text{ mb},\\
  s_M^{pp} = 15.977 \pm 0.075\text{ GeV}^2.                                  
\end{array}\label{eq:result-pdg2016}
\end{equation}

\subsection{Parametrization by Amaldi \etal}
\label{subsec:rise_Regge_Amaldi}

In 1977, Amaldi \etal have proposed~\cite{Amaldi:1977} an \textit{empirical} parametrization for the leading term of $\sigmatot$ in order
to fit the data obtained in the ISR/CERN at the energies $\sqrt{s} = 30.6,\, 44.7,\,52.9,\,62.4$ GeV for $pp$ scattering.
In the fits, they considered data from $pp$ and $\ppbar$ scattering. The full parametrization, with their notation, for the total cross section reads
\begin{align}
 \sigma_{\pm} & =  C_1E^{-\nu_1} \mp C_2E^{-\nu_2} + \sigma_{\infty}\label{eq:par-amaldi1}\\
 \sigma_\infty & =  B_1 + B_2\{\ln(s)\}^\gamma\label{eq:par-amaldi2},
\end{align}

\noindent where $\sigma_{+}$ ($\sigma_{-}$) denotes the total cross section for $pp$ ($\bar{p}p$) scattering, 
$E$ is the energy in the laboratory system and $\gamma$ is a free parameter to be determined in the fit, allowed to assume real values,
and it is this parameter that consists in the novelty of the work. 
We note that this parametrization also considers that the $pp$ and $\ppbar$
cross sections will be equal at high energies, typical of an even contribution, like the Pomeron. 
Although not mentioned by the authors,
a scale fixed at 1 GeV$^2$ in the logarithm is assumed.

The fits were constrained with the $\rho$ parameter data. The connection between the real and imaginary parts of the amplitude
was done by using Integral Dispersion Relations (see Section~\ref{subsec:rise_IDR}). As will be discussed later, the use of integral relations
with a $\ln^\gamma s$ term demands \textit{numerical integration}.

With data in the range 5~GeV~$< \sqrt{s} \leq 62.4$~GeV, Amaldi \etal obtained
\begin{equation}
 \gamma_\text{Amaldi} = 2.10 \pm 0.10.
 \label{eq:result-amaldi}
\end{equation}

Two decades later, the UA4/2 Collaboration~\cite{Augier:1993} performed fits with the same Amaldi \etal parametrization,
but now using an updated dataset ($\sqrt{s_\text{max}}=546$~GeV) and energy scale explicitly fixed in 1~GeV$^2$.
Their result reads
\begin{equation}
 \gamma_\text{UA4/2} = 2.25^{+0.35}_{-0.31}.
 \label{eq:result-ua42}
\end{equation}

Afterwards, a similar study was performed by Bueno and Velasco~\cite{Bueno_Velasco:1996}, in which two parametrizations were considered:
the Amaldi parametrization, Eqs.~\eqref{eq:par-amaldi1} and \eqref{eq:par-amaldi2}, and the Donnachie-Landshoff parametrization~\cite{Donnachie:1992},
with the leading term corresponding to the soft Pomeron, Eq.~\eqref{eq:Pomeron-soft-simple-imag}. Tests on the cutoff ($\sqrt{s_\text{min}}$) were also done.
For $\sqrt{s_\text{min}}=5$ GeV the same value of Eq.~\eqref{eq:result-ua42} was obtained and for $\sqrt{s_\text{min}} = 10$~GeV the result was:
\begin{equation}
 \gamma = 2.64^{+0.50}_{-0.32}.
 \label{eq:result-buenovelasco}
\end{equation}

\noindent Once more, the energy scale was fixed, in an arbitrary way, at $s_0=1$~GeV$^2$.

In 2012, after the first measurement of the total cross section at 7~TeV by the TOTEM 
Collaboration, Fagundes, Menon and Silva (hereafter FMS) developed an amplitude analysis
based on the parametrization introduced by Amaldi \etal for the total cross section~\cite{Fagundes_Menon_Silva:2012a}.
In this work, the data reductions were also limited to $pp$ and $\bar{p}p$ scattering.
This analysis was then developed and extended in~\cite{Fagundes_Menon_Silva:2013a} and further updated 
and discussed by Menon and Silva~\cite{Menon_Silva:2013a,Menon_Silva:2013b},
leading to values of $\gamma$ in the interval 2.2 - 2.4~\cite{Menon_Silva:2013a}.
Including the first measurement at 8~TeV by the TOTEM Collaboration, the simultaneous fit to
$\sigmatot$ and $\rho$ resulted in~\cite{Menon_Silva:2013b}
\begin{equation} 
\gamma = 2.23 \pm 0.11.
\nonumber
\end{equation}

In these works, when the $\rho$ data is predicted from fits to $\sigmatot$~\cite{Fagundes_Menon_Silva:2012a} 
or included in the fits~\cite{Fagundes_Menon_Silva:2013a,Menon_Silva:2013a,Menon_Silva:2013b}, 
the analytic connection between the real and imaginary parts was done using \textit{Derivative Dispersion Relations}
(Section~\ref{subsec:rise_DDR}).

The COMPAS group also mentioned fits with this type of parametrization. They quote
a result obtained with a leading term $\ln^c(s/s_M)$ with $c=1.98\pm 0.01$ without further details of the analysis \cite{PDG:2014,PDG:2016}.

\section{Connecting the Real and Imaginary Parts of Elastic Scattering Amplitude}
\label{sec:rise_RealImagParts}

As commented in the introduction of this chapter, we are interested in simultaneous
analyses of $\sigmatot$ and $\rho$ parameter data. Therefore, we need to connect the imaginary
and real parts of the amplitude. In this section, we present three ways of attaining this:
Integral Dispersion Relations, Derivative Dispersion Relations, and Asymptotic Uniqueness Relations.
Precisely, what we do is to determine the real part from the imaginary part, which, in turn, is
given by the parametrization to $\sigmatot$.

\subsection{Integral Dispersion Relations and the High-Energy Approximation}
\label{subsec:rise_IDR}

Integral Dispersion Relations (IDR) are useful tools when dealing with complex functions.
They are derived from Analyticity and Causality and connect the real and imaginary part
of a complex-valued function by means of a Hilbert integral transform~\cite{Byron_Fuller_book:1970}.

Here, after recalling the analytic result, we discuss the important concept of the \textit{high-energy approximation},
which led to the introduction of an \textit{effective subtraction constant}
in the analyses by Fagundes, Menon and Silva, also employed in this thesis.

In high-energy forward ($t=0$) scattering, these relations are written in terms
of the even/symmetric ($+$) and the odd/antisymmetric ($-$) amplitudes
\begin{equation}
\begin{split}
F_{+}(-s) & = F_{+}(se^{i \pi}) = F_{+}^{*}(s),\\
F_{-}(-s) & = F_{-}(se^{i \pi}) = - F_{-}^{*}(s),
\end{split}
\label{eq:even-odd-amplitudes}
\end{equation}

\noindent where the asterisk (*) denotes complex conjugation and for simplicity we have omitted the argument $t=0$.
The hadronic amplitudes are recovered from the even and odd amplitudes through the relations
\begin{equation}
\begin{split}
F_{pp}(s) & = F_{+}(s) + F_{-}(s),\\
F_{\ppbar}(s) & = F_{+}(s) - F_{-}(s).
\end{split}
\label{eq:hadronic-amplitudes}
\end{equation}

Usually, in order to guarantee the convergence of the integral, 
the analysis of $\sigmatot$ demands singly-subtracted integral dispersion relations, 
with the subtraction present in the even part of the amplitude. 
In terms of the c.m. energy squared, the IDR read \cite{Goldberger:1957,Soding:1964}
\begin{align} 
\frac{\Real F_{+}(s)}{s} & = \frac{K}{s} + \frac{2s}{\pi}\,P\int_{s_\text{th}}^{\infty} ds' \left[\frac{1}{s'^2-s^2}\right] \frac{\Imag F_{+}(s')}{s'},\label{eq:IDR-even}\\[10pt] 
\frac{\Real F_{-}(s)}{s} & =  \frac{2}{\pi}\,P\int_{s_\text{th}}^{\infty}ds' \left[\frac{s'}{s'^2-s^2}\right] \frac{\Imag F_{-}(s')}{s'},\label{eq:IDR-odd}
\end{align}

\noindent where $P$ denotes the principal Cauchy value, $K$ is the \textit{subtraction constant},\footnote{The subtraction constant
corresponds to the real part evaluated at the subtraction point, usually chosen in the non-physical region 
of the scattering amplitude.}
and $s_\text{th}$ denotes the physical threshold for scattering states. Here, we are interested in $pp$ and $\bar{p}p$ scattering,
therefore from $s~=~2~m_p(E + m_p)$ and $E_\text{th} = m_p$ we have:
\begin{equation}
s_\text{th} = 4 m_p^2 \approx 3.521 \mathrm{GeV}^2.
\nonumber
\end{equation}

In the application of IDR and also, as discussed in the next subsection,
in the replacement of IDR by Derivative Dispersion Relations (DDR) in amplitude
analyses, several authors consider the \textit{high-energy approximation}, which consists in taking the
limit 
\begin{equation}
s_\text{th} \rightarrow 0
\nonumber
\end{equation}

\noindent in the above integrals. Although usual, this approximation
is not well justified and some comments are appropriate.

\begin{enumerate}

\item Experimental data of $\sigmatot$ from $pp$ and $\ppbar$ are characterized by two different regions:
below $\sqrt{s} \sim$ 2 GeV the data show narrow peaks, caused by the formation of resonances,
while reaching the scattering region ($s_\text{th} = 4 m_p^2$), $\sigmatot$($s$) presents a smooth energy dependence. 
The total cross section decreases monotonically up to $\sim$ 20 GeV and then starts to rise.
As discussed in the previous section, the region of the smooth decrease is expected to be described by the Reggeon
exchanges. Therefore, the region $s < 4m_p^2$ corresponds to an \textit{unphysical} region for scattering
states. 

\item The usual dataset for amplitude analyses starts at $\sqrt{s_\text{min}}$ = 5 GeV
(the energy cutoff), which is not far above the threshold $\sqrt{s_\text{th}} \sim $ 2 GeV.
Or, in other words, for fits with $\sqrt{s_\text{min}}$ = 5 GeV (as those by COMPETE
and PDG) it seems unreasonable to consider $\sqrt{s_\text{th}} \sim $ 2 GeV as zero.

\item With respect to our parametrizations (FMS models, see Section~\ref{subsec:rise_DDR_models}) and in all the
data reductions developed in this chapter,
the energy scale is assumed at $s_0 = s_\text{th} = 4 m_p^2$ and therefore this scale cannot be considered null.

\end{enumerate}

\subsection*{High-energy Approximation and the Effective Subtraction Constant}

Let us illustrate the application of IDR and the effects of not taking into account the high-energy approximation, 
by considering a parametrization for $\sigmatot$ based on
Reggeon exchanges at low-energies and the simple-pole Pomeron contribution in high energies (Section~\ref{subsec:rise_Regge_simple_pole}).
For simplicity, we consider here the energy scale $s_0 = 1$ GeV$^2$:
\begin{equation}
\sigmatot(s) = a_1 s^{-b_1} + \tau a_2 s^{-b_2} + \delta s^{\epsilon},
\label{eq:rise-sigtot-regge}
\end{equation}

\noindent where $\tau = -1\; (+1)$ for $pp$ ($\ppbar$). Our goal is to determine the real parts
(and consequently $\rho(s)$) by means of the crossing relations, Eq.~\eqref{eq:hadronic-amplitudes},
and the IDR, Eqs.~\eqref{eq:IDR-even} and \eqref{eq:IDR-odd},
with $s_\text{th}\neq 0$ fixed.

From Eq.~\eqref{eq:rise-sigtot-regge}, with the crossing relations Eq.~\eqref{eq:even-odd-amplitudes}, and the optical theorem Eq.~\eqref{eq:rise-optical-theo}, we obtain
\begin{equation}
\begin{split}
\frac{\Imag F_{+}}{s} & = a_1 s^{-b_1} + \delta s^{\epsilon},\\
\frac{\Imag F_{-}}{s} & = - a_2 s^{-b_2}.
\end{split}
\label{eq:even-odd-regge}
\end{equation}

One can write the integrals appearing in Eqs.~\eqref{eq:IDR-even} and \eqref{eq:IDR-odd} as
\begin{equation}
\int_{s_\text{th}}^{\infty}\; ds' = \int_{0}^{\infty} \; ds' - \int_{0}^{s_\text{th}}\; ds'
\label{eq:IDR-separeted-integrals}
\end{equation}

\noindent and substitute Eq.~\eqref{eq:even-odd-regge}. The first integral in the RHS results in trigonometric functions and the second one 
in hyper-geometric functions~\cite{Bertini:1996,Avila_Menon:2004},
that in turn can be written as a series expansion~
\cite{Gradshteyn_etal_book:1980,Abramowitz_Stegun_book:1964}.
We obtain for the even case
\begin{align}
& \frac{\Real F_{+}(s)}{s} = \frac{K}{s} - a_1 \tan \left(\frac{\pi b_1}{2}\right) s^{- b_1} +
\delta \tan \left(\frac{\pi \epsilon}{2}\right) s^{\epsilon} + \Delta^{+},\label{eq:real-IDR-even-regge-complete}
%
\intertext{where}
%
& \Delta^{+} = \frac{2}{\pi} \sum_{j=0}^{\infty} \frac{a_1 s_\text{th}^{-b_1}}{2j + 1 - b_1}\left[\frac{s_\text{th}}{s}\right]^{2j + 1}
           + \frac{2}{\pi} \sum_{j=0}^{\infty} \frac{\delta s_\text{th}^{\epsilon}}{2j + 1 + \epsilon}\left[\frac{s_\text{th}}{s}\right]^{2j + 1},
\label{eq:correction-IDR-even-regge}
\end{align}
\noindent and for the odd case
\begin{align}
& \frac{\Real F_{-}(s)}{s} =  - a_2 \cot \left(\frac{\pi b_2}{2}\right) s^{- b_2}  - \Delta^{-},
 \label{eq:real-IDR-odd-regge-complete}
%
\intertext{where}
%
& \Delta^{-} = \frac{2}{\pi s} \sum_{j=0}^{\infty} 
\frac{a_2 s_\text{th}^{1 -b_2}}{2j + 2 - b_2}\left[\frac{s_\text{th}}{s}\right]^{2j + 1}.
\label{eq:correction-IDR-odd-regge}
\end{align}

Therefore, denoting the ``correction'' term
\begin{equation}
\Delta \equiv \Delta^{+} + \tau \Delta^{-},
\nonumber
\end{equation}

\noindent the real parts of the $pp$ and $\bar{p}p$ amplitudes can be expressed by
\begin{equation}
\frac{\Real F(s)}{s} = \frac{K}{s} - a_1 \tan \left(\frac{\pi b_1}{2}\right) s^{- b_1} + \tau
a_2 \cot \left(\frac{\pi b_2}{2}\right) s^{- b_2} + 
\delta \tan \left(\frac{\pi \epsilon}{2}\right) s^{\epsilon} + \Delta,
\label{eq:real-pp-ppbar-regge-IDR-complete}
\end{equation}

\noindent where
\begin{equation}
\Delta = 
\frac{2}{\pi} \sum_{j=0}^{\infty} \left[\frac{a_1 s_\text{th}^{-b_1}}{2j + 1 - b_1} +
\frac{\delta s_\text{th}^{\epsilon}}{2j + 1 + \epsilon} + \tau \frac{1}{s} \frac{a_2 s_\text{th}^{1 -b_2}}{2j + 2 - b_2}\right]
\left[\frac{s_\text{th}}{s}\right]^{2j + 1}, 
\nonumber
\end{equation}

\noindent again with $\tau = -1 \, (+1)$ for $pp$ ($\bar{p}p$).

Note that if $s_\text{th} \rightarrow 0$, then $\Delta \rightarrow 0$ and we recover
the Regge-Gribov result for $\rho$, Eq.~\eqref{eq:rho-2Reggeons-SoftPomeron}, 
except for the subtraction constant in the form $K/s$ present in Eq.~\eqref{eq:real-pp-ppbar-regge-IDR-complete}.
That is the point we are interested in here. By expanding the correction term, $\Delta$, we obtain
\begin{equation}
\Delta = 
\frac{2}{\pi} \left\{ \left[\frac{a_1 s_\text{th}^{1 -b_1}}{1 - b_1} + \frac{\delta s_\text{th}^{1 + \epsilon}}{1 + \epsilon}\right]\frac{1}{s} +
\tau \left[\frac{a_2 s_\text{th}^{2 -b_2}}{2 - b_2}\right]\frac{1}{s^2} + 
\left[\frac{a_1 s_\text{th}^{3 -b_1}}{3 - b_1} + \frac{\delta s_\text{th}^{3 + \epsilon}}{3 + \epsilon}\right]\frac{1}{s^3} +
\tau \left[\frac{a_2 s_\text{th}^{4 -b_2}}{4 - b_2}\right]\frac{1}{s^4} + ...
\right\}. 
\nonumber
\end{equation}

Expressing the leading term in the above equation
\begin{equation}
\frac{2}{\pi} \left[\frac{a_1 s_\text{th}^{1 -b_1}}{1 - b_1} + \frac{\delta s_\text{th}^{1 + \epsilon}}{1 + \epsilon}\right]
\equiv f(s_\text{th}, a_1, b_1, \delta, \epsilon),
\nonumber
\end{equation}

\noindent we have 
\begin{equation}
\Delta = \frac{f(s_\text{th}, a_1, b_1, \delta, \epsilon)}{s} + \mathcal{O} (1/s^2).
\nonumber
\end{equation}

Therefore, in Eq.~\eqref{eq:real-pp-ppbar-regge-IDR-complete}, this leading term can be absorbed in the subtraction constant, defining 
an \textit{effective subtraction constant}:
\begin{equation}
\frac{K + f(s_\text{th}, a_1, b_1, \delta, \epsilon)}{s} \equiv \frac{K_\text{eff}}{s},
\label{eq:Keff-def}
\end{equation}

\noindent which is the same for $pp$ and $\bar{p}p$ scattering.

With this concept and definition, we can re-express the IDR in the form (note the lower limits):
\begin{align}
\frac{\Real F_{+}(s)}{s} & =\frac{K_\text{eff}}{s} + \frac{2s}{\pi}\,P\int_{0}^{\infty} ds' \left[\frac{1}{s'^2-s^2}\right]\frac{\Imag F_{+}(s')}{s'},
\label{eq:IDR-even-Keff}\\[10pt]
%
\frac{\Real F_{-}(s)}{s} & =  \frac{2}{\pi}\,P\int_{0}^{\infty} ds' \left[\frac{s'}{s'^2-s^2}\right]\frac{\Imag F_{-}(s')}{s'}.
\label{eq:IDR-odd-Keff}
\end{align}

This result deserves some comments and explanations as follows.

\begin{enumerate}

\item If $K_\text{eff} = 0$, the IDR correspond to the high-energy approximation,
namely $s_\text{th} \rightarrow 0$. It is important to stress this point: if in the data reductions
with IDR (or DDR as we shall see), the subtraction constant is omitted (which means to be assumed zero), then
the high-energy approximation is implicit and therefore the unphysical region
from 0 to $s_\text{th}$ is taken into account.

\item When $K_\text{eff}$ is considered as a free fit parameter in data reductions, it has a
clear and important physical meaning as a first-order contribution related to the finite value of the lower limit
(see Eq.~\eqref{eq:Keff-def}). 
Consequently, \textit{it improves the applicability of the formalism in the regions of lower 
and intermediate energies}.

\item According to Eq.~\eqref{eq:Keff-def}, for the case of simple poles $K_\text{eff}$ is connected with
the other free parameters present in the analytic input for the total cross section.
Moreover, this connection involves not only the Reggeon parameters ($a_1, b_1$) but also those related to the Pomeron
contribution ($\delta, \epsilon$). As a consequence,
in data reductions we expect that the subtraction constant as a free fit parameter
is correlated with all the other parameters, including those associated
with any form of the leading Pomeron contribution (high-energy region). 
This can be seen in the analyses by Fagundes, Menon and Silva \cite{Fagundes_Menon_Silva:2013a} 
and Menon and Silva \cite{Menon_Silva:2013a}, 
where the correlation matrices are shown and discussed.

\item The same interpretation of $K_\text{eff}$ can be extended to the replacements of IDR by DDR,
as we shall show in the next subsection.

\end{enumerate}

\subsection{Derivative Dispersion Relations with the Effective Subtraction Constant}
\label{subsec:rise_DDR}

Derivative Dispersion Relations are the method employed in the amplitude analyses by Fagundes, Menon and Silva and play a fundamental
role in this chapter. As we shall show, one of the main points concerns the possibility to evaluate the $\rho$ function associated with the
$\ln^\gamma s$ form ($\gamma$ real) in an \textit{analytical} way (which is not the case with IDR).

Derivative Dispersion Relations (DDR) 
are usually obtained from IDR in the high-energy limit ($s_\text{th}\to0$). 
This is done with a change of variable $s = s_\text{th}\, e^{\xi}$ in
the integrands of Eqs.~\eqref{eq:IDR-even} and \eqref{eq:IDR-odd}, which are then expanded in power series and, after that, integrated 
by parts (see Appendix~\ref{app:DDR} for details). At last, the primitives are evaluated at the upper and lower integration limits.
The derivation can be found in several works (see quoted papers below and references therein).

First, let us focus on the result by Bronzan, Kane, and Sukhatme that is based on the \textit{high-energy approximation}
and does not have any reference to the subtraction constant\footnote{In that work,
the authors consider also an additional parameter, which, in fact, is not necessary 
(see the details of the calculation and critical discussion by \'Avila and Menon in Ref.~\cite{Avila_Menon:2004}).}
\cite{Bronzan_Kane_Sukhatme:1974}.
The results can be written in operational form with derivatives with respect to
the logarithm of the energy in the arguments of trigonometric operators. 
For crossing even and odd amplitudes these singly-subtracted DDR are given by
\begin{align}
\frac{\Real F_{+}(s)}{s} & = \tan\left[\frac{\pi}{2}\frac{d}{d\ln s} \right]\frac{\Imag F_{+}(s)}{s}, \label{eq:DDR-even}\\[10pt]
%
\frac{\Real F_{-}(s)}{s} & = \tan\left[\frac{\pi}{2}\left(1 + \frac{d}{d\ln s}\right) \right] \frac{\Imag F_{-}(s)}{s}.\label{eq:DDR-odd}
\end{align}

In practice,
the trigonometric operators are expanded in power series and the derivatives of the imaginary part are
calculated term by term providing the corresponding real parts by summing the series.

These relations were used, for instance, by the COMPETE Collaboration 
in their amplitude analyses~\cite{COMPETE:2002a},
and also by the COMPAS Group in the results reported by RPP~\cite{PDG:2012}. 

Up to our knowledge, the first results for the DDR taking into account the finite
lower limit (i.e. without the high-energy approximation) and the effect of
the primitive at both upper and lower limits were obtained by \'Avila and Menon in 2005 \cite{Avila_Menon:2006,Avila_Menon:2007a}.
The correction term can be expressed as a double infinite series\footnote{Later, Ferreira and Sesma \cite{Ferreira_Sesma:2008,Ferreira_Sesma:2013}
have shown that it is possible to reduce the result to a single series, using sum rules and the incomplete Gamma function.}.

For our purposes, the main point concerns the fact
that for simple poles, this correction term can be also expressed as inverse powers of $s$, so that the
leading contribution can be absorbed by the subtraction constant \cite{Avila_phdthesis:2009}. 
This is analogous to the IDR calculations and corresponds 
to the introduction of the \textit{effective subtraction constant} as a free fit parameter in data reductions.
The complete practical equivalence in data reductions between the IDR without the high-energy approximation and the DDR with the subtraction constant as
a free fit parameter is demonstrated by \'Avila and Menon in Refs.~\cite{Avila_Menon:2004,Avila_Menon:2007b,Avila_Menon:2004b} and in more detail 
by \'Avila in Ref.~\cite{Avila_phdthesis:2009}.
The replacement of IDR by DDR has been also discussed by Cudell, Martynov and Selyugin
\cite{Martynov:2004,Cudell:2003} and more recently (2017) by Ferreira, Kohara and Sesma \cite{Ferreira:2017a,Ferreira:2017b}.

It is interesting to note that the IDR, Eqs.~\eqref{eq:IDR-even} and \eqref{eq:IDR-odd}, have a \textit{non-local}
character since, in order to obtain the real part of the amplitude, the imaginary part must be known
for all values of the energy. On the other hand, DDR present a nearly local character. 
Moreover, in some cases the use of IDR can demand numerical integration, as was the case 
of the analyses by Amaldi \etal~\cite{Amaldi:1977}, UA4/2 Collab.~\cite{Augier:1993} and
Bueno-Velasco~\cite{Bueno_Velasco:1996} discussed in Section~\ref{subsec:rise_Regge_Amaldi}.
As already mentioned, contrary to this, the use of DDR provides analytical expressions since the imaginary part of the amplitude 
depends only on the variable $\ln s$. In this case, the term $\ln^\gamma s$
considered in the aforementioned analyses has an analytical form for the real part, 
as we show in what follows.

In our analyses, we consider the operator expansion in the form 
introduced by Kang and Nicolescu \cite{Kang_Nicolescu:1975} 
in 1975 and discussed in \cite{Avila_Menon:2004,Avila_Menon:2007a} with the inclusion of \textit{the effective subtraction constant}.
The even and odd relations are given by the operational forms:
\begin{align} 
\frac{\Real F_{+}(s)}{s} & = \frac{K_\text{eff}}{s} + \left[ \frac{\pi}{2} \frac{d}{d\ln s} + \frac{1}{3} \left(\frac{\pi}{2}\frac{d}{d \ln s}\right)^3 +
\frac{2}{15} \left(\frac{\pi}{2}\frac{d}{d \ln s}\right)^5 + \dots \right] \frac{\Imag F_{+}(s)}{s}, \label{eq:DDR-KN-even}\\[10pt]
\frac{\Real F_{-}(s)}{s}  &= - \int \left\{\frac{d}{d\ln s}\left[\cot \left( \frac{\pi}{2}\frac{d}{d\ln s} \right)\right]\frac{\Imag F_{-}(s)}{s} \right\} d\ln s \nonumber \\
                          &= - \frac{2}{\pi}\int \left\{ \left[ 1 - \frac{1}{3} \left(\frac{\pi}{2}\frac{d}{d \ln s}\right)^2 \frac{1}{45} \left(\frac{\pi}{2}\frac{d}{d \ln s}\right)^4
                             - \dots \right] \frac{\Imag F_{-}(s)}{s} \right\} \, d \ln s. \label{eq:DDR-KN-odd}
\end{align} 

As an example of the applicability of the above DDR, consider the parametrization
for $\sigmatot$ given by Eq.~\eqref{eq:rise-sigtot-regge}.
Using Eq.~\eqref{eq:even-odd-amplitudes} to obtain the imaginary part
of even and odd amplitudes and substituting them in Eqs.~\eqref{eq:DDR-KN-even}
and \eqref{eq:DDR-KN-odd}, it is easy to show that the resulting infinite series has a closed form, so that
\begin{align}
 \frac{\Real F_+(s)}{s} & = \frac{K_\text{eff}}{s} - a_1s^{-b_1}\tan\left(\frac{\pi b_1}{2}\right) + \delta s^{\epsilon}\tan\left(\frac{\pi \epsilon}{2}\right),\label{eq:amp-even-DDR-PowerLaw}\\
 \frac{\Real F_-(s)}{s} & = -a_2s^{-b_2}\cot\left(\frac{\pi b_2}{2}\right)\label{eq:amp-odd-DDR-PowerLaw},
\end{align}

\noindent which corresponds to the results obtained with IDR, 
see Eqs.~\eqref{eq:real-IDR-even-regge-complete} and \eqref{eq:real-IDR-odd-regge-complete},
apart from the correction term (here included in $K_\text{eff}$)
and also to the results obtained directly from the Regge-Gribov Formalism (Section~\ref{subsec:rise_Regge_simple_pole})
except for the presence here of the subtraction constant.

Of interest here, let us consider the leading \textit{even} contribution to $\sigmatot$
as introduced by Amaldi \etal \cite{Amaldi:1977} (discussed in Section~\ref{subsec:rise_Regge_Amaldi})
\begin{equation} 
\sigma^\pomeron(s) = \frac{\Imag F^\pomeron(s)}{s} = \alpha + \beta \ln^{\gamma}\left(\frac{s}{s_{0}}\right).
\label{eq:Lgamma-term-sigtot}
\end{equation} 

From Eq.~\eqref{eq:DDR-KN-even}, the corresponding $\rho$ parameter can be expressed as
\begin{equation} 
\frac{\Real F^\pomeron(s)}{s} = \mathcal{A}\,\ln^{\gamma - 1} \left(\frac{s}{s_0}\right) +
\mathcal{B}\,\ln^{\gamma - 3} \left(\frac{s}{s_0}\right) +
\mathcal{C}\,\ln^{\gamma - 5} \left(\frac{s}{s_0}\right) + \dots
\label{eq:Lgamma-term-real-DDR-general}
\end{equation} 
\noindent where
\begin{equation} 
\begin{split}
\mathcal{A} & = \frac{\pi}{2} \, \beta\, \gamma,  
\qquad 
\mathcal{B} = \frac{1}{3} \left[\frac{\pi}{2}\right]^3 \, \beta\, \gamma\, [\gamma - 1][ \gamma - 2], \\
\mathcal{C} & = \frac{2}{15} \left[\frac{\pi}{2}\right]^5 \, \beta\, \gamma\, [\gamma - 1][ \gamma - 2]
[\gamma - 3][ \gamma - 4], \dots
\end{split}
\label{eq:Lgamma-factors-rho-DDR-general}
\end{equation} 

If we consider a double pole ($D$) in the complex angular momentum space, which corresponds to $\gamma$ = 1, 
we have $\mathcal{A} = \pi\,\beta/2$, $\mathcal{B} = \mathcal{C} = \dots =0$ and
%
\begin{equation} 
\frac{\Real F^\pomeron_{D}(s)}{s} = \frac{\pi}{2}\beta,
\end{equation}

\noindent and considering a triple pole ($T$), $\gamma$ = 2, $\mathcal{A} = \pi\,\beta$, $\mathcal{B} = \mathcal{C} = \dots =0$
and
%
\begin{equation} 
\frac{\Real F^\pomeron_{T}(s)}{s} = \pi \beta \ln\left(\frac{s}{s_{0}}\right), 
\end{equation}

\noindent as obtained in the Regge-Gribov formalism, Eqs.~\eqref{eq:Regge-Gribov-double-pole-1storder}
and \eqref{eq:Regge-Gribov-triple-pole-1storder}.
Therefore, Eqs.~\eqref{eq:Lgamma-term-sigtot}-\eqref{eq:Lgamma-factors-rho-DDR-general} constitute a \textit{generalization} of the
$\ln^2s$ and $\ln s$ cases for real (not integer) exponents of the logarithm.

In all analyses performed with the leading term given by Eq.~\eqref{eq:Lgamma-term-sigtot}, 
the data reductions, with $\gamma$ as a real free fit parameter have shown that its
value does not exceed $\sim$ 2.5 \cite{Fagundes_Menon_Silva:2012a,Fagundes_Menon_Silva:2013a,Menon_Silva:2013a,
Menon_Silva:2013b,Fagundes_Menon_Silva:2017a,Fagundes_Menon_Silva:2017b}.
Therefore, the expansion up to third order is enough to ensure the convergence of the fit
and higher orders do not affect the results.
In the parametrization to be considered in the next section, we will truncate the series after the third-order term.
Recall that for a leading contribution in the form $\ln^{\gamma}(s/s_\text{th})$
the IDR can not provide an analytic result, but that is not the case for the DDR, as shown above.

We want to stress two crucial advantages of this approach in amplitude analyses:

\begin{enumerate}

\item It provides analytic results in all cases of interest, which are adequate for data reductions
and allow standard statistical determination of the uncertainties in all free fit
parameters involved (and consequently, analytic propagation of the uncertainties
to the physical quantities). 

\item With the subtraction constant as an additional free fit parameter, related to its \textit{effective character},
the approach is not constrained by the high-energy approximation: its applicability covers, in principle, all the
energies above the physical threshold, without reference to the unphysical region.

\end{enumerate}

\subsection{Asymptotic Uniqueness and the Phragm\'en-Lindel\"off Theorems}
\label{subsec:rise_AU}

Asymptotic Uniqueness, associated with 
Phragm\'en-Lindel\"off theorems, constitute another analytic way
for the determination of the real part of the forward amplitude.
In this subsection we present the ideas and basic theorems osn which this approach is based.
To do that, we follow Section 7.1 of the book by Eden~\cite{Eden_book:1967}.
This subject is also treated in Ref.~\cite{Block:2006} (Sect. 10.3)
and in Ref.~\cite{Block_Cahn:1985} (Sect. IV.D), but with a different approach.

As we shall show, in case of the $\ln^\gamma s$ law with $\gamma$ real, the analytic results
are not equivalent to those obtained through DDR.

\subsubsection{Basic Concepts}

Asymptotic Uniqueness (AU) is based on the concepts of crossing symmetry
and analyticity, associated with the \textit{forward} scattering amplitude in
the complex-$s$ plane. We apply AU for even and odd amplitudes, Eq.~\eqref{eq:even-odd-amplitudes},
from which we recover the hadronic amplitudes with Eq.~\eqref{eq:hadronic-amplitudes}.

Recall that in the polar form, $F(s) = |F(s)| e^{i \theta}$, the phase of the amplitude
is given by
\begin{equation}
\theta = \tan^{-1} \left[ \frac{\Imag F(s)}{\Real F(s)}\right].
\label{eq:phase-amplitude-theta}
\end{equation}

The asymptotic uniqueness constitutes a way to determine
the phase of the amplitudes, once given a \textit{real} function \textit{related} to its imaginary
part as an input. Actually, these asymptotic results provide the crossing even
and odd amplitudes to within a $\pm$ factor. The correct sign is determined by 
physical conditions involved \cite{Eden_book:1967}. Specifically, from Eq.~\eqref{eq:hadronic-amplitudes}  and the optical
theorem, Eq.~\eqref{eq:rise-optical-theo},
\begin{equation}
\frac{\Imag F_+(s)}{s} = \frac{1}{2} \left\{\sigma_{pp} + \sigma_{\bar{p}p} \right\},
\qquad
\frac{\Imag F_-(s)}{s} = \frac{1}{2} \left\{\sigma_{pp} - \sigma_{\bar{p}p} \right\}.
\label{eq:amp-even-odd-sigtot-pp-ppbar}
\end{equation}

Since $\sigma_{pp}, \sigma_{\bar{p}p} > 0$, we have always $\Imag F_+(s) > 0$.
For the Reggeons contribution, once associated with the region where 
$\sigma_{\bar{p}p} > \sigma_{pp}$ (low-energy, see Fig.~\ref{fig:sigmatot_rho_data_pp_ppbar}) we have $\Imag F_-(s) < 0$,
as discussed in Section~\ref{subsec:rise_Regge_simple_pole}.
For the Pomeron contribution, dominating the region where $\sigma_{\bar{p}p} = \sigma_{pp}$,
$\Imag F_-(s) = 0$. 

Following Ref.~\cite{Eden_book:1967}, let us consider the following corollary that provides the essential concept and 
role for the determination of the phase of the amplitude:



\begin{corollary}
``If $f(z)$ is bounded by a polynomial, and $f(z)$ tends to the limits $L_1$ and $L_2$
along the rays $z = x + i0$ as $x \rightarrow + \infty$ and $- \infty$, then we must
have $L_1 = L_2$."
\end{corollary}

The above statement, where $z = x + i0$ denotes the limit from the upper half plane, says that it is possible to
obtain the phase of the function $f(z)$ for all $z$ from its asymptotic limits. 

In our case, $f(z)$ corresponds to the forward elastic scattering amplitude, namely $f(z) = F(s,0)$. Furthermore,
as commented in Section~\ref{sec:basic_FLM_bound}, to have the amplitude bounded by a polynomial is a usual assumption for the 
elastic scattering amplitude. Therefore, the corollary can be applied to determine the forward phase of $F(s) = F(s,0)$.

Below, we illustrate in detail the use of AU by applying the method to simple-poles
contributions to even and odd amplitudes (that corresponds to Power Laws),
triple-pole contribution to even amplitude (Log-squared Law) and to 
Log-raised-to-$\gamma$ Law, Eq.~\eqref{eq:Lgamma-term-sigtot}.

\subsubsection{Power Law (Simple Poles)}

We begin with the simple-pole contribution to the total cross section. From Regge theory,
we know that the simple pole contributes to the total cross section as $\beta s^{\alpha - 1}$, so that
\begin{equation}
\Imag F(s) = \beta s^{\alpha}.
\label{eq:imag-power-law}
\end{equation}

\noindent In order to apply the corollary, let us consider the complex function 
\begin{equation}
\frac{F(s)}{s^{\alpha}},
\nonumber
\end{equation}

\noindent which is bounded by a polynomial. Therefore, if in the limit $s \rightarrow \infty$ we have

\begin{equation}
\frac{F(s)}{s^{\alpha}} \rightarrow L_1 \equiv M e^{i \theta},
\nonumber
\end{equation}

\noindent and
\begin{equation}
\frac{F(se^{i\pi})}{[se^{i \pi}]^{\alpha}} \rightarrow L_2,
\nonumber
\end{equation}

\noindent then, by the Corollary, we must have $L_2 = L_1$ and therefore,
\begin{equation}
F(s) = Ms^{\alpha} e^{i \theta} \quad \text{and} \quad F(se^{i\pi}) = Ms^{\alpha} e^{i (\theta + \pi \alpha)}.
\nonumber
\end{equation}

Considering a \textit{symmetric}/even ($+$) amplitude, using Eq.~\eqref{eq:even-odd-amplitudes} and the above result,
we have
\begin{equation}
Ms^{\alpha_+} e^{i (\theta_+ + \pi \alpha_+)} = + Ms^{\alpha_+} e^{- i \theta_+}.
\nonumber
\end{equation}

Therefore, we can calculate the phase \textit{explicitly}: 
\begin{equation}
\theta_{+} = n\pi - \frac{\pi \alpha_+}{2} \quad (n = 0,\pm 1,\pm 2,\dots),
\nonumber
\end{equation}

Using the optical theorem, Eq.~\eqref{eq:rise-optical-theo}, and Eqs.~\eqref{eq:phase-amplitude-theta}
and~\eqref{eq:imag-power-law} and denoting
$\Imag F_{+}(s) =  \beta_{+} s^{\alpha_{+}}$, $\beta_+ > 0$, the real part of the amplitude reads:
\begin{equation}
\Real F_{+}(s) = \frac{\Imag F_+}{\tan \theta_+} = 
\frac{\beta_{+} s^{\alpha_{+}}}{- \tan (\pi \alpha_+/2)}
= - \beta_{+} \cot \left(\frac{\pi \alpha_+}{2}\right) s^{\alpha_{+}},
\nonumber
\end{equation}

\noindent leading to the \textit{complex symmetric amplitude}:
\begin{equation}
 F_{+}(s) = \beta_{+} \left[ i - \cot \frac{\pi \alpha}{2} \right] s^{\alpha_{+}}.
\label{eq:amp-even-AU-PowerLaw}
 \end{equation}

Now, if we consider an \textit{antisymmetric}/odd ($-$) amplitude, the procedure is analogous. 
From Eq.~\eqref{eq:even-odd-amplitudes}
\begin{equation}
Ms^{\alpha_-} e^{i (\theta_- + \pi \alpha_-)} = - Ms^{\alpha_-} e^{- i \theta_-}.
\nonumber
\end{equation}

With $-1 = e^{i\pi}$, we can obtain again the phase explicitly, but now it reads
\begin{equation}
\theta_{-} = n\pi + \frac{\pi}{2} (1 - \alpha_{-}) \quad (n = 0, \pm 1, \pm 2,\dots).
\nonumber
\end{equation}

As commented above (see Eq.~\eqref{eq:amp-even-odd-sigtot-pp-ppbar}), 
in the odd case we have to consider the solution with the minus sign 
in order to obtain a positive cross section. Therefore, we have
$\mathrm{Im} A_- = - \beta_- s^{-\alpha_-}$,
with $\beta_- > 0$. From Eq.~\eqref{eq:phase-amplitude-theta}, we obtain
\begin{equation}
\Real F_{-}(s) = \frac{\Imag F_-}{\tan \theta_-} = - \frac{\beta_{-} s^{\alpha_{-}}}{\tan (\pi/2 - \pi \alpha/2)}
= - \beta_{-} \tan \left(\frac{\pi \alpha}{2}\right) s^{\alpha_{-}},
\nonumber
\end{equation}

\noindent and the \textit{complex antisymmetric amplitude}:
\begin{equation}
 F_{-}(s) = - \beta_{-} \left[ i + \tan \frac{\pi \alpha_-}{2} \right] s^{\alpha_{-}}.
 \label{eq:amp-odd-AU-PowerLaw}
\end{equation}

We note that the asymptotic results Eqs.~\eqref{eq:amp-even-AU-PowerLaw}
and \eqref{eq:amp-odd-AU-PowerLaw} are exactly the same as those obtained in the
Regge-Gribov formalism for the even and odd amplitudes (Section~\ref{subsec:rise_Regge_simple_pole}) 
and also through dispersion relations, Eqs.~\eqref{eq:amp-even-DDR-PowerLaw}
and \eqref{eq:amp-odd-DDR-PowerLaw}, without the subtraction constant.

Therefore, the full expressions for $\sigmatot$ and $\rho$ for $pp$ and $\bar{p}p$ scattering
associated with an even and an odd contribution for Reggeons and with the simple-pole (even) Pomeron,
are the same as Eqs.~\eqref{eq:sigtot-2Reggeons-SoftPomeron} 
and~\eqref{eq:rho-2Reggeons-SoftPomeron} (including the energy scale):
\begin{align}
\sigmatot(s) &= a_1 s^{-b_1} + \tau a_2 s^{-b_2} + \delta s^{\epsilon}, \nonumber \\ 
\rho(s) &= \frac{1}{\sigmatot(s)} \left\{- a_1 \tan \left(\frac{\pi b_1}{2}\right) s^{- b_1} + 
\tau a_2 \cot \left(\frac{\pi b_2}{2} \right) s^{-b_2}  + \delta \tan \left(\frac{\pi \epsilon}{2}\right) s^{\epsilon}    \right\},
\nonumber
\end{align}

\noindent with $\tau = -1$ for $pp$ and $\tau = +1$ for $\bar{p}p$.

\subsubsection{Log-squared Law (Triple Pole)} 

Let us discuss now the case of a Triple Pole.
Once it represents the Pomeron (even signature), we consider here only
the \textit{symmetric} relation in Eq.~\eqref{eq:even-odd-amplitudes}. 
We will use the index $T$ standing for triple-pole leading contribution at high energies.

Consider $\sigma^{T}(s) = \beta \ln^2{s}$, so that $\Imag F^{T}(s) = \beta s \ln^2(s)$.
As it was done in the case of Power Laws, we use the corollary to obtain the asymptotic behavior
\begin{equation}
F^{T}(s) = Ms\ln^{2}(s)e^{i \theta} \quad \mathrm{and} \quad F^{T}(se^{i\pi}) = Ms\ln^{2}(se^{i\pi}) e^{i (\theta + \pi)}.
\label{eq:even-amp-Log2-asymp-limits}
\end{equation}

Using the symmetric relation of Eq.~\eqref{eq:even-odd-amplitudes} to connect the two limits above
(and omitting the index $+$, for simplicity)
\begin{equation}
\ln^{2}(se^{i\pi}) e^{i (\theta + \pi)} = \ln^{2}(s)e^{-\, i \theta}.
\nonumber
\end{equation}

\noindent One can write 
\begin{equation}
\ln^{2}(se^{i\pi}) = \ln^{2}(s)\,\left[1 + i\frac{\pi}{\ln(s)}\right]^{2},
\nonumber
\end{equation}

\noindent extract the square root from both sides, and, finally, take the complex conjugate to obtain
\begin{equation}
e^{i\theta} = \pm \left[\frac{\pi}{\ln s} + i\right].
\nonumber
\end{equation}

Contrary to what happens in the case of Power Laws, here we cannot determine the phase \textit{explicitly},
since we have the $e^{i \pi}$ factor in the argument of the logarithm.
In this case, from $F^{T}(s)$ in Eq.~\eqref{eq:even-amp-Log2-asymp-limits}, denoting $M = \beta$ and choosing
the $+$ sign in order to have $\Imag F^{T}(s) >0$, we obtain the complex amplitude
\begin{equation}
\frac{F^{T}(s)}{s} = \beta [ \pi \ln(s) + i \ln^2(s)].
\label{eq:amp-Pomeron-Triple-Pole-AU}
\end{equation}
 
Therefore, as obtained in the Regge-Gribov and dispersion formalisms (Sections~\ref{sec:rise_ReggeGribov}
and~\ref{sec:rise_RealImagParts}), the Pomeron contribution associated with a triple pole reads
\begin{equation}
\sigma^{T} = \beta \ln^2(s), \qquad \rho^{T} = \frac{1}{\sigma^{T}_{P}}[\beta \pi \ln(s)] = \frac{\pi}{\ln(s)}.
\end{equation}

\subsubsection{Log-raised-to-$\gamma$ Law}

Now we turn to the main focus of this work: 
$\Imag F(s) = \beta s \ln^{\gamma}(s)$, with $\gamma$ a real number.
For simplicity, we shall omit any index in the amplitude.

We first present the derivation of an exact result (introduced in Ref.~\cite{Fagundes_Menon_Silva:2017b}),
which is used in the data reductions developed here (Sects.~\ref{sec:rise_analytic_models} and \ref{sec:rise_results}). 
After that we discuss a high-energy
approximation and comparison with other results, followed by the particular case in which
$\gamma = 2$ and another analytic result obtained by means of a binomial expansion.

\begin{itemize}

\item \textit{Exact Result} 

We follow the same steps used above and we use the corollary to write
\begin{equation}
F(s) = Ms\ln^{\gamma}(s)e^{i \theta} \quad \mathrm{and} \quad F(se^{i\pi}) = Ms\ln^{\gamma}(se^{i\pi}) e^{i (\theta + \pi)}.
\label{eq:even-amp-Loggamma-asymp-limits}
\end{equation}

Again, since this term corresponds to an even contribution to the amplitude, we use the symmetric relation in Eq.~\eqref{eq:even-odd-amplitudes}, 
\begin{equation}
\ln^{\gamma}(se^{i\pi}) e^{i (\theta + \pi)} = \ln^{\gamma}(s)e^{-\, i \theta}.
\end{equation}

As before, we can write
\begin{equation}
\ln^{\gamma}(se^{i\pi}) =
\ln^{\gamma}(s)\,\left[1 + i\frac{\pi}{\ln(s)}\right]^{\gamma},
\nonumber
\end{equation}

\noindent and we obtain
\begin{align}
e^{-i(2\theta + \pi)} & = \left[1 + i\frac{\pi}{\ln(s)}\right]^{\gamma},\label{eq:Lgamma-exp-phase-AU-A}
%
\intertext{or}
%
e^{-i(2\theta + \pi)} & = \left[1 + \frac{\pi^2}{\ln^2(s)}\right]^{\gamma/2} e^{i \gamma \phi(s)},\label{eq:Lgamma-exp-phase-AU-B}
\end{align}

\noindent where
\begin{equation}
\boxed{\phi(s) = \tan^{-1}\left[ \frac{\pi}{\ln(s)}\right].}
\label{eq:phi-AU-theory}
\end{equation}

By extracting the square root,
\begin{equation}
e^{-i(\theta + \pi/2)} = \pm \frac{1}{\ln^{\gamma/2}(s)}
 \left[\ln^2(s) + \pi^2 \right]^{\gamma/4} e^{i \gamma \phi /2},
\nonumber
\end{equation}

\noindent we obtain
\begin{equation}
\ln^{\gamma}(s) e^{i\theta} = \pm \ln^{\gamma/2}(s)
 \left[\ln^2(s) + \pi^2 \right]^{\gamma/4} e^{- i [\gamma \phi + \pi]/2}.
\nonumber
\end{equation}

Now, from Eq.~\eqref{eq:even-amp-Loggamma-asymp-limits} we have asymptotically $F(s) = Ms\ln^{\gamma}(s)e^{i \theta}$.
Denoting $M=\beta > 0$, and under the condition $\Imag F(s) > 0$, the above
equation provides the \textit{exact} result for the complex amplitude:
\begin{equation}
\boxed{\frac{F(s)}{s} =  \beta \ln^{\gamma/2}(s) \left[ \ln^2(s) + \pi^2 \right]^{\gamma/4} 
\left[ \sin \left(\frac{\gamma \phi}{2} \right) + i \cos \left(\frac{\gamma \phi}{2} \right) \right],}
\label{eq:amp-Lgamma-AU-exact}
\end{equation}

\noindent where $\phi = \phi(s)$ is given by Eq.~\eqref{eq:phi-AU-theory}.

\item \textit{Approximate Result} 

At sufficiently high energies and since $\gamma < 3$ from the data 
reductions~\cite{Fagundes_Menon_Silva:2012a,Fagundes_Menon_Silva:2013a,Menon_Silva:2013a},
we can approximate
\begin{equation}
\tan \phi = \frac{\pi}{\ln(s)} \approx \phi, \qquad \sin\left(\frac{\gamma\phi}{2}\right)\approx \frac{\gamma\phi}{2} =
\frac{\gamma \pi}{2 \ln(s)}, \qquad \cos\left(\frac{\gamma\phi}{2}\right)\approx 1
\nonumber
\end{equation}

\noindent and 
\begin{equation}
\left[ \ln^2(s) + \pi^2 \right]^{\gamma/4} = \ln^{\gamma/2}(s) 
\left[1 +\frac{\pi^2}{\ln^2 (s)}\right]^{\gamma/4}\approx \ln^{\gamma/2}(s).
\nonumber
\end{equation}

Substituting in Eq.~\eqref{eq:amp-Lgamma-AU-exact} we obtain the \textit{approximate} result
\begin{equation}
\frac{F(s)}{s} \approx \beta \ln^{\gamma}(s) \left[ \frac{\gamma \pi}{2\ln(s)} + i\ \right],
\label{eq:amp-Lgamma-AU-approx}
\end{equation}

\noindent which corresponds to the first-order result obtained with DDR, 
Eqs.~\eqref{eq:Lgamma-term-real-DDR-general} and \eqref{eq:Lgamma-factors-rho-DDR-general}.

\item \textit{AU Result for $\gamma=2$.}

Now we turn to the comparisons between the results obtained with DDR and AU. 
Let us first consider the exact AU result given by Eq.~\eqref{eq:amp-Lgamma-AU-exact}.
For $\gamma$ = 2, we obtain
\begin{equation}
\frac{F^{\gamma=2}(s)}{s} = \beta \ln(s)\, \left[ \ln^2(s) + \pi^2 \right]^{1/2} \left[ \sin \phi + i \cos \phi \right],
\label{eq:amp-Lgamma2-AU-exact}
\end{equation}

\noindent which does not correspond to the triple-pole contribution, Eq.~\eqref{eq:amp-Pomeron-Triple-Pole-AU}.

On the other hand, from the high-energy approximate result~\eqref{eq:amp-Lgamma-AU-approx}
for $\gamma = 2$, we obtain the triple-pole contribution~\eqref{eq:amp-Pomeron-Triple-Pole-AU}.
Moreover, as commented above, Eq.~\eqref{eq:amp-Lgamma-AU-approx} in the general case
corresponds to the first-order expansion for $\rho(s)$ obtained in the DDR approach.

\item \textit{Binomial Expansion} 

Next, we present another result for the amplitude, without the 
explicit determination of the phase, but by using a binomial
expansion. We also compare the results obtained with AU and DDR.

Returning to Eq.~\eqref{eq:Lgamma-exp-phase-AU-A} and extracting the square root, we can express
\begin{equation}
e^{i\theta} = \pm \left[1 - i\frac{\pi}{\ln(s)}\right]^{\gamma/2}.
\nonumber
\end{equation}

Consider now the binomial expansion,
\begin{equation}
(1 + x)^p = 1 + \sum_{k=1} \frac{1}{k!} p (p-1) (p-2) \cdots (p-[k-1]) x^k
\nonumber
\end{equation}

\noindent in the variable       
\begin{equation}
x = -i\,\frac{\pi}{\ln s}.
\nonumber
\end{equation}

Since from Eq.~\eqref{eq:even-amp-Loggamma-asymp-limits} $F(s)/s = \beta \ln(s) e^{i \theta}$, for
$\Imag F(s) > 0$, we obtain the complex amplitude in the form
of a series expansion:
\begin{equation}
\begin{split}
\frac{F(s)}{s} &= \beta \left\{\frac{\gamma}{1!} \left[\frac{\pi}{2}\right] \ln^{\gamma-1}(s) -
\frac{\gamma (\gamma-2) (\gamma-4)}{3!}  \left[\frac{\pi}{2}\right]^3  \ln^{\gamma-3}(s)  \right. \\
 & \quad + \left. \frac{\gamma (\gamma-2) (\gamma-4) (\gamma - 6) (\gamma -8)}{5!}  \left[\frac{\pi}{2}\right]^5  \ln^{\gamma-5}(s)
+ \dots \right\} \\
 & \quad + i\, \beta \left\{\ln^{\gamma}(s) -
\frac{\gamma (\gamma-2)}{2!}  \left[\frac{\pi}{2}\right]^2  \ln^{\gamma-2}(s)  \right. \\
 & \quad + \left. \frac{\gamma (\gamma-2) (\gamma-4) (\gamma - 6)}{4!}  \left[\frac{\pi}{2}\right]^4  \ln^{\gamma-4}(s)
+ \dots \right\}.
\end{split}
\label{eq:amp-Lgamma-AU-binomial-exact}
\end{equation}

We note that for $\gamma$ = 2, we recover 
\begin{equation}
\frac{\Imag F^\pomeron(s)}{s} = \beta \ln^2(s), \qquad \frac{\Real F^\pomeron(s)}{s} = \pi \beta\ln(s).
\nonumber
\end{equation}

Comparing Eq.~\eqref{eq:amp-Lgamma-AU-binomial-exact} with the leading contribution obtained in the DDR approach,
Eqs.~\eqref{eq:Lgamma-term-sigtot}-\eqref{eq:Lgamma-factors-rho-DDR-general}, the results are the same only in first order
\begin{equation}
\frac{\Imag F^\pomeron(s)}{s} \sim \beta \ln^{\gamma}(s),
\qquad
\frac{\Real F^\pomeron(s)}{s} = \beta \gamma \frac{\pi}{2} \ln^{\gamma - 1}(s).
\nonumber
\end{equation}

\end{itemize}

We summarise in Table~\ref{tab:Lgamma_analytical_results}
all analytical results of interest obtained for the L2 and L$\gamma$ laws through DDR and AU.
From the Table we see that in case of the L$\gamma$ law, the analytic results through DDR and AU are distinct,
including the imaginary parts (total cross section). We shall return to this important point in Section~\ref{sec:rise_discussion_comments}.
In the next section, we define the analytic models considered in the data reductions.
 \renewcommand{\arraystretch}{1.25}
\begin{table}[htb]
 \centering
 \caption{\label{tab:Lgamma_analytical_results}Summary of results obtained with DDR and AU for the L2 and L$\gamma$ law. 
 The constant factors $\mathcal{A}$, $\mathcal{B}$ and $\mathcal{C}$ are given in Eq.~\eqref{eq:Lgamma-factors-rho-DDR-general} 
 and the function $\phi=\phi(s)$ is given by Eq.~\eqref{eq:phi-AU-theory}. The energy scale $s_0$ has been omitted for clarity.}
 \subfloat[\label{subtab:Lgamma_Imag_Part}Imaginary Part]{
 \begin{tabular}{c|c|c|c}\hline\hline
        & L$2$ & L$\gamma$ & L$\gamma$ with $\gamma=2$\\\hline
 DDR    & $\beta\ln^2 s$  & $\beta\ln^\gamma s$ & $\beta\ln^2 s$\\\hline 
 AU     & $\beta\ln^2 s$  & $\beta\ln^{\gamma/2}(s) [\ln^2 s + \pi^2]^{\gamma/4}\cos\left(\gamma\phi/2\right)$ & $\beta\ln(s)[\ln^2 s + \pi^2]^{1/2}\cos\left(\phi\right)$ \\\hline\hline
 \end{tabular}}\\
 \subfloat[\label{subtab:Lgamma_Real_Part}Real Part]{
 \begin{tabular}{c|c|c|c}\hline\hline
        & L$2$ & L$\gamma$ & L$\gamma$ with $\gamma=2$\\\hline
 DDR    & $\pi\beta\ln s$  & ${\mathcal{A}}\ln^{\gamma-1}s+{\mathcal{B}}\ln^{\gamma-3}s+{\mathcal{C}}\ln^{\gamma-5}s$ & $\pi\beta\ln s$\\\hline 
 AU     & $\pi\beta\ln s$  & $\beta\ln^{\gamma/2}(s) [\ln^2 s + \pi^2]^{\gamma/4}\sin\left(\gamma\phi/2\right)$ & $\beta\ln  (s) [\ln^2 s + \pi^2]^{1/2}\sin\left(\phi\right)$ \\\hline\hline
 \end{tabular}}
\end{table}

\renewcommand{\arraystretch}{1.0}

\section{Analytic Models}
\label{sec:rise_analytic_models}

In this section we introduce the analytic parametrizations for
$\sigmatot(s)$ and $\rho(s)$ that will be used in our data reductions.
They are based on the analytic results discussed in the previous section.

We start with the definition of a useful notation for models (Sect.~\ref{subsec:rise_notations}).
Next we present the analytic  parametrizations for $\sigmatot(s)$ and $\rho(s)$ constructed through
two methods: the DDR approach (Sect.~\ref{subsec:rise_DDR_models}) and  the AU approach (Sect.~\ref{subsec:rise_AU_models}).
The data reductions with these models are the subject of Sect.~\ref{sec:rise_results}.
For convenience, we repeat some equations already presented in the previous sections.

\subsection{Notation}
\label{subsec:rise_notations}

The parametrizations of interest have a structure
similar to that selected by COMPETE and following their notation,
could be represented by RRPL2 and RRPL$\gamma$, where R stands for the Regge contribution,
P for the constant Pomeranchuk term (critical Pomeron), L2 for the $\ln^2 s$ contribution, i.e.
the triple-pole Pomeron, and L$\gamma$ for the generalization $\ln^\gamma s$ proposed by Amaldi \etal
(see Section~\ref{subsec:rise_Regge_Amaldi}).
Now, for simplicity, we can omit from the notation the RR and P, since the Regge and constant
Pomeranchuk contributions are the same in all cases.
On the other hand, as discussed in the previous section,
in case of L$\gamma$, the way to construct the parametrizations  
for $\sigmatot(s)$ and mainly the connections with $\rho(s)$ are different in the DDR and AU
approaches and the analytic expressions. Moreover, the relations between 
the L2 and the L$\gamma$ terms are also not
the same in these two methods. 

Based on these details, we consider two aspects related to the DDR and
AU approaches for the definition of a notation:

\begin{enumerate}
 \item Since the previous FMS analyses with the L$\gamma$ 
term~\cite{Fagundes_Menon_Silva:2012a,Fagundes_Menon_Silva:2013a,Menon_Silva:2013a,Menon_Silva:2013b}
are based on DDR  with some specific assumptions, namely to consider the DDR series up to the third order
and the inclusion of the effective subtraction constant, we shall adopt
here these FMS parametrizations as \textit{representative} of the DDR approach (in what concerns the leading L$\gamma$). 
For that reason, the DDR parametrizations will be denoted as FMS-L$\gamma$ model and FMS-L2 model.

 \item As discussed in Sect.~\ref{subsec:rise_DDR}, within the DDR approach, the FMS-L2 model is nothing more than
the FMS-L$\gamma$ model for $\gamma = 2$, i.e. a particular case. However, setting $\gamma=2$ in the result for L$\gamma$ within AU
does not correspond to the L2 result, as can be seen in Eq.~\eqref{eq:amp-Lgamma2-AU-exact} and Table~\ref{tab:Lgamma_analytical_results}.
Therefore, in order to refer to a specific
model in the AU approach we need to distinguish these three cases, namely
L$\gamma$, L$\gamma$ for $\gamma = 2$ and L2. To this end,
we will adopt the following short notation: AU-L$\gamma$ model, AU-L$\gamma$=2 model and AU-L2
model, respectively.
\end{enumerate}

All these analytical parametrizations can are summarized as follows:
\begin{equation*}
\text{DDR Approach} \begin{cases}
 \text{FMS-L$\gamma$ Model} \\ 
 \text{FMS-L2 Model}
\end{cases}
\qquad
\text{AU\ Approach} \begin{cases}
 \text{AU-L2 Model} \\
 \text{AU-L$\gamma$ Model} \\
 \text{AU-L$\gamma$=2 Model}
\end{cases} 
\end{equation*}

The corresponding formulas, displayed in the next sections,
are constructed with the results discussed in the previous sections
together with the optical theorem, Eq.~\eqref{eq:rise-optical-theo}, and the
definition of $\rho$ parameter, Eq.~\eqref{eq:rise-rho-def}.

\subsection{Derivative Dispersion Relation Approach}
\label{subsec:rise_DDR_models}

\subsubsection{FMS-L$\gamma$ Model}
\label{subsubsec:rise_FMS_Lgamma}


The total cross section is given by the
parametrization introduced by Amaldi \etal~\cite{Amaldi:1977}.
Following the notation of Refs.~\cite{Fagundes_Menon_Silva:2012a,Fagundes_Menon_Silva:2013a}, we write:
\begin{equation}
\sigmatot(s) = a_1\, \left[\frac{s}{s_0}\right]^{-b_1} + 
\tau\, a_2\, \left[\frac{s}{s_0}\right]^{-b_2}
+  \alpha + \beta \ln^{\gamma}\left(\frac{s}{s_0}\right),
\label{eq:sigtot-FMS-Lgamma}
\end{equation}

\noindent where $\tau$ = -1 (+1) for $pp$ ($\bar{p}p$) scattering, 
while $a_1$, $b_1$, $a_2$, $b_2$, $\alpha$, $\beta$, $\gamma$ are real free fit parameters.
The energy scale is \textit{fixed} at the physical threshold for scattering states 
\begin{equation}
s_0 = 4m_p^2,
\label{eq:rise-energy-scale}
\end{equation}

\noindent with $m_p$ the proton mass. In Section~\ref{subsec:rise_fit_procedures}, 
we discuss this choice for the energy scale.

The $\rho(s)$ dependence is analytically determined through DDR
using the Kang and Nicolescu representation~\cite{Kang_Nicolescu:1975} 
(see Sect.~\ref{subsec:rise_DDR} for details):
\begin{equation}
\rho(s) = \frac{1}{\sigmatot(s)}
\left\{ \frac{K_\text{eff}}{s} + T^{\reggeon}(s) + T^{\pomeron}(s) \right\},
\label{eq:rho-FMS-Lgamma}
\end{equation}

\noindent where $K_\text{eff}$ is the \textit{effective subtraction constant}
and the terms ($T$) associated with Reggeon ($\reggeon$) and Pomeron ($\pomeron$) contributions read
\begin{align}
T^{\reggeon}(s) & =
- a_1\,\tan \left( \frac{\pi\, b_1}{2}\right) \left[\frac{s}{s_0}\right]^{-b_1} +
\tau \, a_2\, \cot \left(\frac{\pi\, b_2}{2}\right) \left[\frac{s}{s_0}\right]^{-b_2} 
\label{eq:rho-FMS-Lgamma-Reggeon}\\[10pt]
T^{\pomeron}(s) & =
\mathcal{A}\,\ln^{\gamma - 1} \left(\frac{s}{s_0}\right) +
\mathcal{B}\,\ln^{\gamma - 3} \left(\frac{s}{s_0}\right) +
\mathcal{C}\,\ln^{\gamma - 5} \left(\frac{s}{s_0}\right),
\label{eq:rho-FMS-Lgamma-Pomeron}
\end{align}

\noindent where
\begin{equation} 
\begin{split}
\mathcal{A} & = \frac{\pi}{2} \, \beta\, \gamma,  
\quad 
\mathcal{B} = \frac{1}{3} \left[\frac{\pi}{2}\right]^3 \, \beta\, \gamma\, [\gamma - 1][ \gamma - 2], \\
\mathcal{C} & = \frac{2}{15} \left[\frac{\pi}{2}\right]^5 \, \beta\, \gamma\, [\gamma - 1][ \gamma - 2]
[\gamma - 3][ \gamma - 4]. 
\end{split}
\label{eq:rho-FMS-Lgamma-factors}
\end{equation} 


\subsubsection{FMS-L2 Model}
\label{subsubsec:rise_FMS_L2}

In the particular case of $\gamma = 2$, from Eq.~\eqref{eq:rho-FMS-Lgamma-factors}, 
we have $\mathcal{A} = \pi \beta$, $\mathcal{B} = \mathcal{C} = 0$
and through Eqs.~\eqref{eq:sigtot-FMS-Lgamma}-\eqref{eq:rho-FMS-Lgamma-Pomeron}:
\begin{align}
\sigmatot(s) & = 
a_1\, \left[\frac{s}{s_0}\right]^{-b_1} + 
\tau\, a_2\, \left[\frac{s}{s_0}\right]^{-b_2}
+ \alpha + \beta \ln^{2}\left(\frac{s}{s_0}\right),
\label{eq:sigtot-FMS-L2}\\[10pt]
%
\rho(s) & = \frac{1}{\sigmatot(s)}
\left\{\frac{K_\text{eff}}{s}
- a_1\,\tan \left( \frac{\pi\, b_1}{2}\right) \left[\frac{s}{s_0}\right]^{-b_1}\!\!\! +
\tau \, a_2\, \cot \left(\frac{\pi\, b_2}{2}\right) \left[\frac{s}{s_0}\right]^{-b_2} \!\!\! 
+ \pi \beta \ln\left(\frac{s}{s_0}\right)\right\}.
\label{eq:rho-FMS-L2}
\end{align}

These analytic expressions for $\sigmatot(s)$ and $\rho(s)$, Eqs.~\eqref{eq:sigtot-FMS-L2} and \eqref{eq:rho-FMS-L2},
have solid basis on the Regge-Gribov formalism  and Eq~\eqref{eq:rho-FMS-L2} can be directly obtained 
from Eq.~\eqref{eq:sigtot-FMS-L2} using DDR.

\subsubsection{PDG-L2 Model}

We recall that Eqs.~\eqref{eq:sigtot-FMS-L2} and \eqref{eq:rho-FMS-L2} have also
the same analytic structure as those selected by
the COMPETE Collaboration and used in the successive editions by the PDG, except for modifications in the
energy scale and the pre-factor $\beta$, as discussed in Section~\ref{subsec:rise_Regge_COMPETE_PDG}.
We shall refer as PDG-L2 model the L2 model without the subtraction constant and with the two
constraints given by Eqs.~\eqref{eq:pdg2012-energy-scale} and \eqref{eq:pdg2012-B-parameter}.

\subsection{Asymptotic Uniqueness Approach}
\label{subsec:rise_AU_models}

\subsubsection{AU-L2 Model}
\label{subsubsec:rise_AU_L2}

From Section~\ref{subsec:rise_AU}, see Eq.~\eqref{eq:amp-Pomeron-Triple-Pole-AU}, this model has the
same structure obtained with the Regge-Gribov 
and DDR
formalisms (except, in the last case, for the subtraction constant).
Although the formulas correspond to Eqs.~\eqref{eq:sigtot-FMS-L2} and \eqref{eq:rho-FMS-L2}, 
without $K_\text{eff}$, we display the results for future reference:
\begin{align}
\sigmatot(s) & = 
a_1\, \left[\frac{s}{s_0}\right]^{-b_1} + 
\tau\, a_2\, \left[\frac{s}{s_0}\right]^{-b_2}
+ \alpha + \beta \ln^{2}\left(\frac{s}{s_0}\right),
\label{eq:sigtot-AU-L2}\\[10pt]
%
\rho(s) & = \frac{1}{\sigmatot(s)}
\left\{
- a_1\,\tan \left( \frac{\pi\, b_1}{2}\right) \left[\frac{s}{s_0}\right]^{-b_1} +
\tau \, a_2\, \cot \left(\frac{\pi\, b_2}{2}\right) \left[\frac{s}{s_0}\right]^{-b_2}  
+ \pi \beta \ln\left(\frac{s}{s_0}\right)\right\},
\label{eq:rho-AU-L2}
\end{align}
\noindent where, as before, $\tau$ = -1 (+1) for $pp$ ($\bar{p}p$) scattering. 

\subsubsection{AU-L$\gamma$ Model}
\label{subsubsec:rise_AU_Lgamma}

From Section~\ref{subsec:rise_AU}, the AU approach applied to Reggeons and the $\ln^\gamma$ term leads to:
\begin{equation}
\sigmatot(s) = 
a_1\, \left[\frac{s}{s_0}\right]^{-b_1} + 
\tau\, a_2\, \left[\frac{s}{s_0}\right]^{-b_2} 
+ \alpha + 
\beta \cos \left(\frac{\gamma \phi}{2}\right) \ln^{\gamma/2}\left(\frac{s}{s_0}\right)
\left[\ln^2\left(\frac{s}{s_0}\right) + \pi^2\right]^{\gamma/4},
\label{eq:sigtot-AU-Lgamma}
\end{equation}

\begin{multline}
\rho(s) =\frac{1}{\sigmatot(s)}
 \left\{ - a_1\,\tan \left( \frac{\pi\, b_1}{2}\right) \left[\frac{s}{s_0}\right]^{-b_1} +
\tau \, a_2\, \cot \left(\frac{\pi\, b_2}{2}\right) \left[\frac{s}{s_0}\right]^{-b_2}   
\right. \\
\left. \, + \,  \beta \sin \left(\frac{\gamma \phi}{2}\right)
\ln^{\gamma/2}\left(\frac{s}{s_0}\right)
\left[\ln^2\left(\frac{s}{s_0}\right) + \pi^2\right]^{\gamma/4}\right\},
\label{eq:rho-AU-Lgamma}
\end{multline}

\noindent where 
\begin{equation}
\phi = \phi(s) = \tan^{-1}\left(\frac{\pi}{\ln(s/s_0)}\right).
\label{eq:phi-AU}
\end{equation}

\subsubsection{AU-L$\gamma$=2 Model}
\label{subsubsec:rise_AU_Lgamma2}

For $\gamma = 2$, Eqs.~\eqref{eq:sigtot-AU-Lgamma} and \eqref{eq:rho-AU-Lgamma} read
\begin{equation}
\sigmatot(s) = 
a_1\, \left[\frac{s}{s_0}\right]^{-b_1} + 
\tau\, a_2\, \left[\frac{s}{s_0}\right]^{-b_2} 
+ \alpha + 
\beta \cos \left(\phi\right) \ln\left(\frac{s}{s_0}\right)
\left[\ln^2\left(\frac{s}{s_0}\right) + \pi^2\right]^{1/2},
\label{eq:sigtot-AU-Lgamma2}
\end{equation}

\begin{multline}
\rho(s) = \frac{1}{\sigmatot(s)}
\left\{
- a_1\,\tan \left( \frac{\pi\, b_1}{2}\right) \left[\frac{s}{s_0}\right]^{-b_1} +
\tau \, a_2\, \cot \left(\frac{\pi\, b_2}{2}\right) \left[\frac{s}{s_0}\right]^{-b_2}   
\right. \\
 \left. \, + \, \beta \sin \left(\phi\right)
\ln\left(\frac{s}{s_0}\right)
\left[\ln^2\left(\frac{s}{s_0}\right) + \pi^2\right]^{1/2}\right\},
\label{eq:rho-AU-Lgamma2}
\end{multline}

\noindent with $\phi = \phi(s)$ given by Eq.~\eqref{eq:phi-AU}.

We stress that, differently from the FMS-L$\gamma$ and FMS-L2 models,
this AU-L$\gamma$=2 model does not correspond to the AU-L2 model,
given by Eqs.~\eqref{eq:sigtot-AU-L2}-\eqref{eq:rho-AU-L2}.
In Sect.~\ref{subsec:rise_analytic_differences} the \textit{analytic}
similarities and main differences between the AU and DDR approaches,
related to L2 and L$\gamma$ models are discussed in more detail.

\section{Experimental Data}
\label{sec:rise_expdata}

In this analysis we are interested in comparing two approaches connecting the real and imaginary parts
of the amplitude, and in studying the behavior of $\sigmatot$ and $\rho$ at high and asymptotic energies.
To do that, we will consider reactions that have available data in the largest energy
range: $pp$ and $\ppbar$ scattering, as done in 
Refs.~\cite{Fagundes_Menon_Silva:2012a,Fagundes_Menon_Silva:2013a,Menon_Silva:2013a,Menon_Silva:2013b}.
This choice allows the investigation of possible high-energy
effects that may be unrelated to the trends of the lower-energy data
of other reactions.

\begin{figure}[htb!]
 \centering
 \subfloat[\label{subfig:sigmatot_rho_data_pp_ppbar_panel_a}$\sigmatot$]{\includegraphics[scale=0.39]{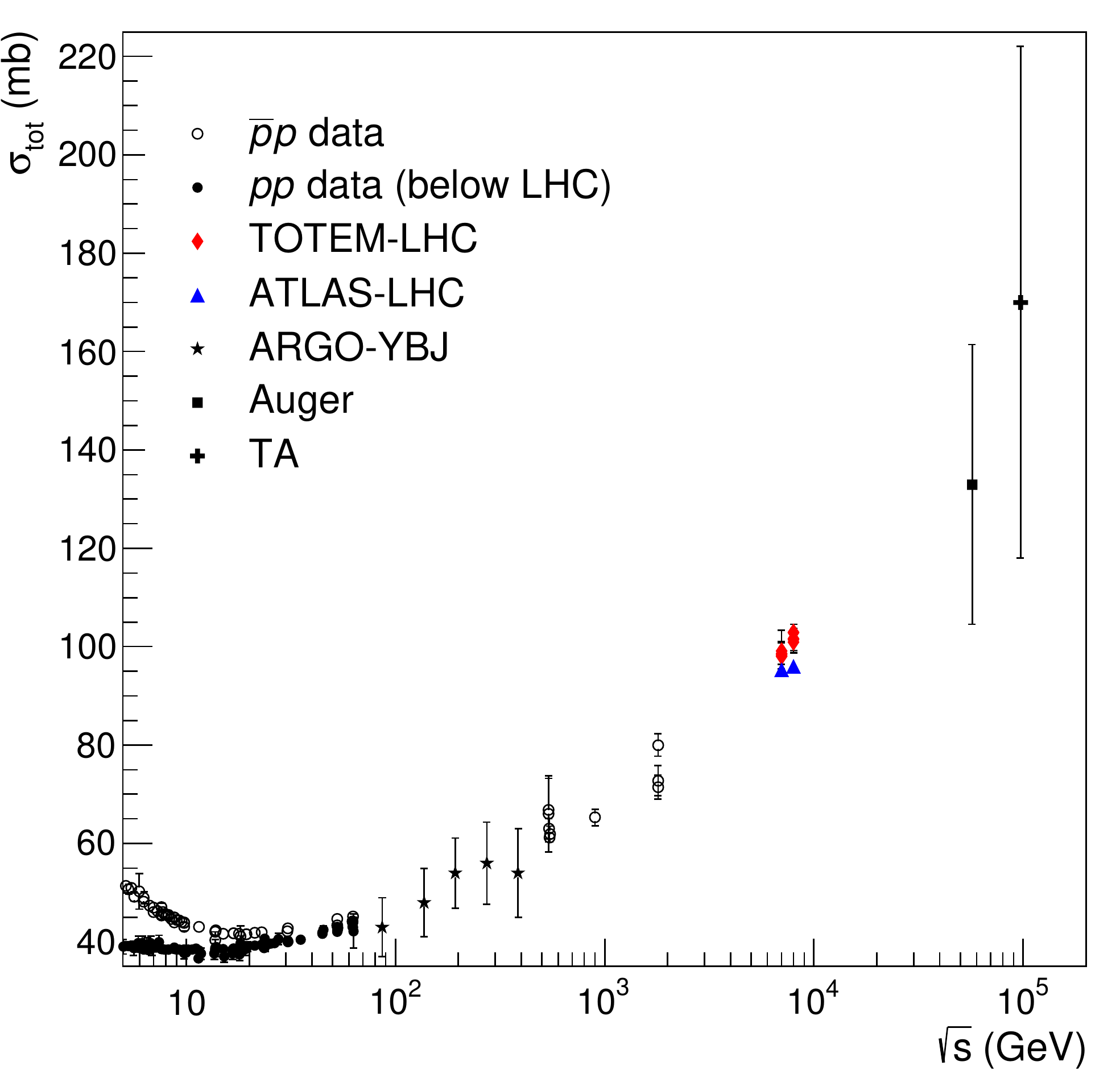}}\hfill
 \subfloat[\label{subfig:sigmatot_rho_data_pp_ppbar_panel_b}$\rho$]{\includegraphics[scale=0.39]{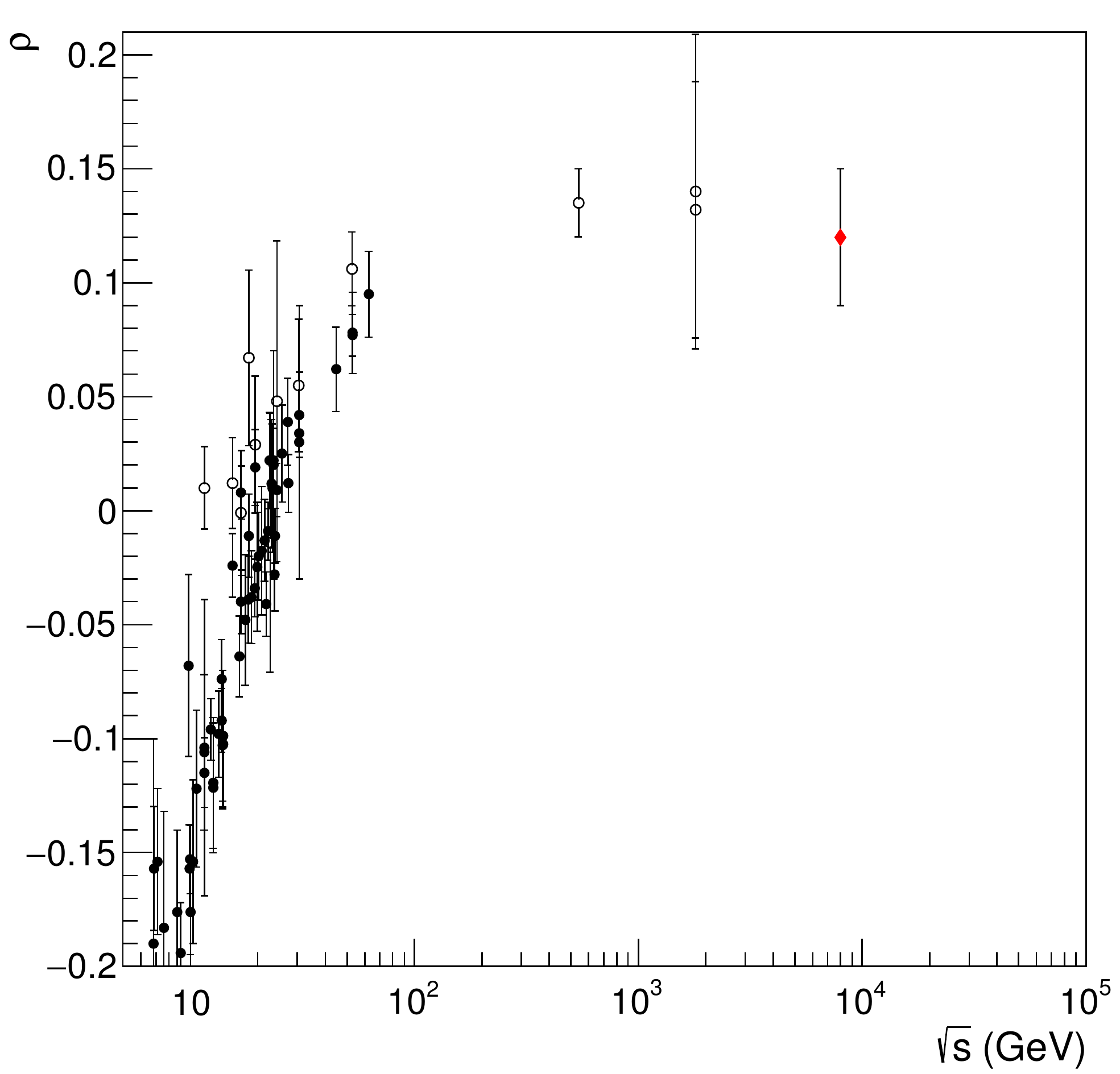}}
 \caption{\label{fig:sigmatot_rho_data_pp_ppbar}Accelerator data on \subref{subfig:sigmatot_rho_data_pp_ppbar_panel_a} $\sigmatot$ and 
 \subref{subfig:sigmatot_rho_data_pp_ppbar_panel_b} $\rho$ from $pp$ and $\bar{p}p$ 
scattering used in this analysis. Some estimations for the
$pp$ total cross section from cosmic-ray experiments (ARGO-YBJ, Auger and TA) are also displayed.
The symbols here defined are assumed in all figures.}
\end{figure}

Our dataset on $\sigmatot$ and $\rho$ comprises all the accelerator data from $pp$ and $\bar{p}p$
elastic scattering above 5 GeV~\cite{PDG_data_website} (same cutoff used in the COMPETE and PDG analyses),
including all published results at the energies 7 and 8~TeV by the TOTEM and ATLAS Collaborations 
(see Chapter~\ref{chapt:data_fits}).  
The data on $\sigmatot$ at the highest-energy region are displayed in Table~\ref{tab:sigmatot_data_pp_ppbar_TeV}
together with information about the uncertainties associated.
The recent measurement of $\rho$ at 8 TeV by the TOTEM
Collaboration~\cite{TOTEM:2016} is also included in the dataset. 
All these data and information on
$\sigmatot$ and $\rho$ are shown in Figure~\ref{fig:sigmatot_rho_data_pp_ppbar},
where we can see the incompatibility between TOTEM and ATLAS measurements at 8~TeV.

Although not taking part in the data reductions,
we  display in the figures, as illustration, some estimations of the $\sigmatot$ from cosmic-ray experiments:
ARGO-YBJ results at $\sim$ 100 - 400~GeV~\cite{ARGO_YBJ:2009}, Auger result at 57~TeV~\cite{PierreAuger:2012} 
and Telescope Array (TA) result at 95~TeV~\cite{TA:2015}. 

Note that at 8 TeV, from Table~\ref{tab:sigmatot_data_pp_ppbar_TeV}, from~\cite{TOTEM:2016} (TOTEM) and \cite{ATLAS:2016}
(ATLAS)
\begin{align}
\frac{\sigmatot^{\text{TOTEM}} - \sigmatot^{\text{ATLAS}}}{\Delta \sigmatot^{\text{TOTEM}}} & = 
\frac{103 - 96.07}{2.3} = 3.0\\
\intertext{and}
\frac{\sigmatot^{\text{TOTEM}} - \sigmatot^{\text{ATLAS}}}{\Delta \sigmatot^{\text{ATLAS}}} & = 
\frac{103 - 96.07}{0.92} = 7.5.
\end{align}

In order to explore the different scenarios given by TOTEM and ATLAS data,
we will consider three \textit{ensembles} of experimental data in our fits.
They have in common the accelerator data above 5 GeV and below the LHC energies,
and differ by the LHC data included, following the notation defined below.

\begin{itemize}
 \item \textbf{Ensemble T}: accelerator data above 5 GeV and below the LHC including only the TOTEM data.

 \item \textbf{Ensemble A}: same as above but now including the ATLAS data in place of TOTEM data.
 
 \item \textbf{Ensemble T+A}: all data available, i.e. accelerator data below LHC with TOTEM and ATLAS data.
\end{itemize}

\begin{table}[htb!]
\caption{\label{tab:sigmatot_data_pp_ppbar_TeV}Experimental information on measurements of the $\bar{p}p$ and $pp$ total cross section at
the highest energies from collider experiments (CERN-SPS, Fermilab-Tevatron and CERN-LHC): central value ($\sigmatot$), statistical uncertainties
($\Delta\sigmatot^\text{stat.}$), systematic uncertainties
($\Delta\sigmatot^\text{syst.}$), total uncertainty from quadrature
($\Delta\sigmatot$) and total relative uncertainty ($\Delta\sigmatot/\sigmatot$).}
\small
\begin{tabular}{c c c c  c c c c }
\hline\hline
 reaction    &$\sqrt{s}$&$\sigmatot$ &$\Delta\sigmatot^\text{stat.}$&
$\Delta\sigmatot^\text{syst.}$&$\Delta\sigmatot$& $\Delta\sigmatot/\sigmatot$ & Collaboration\\
(collider) & (TeV)& (mb) & (mb) & (mb) & (mb) & ($\times$ 100) \% & [reference] \\
\hline
$\bar{p}p$ (SPS)& 0.546& 61.9 & 1.5 & 1.0  &1.8 &2.9 & UA4 \cite{UA4:1984} \\
$\bar{p}p$ (Tevatron)&0.546 & 61.26 & 0.93 & -  & 0.93& 1.5 & CDF \cite{CDF:1994a} \\
$\bar{p}p$ (SPS)& 0.900& 65.3 & 0.7 & 1.55  &1.66 & 2.5& UA5 \cite{UA5:1986a}\\
$\bar{p}p$ (Tevatron)& 1.80& 72.8 & 3.1 & -  &3.1 & 4.3& E710 \cite{E710:1992}\\
 $\bar{p}p$ (Tevatron)& 1.80& 80.03& 2.24& -  &2.24& 2.8& CDF \cite{CDF:1994a}\\
$\bar{p}p$ (Tevatron)& 1.80& 71.42& 1.55& 2.6&3.03& 4.2& E811 \cite{E811:1999}\\ 
\hline
 $pp$ (LHC) & 7.0 & 98.3  & 0.2   & 2.8  & 2.8& 2.9 & TOTEM \cite{TOTEM:2011b}\\
 $pp$ (LHC) & 7.0 & 98.6  &  -    & 2.2  & 2.2& 2.2 & TOTEM \cite{TOTEM:2013a}\\
 $pp$ (LHC) & 7.0 & 98.0  &  -    & 2.5  & 2.5& 2.6 & TOTEM \cite{TOTEM:2013c}\\
 $pp$ (LHC) & 7.0 & 99.1  &  -    &  4.3 & 4.3& 4.3& TOTEM \cite{TOTEM:2013c}\\
 $pp$ (LHC) & 7.0 & 95.35 & 0.38  &1.304 & 1.36& 1.4& ATLAS \cite{ATLAS:2014}\\
 $pp$ (LHC) & 8.0 & 101.7 & -     & 2.9  & 2.9 & 2.9& TOTEM \cite{TOTEM:2013d}\\
 $pp$ (LHC) & 8.0 & 101.5 & -     & 2.1  & 2.1  & 2.1 & TOTEM \cite{TOTEM:2015}\\
 $pp$ (LHC) & 8.0 & 101.9 & -     & 2.1  & 2.1  & 2.1 & TOTEM \cite{TOTEM:2015}\\
 $pp$ (LHC) & 8.0 & 102.9 & -     & 2.3  & 2.3  & 2.2 & TOTEM \cite{TOTEM:2016}\\
 $pp$ (LHC) & 8.0 & 103.0 & -     & 2.3  & 2.3  & 2.2 & TOTEM \cite{TOTEM:2016}\\
 $pp$ (LHC) & 8.0 & 96.07 & 0.18  & 0.85 $\pm$ 0.31  & 0.92 & 1.0 & ATLAS \cite{ATLAS:2016}\\
\hline\hline
  \end{tabular}
  \normalsize
\end{table}

\section{Data Reductions and Results}
\label{sec:rise_results}

In order to confront the DDR and AU approaches with L2 and L$\gamma$ models,
four parametrizations for $\sigmatot$ and $\rho$ are considered: 
FMS-L2 and FMS-L$\gamma$ models, Eqs.~\eqref{eq:sigtot-FMS-Lgamma}-\eqref{eq:rho-FMS-Lgamma-factors}
(since L2 is the particular case of L$\gamma$ for $\gamma$ = 2), 
AU-L$\gamma$=2 model, Eqs.~\eqref{eq:sigtot-AU-Lgamma2}-\eqref{eq:rho-AU-Lgamma2} 
and AU-L$\gamma$ model, Eqs.~\eqref{eq:sigtot-AU-Lgamma}-\eqref{eq:phi-AU}
(since the L2 is not a particular case of L$\gamma$ for $\gamma$ = 2).

For each model, we develop fits to $pp$ and $\ppbar$ data with the three ensembles 
defined in the previous section: T, A and T+A.

\subsection{Fit Procedures}
\label{subsec:rise_fit_procedures}

An important aspect of the parametrizations considered in this study is the presence
of nonlinearity in some parameters. This aspect demands a methodology for the choice of the
initial values (IV) of the free parameters.
We choose as initial values
the central values reported by the PDG in their most recent
data reductions. Here we use the values of the parameters published
in the 2016 edition, which for the $pp$ and $\bar{p}p$
scattering read~\cite{PDG:2016}:
\begin{equation}
\begin{split}
a_1 &= 13.07 \pm 0.17 \text{ mb}, \quad b_1 = 0.4473 \pm 0.0077, \quad a_2 = 7.394 \pm 0.081 \text{ mb},       \\
b_2 &= 0.5486 \pm 0.0049, \quad \alpha = 34.41 \pm 0.13 \text{ mb}, \quad \beta = 0.2720 \pm 0.0024 \text{ mb},\\
s_0 &= 15.977 \pm 0.075 \text{ GeV}^2.
\end{split}
\label{eq:IV-PDG2016}
\end{equation}

\begin{figure}[htb!]
 \centering
 \subfloat[\label{subfig:res_PDG2016_panel_a}$\sigmatot$]{\includegraphics[scale=0.39]{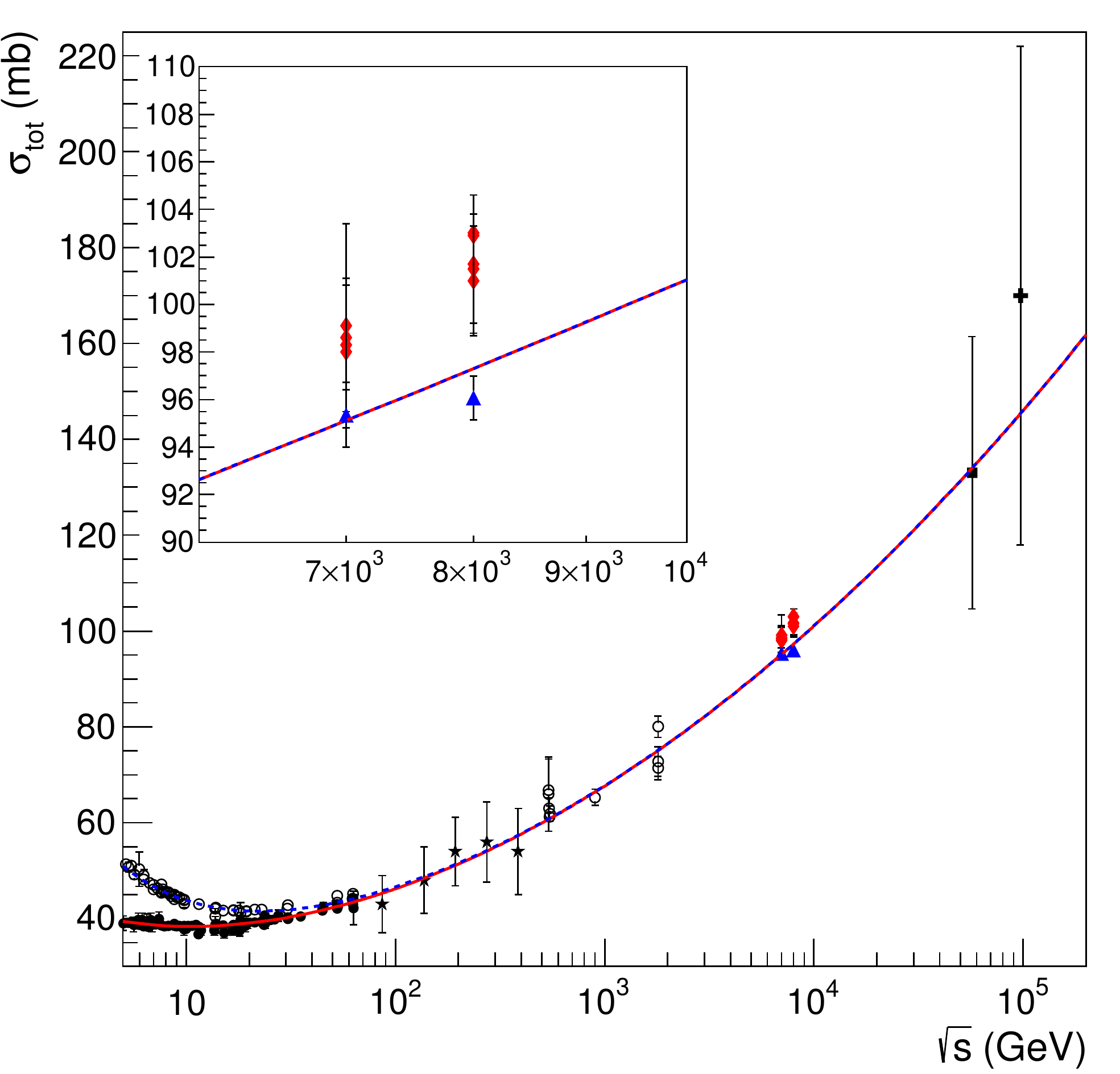}}\hfill
 \subfloat[\label{subfig:res_PDG2016_panel_b}$\rho$]{\includegraphics[scale=0.39]{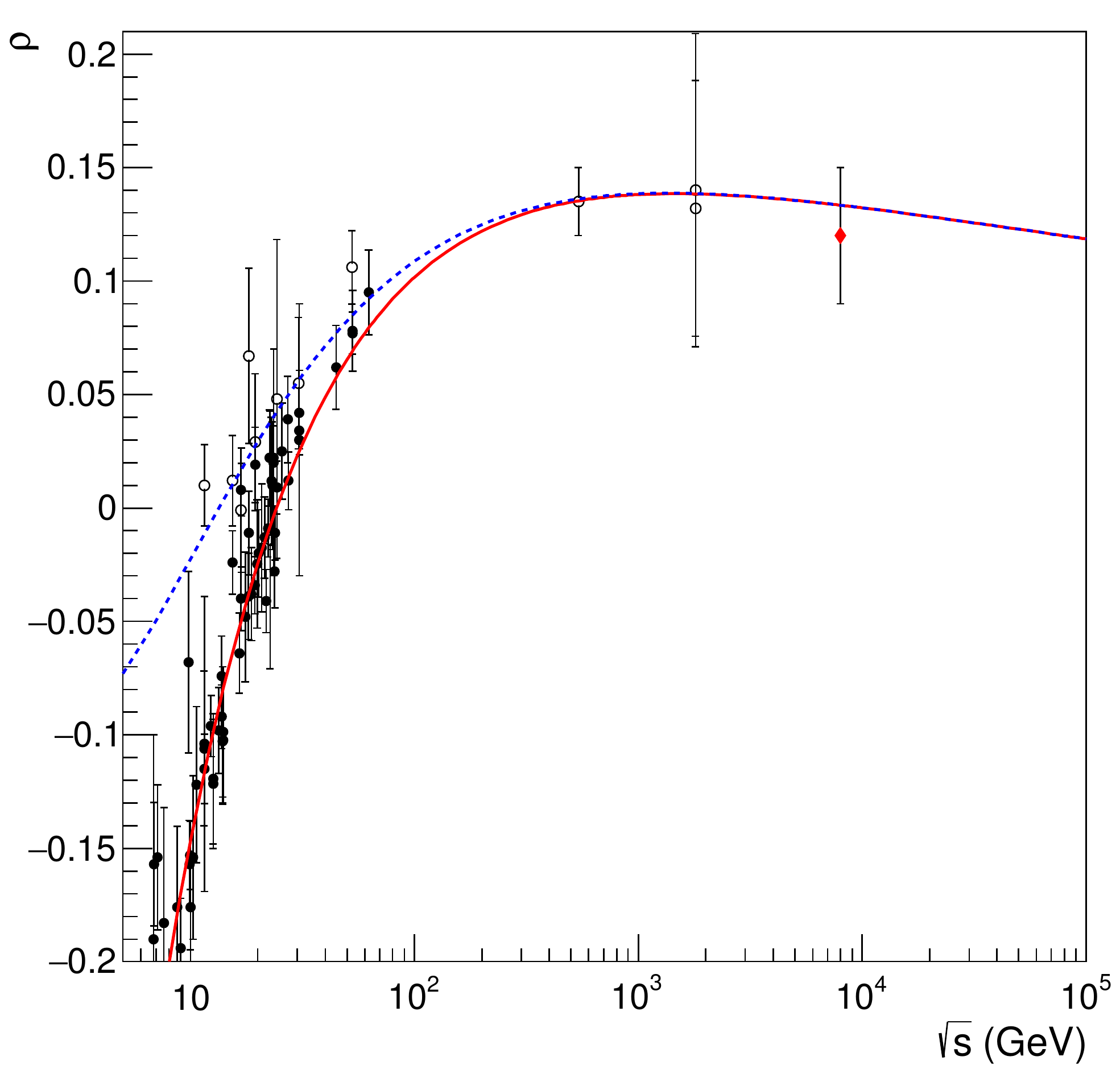}}
 \caption{\label{fig:res_PDG2016}PDG-L2 results from the PDG 2016 \cite{PDG:2016}, parameters from Eq.~\eqref{eq:IV-PDG2016}, compared with 
$pp$ and $\bar{p}p$ data. The insert shows the $\sigmatot$ data and curve
in the region 6 - 10 TeV.}
\end{figure}

Recall that these results were obtained with the PDG-L2 model (Sect.~\ref{subsubsec:rise_FMS_L2})
from fits to data comprising several reactions
and not only to $pp$ and $\bar{p}p$ data. Also, the dataset did not include 
the latest TOTEM measurements and the ATLAS result at 8 TeV.
For future discussion, we display in Figure~\ref{fig:res_PDG2016} the 
corresponding results for $\sigmatot(s)$ and $\rho(s)$ with the PDG-L2 model (2016), 
for $pp$ and $\bar{p}p$ scattering \cite{PDG:2016}, together with the experimental
data used here.

Specifically, we used the PDG2016 results (central values) as initial values for the fits with $\gamma=2$ fixed 
(FMS-L2 and AU-L$\gamma=2$). Since the subtraction constant is absent in the PDG parametrization,
we have considered the initial value 0 for $K_\text{eff}$.
The next step is to let $\gamma$ be a free parameter.
For that, we use in each case the resulting central values of the free parameters
as initial values for data reductions with the corresponding L$\gamma$ models.
These fit procedures are summarized in the scheme below
$$
 \mbox{PDG-L2, Eq.~\eqref{eq:IV-PDG2016}}\ 
\xrightarrow{\text{IV}}
\quad
\begin{cases}
 \mbox{fit\ FMS-L2 model}\ \xrightarrow{\text{IV}}\ \mbox{fit\ FMS-L$\gamma$ model} \\
 \mbox{fit\ AU-L$\gamma$=2 model}\ \xrightarrow{\text{IV}}\ \mbox{fit\ AU-L$\gamma$ model}.
\end{cases}
$$
These procedures have been used with each one of the three ensemble: T, A and T+A. 
Statistical parameters and details of the minimization code are discussed in Chapter~\ref{chapt:data_fits}.

\subsubsection{Energy Scale}
\label{subsubsec:rise_energy_scale}

Besides the nonlinearity of the parametrizations, the energy scale appearing in the power
and logarithmic laws also demands some comments. In this analysis, following Refs.~\cite{Menon_Silva:2013a,Menon_Silva:2013b}
we choose to fix this parameter at the \textit{physical} energy threshold, Eq.~\eqref{eq:rise-energy-scale},
\begin{equation}
s_0 = 4m_p^2 \sim \mathrm{3.521\ GeV}^2,
\nonumber
\end{equation}

\noindent instead of fixing it at arbitrary values, for instance 1 GeV$^2$, or even to fix it at the
cutoff energy, $s_\text{min}$.

Of course, we could also consider $s_0$ as a free parameter. However, letting $s_0$ be free adds additional
nonlinearity in the fits that, at a first moment, we prefer to avoid. In Refs.~\cite{Menon_Silva:2013a,Menon_Silva:2013b}, 
fits with a free energy scale were considered and the implications on the parameter correlations are discussed.

Next, we present the fit results with
the DDR Approach (Sect.~\ref{subsec:rise_results_FMS_models}) followed by those with the AU approach
(Sect.~\ref{subsec:rise_results_AU_models}). A detailed discussion on all these results is the content
of Sect.~\ref{subsec:rise_discussion_fit_results}.

\subsection{FMS-L2 and FMS-L$\gamma$ models}
\label{subsec:rise_results_FMS_models}

\begin{figure}[htb!]
 \centering
 \subfloat[\label{subfig:res_FMS_L2_T_A_TA_panel_a}$\sigmatot$]{\includegraphics[scale=0.39]{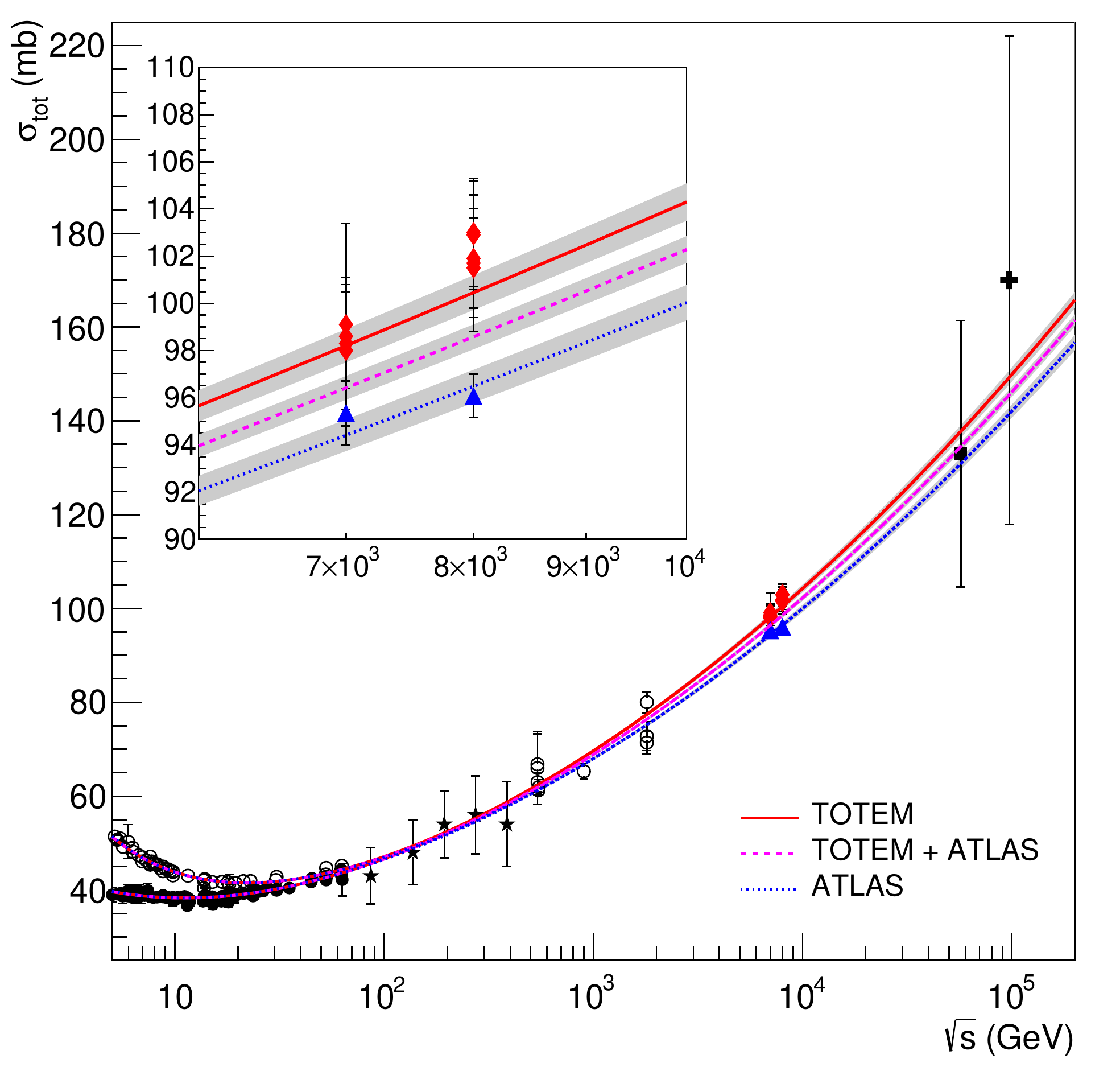}}\hfill
 \subfloat[\label{subfig:res_FMS_L2_T_A_TA_panel_b}$\rho$]{\includegraphics[scale=0.39]{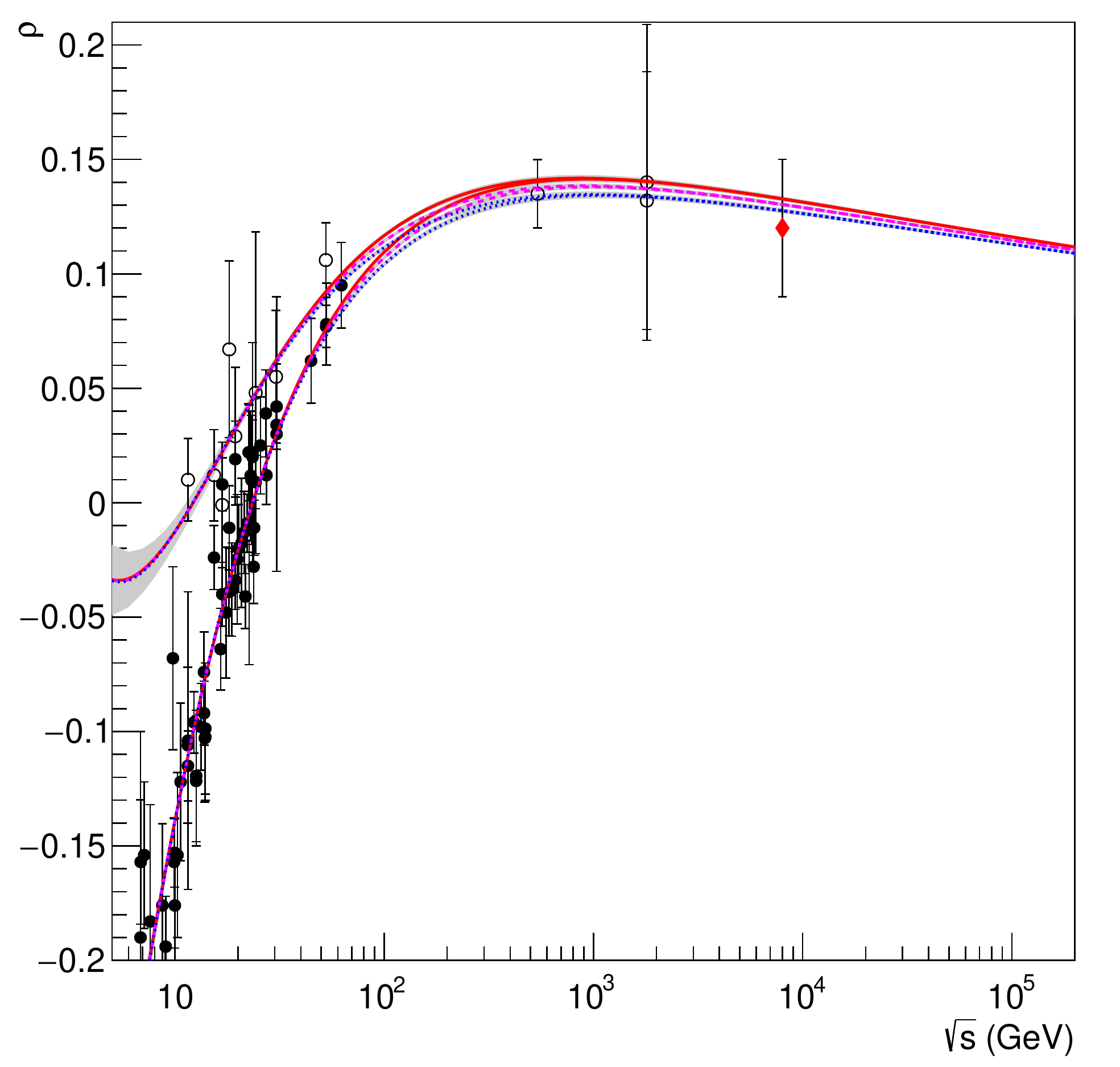}}
 \caption{\label{fig:res_FMS_L2_T_A_TA}Fit results with the FMS-L2 model to ensembles T, A and T+A,
Eqs.~\eqref{eq:sigtot-FMS-Lgamma}-\eqref{eq:rho-FMS-Lgamma-factors}), Table~\ref{tab:res_FMS_L2_Lgamma_T_A_TA}.}
\end{figure}

\begin{figure}[htb!]
 \centering
 \subfloat[\label{subfig:res_FMS_Lgamma_T_A_TA_panel_a}$\sigmatot$]{\includegraphics[scale=0.39]{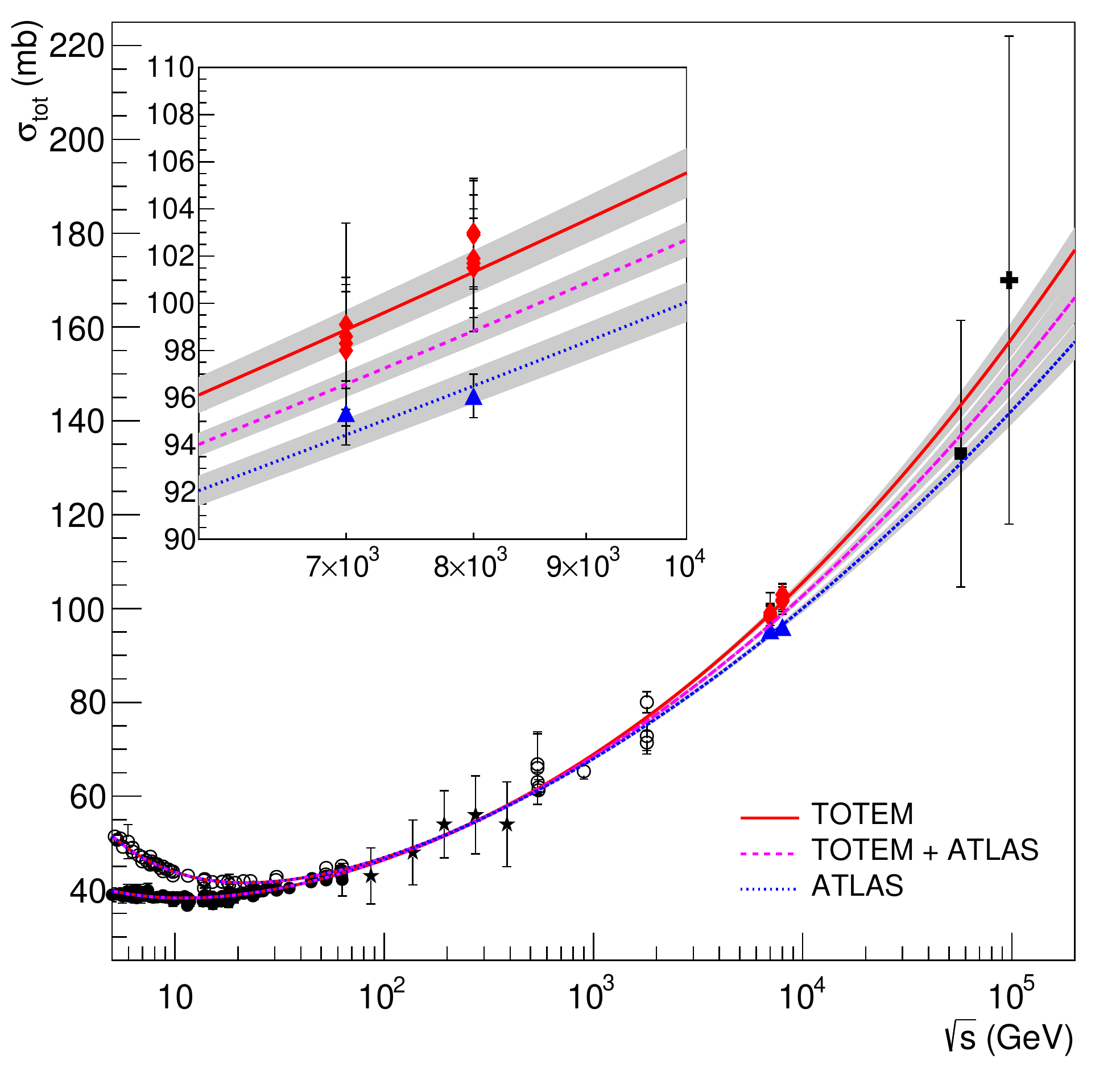}}\hfill
 \subfloat[\label{subfig:res_FMS_Lgamma_T_A_TA_panel_b}$\rho$]{\includegraphics[scale=0.39]{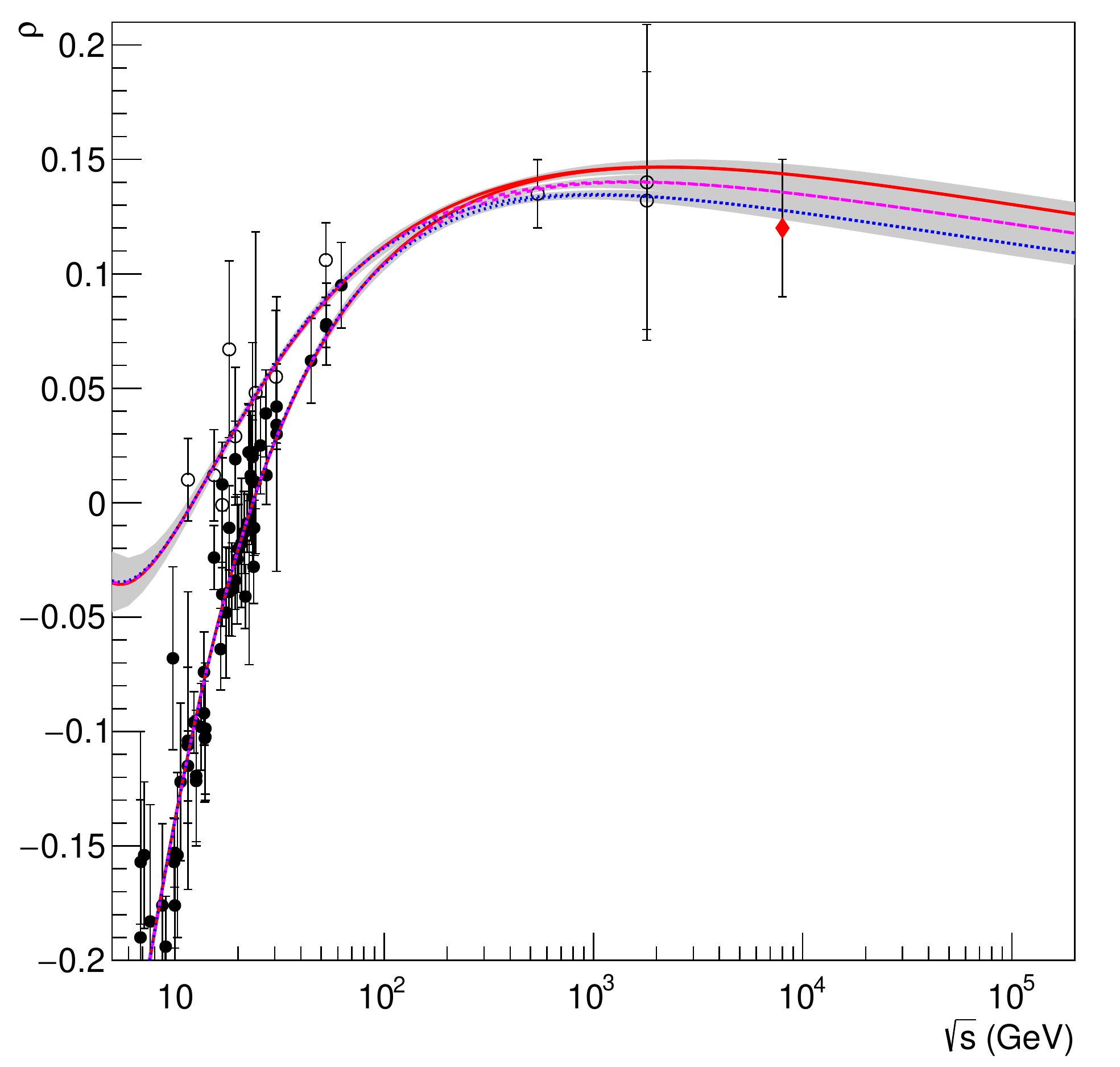}}
 \caption{\label{fig:res_FMS_Lgamma_T_A_TA}Fit results with the FMS-L$\gamma$ model to ensembles T, A and T+A,
Eqs.~\eqref{eq:sigtot-FMS-Lgamma}-\eqref{eq:rho-FMS-Lgamma-factors}, Table~\ref{tab:res_FMS_L2_Lgamma_T_A_TA}.}
\end{figure}

The fit results obtained with models FMS-L2 and FMS-L$\gamma$
are displayed in Table~\ref{tab:res_FMS_L2_Lgamma_T_A_TA}, including statistical information
on the quality of the fit. The corresponding results for $\sigmatot(s)$ and $\rho(s)$
with ensembles T, A and T+A, together with the experimental data analyzed,
are shown in Fig.~\ref{fig:res_FMS_L2_T_A_TA} with the FMS-L2 model and in Fig.~\ref{fig:res_FMS_Lgamma_T_A_TA} with
the FMS-L$\gamma$ model.

\begin{table}[htb!]
\centering
\caption{\label{tab:res_FMS_L2_Lgamma_T_A_TA}Fit results with the FMS-L2 and FMS-L$\gamma$ models,
Eqs.~\eqref{eq:sigtot-FMS-Lgamma}-\eqref{eq:rho-FMS-Lgamma-factors}, 
to ensembles T, A and T+A. Energy scale fixed, $s_0 = 4m_p^2 =$ 3.521 GeV$^2$.
Also displayed is the statistical information on the quality of the fits 
(chi-squared per degree of freedom and integrated probability).
Parameters $a_1$, $a_2$, $\alpha$ and $\beta$ are given in mb, 
$K_\text{eff}$ in mbGeV$^2$ and $b_1$, $b_2$, $\gamma$ are dimensionless.
}
\begin{tabular}{@{}*{7}{c}}
\hline\hline
Ensemble: &\multicolumn{2}{c}{TOTEM}&\multicolumn{2}{c}{ATLAS} & \multicolumn{2}{c}{TOTEM + ATLAS}\\
\cmidrule(lr){2-3} \cmidrule(lr){4-5} \cmidrule(lr){6-7} 
Model:          &    L2      & L$\gamma$ & L2         & L$\gamma$  & L2         & L$\gamma$          \\\hline
$a_1$           & 32.11(60)  & 31.5(1.3) & 32.39(86)  & 32.4(1.0)  & 32.16(67)  & 31.60(98)          \\
$b_1$           & 0.381(17)  & 0.528(57) & 0.435(19)  & 0.438(57)  & 0.406(16)  & 0.484(84)          \\
$a_2$           & 16.98(72)  & 17.10(74) & 17.04(72)  & 17.04(72)  & 17.01(72)  & 17.07(73)          \\
$b_2$           & 0.545(13)  & 0.546(13) & 0.545(13)  & 0.545(13)  & 0.545(13)  & 0.546(13)          \\
$\alpha$        & 29.25(44)  & 34.0(1.1) & 30.88(35)  & 31.0(2.1)  & 30.06(34)  & 32.8(2.2)          \\
$\beta$         & 0.2546(39) & 0.103(29) & 0.2347(35) & 0.231(83)  & 0.2451(28) & 0.151(71)          \\
$\gamma$        & 2 (fixed)  & {\bf2.301(98)} & 2 (fixed)  & {\bf2.01(12)}   & 2 (fixed)  & {\bf2.16(16)}           \\
$K_\text{eff}$  & 50(17)     & 109(36)   & 74(20)     & 75(27)     & 61(17)     & 90(42)             \\\hline 
$\nu$           & 242        & 241       & 235        & 234        & 244        & 243                \\
$\chi^2/\nu$    & 1.09       & 1.07      & 1.08       & 1.09       & 1.15       & 1.14               \\
$P(\chi^2)$     & 0.150      & 0.213     & 0.177      & 0.166      & 0.059      & 0.063              \\
\hline\hline
\end{tabular}
\end{table}

\subsection{AU-L2 and AU-L$\gamma$ Models}
\label{subsec:rise_results_AU_models}

\begin{figure}[htb!]
 \centering
 \subfloat[\label{subfig:res_AU_Lgamma2_T_A_TA_panel_a}$\sigmatot$]{\includegraphics[scale=0.39]{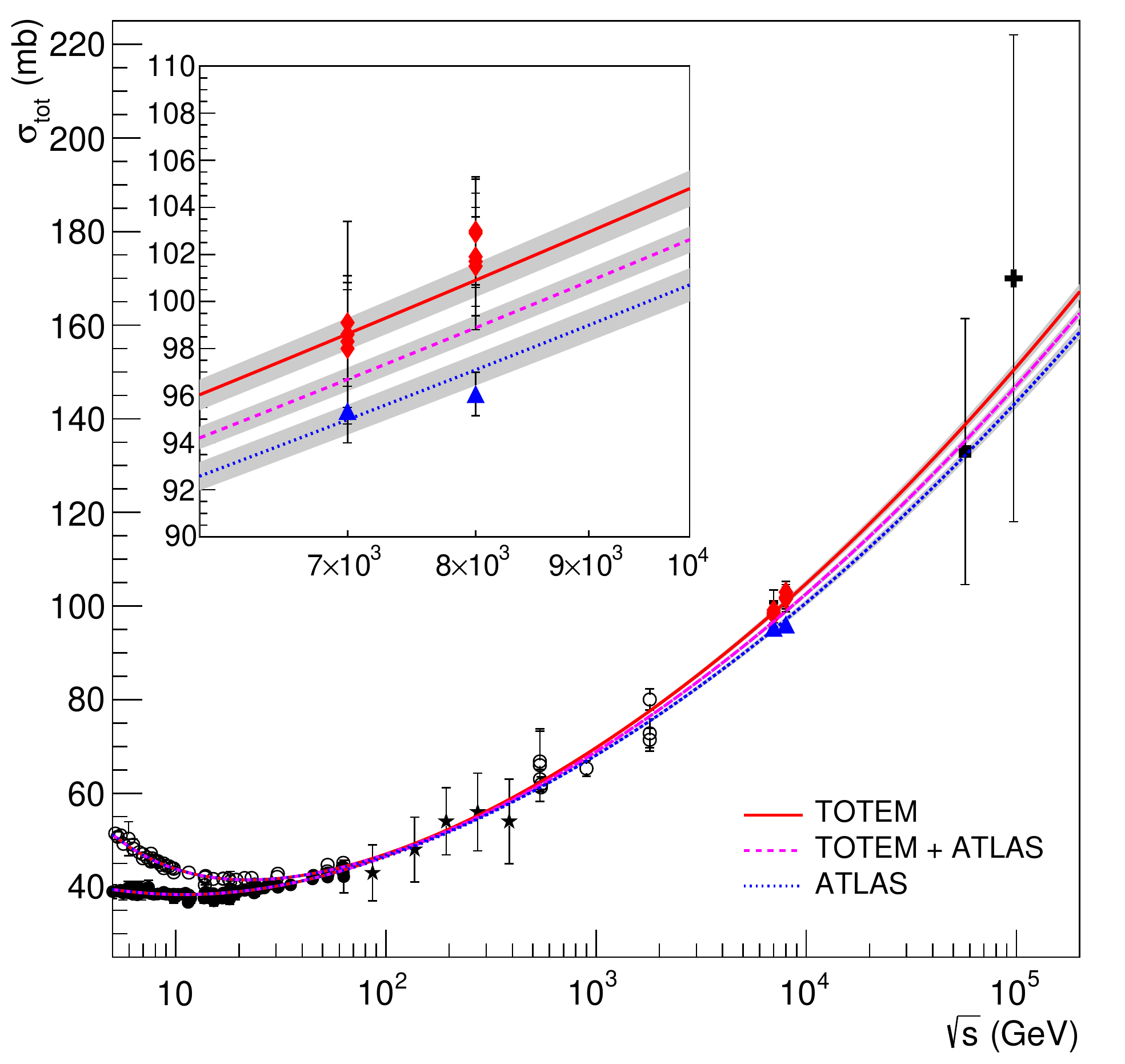}}\hfill
 \subfloat[\label{subfig:res_AU_Lgamma2_T_A_TA_panel_b}$\rho$]{\includegraphics[scale=0.39]{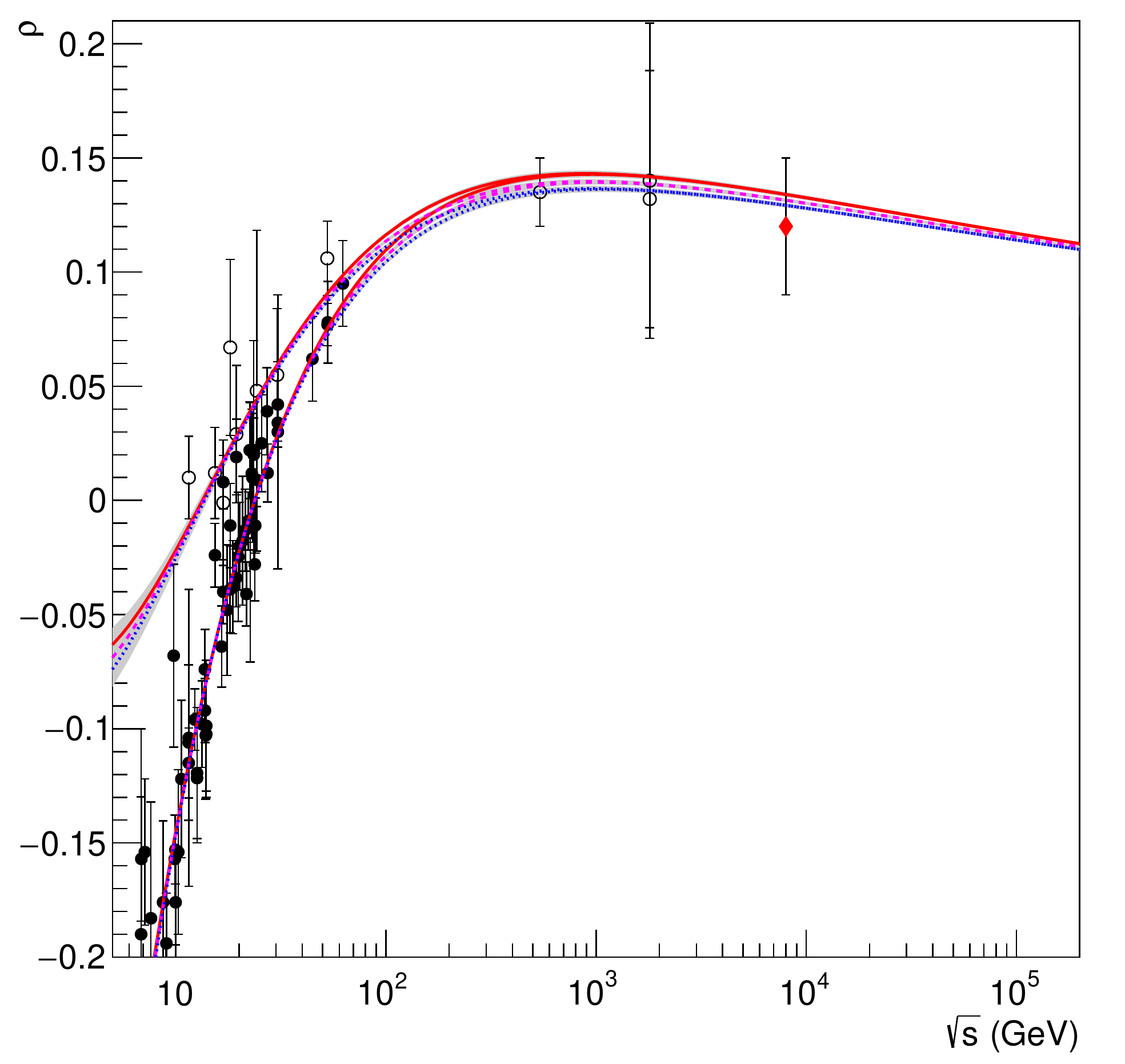}}
 \caption{\label{fig:res_AU_Lgamma2_T_A_TA}Fit results with the AU-L$\gamma$=2 model to ensembles T, A and T+A.
Eqs.~\eqref{eq:sigtot-AU-Lgamma2}-\eqref{eq:rho-AU-Lgamma2}, Table~\ref{tab:res_AU_Lgamma2_Lgamma_T_A_TA}.}
\end{figure}

\begin{figure}[htb!]
 \centering
 \subfloat[\label{subfig:res_AU_Lgamma_T_A_TA_panel_a}$\sigmatot$]{\includegraphics[scale=0.39]{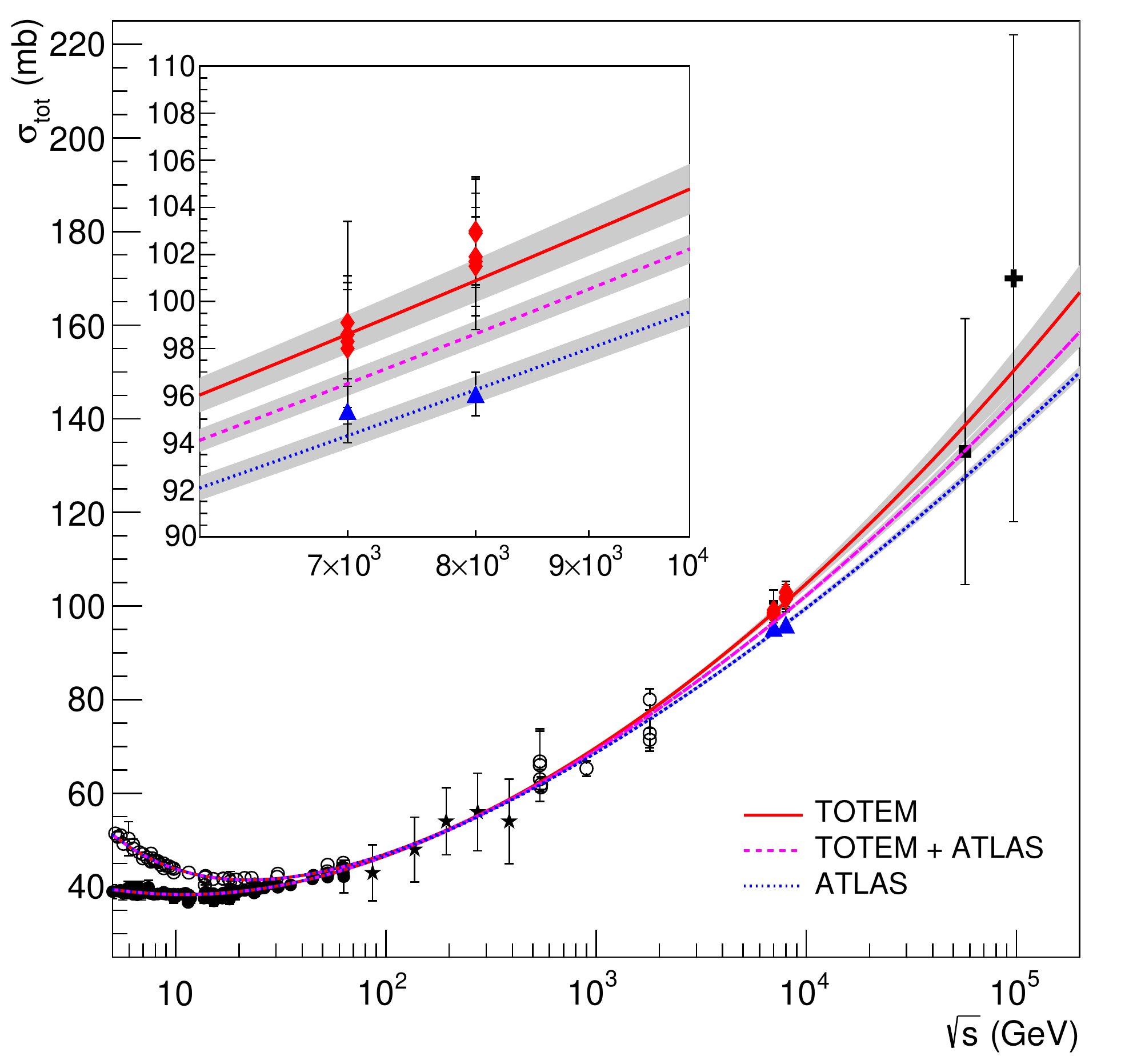}}\hfill
 \subfloat[\label{subfig:res_AU_Lgamma_T_A_TA_panel_b}$\rho$]{\includegraphics[scale=0.39]{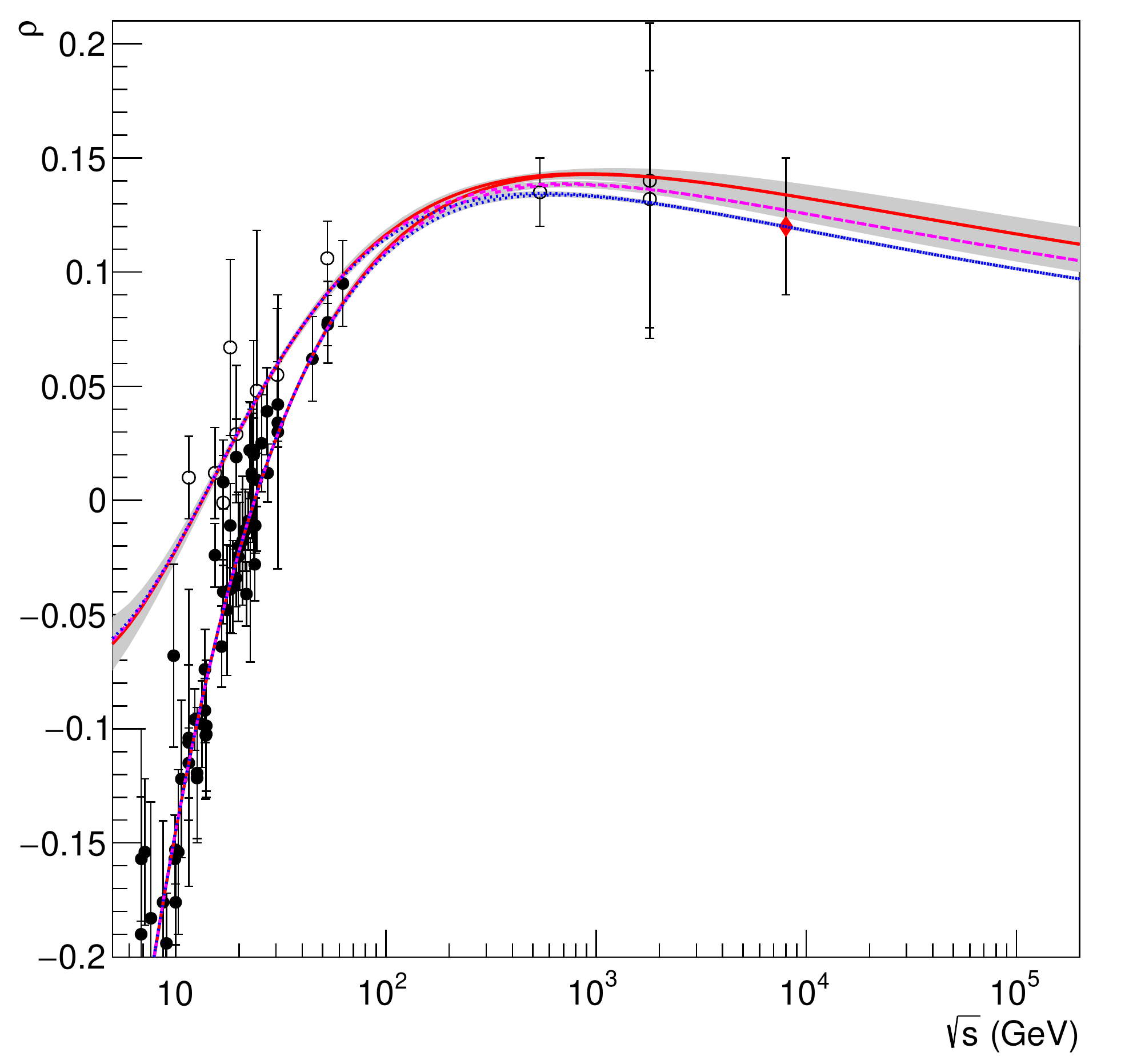}}
 \caption{\label{fig:res_AU_Lgamma_T_A_TA}Fit results with the AU-L$\gamma$ model to ensembles T, A and T+A.
Eqs.~\eqref{eq:sigtot-AU-Lgamma}-\eqref{eq:phi-AU}, Table~\ref{tab:res_AU_Lgamma2_Lgamma_T_A_TA}.}
\end{figure}

The fit results are displayed in Table~\ref{tab:res_AU_Lgamma2_Lgamma_T_A_TA}
and the corresponding results for $\sigmatot(s)$ and $\rho(s)$ with ensembles T, A and T+A,
together with the experimental data analyzed, are shown in Fig.~\ref{fig:res_AU_Lgamma2_T_A_TA} with the AU-L$\gamma=2$ model and in Fig.~\ref{fig:res_AU_Lgamma_T_A_TA} with the AU-L$\gamma$ model.

\begin{table}[htb!]
\centering
\caption{\label{tab:res_AU_Lgamma2_Lgamma_T_A_TA}Data reductions with the AU-L$\gamma$=2 model, 
Eqs.~\eqref{eq:sigtot-AU-Lgamma2}-\eqref{eq:rho-AU-Lgamma2} 
and AU-L$\gamma$ model, Eqs.~\eqref{eq:sigtot-AU-Lgamma}-\eqref{eq:phi-AU}, 
to ensembles T, A and T+A. Energy scale fixed, $s_0 = 4m_p^2 =$ 3.521 GeV$^2$.
For the parameter dimensions, see Table~\ref{tab:res_FMS_L2_Lgamma_T_A_TA}. 
In the 5th column, n.p.d. stands for non-positive definite error matrix. See text for further comments.}
\small
\begin{tabular}{@{}*{7}{c}}
\hline\hline
Ensemble:   & \multicolumn{2}{c}{TOTEM}      & \multicolumn{2}{c}{ATLAS} & \multicolumn{2}{c}{TOTEM + ATLAS}\\
\cmidrule(lr){2-3} \cmidrule(lr){4-5} \cmidrule(lr){6-7} 
Model:      & AU-L$\gamma=2$ & AU-L$\gamma$  & AU-L$\gamma=2$ & AU-L$\gamma$ [n.p.d.] & AU-L$\gamma=2$ & AU-L$\gamma$       \\\hline
$a_1$       & 31.42(47)      & 31.5(3.1)     & 30.87(54)      & 38.09(38)    & 31.10(50)      & 34.0(3.1)      \\
$b_1$       & 0.355(14)      & 0.353(60)     & 0.394(14)      & 0.2944(46)   & 0.376(13)      & 0.326(44)      \\
$a_2$       & 17.30(72)      & 17.30(73)     & 17.45(73)      & 17.28(62)    & 17.38(72)      & 17.28(72)      \\
$b_2$       & 0.553(13)      & 0.553(13)     & 0.557(13)      & 0.552(11)    & 0.555(13)      & 0.553(13)      \\
$\alpha$    & 28.61(43)      & 28.5(3.9)     & 30.14(33)      & 21.60(24)    & 29.46(32)      & 25.9(3.7)      \\
$\beta$     & 0.2584(38)     & 0.26(14)      & 0.2395(33)     & 0.661(17)    & 0.2483(27)     & 0.39(16)       \\
$\gamma$    & 2 (fixed)      & {\bf1.99(17)} & 2 (fixed)      & {\bf1.6763(97)}   & 2 (fixed)      & {\bf 1.85(13)} \\\hline
$\nu$       & 243            & 242           & 236            & 235          & 245            & 244            \\
$\chi^2/\nu$& 1.125          & 1.130         & 1.152          & 1.128        & 1.191          & 1.188          \\
$P(\chi^2)$ & 0.0878         & 0.0809        & 0.0538         & 0.0872       & 0.0217         & 0.0232         \\
\hline\hline
  \end{tabular}
  \normalsize
\end{table}

\section{General Discussion and Comments} 
\label{sec:rise_discussion_comments}

In this section, we critically discuss all the results obtained in this analysis taking into account
the analytic and conceptual differences between
the DDR and AU approaches related to L2 and L$\gamma$ models (Sect.~\ref{subsec:rise_analytic_differences}) and 
the corresponding fit results with ensembles
T, A and T+A (Sect.~\ref{subsec:rise_discussion_fit_results}). We proceed presenting our
partial conclusions (Sect.~\ref{subsec:rise_partial_conclusions}) and some further comments
on the log-raised-to-$\gamma$ law (Sect.~\ref{subsec:rise_further_comments_Lgamma}).

\subsection{Analytic and Conceptual Differences}
\label{subsec:rise_analytic_differences}

Let us compare the analytic results presented in Sect.~\ref{sec:rise_analytic_models}, namely
the FMS-L2, FMS-L$\gamma$ models, Eqs.~\eqref{eq:sigtot-FMS-Lgamma}-\eqref{eq:rho-FMS-Lgamma-factors} and the
AU-L$\gamma$=2, Eqs.~\eqref{eq:sigtot-AU-Lgamma2}-\eqref{eq:rho-AU-Lgamma2}, 
AU-L$\gamma$, Eqs.~\eqref{eq:sigtot-AU-Lgamma}-\eqref{eq:phi-AU} models.
Our focus here concerns the analytic and conceptual differences
among them. 

First, we note that all models present the same Reggeon contributions
(related to the parameters $a_1$, $b_1$ and $a_2$, $b_2$) and the same critical
Pomeron contribution ($\alpha$). 
Beyond the presence of the effective subtraction constant in the FMS models (not in the
AU cases), a central point in these parametrizations concerns the
log-raised-to-$\gamma$ term and the corresponding connection between $\sigmatot$ and $\rho$.

\subsubsection{DDR Approach}
\label{subsubsec:rise_analytic_differences_DDR}

The general analytic expressions of the FMS-L$\gamma$ model are given 
by Eqs.~\eqref{eq:sigtot-FMS-Lgamma}-\eqref{eq:rho-FMS-Lgamma-factors}.
As we showed in Sect.~\ref{subsubsec:rise_FMS_L2}, the FMS-L2 model is a particular case for
$\gamma = 2$. This specific case has the same analytic structure as the
PDG-L2 model and the COMPETE parametrizations RRPL2 for $\sigmatot$ and $\rho$,
except for the absence of a constraint connecting $\beta$ and $M$ (PDG), the fixed scale $s_0 = 4m_p^2$,
and for the presence of the effective subtraction constant, $K_\text{eff}$.

We recall that this approach is based on the use of DDR without the high-energy approximation 
since $K_\text{eff}$ is considered as a free parameter and it takes into account, at least in first order,
the correction term for not taking the aforementioned limit.

We stress that these L2 models (both PDG and FMS) are constructed in accordance with the
Regge-Gribov theory (as shown in Section~\ref{sec:rise_ReggeGribov}). In this context,
the parameters $a_1$, $a_2$, $\alpha$ and $\beta$ correspond to the strengths of the
Reggeons and of the Pomerons (critical and triple pole) and
they are constant factors at $t=0$ (independent of the energy).

\subsubsection{AU Approach}
\label{subsubsec:rise_analytic_differences_AU}

Now, let us turn our attention to the models L2, L$\gamma$=2, and L$\gamma$ 
obtained with the AU approach, showed in Section~\ref{subsec:rise_AU_models} and derived in Section~\ref{subsec:rise_AU}.
Although the AU-L2 model can also be deduced in this context, a crucial
point is the fact that for $\gamma = 2$ the
L$\gamma$ model does not correspond to the L2. The essential difference appears in 
the Pomeron contributions to $\sigmatot$ and $\rho$. 

In the AU-L2 model, the Pomeron contributions ($\pomeron$) to the imaginary and real parts of the
amplitude are given by Eqs.~\eqref{eq:sigtot-AU-L2} and \eqref{eq:rho-AU-L2} 
(omitting the arguments $s$ and $t=0$ of the amplitude),
\begin{align}
\frac{\Imag F_{L2}^{\pomeron}}{s} & = \alpha + \beta \ln^{2}\left(\frac{s}{s_0}\right),
\label{eq:Im-AU-L2-Pomeron}\\
%
\frac{\Real F_{L2}^{\pomeron}}{s} & = \pi \beta \ln\left(\frac{s}{s_0}\right).
\label{eq:Re-AU-L2-Pomeron}
\end{align}
\noindent With the AU-L$\gamma$ model for $\gamma = 2$ (L$\gamma$=2 model), the contributions
from Eqs.~\eqref{eq:sigtot-AU-Lgamma2} and \eqref{eq:rho-AU-Lgamma2} read
\begin{align}
\frac{\Imag F_{L\gamma=2}^{\pomeron}}{s} & =
\alpha + \beta \cos (\phi)
\ln\left(\frac{s}{s_0}\right)
\left[\ln^2\left(\frac{s}{s_0}\right) + \pi^2\right]^{1/2},
\label{eq:Im-AU-Lgamma2-Pomeron}\\[5pt]
%
\frac{\Real F_{L\gamma=2}^{\pomeron}}{s} & =
\beta \sin (\phi)
\ln\left(\frac{s}{s_0}\right)
\left[\ln^2\left(\frac{s}{s_0}\right) 
+ \pi^2 \right]^{1/2}.
\label{eq:Re-AU-Lgamma2-Pomeron}
\end{align}
The essential difference between Eqs.~\eqref{eq:Im-AU-L2-Pomeron} and \eqref{eq:Re-AU-L2-Pomeron}
and Eqs.~\eqref{eq:Im-AU-Lgamma2-Pomeron} and \eqref{eq:Re-AU-Lgamma2-Pomeron} 
is the presence in the latter
of 
trigonometric functions, which \textit{depend on the energy} through
$\phi = \phi(s)$, as given by Eq.~\eqref{eq:phi-AU}. Although in the asymptotic
limit ($s \rightarrow \infty$),
\begin{equation}
\phi \rightarrow 0, \qquad \cos(\phi) \rightarrow 1, \qquad \sin(\phi) \rightarrow 0,
\nonumber
\end{equation}

\noindent that is not the case in the \textit{finite energy-interval investigated} (5 GeV - 8 TeV), in which,
even if limited to the interval $[-1, 1]$, both cosine and sine can, in principle, take on negative,
null and positive values (depending on the ratio $s/s_0$ in Eq.~\eqref{eq:phi-AU}).

However, up to our knowledge, this analytic dependence on the energy does not have an interpretation or 
justification in Regge-Gribov context. In special, it is not clear how we could, for instance,
associate the $\cos\phi$ factor with $\beta$ giving rise to an energy-dependent Pomeron strength 
in~\eqref{eq:Im-AU-Lgamma2-Pomeron}. The same applies to the
real part of the amplitude (note also the $\ln{s}$ dependence in~\eqref{eq:Re-AU-L2-Pomeron}
contrasting with the $\ln^2{s}$ in~\eqref{eq:Re-AU-Lgamma2-Pomeron}), as well as to the more general
AU-L$\gamma$ model.

\begin{figure}[htb!]
 \centering
 \subfloat[\label{subfig:cont_AU_Lgamma2_Lgamma_TA_panel_a}AU-L$\gamma=2$]{\includegraphics[scale=0.39]{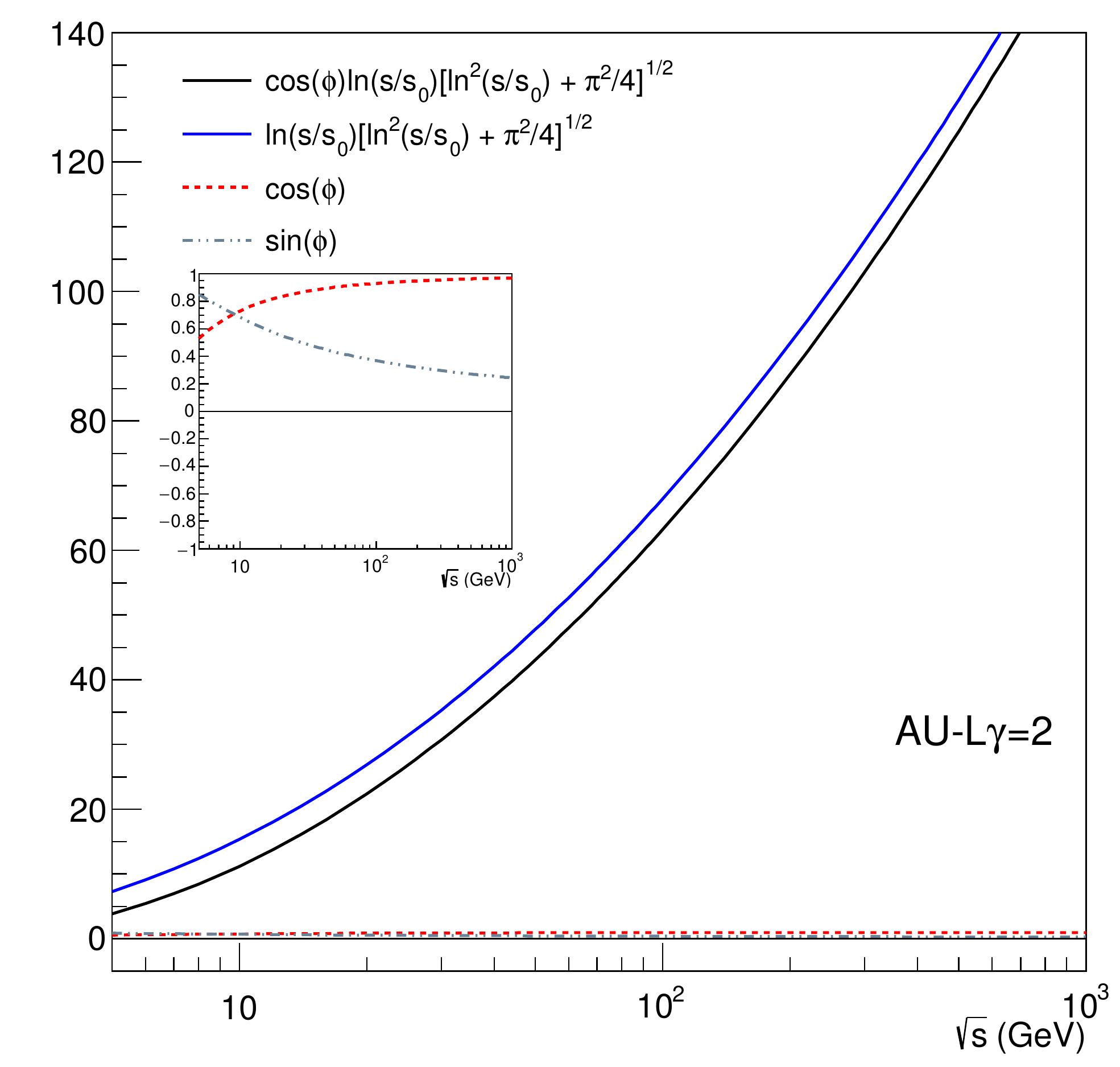}}\hfill
 \subfloat[\label{subfig:cont_AU_Lgamma2_Lgamma_TA_panel_b}AU-L$\gamma$]{\includegraphics[scale=0.39]{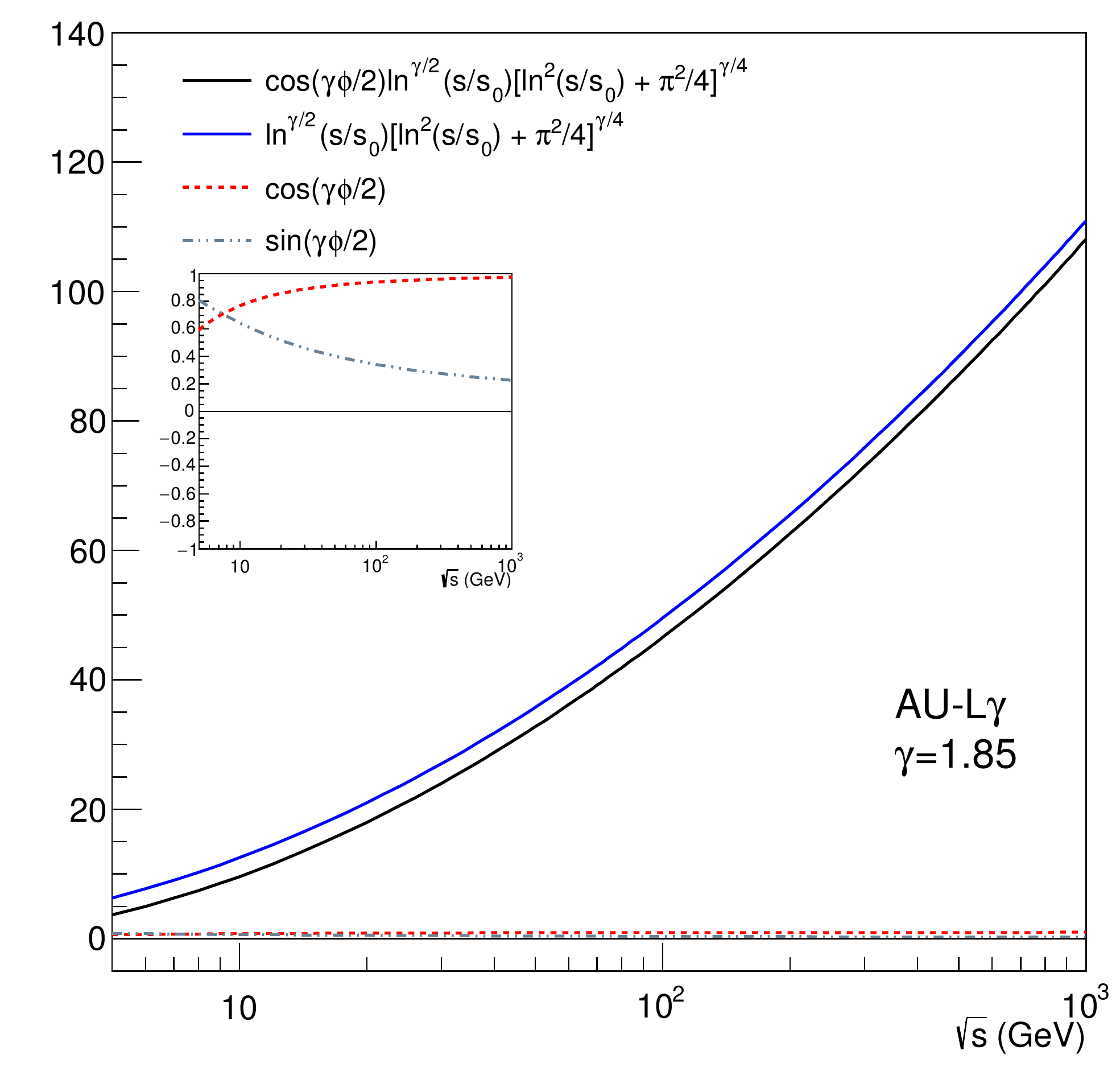}}
 \caption{\label{fig:cont_AU_Lgamma2_Lgamma_TA}Dimensionless contributions to the leading Pomeron component in
 \subref{subfig:cont_AU_Lgamma2_Lgamma_TA_panel_a}~AU-L$\gamma$=2 and \subref{subfig:cont_AU_Lgamma2_Lgamma_TA_panel_b}~AU-L$\gamma$ models, from fits to ensemble T+A,
Table~\ref{tab:res_AU_Lgamma2_Lgamma_T_A_TA}.}
\end{figure}

We illustrate this effect in Fig.~\ref{fig:cont_AU_Lgamma2_Lgamma_TA}, using the AU-L$\gamma$=2
and AU-L$\gamma$ models
from the fits to ensemble T+A (Table~\ref{tab:res_AU_Lgamma2_Lgamma_T_A_TA}). 
In this figure, 
three dimensionless terms associated with the Pomeron component to $\sigmatot$ are shown.
Note that in this case, the energy scale is fixed at the
physical threshold $s_0 = 4m_p^2$, a relatively small value which attenuates the
oscillation. That, however, is not the case for larger values of $s_0$,
as can be easily verified.

Therefore the AU approach for L$\gamma$ and L$\gamma$=2 models
introduces energy-dependent functions in the parametrization for $\sigmatot(s)$
which are not present in the original input, Eq.~\eqref{eq:sigtot-FMS-Lgamma}.

\subsection{Fit Results}
\label{subsec:rise_discussion_fit_results}

We have developed data reductions using four models
(FMS-L2, FMS-L$\gamma$, AU-L$\gamma$=2 and AU-L$\gamma$) and three
ensembles (T, A and T+A). The fit results with the FMS models
are presented in Table~\ref{tab:res_FMS_L2_Lgamma_T_A_TA} and Figures~\ref{fig:res_FMS_L2_T_A_TA} and \ref{fig:res_FMS_Lgamma_T_A_TA}
and those with the AU models in Table~\ref{tab:res_AU_Lgamma2_Lgamma_T_A_TA} and Figures~\ref{fig:res_AU_Lgamma2_T_A_TA} and \ref{fig:res_AU_Lgamma_T_A_TA}.
We call the attention to the fact that the fit performed with AU-L$\gamma$ model to ensemble A
showed in Table~\ref{tab:res_AU_Lgamma2_Lgamma_T_A_TA} has a \textit{non-positive definite error matrix}. 
Therefore it does not constitute a reliable result as discussed in Chapter~\ref{chapt:data_fits}.
In what follows, \textit{we shall not include this result in our discussion}.

Let us discuss all the \textit{fit results}, by comparing
separately the following aspects: 
ensembles T, A and T+A, 
L2 and L$\gamma$ models with the DDR approach,
L$\gamma$ models with the DDR and AU approaches.

\subsubsection{Ensembles T, A and T+A}
\label{subsubsec:rise_discussion_fit_results_T_A_TA}


Within all models, the goodness-of-fit is slightly better with ensemble T
than with A or T+A: $\chi^2/\nu \sim 1.07 - 1.13$ (T), $\chi^2/\nu \sim 1.08 - 1.15$ (A) and 
$\chi^2/\nu \sim 1.14 - 1.19$ (T+A),
where the smallest $\chi^2/\nu$ values are associated with FMS models.

From the figures, all TOTEM data are quite well
described with ensemble T, but in case of ensemble T+A 
all curves lie between the data points, barely reaching
the extrema of the uncertainty bars.
ATLAS data are also well described with ensemble A.

Ensemble T indicates a rise of the total cross section faster than
ensembles A and T+A, as shown by the extrapolated curves and, for example, by
the $\gamma$ values with the FMS-L$\gamma$ model:
$\gamma \sim 2.30 \pm 0.10$ (T), $\gamma \sim 2.01 \pm 0.12$ (A) and $\gamma \sim 2.16 \pm 0.16$ (T+A). 

The ATLAS datum at 8 TeV is not described by any fit result to ensembles T and T+A:
all curves lie above this point within the uncertainties. 

We recall that the result AU-L$\gamma$ with ensemble A has a non-positive definite error matrix and
we will not include it in our discussions.


\subsubsection{FMS-L2 and FMS-L$\gamma$ Models}
\label{subsubsec:rise_discussion_fit_results_FMS}

 
Taking into account the distinct characteristics of ensembles T, A and T+A,
both models present agreement with the experimental data analyzed and
cannot be distinguished on statistical grounds: with ensemble T,
$\chi^2/\nu$ = 1.09 (L2) and 1.07 (L$\gamma$), with ensemble A
$\chi^2/\nu$ = 1.08 (L2) and 1.09 (L$\gamma$) and with ensemble T+A,
$\chi^2/\nu$ = 1.15 (L2) and 1.14 (L$\gamma$).

With ensemble T, the L$\gamma$ results confirm our previous determination
of the parameter $\gamma$. The slightly high value, $\gamma \sim 2.30 \pm 0.10$
(as compared with the previous $2.23 \pm 0.11$ \cite{Menon_Silva:2013b}) is a consequence
of the latest TOTEM data at 8 TeV. On the other hand, with ensemble A we have a lower value for $\gamma$,
namely $2.01 \pm 0.12$, indicating a slower rise of $\sigmatot$ compared to the result with ensemble T.

We also note the anti-correlation between the parameters $\beta$ and $\gamma$:
$\beta \sim 0.25$ mb for $\gamma = 2$ and 
$\beta \sim 0.10 $ mb for $\gamma \sim 2.3$ (T)
(see discussion in the next section).

%

\subsubsection{FMS-L$\gamma$ and AU-L$\gamma$ Models}
\label{subsubsec:rise_discussion_fit_results_AU}


With respect to fits with ensembles T and T+A, 
the resulting $\gamma$-values with FMS are higher than with AU:
within ensemble T, $\gamma \sim 2.3 \pm 0.1$ (FMS) and $\sim 2.0 \pm 0.2$ (AU)
and within ensemble T+A, $\gamma \sim 2.2 \pm 0.2$ (FMS) and $\sim 1.9 \pm 0.1$ (AU).
With ensemble A and FMS model, we get a $\gamma$ value compatible with 2, $\gamma\sim 2.0\pm 0.1$.

Moreover, taking into account the uncertainties of all $\gamma$ values determined in the fits,
we may infer a range of possible values of this parameter for the present energies. 
From fits with ensembles T, A and T+A with the FMS model, we have $ 1.9 \lesssim \gamma \lesssim 2.4$,
and from fits with ensembles T and T+A with AU model, $ 1.7\lesssim \gamma \lesssim 2.2$.


\subsection{Partial Conclusions}
\label{subsec:rise_partial_conclusions}

On the basis of the discussions in Sects.~\ref{subsec:rise_analytic_differences} 
and \ref{subsec:rise_discussion_fit_results}, we are led to the 
partial conclusions that follow.

\begin{enumerate}

\item
In what concerns the
\textit{L$\gamma$ models} (and the energy-interval investigated), the DDR approach
is consistent with the Regge-Gribov theory: (1) the derivative
dispersion relations
apply to any function of interest in
amplitude analyses (power or logarithmic laws); (2) the triple-pole
contribution is nothing more than a particular case of L$\gamma$ for
$\gamma = 2$; (3) with the FMS models, given a parametrization for $\sigmatot(s)$, the determination of
$\rho(s)$ does not involve the high-energy approximation 
(due to the \textit{effective subtraction constant}) and therefore,
most importantly, it is not associated with the asymptotic condition
($s \rightarrow \infty$) being adequate for the finite energy interval investigated.

On the other hand, the AU approach leads to analytic results for \textit{both} $\sigmatot(s)$ and $\rho(s)$
with oscillatory energy-dependent factors in the leading term, which do not have justifications
in the Regge-Gribov context. Moreover, the AU-L$\gamma$ does not reproduce the
AU-L2 model for $\gamma~=~2$ and this model, AU-L$\gamma$=2, has also oscillatory
terms.

Therefore, we conclude that, regarding L$\gamma$,
\textit{the FMS models are more consistent in the formal context 
and adequate for the energy interval
investigated than the AU models}.

\item
Taking into account all the experimental data presently available (ensemble T+A),
the discrepancy between the TOTEM and ATLAS data does not allow a high-quality
fit on statistical grounds. 

\item
Both FMS-L2 and FMS-L$\gamma$ models present agreement with the experimental data
analyzed and cannot be discriminated on statistical grounds.

\item
The fit results indicate that
the TOTEM data and the ATLAS data favor different scenarios for
the asymptotic rise of the total cross section.

\end{enumerate}

Predictions of the FMS-L2 and FMS-L$\gamma$ models (ensembles T, A and T+A) for $\sigmatot(s)$
and $\rho(s)$ at 13, 14, 57 and 95 TeV are presented in Tables~\ref{tab:pred_FMS_L2_T_A_TA} and \ref{tab:pred_FMS_Lgamma_T_A_TA},
respectively.

\begin{table}[htb!]
\centering
\caption{\label{tab:pred_FMS_L2_T_A_TA}Predictions for $\sigmatot$ and $\rho$ with the FMS-L2 model.} 
\small
\begin{tabular}{c c c c c c c}
\hline\hline
 Ensemble: &\multicolumn{2}{c}{TOTEM}&\multicolumn{2}{c}{ATLAS}&\multicolumn{2}{c}{TOTEM + ATLAS} \\
 \cmidrule(lr){2-3} \cmidrule(lr){4-5} \cmidrule(lr){6-7} 
 $\sqrt{s}$ (TeV) & $\sigmatot$ (mb)   & $\rho$     & $\sigmatot$ (mb)  & $\rho$      & $\sigmatot$ (mb)  & $\rho$     \\
\hline
13               & 108.94(86)  & 0.1296(10) & 104.31(81) & 0.12489(92) & 106.76(61) & 0.12740(75)\\
14               & 110.28(88)  & 0.12915(96)& 105.55(83) & 0.12447(91) & 108.04(63) & 0.12694(74)\\
57               & 137.8(1.3)  & 0.11980(73)& 130.9(1.2) & 0.11625(69) & 134.51(92) & 0.11813(57)\\
95               & 148.8(1.5)  & 0.11646(66)& 141.0(1.4) & 0.11324(63) & 145.1(1.0) & 0.11494(52)\\
\hline\hline
\end{tabular}
\normalsize
\end{table}

\begin{table}[htb!]
\centering
\caption{\label{tab:pred_FMS_Lgamma_T_A_TA}Predictions for $\sigmatot$ and $\rho$ with the FMS-L$\gamma$ model.} 
\small
\begin{tabular}{c c c c c c c}
\hline\hline
 Ensemble: &\multicolumn{2}{c}{TOTEM}&\multicolumn{2}{c}{ATLAS}&\multicolumn{2}{c}{TOTEM + ATLAS} \\
 \cmidrule(lr){2-3} \cmidrule(lr){4-5} \cmidrule(lr){6-7} 
 $\sqrt{s}$ (TeV) & $\sigmatot$ (mb)   & $\rho$     & $\sigmatot$ (mb)  & $\rho$      & $\sigmatot$ (mb)  & $\rho$     \\
\hline
13                & 110.7(1.2)  & 0.1417(47) & 104.34(97) & 0.1251(44) & 107.39(94) & 0.1333(63)\\
14                & 112.1(1.3)  & 0.1413(47) & 105.6(1.0) & 0.1247(44) & 108.8(1.0) & 0.1329(64)\\
57                & 143.4(2.8)  & 0.1337(52) & 131.0(2.2) & 0.1165(51) & 137.0(2.8) & 0.1251(73)\\
95                & 156.2(3.6)  & 0.1306(52) & 141.2(2.8) & 0.1135(52) & 148.4(3.7) & 0.1221(75)\\
\hline\hline
\end{tabular}
\normalsize
\end{table}

\subsection{Further Comments on the Log-raised-to-$\gamma$ Law}
\label{subsec:rise_further_comments_Lgamma}

Presently, we do not have yet a clear and direct justification for the $\ln^\gamma s$ term within the Regge-Gribov formalism.
However, this function has an empirical motivation in the sense that it can be used to check, from fits to data,
how close the rise of $\sigmatot$ is from the FLM bound.
Perhaps, if not speculative, this real exponent could be seen as a
kind of effective contribution, similar to an effective exponent in the
power law associated with the simple-pole Pomeron.
Despite this limitation, to treat the exponent $\gamma$ as a free fit parameter
leads to some interesting consequences and useful results, including a possible mathematical connection with a branch point singularity.
We discuss these aspects in what follows.

\begin{enumerate}

\item In the general case, data reductions with this term involve three free parameters,
$\beta$, $\gamma$ and $s_0$, which are strongly correlated as demonstrated in the Appendix of Ref. \cite{Fagundes_Menon_Silva:2013a} and
also discussed in \cite{Menon_Silva:2013a} (see Sect. 4.2 and Table 6). For $s_0$ fixed, as assumed in this analysis,
$\beta$ and $\gamma$ are anti-correlated.
For example, from Tables~\ref{tab:res_FMS_L2_Lgamma_T_A_TA} and \ref{tab:res_AU_Lgamma2_Lgamma_T_A_TA}:
\begin{equation}
\begin{split}
\gamma &\sim 2.3\, \text{ (FMS)}\ \iff \beta \sim 0.10\, \text{ mb}\, \text{ (T)}  \\ 
\gamma &\sim 2.2\, \text{ (FMS)}\ \iff \beta \sim 0.15\, \text{ mb}\, \text{ (T+A)} \\
\gamma &= 2.0\, \text{ (FMS, AU)}\ \iff \beta \sim 0.26\, \text{ mb}\, \text{ (T)},\, \beta \sim 0.25\, \text{ mb}\, \text{ (T+A)} \\
\gamma &\sim 1.85\, \text{ (AU)}\ \iff \beta \sim 0.39\, \text{ mb}\, \text{ (T+A)}
\end{split}
\label{eq:gamma-beta-FMS-AU}
\end{equation}

These different values associated with $\gamma$ and $\beta$ may have some connections
with recent phenomenological and theoretical ideas and results, as
discussed in the next item.

\item In the phenomenological context, a real (not-integer) exponent in the interval
$1 < \gamma < 2$ is predicted in the QCD mini-jet model with soft gluon re-summation
\cite{Grau_etal:2012,Fagundes:2015}. 
As commented in \cite{PDG:2014,PDG:2016}, a rise slower than L2 is also
predicted in the Cheng and Wu formalism \cite{Cheng_Wu:1970}.

We may also associate the effective real exponent character to the
presence of sub-leading contributions, beyond the leading log-squared component.

As commented in Chapter~\ref{chapt:intro},
a result on the rise of total cross section in the asymptotic limit has been recently
obtained in a nonperturbative approach to soft scattering by 
Giordano and Meggiolaro \cite{Giordano_Meggiolaro:2014,Giordano_Meggiolaro:2015}.
This topic will be our focus in Chapter~\ref{chapt:SLT_sigmatot}, where we will discuss the details.
For now, we only recall the points of interest here.
In Ref.~\cite{Giordano_Meggiolaro:2014}, the hadronic total cross section behaves,
under some specific assumptions, as
\begin{equation}
 \sigmatot(s) \sim B \ln^2 s + C\ln s \cdot \ln\ln s \quad (s\to\infty),
 \nonumber
\end{equation}

\noindent where the coefficients $B$ and $C$ are universal (independent of the properties of the scattered particles).
Instead, this coefficients are related to the QCD spectrum. Of interest here, by considering the spectrum of stable
particles in the \textit{unquenched} QCD, the $B$ factor reads,
\begin{equation}
B_\text{th} \simeq 0.22 \text{ mb},
\nonumber
\end{equation}

\noindent while in the case of \textit{quenched} QCD ($q$), the estimated lower value (among all possibilities) reads
\begin{equation}
B_\text{th}^{q} \simeq 0.42 \text{ mb}.
\nonumber
\end{equation}

\noindent These two results indicate that the inclusion of dynamical
fermions (full QCD) has a pronounced influence in the (universal) value of 
$B$, a result in contrast with the
usual phenomenological models which consider that this asymptotic behavior is governed
by the gluonic sector of QCD.

From Tables~\ref{tab:res_FMS_L2_Lgamma_T_A_TA} and  \ref{tab:res_AU_Lgamma2_Lgamma_T_A_TA}
(and estimates in Eq.~\eqref{eq:gamma-beta-FMS-AU}), the $\beta$ results from models with $\gamma = 2$ (FMS and AU)
are in agreement with the full QCD 
prediction (as well as the PDG-L2 result, Eq.~\eqref{eq:IV-PDG2016}),
while the AU-L$\gamma$ result favors the quenched case. The smallest $\beta$ value (0.10 mb) 
was obtained from FMS-L$\gamma$ with ensemble T,
and corresponds to a mass
\begin{equation}
M = \sqrt{\frac{\pi}{\beta}} \sim 3.5\ \mathrm{GeV}.
\nonumber
\end{equation}

\item An important and interesting aspect
related to a leading L$\gamma$ component of $\sigmatot$ (DDR approach),
%
\begin{equation}
 \sigma^\pomeron(s) = \alpha + \beta \ln^{\gamma}\left(\frac{s}{s_0}\right),
 \label{eq:sigtot-Pomeron-Lgamma-DDR}
\end{equation}
\noindent concerns the \textit{slope} ($S$) and \textit{curvature} ($C$) of
$\sigma^\pomeron$ \textit{in terms of the variable} $\ln{s}$. Indeed, in the particular case of a L2 model
(or in the asymptotic FLM bound), we have
\begin{align}
S_{L2}(\ln s) & = 2 \beta \ln \left(\frac{s}{s_0}\right) \, \text{(linear)}, \label{eq:slope-sigtot-L2} \\
C_{L2}(\ln s) & = 2 \beta \, \text{(constant)},\label{eq:curvature-sigtot-L2}
\end{align}

\noindent while in the general case ($\gamma$ real),
\begin{align}
S_{L\gamma}(\ln s) & = \beta \gamma \ln^{\gamma - 1} \left(\frac{s}{s_0}\right),\label{eq:slope-sigtot-Lgamma}\\
C_{L\gamma}(\ln s) & = \beta\,\gamma (\gamma - 1) \ln^{\gamma - 2} \left(\frac{s}{s_0}\right),\label{eq:curvature-sigtot-Lgamma}
\end{align}

\noindent both, therefore, energy dependent. This means that
any deviation from a linear rate of change of $\sigmatot(\ln s)$ and from a
constant curvature ($2 \beta$) can be ``detected"  by a L$\gamma$ model, Eq.~\eqref{eq:sigtot-Pomeron-Lgamma-DDR}. 
Moreover, Eqs.~\eqref{eq:slope-sigtot-Lgamma} and \eqref{eq:curvature-sigtot-Lgamma} allows to quantify the differences in the
rate of rise of $\sigmatot$ from the TOTEM and ATLAS data. 
For example, at 8 TeV and with the L$\gamma$ model, the fits with ensembles
T and A (Table~\ref{tab:res_FMS_L2_Lgamma_T_A_TA}) predict: $S \sim 9.3 \pm 0.2$ mb (T) 
and $S \sim 7.9 \pm 0.3$ mb (A), indicating, therefore, a faster
rise of $\sigma_{tot}(s)$ from the TOTEM data than from the ATLAS data. 

We note that in case of a AU approach, the analytic structure and oscillatory terms do not
allow this simple interpretation.

\item The dependence of the parametrization~\eqref{eq:sigtot-Pomeron-Lgamma-DDR} with the energy scale $s_0$ is another
interesting aspect that deserves some comments. 
The presence of a \textit{real} (not-integer) exponent of $\ln(s/s_0)$ in the total cross section
puts some constraints on the definition of this parametrization. 
In order to obtain positive values of $\ln (s/s_0)$ and consequently real (physical) values 
for the total cross section, we must have $s \geq s_0$. Otherwise, we would get a free power of a negative number and, consequently,
a complex value for $\sigmatot$, which has no physical mean.

Therefore, Eq.~\ref{eq:sigtot-Pomeron-Lgamma-DDR} ``starts'' at $s=s_0$ (and here we chose the threshold for scattering states, $4m_p^2$)
with
\begin{equation}
\sigma^\pomeron(s_0) = \alpha,
\nonumber
\end{equation}

\noindent and from this point on, $\sigma^\pomeron(s)$ increases as the energy increases according to $\alpha+\beta\ln^\gamma(s/s_0)$ and 
in accordance with the concept of a super-critical Pomeron.

In case of the same energy scale for the Reggeon components,
as in the FMS-L$\gamma$ model, Eq.~\eqref{eq:sigtot-FMS-Lgamma}, at $s=s_0$ the contributions
to the $pp$ and $\bar{p}p$ total cross section read
\begin{equation}
\sigma_{pp}^\reggeon(s_0) = a_1 - a_2
\qquad
\sigma_{\bar{p}p}^\reggeon(s_0) = a_1 + a_2
\nonumber
\end{equation}

\noindent and from this point on, $\sigma_{pp}^\reggeon(s)$ and  $\sigma_{\bar{p}p}^\reggeon(s)$
decrease as the energy increases.  

It is interesting to investigate these quantities at the threshold points
as well as their evolution with the energy. For example, from our fit
results with the FMS-L$\gamma$ model to ensemble T+A (all data presently available), the numerical results
are (Table~\ref{tab:res_FMS_L2_Lgamma_T_A_TA}):
\begin{equation}
\sigma^\pomeron(s_0) \sim 32.8 \text{ mb},
\qquad
\sigma_{pp}^\reggeon(s_0) \sim 14.5 \text{ mb},
\qquad
\sigma_{\bar{p}p}^\reggeon(s_0) \sim 48.7 \text{ mb}.
\nonumber
\end{equation}

\noindent Therefore, at $s=s_0$ 
$\sigma_{pp}^\reggeon(s_0) < \sigma^\pomeron(s_0) < \sigma_{\bar{p}p}^\reggeon(s_0)$
and above $s_0$, the Reggeon contributions decrease and the Pomeron
one increases as $s$ increases. 
Therefore, besides being a leading contribution at high energies,
the Pomeron component is also
significant at low energies,
as illustrated in Fig.~\ref{fig:comp_reggeon_pomeron_FMS_Lgamma_TA}.
Note that $\sigma^\pomeron(s)$ is the leading contribution for all $s$ above the energy cutoff, $\sqrt{s_\text{min}}$.
From Tables~\ref{tab:res_FMS_L2_Lgamma_T_A_TA} and \ref{tab:res_AU_Lgamma2_Lgamma_T_A_TA}, 
this effect is also present in all fit results, independently of model or ensemble.

\begin{figure}[htb!]
 \centering
 \includegraphics[scale=0.55]{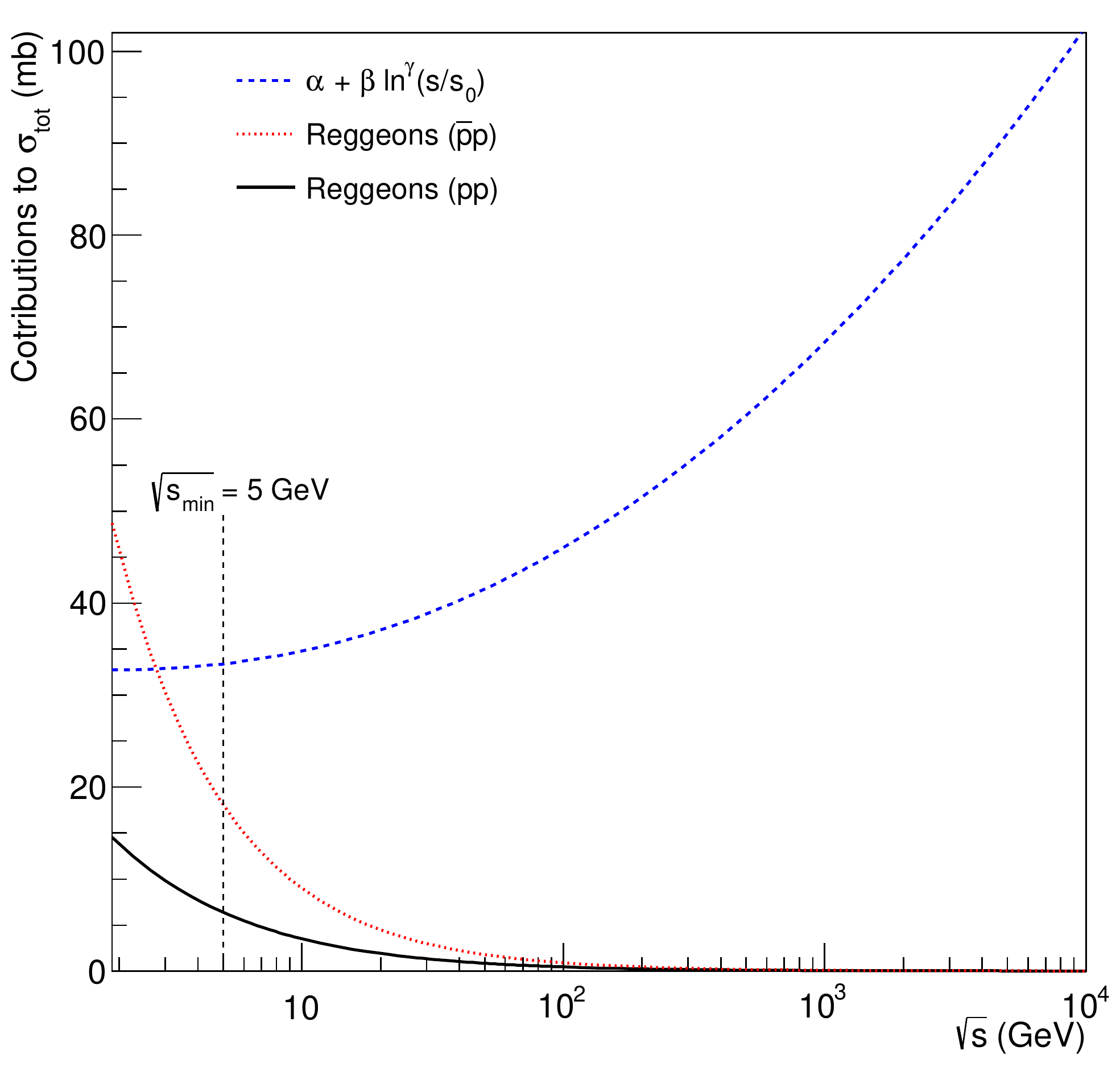}
 \caption{\label{fig:comp_reggeon_pomeron_FMS_Lgamma_TA}Reggeon components, $\sigma_{pp}^\reggeon(s)$, $\sigma_{\bar{p}p}^\reggeon(s)$ and
 Pomeron component,  $\sigma^\pomeron(s)$, of $\sigmatot(s)$ for $pp$ and $\bar{p}p$
 scattering, above the physical threshold $\sqrt{s_0} = 2m_p \sim $ 1.9 GeV$^2$. Results obtained with
 the FMS-L$\gamma$ model and ensemble T+A (energy cutoff at $\sqrt{s_{\mathrm{min}}}$ = 5 GeV).}
\end{figure}

\item Even if we do not have a complete interpretation of $\ln^\gamma s$ within the Regge-Gribov approach, a natural question is
what kind of singularity in the complex angular-momentum $\ell$-plane could be associated with this asymptotic contribution.
This can be investigated, in a mathematical context, by means of the Mellin transform \cite{Bertrand:2000},
which can correlate the asymptotic behaviour of a real function of a real variable with the singularities of a real function
of a complex variable.

As demonstrated and discussed in Appendix ~\ref{app:mellin}, 
the Mellin transform of $\ln^\gamma s$ reads
\begin{equation}
 \frac{\Gamma(\gamma+1)}{(\ell-1)^{\gamma+1}}
 \label{eq:Lgamma-Mellin}
\end{equation}

\noindent for $\gamma > -1$, $\Real[\ell-1]>0$, where $\Gamma$ is the Euler gamma function. 
The above equation indicates that for $\gamma$ real (not-integer), the $\ln^\gamma s$ is associated with a branch point at $\ell=1$
in the complex $\ell$ plane. For $\gamma=2$ and $\gamma=1$, the singularities reduce to a triple and double pole, respectively, as expected
from Regge theory.

At this point, it is worth noting that branch points with the attached cut also appear in Regge Theory. However, in this case
the branch point is associated with the exchange of two or more Reggeons (or Pomerons) 
and the contribution to the total cross-section, in case of $N$ exchanges, reads \cite{Barone_Predazzi_book:2002}
\begin{equation}
 \sigma^N_\text{cut}(s) \sim \frac{s^{\alpha_\text{cut}(t=0)-1}}{\ln^{N-1} (s)}, \quad \text{with } \gamma > 0,\nonumber
\end{equation}

\noindent where $\alpha_\text{cut}(t)$ is an effective trajectory associated with the exchanged objects. 
The result above is obtained with the Gribov Calculus, which is based on perturbative techniques. 
We note that, since $N\geq2$, 
the logarithmic term always appears in the denominator and, at $t=0$, this contribution tends to tame the power rise of $\sigmatot$. 
Contrasting with this behaviour, in the L$\gamma$ law we have $\gamma>0$ and the term $\ln^\gamma s$ in the numerator explicitly rise with the energy.
Therefore, 
the branch point associated with L$\gamma$ cannot be related to the exchange of two or more Pomerons.

On the other hand, given the nonperturbative character of $\sigmatot(s)$, it seems reasonable to think that the L$\gamma$ law
and associated branch point might be related to some (unknown) nonperturbative effect.

At last, we note that the analytic connection between the L$\gamma$ law and the associated branch point, can be demonstrated
through generalizations of Eqs.~\eqref{eq:higher-order-pole-l-plane} and \eqref{eq:higher-order-pole-power-law}
from \textit{integer order} derivatives to \textit{real order} derivatives.
The approach, based on Fractional Calculus, employs the non-local Caputo fractional derivative and is discussed in Ref.~\cite{CapelasdeOliveira:2017}.

\item Recent fits with the L$\gamma$ parametrization to \textit{only total cross section data} 
within ensemble T+A (all data presently available) and fixed energy scale ($s_0 = 4m_p^2$)
indicate $\gamma=2.21\pm0.14$~\cite{CapelasdeOliveira:2017}, which
is $1.5\sigma$ above the value $\gamma=2$.

\end{enumerate}

\section{Summary and Conclusions}
\label{sec:rise_conclusions}

We presented in this chapter a comparative study on some parametrizations for the forward
elastic scattering amplitude related to methods of connecting the real and imaginary parts, together with
fits to $pp$ and $\ppbar$ data in the energy region 5 GeV - 8 TeV.

The analytic parametrizations for the imaginary part of the amplitude ($\sigmatot \sim \Imag F(s,t=0)$)
were based on the Regge-Gribov formalism, in which the low-energy behaviour of $\sigmatot$
is described by the exchange of Reggeons, while the high-energy domain is determined by Pomeron exchange. 
Two leading terms were considered: the triple-pole Pomeron (log-squared term) and the empirical ansatz 
introduced by Amaldi \etal in the 1970s \cite{Amaldi:1977} given by
a log-raised-to-$\gamma$ term, where $\gamma$ is a real free parameter.
We denote the two terms by L2 and L$\gamma$, respectively.

Alongside, two analytical ways of relating the real and imaginary parts of 
the elastic scattering amplitude in the forward direction were explored: the Derivative Dispersion Relations (DDR)
and the Asymptotic Uniqueness (AU), the latter based on the Phragmén-Lindelöff theorems. 

DDR are derived from the Integral Dispersion Relations (IDR) in the so-called high-energy limit.
However, the consideration of the \textit{effective subtraction constant} ($K_\text{eff}$),
as introduced in Section~\ref{subsec:rise_DDR}, takes into account the corrections to this limit, at least in first order.
Applying DDR in the L$\gamma$ model gives us analytical results for the real part\footnote{We recall that 
Amaldi \etal \cite{Amaldi:1977} have considered IDR, which demand numerical integration for functions like $\ln^\gamma s$.}, which
recover the L2 equations when $\gamma$ is fixed to the value 2. Therefore, we can understand the L$\gamma$ model in the DDR approach as
a generalization of the L2 model for a real exponent of $\ln s$. Given the introduction of the $K_\text{eff}$, we denote these models as 
FMS-L2 and FMS-L$\gamma$. 

On the other hand, even if the AU approach allows for an analytic determination of the real
and imaginary parts from a real-valued function,
this method is based on the behaviour of the chosen function in the asymptotic limit ($s\to\infty$). 
Moreover, the resulting amplitude from a $\ln^\gamma s$ function has oscillatory energy-dependent factors appearing 
in front of the leading term $\ln^\gamma s$, which do not have an interpretation within Regge-Gribov formalism.
Another drawback is the fact that setting $\gamma=2$ in the L$\gamma$ model does not recover the L2 result in this approach,
resulting in a conceptual difference between AU and DDR approaches (in what concerns L$\gamma$ law). 
Since fixing $\gamma=2$ in the L$\gamma$ model does not recover the L2 model,
we have three models in this approach that we denote AU-L2, AU-L$\gamma$ and AU-L$\gamma=2$. 
In what concerns the fit to experimental data and the AU approach,
we have considered only the last two models.

Based on these considerations, we conclude that the DDR approach with the effective subtraction constant is, at least from the
analytical point of view and considering the Regge-Gribov interpretation,
the appropriate way to connect real and imaginary parts of the forward amplitude.

We performed simultaneous fits to $\sigmatot$ and $\rho$ data including recent data obtained in the LHC by TOTEM and ATLAS Collaborations.
We also took into account the present discrepancies in the TOTEM and ATLAS $\sigmatot$ measurements at 7 and 8 TeV by means of different
ensembles of data. Up to the Tevatron energy (1.8 TeV), all ensembles comprise the same data (collected from the PDG website). 
Three ensembles were considered:
one including only TOTEM data (denoted T), one including only ATLAS data (denoted A)
and another one with all available data included (denoted T+A).

All data reductions led to consistent description of the experimental data analysed, 
with reduced chi-squared smaller for FMS models than those for AU models. 
Interesting enough, all fits with FMS models have $\gamma$ central values $\geq 2$, 
while fits with AU models give $\gamma$ central values $<2$. Of course, considering the associated uncertainties,
the results obtained with FMS-L$\gamma$ with ensembles A and T+A and AU-L$\gamma$ with ensemble T are compatible with 2.
Another interesting fact is the similarity between the result from FMS-L$\gamma$ with ensemble T+A ($\gamma = 2.16\pm 0.16$)
and the result obtained by Amaldi \etal ($2.10\pm0.10$). At last, we may infer a range of possible values for the $\gamma$ parameter:
$1.9\lesssim \gamma \lesssim 2.4$ in the FMS approach and $1.7\lesssim\gamma \lesssim2.2$ in AU approach.

Another advantage of FMS-L$\gamma$ model is to have the slope and curvature (in respect of the variable $\ln s$) energy dependent.
Using the parameters obtained in the fit, it is clear that TOTEM data points to a faster rise of $\sigmatot$ than ATLAS data.

We also mention that even if the $\ln^\gamma s$ term with $\gamma>2$ may indicate a faster rise than allowed by the FML bound,
this theoretical bound is for asymptotic energies ($s\to\infty$) and 
here we analysed energies up to 8 TeV. This rise faster than $\ln^2 s$ may be a local effect, and the $\ln^2 s$
behaviour may be recovered as energy get larger. 
Nonetheless,
we recall that even with $\gamma>2$, the values of $\sigmatot$ are less than
the numerical limit given by Eq.~\eqref{eq:rise-FLM-bound-coeff}.

At last, we note that most analyses with the L$\gamma$ law, with $\gamma$ as a free fit parameter, favor \textit{real} (\textit{not integer})
$\gamma$ values (above or below 2). If that is really the case, the associated singularity cannot be a triple pole ($\gamma=2$),
but, as we have shown, a branch point singularity.

In the next chapter, we will focus on the subleading terms to the total cross section based on recent
nonperturbative QCD results by Giordano and Meggiolaro~\cite{Giordano_Meggiolaro:2014}.

%% file: ch_SLT_sigmatot_f.tex
%
%
%
\cleardoublepage



\chapter[Studies on Subleading Contributions to the Hadronic $\sigmatot$]{Studies on Subleading Contributions to the Hadronic Total Cross-Section}
\label{chapt:SLT_sigmatot}

\section{Introduction}
\label{sec:slt_intro}

In the previous chapter, we focused on the \textit{leading} behaviour of the total cross section
at high energies and two approaches to connect the real and imaginary parts of the amplitude
by means of fits to experimental data. Two leading terms were considered: a log-squared term ($\ln^2s$) 
and a log-raised-to-$\gamma$ term ($\ln^\gamma s)$. The parametrizations were based on the Regge-Gribov Phenomenology.

As discussed along the text, the energy dependence of $\sigmatot$ is an intrinsically
nonperturbative QCD problem, since it is connected to the elastic scattering amplitude 
at zero transferred momentum through the optical
theorem, Eq.~\eqref{eq:rise-optical-theo}.
In this chapter, based on a nonperturbative QCD approach for the
elastic scattering at asymptotic energies, our focus concerns the \textit{subleading}
contributions to the hadronic total cross sections.
By means of fits to experimental data, we will estimate and compare the contribution, at the present energies,
of two subleading terms predicted in the aforementioned approach. 
The chapter is based on the research presented in Ref.~\cite{Giordano_Meggiolaro_Silva:2017}.

This study is based on the nonperturbative approach proposed
by Nachtmann~\cite{Nachtmann:1991} to describe elastic scattering of quarks.
This approach was extended by several authors~\cite{Dosch_etal:1994,Nachtmann:1997_1,Berger_Nachtmann:1999,Shoshi_etal:2002} 
to describe elastic scattering of composite particles (hadrons). 
Specifically, we start from the recent results obtained by Giordano and Meggiolaro~\cite{Giordano_Meggiolaro:2014},
which connect the coefficients of the asymptotic leading ($\sim \ln^2 s$) and 
subleading ($\sim \ln s \ln\ln s$) terms with the QCD spectrum. As will be show later, 
these two coefficients are \textit{universal}, in the sense that they do not depend on the properties of the particles involved in the
scattering process. On the other hand, a second subleading term is obtained ($\sim \ln s$). However, this contribution is
\textit{reaction dependent}. 

In order to compare the two subleading terms, we performed fits to $\sigmatot$ 
data considering several variants of the parametrizations, in what concerns
the presence or not of these terms. As the leading term, we considered only the $\ln^2 s$ contribution which
emerges in a QCD approach under specific assumptions. Two different datasets were also considered.
First, only fits do $pp$ and $\ppbar$ data were taken into account. 
After that, to test the universality of some terms, we included data from other meson-baryon
and baryon-baryon scattering.

The chapter is organized as follows. In Section~\ref{sec:slt_nonperturbative_approach}
we present a summary of theoretical results of interest for this work, 
namely a nonperturbative QCD approach to elastic hadronic scattering.
In Section~\ref{sec:slt_parametrization} we discuss our choice to parametrize
$\sigmatot$ followed by a description of our dataset and our methodology 
in Section~\ref{sec:slt_dataset_methodology}. The results of fits to experimental data
are presented and discussed in Section~\ref{sec:slt_fit_results}. Finally, the conclusions
are the contents of Section~\ref{sec:slt_conclusions}.

\section{A Nonperturbative Approach to Hadronic Elastic Scattering}
\label{sec:slt_nonperturbative_approach}

\subsection{Hadron-hadron scattering}
\label{subsec:slt_hh}

Otto Nachtmann proposed, in his seminal paper \cite{Nachtmann:1991} back in 1991,
a nonperturbative approach in the framework of QCD, using functional integral techniques
to describe quark-quark elastic scattering in the large $s$ and small $|t| \ll s$ regime
(eikonal approximation). In this case, the quarks may be described as traveling on their
classical and almost unperturbed trajectories and the amplitude describing the elastic 
scattering depends on the properly normalized correlation function of two infinite 
lightlike Wilson lines that, in turn, run along the above-mentioned classical trajectories
\cite{Verlinde_Verlinde:1993,Korchemsky:1994,Korchemskaya_Korchemsky:1995,Meggiolaro:1996,Meggiolaro:2001}.

This formalism was pushed forward 
to describe meson-meson elastic scattering. 
In this case, the amplitude depends on the correlation function of two Wilson loops 
\cite{Dosch_etal:1994,Nachtmann:1997_1,Berger_Nachtmann:1999,Shoshi_etal:2002} 
that run along the trajectories of the colliding hadrons. Specifically, these Wilson 
loops correspond to two colour dipoles of fixed transverse size $\mathbf{R}_{i\bot}$ 
and fixed longitudinal-momentum fractions $f_i$ $(i=1,\, 2)$. The correlation function 
gives us the amplitude describing the scattering of two dipoles, $F^{dd}(s,t;\nu_1,\nu_2)$, 
where $\nu_i \equiv (\mathbf{R}_{i\bot},f_i)$ is a short notation for the dipole variables.
We then need to fold the dipole-dipole scattering amplitude with the meson wave functions 
$\psi_i(\mathbf{R}_{i\bot},f_i)$ in order to obtain the hadron-hadron scattering amplitude
\begin{equation}
\begin{split}
 F^{ab}(s,t) & = \int d^2 \mathbf{R}_{1\bot}\int_0^1 df_1|\psi_1(\mathbf{R}_{1\bot},f_1)|^2 
 \int d^2 \mathbf{R}_{2\bot}\int_0^1 df_2|\psi_2(\mathbf{R}_{2\bot},f_2)|^2 F^{dd}(s,t;\nu_1,\nu_2)\\
             & \equiv \langle\langle F^{dd} (s,t;\nu_1,\nu_2) \rangle\rangle,
\end{split}
\label{eq:elaticamplitude-general}
\end{equation}

\noindent where $\langle\langle \; \rangle\rangle$ denotes the integral with the wave functions in the 
dipole variables $\mathbf{R}_i$ and $f_i$ . We also consider normalized wave functions
\begin{equation}
 \int d^2 \mathbf{R}_{i\bot}\int_0^1 df_i|\psi_i(\mathbf{R}_{i\bot},f_i)|^2 = 1 \quad (i=1,\, 2), 
 \text{ so that } \langle\langle 1 \rangle\rangle = 1.
 \label{eq:normalization-wavefunction}
\end{equation}

The Wilson loops are defined in Minkowski space and run in the paths formed by the classical 
trajectories of the quark and antiquark that constitute the mesons. These loops form a hyperbolic 
angle in the longitudinal plane
\begin{equation}
\chi \simeq \ln(s/m_a m_b). \nonumber
\end{equation}

\begin{figure}[h!]
 \centering
 \includegraphics[scale=0.4]{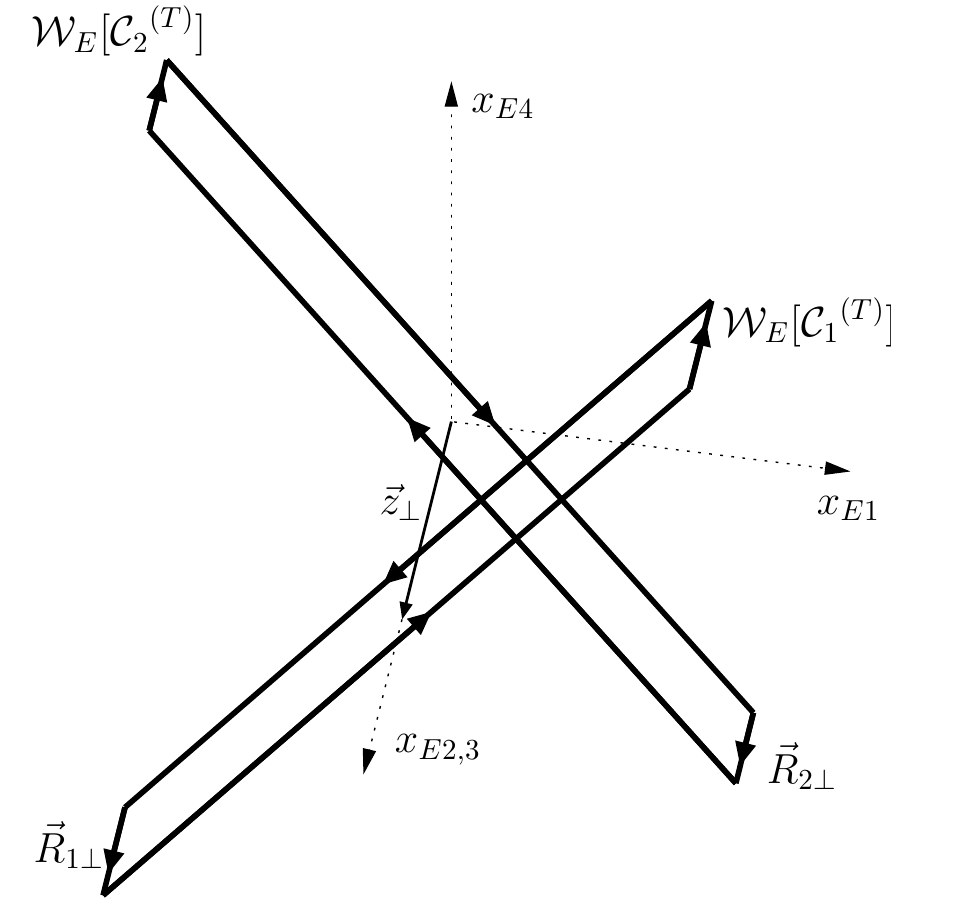}
 \caption{\label{fig:Wilson_loops} The relevant Wilson loops (in Euclidean space) 
 to the elastic scattering amplitude. Figure taken from Ref.~\cite{Giordano_Meggiolaro:2014}.}
\end{figure}

Cuts in the paths at proper times $\pm T$ are introduced in order to regularize infrared divergences and
in these cuts the loop is closed by 
straight-lines to preserve gauge invariance. At some point, we take the limit $T\to\infty$. 
Meggiolaro and  {collaborators} have showed that the Minkowskian correlation function 
$\mathcal{G}_M (\chi;T;\mathbf{z}_\bot; \nu_1,\nu_2)$ can be obtained from the Euclidean
correlation function of two Euclidean Wilson loops by means of analytic continuation 
\cite{Meggiolaro:1997,Meggiolaro:1998,Meggiolaro:2002,Meggiolaro:2005,Giordano_Meggiolaro:2009}. 
This correlation function reads
\begin{equation}
 \mathcal{G}_E(\theta;T;\mathbf{z}_\bot;\nu_1,\nu_2) \equiv \frac{\langle \mathcal{W}_E[\mathcal{C}_1^{(T)}]\mathcal{W}_E[\mathcal{C}_2^{(T)}]\rangle_E}
 {\langle\mathcal{W}_E[\mathcal{C}_1^{(T)}]\rangle_E \langle\mathcal{W}_E[\mathcal{C}_2^{(T)}]\rangle_E} -1,
 \label{eq:correlator-WilsonLoops-Euclidean}
\end{equation}

\noindent where $\langle \quad \rangle_E$ means the average in the sense of the Euclidean QCD 
functional integral and $\mathbf{z}_\bot$ is the impact parameter. In turn, the Euclidean Wilson
loop along a path $\mathcal{C}$ is defined by
\begin{equation}
 \mathcal{W}_E[\mathcal{C}]\equiv \frac{1}{N_c} \Trace \hat{P} \exp\left\{-ig \oint_\mathcal{C} A_{E\mu}(x_E)dx_{E\mu}\right\},
 \label{eq:def-WilsonLoop-Euclidean}
\end{equation}

\noindent with $N_c$ denoting the number of colours, $\hat{P}$ the path-ordering operator
(with larger values of the path parameters on the left), $A_E(x_E)$ the gauge field in the
Euclidean space and $x_E$ the coordinates in Euclidean space. The relevant Wilson loops 
that contribute to the correlation function are shown in Figure~\ref{fig:Wilson_loops}.
For more details about how the paths $\mathcal{C}_i$ are defined see Ref.~\cite{Giordano_Meggiolaro:2014}.

We denote the correlators, after taking the limit $T\to\infty$, as
\begin{align}
  \mathcal{C}_E(\theta,\mathbf{z}_\bot;\nu_1,\nu_2) & \equiv \lim_{T\to\infty}\mathcal{G}_E(\theta;T;\mathbf{z}_\bot;\nu_1,\nu_2), \label{eq:def-Correlators-limits-E}\\
  \mathcal{C}_M(\chi,\mathbf{z}_\bot;\nu_1,\nu_2) & \equiv \lim_{T\to\infty}\mathcal{G}_M(\chi;T;\mathbf{z}_\bot;\nu_1,\nu_2). \label{eq:def-Correlators-limits-M}
\end{align}

The angle $\theta$ appearing in the Euclidean correlation function is inside the range 
$\theta \in (0,\pi)$ and the analytic continuation consists in taking the limit of 
$\theta$ going to complex values, $\theta~\to~-i\chi$, with $\chi \simeq \ln(s/m_am_b) \in \mathds{R}^+$. 
Therefore, we have $\mathcal{C}_M(\chi)~=~\mathcal{C}_E(\theta\to -i\chi)$. 

Finally, the dipole-dipole scattering amplitude in terms of the correlation function reads
\begin{equation}
 F^{dd}(s,t) \equiv -2is \int d^2\mathbf{z}_\bot e^{i\mathbf{q}_\bot \cdot \mathbf{z}_\bot} \mathcal{C}_M(\chi;\mathbf{z}_\bot;\nu_1,\nu_2),
 \label{eq:amplitude-dipole-scattering}
\end{equation}

\noindent where $\mathbf{q}_\bot$ is the transferred momentum ($t = -|\mathbf{q}_\bot|^2$).

Taking advantage of invariance under rotation of Minkowskian theory, we get for the
hadron-hadron scattering amplitude \cite{Giordano_Meggiolaro:2014}
\begin{equation}
 F^{ab}(s,t) = -4\pi i s\langle\langle \int_0^\infty bdb J_0(b\sqrt{-t})\mathcal{C}_M(\chi;\mathbf{b}_\bot;\nu_1,\nu_2) \rangle\rangle_0,
 \label{eq:elastic-amplitude-correlator}
\end{equation}

\noindent where the subscript $0$ indicates the assumption that the wave functions
are rotational invariant and $\mathbf{b}_\bot = (b,0)$ with $b=|\mathbf{z}_\bot|$ is 
the impact parameter rotated to coincide with the Euclidean ``time''. Comparing with 
Eq.~\eqref{eq:amplitude-profile-FBtransform} we see that, apart from the folding with 
the mesons wave functions, the equation above is the Fourier-Bessel transform relating 
the amplitude with the profile function. Therefore, the Wilson loop correlation function
is related to the profile function.

\subsection{Connecting $\sigmatot(s)$ to the QCD Spectrum}
\label{subsec:slt_connection_QCDspectrum}

In Ref.~\cite{Giordano_Meggiolaro:2014}, Giordano and Meggiolaro have
connected the asymptotic behaviour of the total cross section with the QCD spectrum.
In this section, we summarize the results obtained in this approach.

To be able to connect the formalism discussed in the previous section, we need to
consider the Wilson loop \textit{operator} $\hat{\mathcal{W}}_E$ in Euclidean space,
defined by
\begin{equation}
 \hat{\mathcal{W}}_E[\mathcal{C}]\equiv \frac{1}{N_c} \Trace \hat{T} \hat{P} \exp\left\{-ig \oint_\mathcal{C} \hat{A}_{E\mu}(x_E)dx_{E\mu}\right\},
 \label{eq:def-WilsonLoop-Euclidean-Operator}
\end{equation}
 
\noindent where $\hat{T}$ represents the time-ordering operator (acting on the Euclidean time)
and the hat symbol indicates an operator. Now, the average of one or more Wilson 
loops appearing in the correlation function is rewritten  {as a vacuum expectation value of 
the Wilson loop operators}
\begin{equation}
 \langle \mathcal{W}_E[\mathcal{C}_1] \dots \mathcal{W}_E[\mathcal{C}_n] \rangle_E =  { \langle 0 | \hat{T}\left\{ \hat{\mathcal{W}}_E[\mathcal{C}_1] \dots \hat{\mathcal{W}}_E[\mathcal{C}_n] \right\} |0\rangle. }
 \label{eq:MeanValue-Functional-Operator}
\end{equation}

This relation works as a bridge between the functional-integral formalism and the operator
formalism. Finally, the correlation function reads
\begin{equation}
 \mathcal{G}_E(\theta;T;b;\nu_1,\nu_2) \equiv \frac{\langle 0| \hat{T}\left\{ \hat{\mathcal{W}}_E[\mathcal{C}_1^{(T)}]
 \hat{\mathcal{W}}_E[\mathcal{C}_2^{(T)}]\right\}|0\rangle_E}
 {\langle0|\hat{\mathcal{W}}_E[\mathcal{C}_1^{(T)}]|0\rangle_E \langle0|\hat{\mathcal{W}}_E[\mathcal{C}_2^{(T)}]|0\rangle_E} -1.
 \label{eq:correlator-WilsonLoops-Euclidean-Operator}
\end{equation}

The time-ordering operator appearing in the equation above may be dropped if one considers
loops that do not overlap in Euclidean time. This will be the case if the impact parameter is
larger than some $b_0 (\nu_1,\nu_2)$ that depends only on the dipoles variables. 
In Ref.~\cite{Giordano_Meggiolaro:2014} this assumption was made, therefore, in what follows,
the time-ordering operator will not be necessary anymore.

To calculate the vacuum-expectation values, we introduce a complete set of states 
in Eq.~\eqref{eq:correlator-WilsonLoops-Euclidean-Operator}. Consider the projector
for the $n$-particle states (in a representative form, for more detail see Ref.~\cite{Giordano_Meggiolaro:2014})
\begin{equation}
 |n\rangle\langle n| \equiv \frac{1}{n!}\sum_\alpha \delta_{\mathcal{N}_{\alpha},n} 
 \mathcal{P}_\alpha \sum_{\{s\}_\alpha} \int d\Omega_\alpha |\alpha,\{\mathbf{p}\}_\alpha,\{s_3\}_\alpha\rangle \langle\alpha,\{\mathbf{p}\}_\alpha,\{s_3\}_\alpha|,
 \label{eq:n-states-Projector}
\end{equation}

\noindent where $|\alpha,\{\mathbf{p}\}_\alpha,\{s_3\}_\alpha\rangle$ represents the asymptotic
states of the theory, which contains any number of particles (including bound states)
with non-zero mass. The index $\alpha$ contains information about the particle content of the state, 
$\{\mathbf{p}\}_\alpha$ and $\{s_3\}_\alpha$ represent the sets of all momenta and all third 
component of spin of the particles in the state, respectively. We also have in Eq.~\eqref{eq:n-states-Projector}
a sum over the particle content of state $|\alpha,\dots\rangle$, the total number of particles 
$\mathcal{N}_\alpha$ in state $|\alpha,\dots\rangle$, a combinatorial factor $\mathcal{P}_\alpha$, 
the sum over the spins of the particles of the state $|\alpha,\dots\rangle$ and an integration over
the phase space $d\Omega_\alpha$. 

It follows that (noting that the $n=0$ term cancels with the $-1$ appearing in Eq.~\eqref{eq:correlator-WilsonLoops-Euclidean-Operator})
\begin{equation}
 \mathcal{G}_E(\theta;T;b;\nu_1,\nu_2) = \sum_{n=1}^\infty \frac{\langle 0| \hat{\mathcal{W}}_E[\mathcal{C}_1^{(T)}]|n\rangle}{\langle0|\hat{\mathcal{W}}_E[\mathcal{C}_1^{(T)}]|0\rangle}
 \frac{\langle n| \hat{\mathcal{W}}_E[\mathcal{C}_2^{(T)}]|0\rangle}{\langle0|\hat{\mathcal{W}}_E[\mathcal{C}_2^{(T)}]|0\rangle} = \sum_{n=1}^\infty \frac{1}{n!} G_n(\theta;T;b;\nu_1,\nu_2),
\end{equation}

\noindent where $G_n(\theta;T;b;\nu_1,\nu_2)$ contains all the information about the Wilson-loop matrix elements
\begin{eqnarray}
 W_\alpha^{(T)}(\{\mathbf{p}\}_\alpha,\{s_3\}_\alpha;\nu_i) = \frac{\langle 0|\hat{\mathcal{W}}_E[\mathcal{C}_i^{(T)}]|\alpha,\{\mathbf{p}\}_\alpha,\{s_3\}_\alpha\rangle}
 {\langle 0| \hat{\mathcal{W}}_E[\mathcal{C}_i^{(T)}]|0\rangle},\\
 \overline{W}_\alpha^{(T)}(\{\mathbf{p}\}_\alpha,\{s_3\}_\alpha;\nu_i) = \frac{ \langle\alpha,\{\mathbf{p}\}_\alpha,\{s_3\}_\alpha| \hat{\mathcal{W}}_E[\mathcal{C}_i^{(T)}] |0\rangle}
 {\langle 0| \hat{\mathcal{W}}_E[\mathcal{C}_i^{(T)}]|0\rangle}.
\end{eqnarray}

The next important step is to continue analytically the correlator from Euclidean to Minkowskian space,
with the limit $T\to\infty$ taken. For this, two crucial assumptions are made:

\begin{itemize}
 \item the analytic continuation can be done term by term, i.e., for each state contribution separately;
 
 \item the Wilson loop matrix elements are analytic in $\theta$, in a complex domain 
 that includes the real segment $(0,\pi)$ and the negative imaginary axis.
\end{itemize}

Once the analytic continuation is done and the large $b$ and large $s$ limits are taken, 
the Wilson loop correlator can be written in a general way as
\begin{equation}
 \mathcal{C}_M(\chi;b;\nu_1,\nu_2) \sim g(w;\nu_1,\nu_2) -1 \quad (s\to\infty),
\end{equation}

\noindent where $w$ is a specific combination of the hyperbolic angle $\chi = \chi(s)\sim \ln(s/m_am_b)$ and $b$,
\begin{equation}
 w(\chi,b) \equiv \frac{e^{(\tilde{s}-1)\chi}e^{-\tilde{m}b}}{\sqrt{\tilde{m}b}}.
\end{equation}

\noindent In the above equation, $\tilde{s}$ and $\tilde{m}$ are respectively the spin and mass 
of the state that maximizes the following ratio involving the spin $s$ and mass $m$
\begin{equation}
 l \equiv \frac{s-1}{m}
\label{eq:ratio-s-m}
\end{equation}

\noindent considering the QCD-stable spectrum\footnote{By QCD stable, we mean all 
particles that are stable under strong interactions.} (excluding, however, 
particles with spin $0$ and $1$ and zero mass). At last, $g$ is a power 
series of $w$ whose coefficients depend on the dipole variables.
From unitarity we also have $|g| \leq 1$.

To obtain the dependence with $s$ of the elastic scattering amplitude and then
calculate the asymptotic limit of $\sigmatot$, some assumptions on the asymptotic
behaviour of this function $g$ are necessary. Below, we describe the cases considered 
in Ref.~\cite{Giordano_Meggiolaro:2014} and the corresponding result obtained for the total cross section.

\begin{enumerate}
 \item $g \to 0$ or $g$ oscillates as its argument goes to $\infty$. Both cases lead to the following behaviour
\begin{equation}
 \sigmatot^{ab}(s) \to \frac{2\pi}{\tilde{m}^2} \eta^2 + \mathcal{O}(\eta),
 \label{eq:sigmatot-case1}
\end{equation}

\noindent with 
%
%
\begin{equation}
\eta = \frac{1}{2}W(2e^{2(\tilde{s}-1)\chi}) = (\tilde{s}-1)\chi
 -\frac{1}{2} \ln[(\tilde{s}-1)\chi] + 
 \frac{\ln[(\tilde{s}-1)\chi]}{4(\tilde{s}-1)\chi} + \dots \,,
 \label{eq:def-eta}
\end{equation}

\noindent where $W$ is the Lambert $W$ function~\cite{LambertW} 
and the $\mathcal{O}(\eta)$ depends on the dipole variables.

Therefore, this result has a leading term $\ln^2(s/m_am_b)$ which is in
agreement with Froissart-Martin bound, and its coefficient is universal, in the sense that it
does not depend on the scattered particles (since it does not depend on $\nu_i$).

\item $g\to g_\infty(\nu_1,\nu_2)$ as its argument goes to $\infty$, i.e., 
$g$ goes to a constant value that in principle can depend on $\nu_i$. From 
the bound $|g|\leq1$, we can write $g_\infty(\nu_1,\nu_2) = e^{-\rho_\infty(\nu_1,\nu_2)}e^{i\phi_\infty(\nu_1,\nu_2)}$,
with $\rho_\infty(\nu_1,\nu_2) > 0$. From analyticity and crossing 
symmetry of the amplitude, we have $\phi_\infty(\nu_1,\nu_2) = 0$ or $\pi$, 
therefore independent of the dipole variables.  
The total cross section then reads 
\begin{equation}
 \sigmatot^{ab}(s) \to \frac{2\pi}{\tilde{m}^2}\kappa^{ab}\eta^2  + \mathcal{O}(\eta),
 \label{eq:sigmatot-case2}
\end{equation}

\noindent where $\kappa^{ab} \equiv 1 \mp \langle\langle e^{-\rho_\infty(\nu_1,\nu_2)}\rangle \rangle$. 
This case breaks the universality of the previous cases since, in principle, $\kappa^{ab}$ may depend on 
the colliding particles. The universality would be preserved if, at the same time, 
$\rho_\infty(\nu_1,\nu_2) = \bar{\rho}_\infty$, independently of the dipole variables $\nu_i$. 
Therefore, $\kappa^{ab} \to \kappa \equiv 1 \mp e^{-\bar{\rho}_\infty}$, independent of $\nu_i$. 
The cross section then reads (using Eq.~\eqref{eq:def-eta} to expand $\eta$)
\begin{equation}
 \sigmatot^{ab}(s) \to \kappa \left[ B_\text{th}\ln^2\left(\frac{s}{m_am_b}\right) + C_\text{th} 
 \ln\left(\frac{s}{m_am_b}\right) \ln\ln\left(\frac{s}{m_am_b}\right) \right]  + \mathcal{O}(\eta),
 \label{eq:sigmatot-th}
\end{equation}

\noindent with
\begin{eqnarray}
 B_\text{th} = 2\pi\frac{(\tilde{s}-1)^2}{\tilde{m}^2},\label{eq:Bth}\\
 C_\text{th} = - 2\pi\frac{(\tilde{s}-1)}{\tilde{m}^2}\label{eq:Cth}
\end{eqnarray}

\noindent and ratio
\begin{equation}
 \frac{B_\text{th}}{C_\text{th}} = 1-\tilde{s}.
 \label{eq:ratio-BC}
\end{equation}

Therefore, we have a leading term and a subleading term, both with universal coefficients,
to the total cross section for $s\to\infty$. The equation above reduces 
to the one obtained in case 1 when $\kappa = 1$. In any case,
analyticity requires $\kappa$ to be real, and unitarity requires that $\kappa\in[0,2]$.

\end{enumerate}

We also want to mention another subleading term of order $\mathcal{O}(\eta)$, 
i.e., $Q^{ab}\ln (s/m_am_b)$ found in Ref.~\cite{Giordano_Meggiolaro:2014} (see Eq.~\eqref{eq:sigmatot-th}).
Contrary to the leading term and the subleading term discussed above, this subleading term has
a coefficient $Q^{ab}$ that depends on the colliding particles,
since it is related to the integral of the correlator over the dipole variables \cite{Giordano_Meggiolaro:2014}, 
even if our assumption on $\kappa$ is met.

Summarizing, we have the following general result for the leading and two subleading components at \textit{asymptotic energies}
\begin{equation}
 \sigmatot^{ab}(s) \sim \kappa \left[ B_\text{th}\ln^2\left(\frac{s}{m_am_b}\right) + C_\text{th} 
 \ln\left(\frac{s}{m_am_b}\right) \ln\ln\left(\frac{s}{m_am_b}\right) \right] + Q^{ab}\ln\left(\frac{s}{m_am_b}\right)
\end{equation}

\subsubsection{Values of $B_\text{th}$ and $C_\text{th}$}

Considering the QCD-stable spectrum, the particle with spin greater than 1 
that maximizes Eq.~\eqref{eq:ratio-s-m} is the $\Omega^\pm$ baryon, 
with mass $m_{\Omega^{\pm}} \approx 1.67$ GeV and spin $3/2$, that gives 
\begin{equation}
 B_\text{th}^{\Omega} = 0.22 \text{ mb} \quad \text{and} \quad C_\text{th}^{\Omega} = -2B_\text{th}^\Omega = -0.44 \text{ mb}.
 \label{eq:B-Cth-Omega}
\end{equation}

Now, if one considers the \textit{quenched} limit of the theory 
(or the limit of large number of colours $N_c$), only the \textit{glueball}
states need to be considered. Of interest here (for details see 
ref.~\cite{Giordano_Meggiolaro:2014}) are the glueball states $2^{++}$,
with mass $m_{g2^{++}} \approx 1.40$ GeV, and $3^{+-}$,
with mass $m_{g3^{+-}} \approx 3.55$ GeV (both calculated in the \textit{quenched} 
approximation), for which one finds:
\begin{eqnarray}
 B_\text{th}^{g2^{++}} = 0.42 \text{ mb} \quad \text{and} \quad C_\text{th}^{g2^{++}} = -B_\text{th}^{g2^{++}} = -0.42 \text{ mb},\label{eq:B-Cth-g2pp}\\
 B_\text{th}^{g3^{+-}} = 0.78 \text{ mb} \quad \text{and} \quad C_\text{th}^{g3^{+-}} = -B_\text{th}^{g3^{+-}}/2 = -0.39 \text{ mb}.\label{eq:B-Cth-g3pm}
\end{eqnarray}

The value calculated for the $\Omega^{\pm}$ baryon is the closest to the value 
determined in the fits to forward quantities (with $\sqrt{s}\geq 5$ GeV) 
published in the Review of Particle Physics by the Particle Data Group in 2014:
$B_{PDG2014}~=~0.2704~\pm~0.0038$ mb \cite{PDG:2014}. 
A fit with $\sqrt{s} \geq 7$ GeV  was also performed and the 
result is $B_{PDG2014}~=~0.2838~\pm~0.0045$ mb.

\subsubsection{Further comments on $\kappa$}

With the formalism discussed above, we can also obtain the leading
behaviour of the elastic scattering amplitude in the limit $s\to\infty$.
In the cases considered above, we have
\begin{equation}
 F^{ab} (s,t) \sim 4\pi i s \kappa \left(\frac{\eta}{\tilde{m}}\right)^2 \frac{J_1(x)}{x}, \quad \text{with } x = \eta\sqrt{-t}/\tilde{m} \quad (s\to\infty).
 \label{eq:elastic-amplitude-Meggiolaro}
\end{equation}

Here, we have assumed the universality of the leading term, 
hence $\kappa$ does not depend on the scattering particles and 
this result corresponds to the case 2 discussed above. 
If $\kappa = 1$, we recover the case 1. The equation above 
has a very familiar form: it represents the black-disk 
behaviour (if $\kappa = 1$) or a grey disk (if $\kappa < 1$).

If we integrate Eq.~\eqref{eq:elastic-amplitude-Meggiolaro} in $t$, 
we obtain (assuming that the small-$t$ region gives the dominant contribution)
the asymptotic integrated elastic cross section ($\sigmael$) and,
using the result obtained for the total cross section, we get for 
the asymptotic value for the ratio $\sigmael/\sigmatot$
\begin{equation}
 \frac{\sigmael}{\sigmatot} \sim \frac{\kappa}{2} \quad (s\to\infty).
 \label{eq:ratio-eltot-kappa}
\end{equation}

Therefore, recalling our discussion in Section~\ref{sec:ratioX_asymp_values},
if $\kappa=1$ we have indeed a black disk ($\sigmael/\sigmatot = 1/2$), 
if $\kappa<1$ a grey disk ($\sigmael/\sigmatot < 1/2$) and if $\kappa > 1$ we have 
the antishadowing regime ($\sigmael/\sigmatot >1/2$)~\cite{Troshin_Tyurin:1993,Troshin_Tyurin:2007}.

Moreover, following the discussion in Section~\ref{sec:basic_profile_models} 
and comparing the equation above with Eq.~\eqref{eq:ratio-eltot-grey-gaussian} 
for the grey-disk model, we identify $\kappa\leftrightarrow \Gamma_0$, i.e. 
$\kappa$ \textit{is connected to the central value of the profile function} and, 
at least in first order, to the central opacity of the colliding particles.

Along the chapter, we will \textit{assume} that $\kappa$ 
is independent of the  dipole variables $\nu_i$.
A more general analysis should have this parameter to be particle dependent and, moreover,
to be a free parameter. In this case, the universality, if true, should be present in the fit results.
However, to have $\kappa^{ab}$ free would increase the number of free parameter.
Therefore, we shall consider $\kappa$ to be a universal parameter.

\subsection{Subleading Term of $\sigmatot$ from Other Analyses}
\label{subsec:slt_otheranalysis}

A subleading term to $\sigmatot$ in the form $\ln s \ln \ln s$ has
appeared before in other analyses. In this section we list these
works and present some differences and connections with the result
discussed in the previous sections.

\begin{itemize}
 \item Martin and Roy (MR) published in 2014 a bound to the mean value of
$\sigmatot$~\cite{Martin_Roy:2014}
\begin{equation}
 \bar{\sigma}_\text{tot}(s,\infty) \equiv s \int_s^\infty ds' \frac{\sigmatot(s')}{s'^2} \leq \frac{\pi}{m_\pi^2}[\ln(s/s_0) + \frac{1}{2}\ln\ln(s/s_0) +1]^2
\end{equation}

\noindent with $s_0 = m_\pi^2/(17\pi \sqrt{\pi/2}) \approx 2.72 \times 10^{-4}$ GeV$^2$
for $\pi^0\pi^0$ scattering. Therefore
\begin{equation}
 \bar{\sigma}_\text{tot}(s,\infty) \leq \frac{\pi}{m_\pi^2}[\ln^2(s/s_0) + \ln(s/s_0)\ln\ln(s/s_0) + \dots]
 \label{eq:sigtot-MR}
\end{equation}

\noindent being, in respect to our notation,
$B_\text{MR} = C_\text{MR} = \pi/m_\pi^2$ and
\begin{equation}
 \frac{B_\text{MR}}{C_\text{MR}} = 1.
\end{equation}

\item In a recent work based in AdS/CFT, D\'iez \etal~\cite{Diez_etal:2015} (AdS/CFT)
presented an improvement in the Froissart bound that reads
\begin{equation}
 \sigmatot^\text{AdS/CFT}(s) \leq \frac{\pi}{M_1^2}\left[\frac{1}{4}\ln^2 (s/s_0) + \beta \ln(s/s_0) - \ln(s/s_0)\ln\ln(s/s_0) \right],
 \label{eq:sigtot-Ads/CFT}
\end{equation}

\noindent where $M_1$ and $\beta$ depend on the parameters of the model
and $s_0 = 2m_p^2$. In this case we have $B_\text{AdS/CFT} = \pi/4M_1^2$
and $C_\text{AdS/CFT}~ =~ -\pi/M_1^2$, therefore
\begin{equation}
\frac{B_\text{AdS/CFT}}{C_\text{AdS/CFT}} = -\frac{1}{4}.
\end{equation}

It is important to note that the authors do not address the 
dependence of the leading and subleading terms on the colliding particles.

\item Nastase and Sonnenschein (NS) have published~\cite{Nastase_Sonnenschein:2015}
a revision of Heisenberg's model~\cite{Heisenberg:1952} that admits the Froissart bound
to be saturated. In their calculations, they determine a correction for the 
leading term of $\sigmatot$ at asymptotic energies
\begin{equation}
 \sigmatot^\text{NS}(s) \sim \frac{\pi}{4m_\pi^2}\left[\ln^2(s/s_0) - 2\ln(s/s_0)\ln\ln(s/s_0) + \ln^2(\ln(s/s_0))\right],
 \label{eq:sigtot-NS}
\end{equation}

\noindent  {giving $B_\text{NS} = \pi/4m_\pi^2$, $C_\text{NS} = -\pi/2m_\pi^2$} and
\begin{equation}
\frac{B_\text{NS}}{C_\text{NS}} = -\frac{1}{2}.
\end{equation}

We would like to point out that the $\pi/4m_\pi^2$ factor appearing in $\sigmatot^\text{NS}(s)$
is four times smaller than the maximum value determined by Lukaszuk and Martin~\cite{Lukaszuk_Martin:1967}.
In the NS paper, $s_0 $ is equal to $\langle k_{0,\pi}\rangle$, the average emitted energy per pion
~\cite{Nastase_Sonnenschein:2015}, calculated in the context of Heisenberg's model.

\end{itemize}

Regarding the result obtained by Giordano and Meggiolaro (GM), the ratio $B_\text{th}/C_\text{th}$,
Eq.~\eqref{eq:ratio-BC}, depends on the spin $\tilde{s}$ of the QCD-stable particle that maximizes
the ratio of Eq.~\eqref{eq:ratio-s-m}. To be able to reproduce the above-reported ratios,
we would need $\tilde{s} = 0$ for the MR case and this is not supported by GM calculations
since particles with spin 0 and 1 are excluded. Also the AdS/CFT case is not supported by 
GM calculations, since it would require the unphysical value $\tilde{s} = 5/4$. Instead, the NS case 
is consistent with the GM calculations, provided that $\tilde{s} = 3/2$ (the spin of the $\Omega^\pm$ baryon).
However, the coefficients multiplying the $\ln^2s$ and $\ln s \ln\ln s$ terms are different, 
since in GM the mass of the $\Omega^\pm$ baryon is used, while in NS appears the pion mass.

\section{Parametrization for the Total Cross Section}
\label{sec:slt_parametrization}

In this section, we present the parametrization used in the fits to the total cross section data.

For clarity and future reference, we will divide the total cross section in two parts:
the contribution to low energies (LE) and to high energies (HE), so that
\begin{equation}
 \sigmatot(s) = \sigma_\text{LE}(s)+ \sigma_\text{HE}(s).
 \label{eq:sigtot-par-notation}
\end{equation}

Similar to Chapter~\ref{chapt:rise_sigmatot}, we parametrize the low-energy dependence of $\sigmatot(s)$ 
in the scattering region by means of a \textit{Reggeon} exchange in the $t-$channel, 
following Regge Theory \cite{Collins_book:1977,Barone_Predazzi_book:2002}.
From our discussion in Section~\ref{sec:rise_ReggeGribov} and with a suitable notation here, 
this contribution reads
\begin{equation}
 R_i^{ab}(s) = A_i^{ab}\left(\frac{s}{s^{ab}_0}\right)^{-b_i} \quad(i=1,\,2),
\end{equation}

\noindent where $A_i^{ab}$ is associated to the residue function and $b_i$ to the intercept 
of the Reggeon trajectory $\alpha_i(t)$, i.e. $b_i = 1- \alpha_i(0)$. 
We consider two Reggeon contributions: one corresponding to a trajectory with even signature ($i=1$)
and another one with odd signature ($i=2$).
The latter contributes with a \textit{minus} sign to $ab$ scattering and with a \textit{plus}
sign to the crossed channel, $\bar{a}b$. Summarizing, the low-energy parametrization reads\footnote{In LHS of the equations,
$a^+ \equiv a$, representing a positive charged particle, 
and $a^-\equiv \bar{a}$ corresponds to its antiparticle.}
\begin{equation}
 \sigma_\text{LE}^{a^\pm b}(s) = A_1^{ab}\left(\frac{s}{s_0^{ab}}\right)^{-b_1} \mp A_2^{ab}\left(\frac{s}{s_0^{ab}}\right)^{-b_2}.
 \label{eq:sigtot-le}
\end{equation}
 
The high-energy contribution is parametrized by a \textit{Pomeron} exchange. 
We consider a critical Pomeron with $\alpha_\pomeron(0)=1$, which contributes as a constant, and 
a triple pole with $\alpha_\pomeron(0)=1$, 
giving a log-squared contribution to $\sigmatot(s)$ 
that is in accordance with the leading term obtained in~\cite{Giordano_Meggiolaro:2014} 
and with the Froissart-Martin upper bound for the total cross section (see Section~\ref{sec:basic_FLM_bound}). 
We include the two subleading terms discussed in Section~\ref{subsec:slt_connection_QCDspectrum}: 
$\sim \ln s \cdot \ln \ln s$ and $\sim \ln s$.
Therefore, the high-energy term reads 
 \begin{equation}
 \sigma_\text{HE}^{a^\pm b}(s) = \AP^{ab} + \kappa\left[B\ln^2\left(\frac{s}{s_0^{ab}}\right) + C\ln\left(\frac{s}{s_0^{ab}}\right)\ln\ln\left(\frac{s}{s_0^{ab}}\right)\right] + Q^{ab}\ln\left(\frac{s}{s_0^{ab}}\right)
 \label{eq:sigtot-he}
 \end{equation}
 
\noindent where, following the analysis of Ref.~\cite{Giordano_Meggiolaro:2014}
(see Sect.~\ref{sec:slt_nonperturbative_approach}), $B$ and $C$ are treated as
\textit{universal} parameters, while $Q^{ab}$,
as well as $A_1^{ab}$, $A_2^{ab}$ and $\AP^{ab}$,
are \textit{reaction-dependent}.
As already said at the end of Sect.~\ref{subsec:slt_connection_QCDspectrum}, we will also assume
(as in Ref.~\cite{Giordano_Meggiolaro:2014}) that $\kappa$
is independent of the properties of the scattering particles.
The energy scale is a fixed parameter and depends
only on the masses of the scattering particles\footnote{Note that
this scale is different from what was used in the previous chapter. 
Here, we followed the scale used in \cite{Giordano_Meggiolaro:2014}.}
\begin{equation}
 s_0^{ab}=m_am_b.
\end{equation}

Moreover, $Q^{ab}$ is taken to be crossing-symmetric, i.e.,
$Q^{\bar{a}b}=Q^{ab}$. This means that we are neglecting here
a possible Odderon contribution\footnote{The Odderon is the odd ``version'' of the Pomeron. 
It carries the quantum numbers of the vacuum but has an odd signature.}
to the total cross sections.
Notice that also the PDG and the highest-rank COMPETE parametrizations of
the total cross sections are crossing-symmetric.

Summarizing, the parameters $A_i^{ab}$ (mb), $b_i$ (dimensionless) [$i=1,\,2$] and $\AP^{ab}$ (mb)
are always free parameters to be determined in the fits. The parameters $B$ (mb), 
$C$ (mb), $Q^{ab}$ (mb) and $\kappa$~(dimensionless) can be either fixed or free, 
as detailed below in the descriptions of our variants of fits.

The names of the variants are written using the following notation:

\begin{itemize}
 \item LT stands for \textbf{L}eading \textbf{T}erm and SLT for \textbf{S}ub-\textbf{L}eading 
 \textbf{T}erm;

\item the subscript ``th'' refers to the
case where we fix $B$ ($\LTth$) or both $B$ and $C$ ($\SLTth$)
to the theoretical values discussed in Section~\ref{subsec:slt_connection_QCDspectrum};

\item the subscript ``$\kappa$'', after the first subscript and separated by a comma, indicates that the parameter $\kappa$ is free
 (for instance, $\SLTthk$);

\item by default, the coefficient $Q^{ab}$ of the logarithmic term is \textit{fixed to zero}:
in those cases in which a nonzero logarithmic term is considered
($Q^{ab}$ free), we shall add a ``Q'' in front of the variant name (for instance, $\QLTth$).
\end{itemize}

The main variants considered here are described below.

 \begin{itemize}
  \item $\LT$: $\kappa = 1$, $C=0$ and $Q=0$ are fixed parameters, while $B$ is
    free. This case (in which the subleading terms are absent) corresponds
    analytically to the parametrization used by the PDG in their analysis of
    forward data~\cite{PDG:2016} and to the highest-rank result
    obtained by the COMPETE Collaboration~\cite{COMPETE:2002a,COMPETE:2002b} (see
    below for more details).
    As mentioned above, when $Q$ is included as a free parameter,
    we shall denote this variant as $\QLT$.
    The same rule also applies to the other variants.
  
  \item $\SLT$: $\kappa = 1$ and $Q=0$ are fixed parameters, while $B$ and $C$
    are free parameters. {This case corresponds to the previous parametrization
    with the inclusion of the subleading term.}
  
  \item $\LTth$: $\kappa=1$, $B=B_\text{th}$, $C=0$ and $Q=0$ are fixed
    parameters. This variant has $B$ fixed to the theoretical values discussed
    in Sect.~\ref{subsec:slt_connection_QCDspectrum} and no subleading term is included.
  
  \item $\SLTth$: $\kappa=1$, $B=B_\text{th}$, $C=C_\text{th}$ and $Q=0$
    are fixed parameters. This variant has both $B$ and $C$ fixed to the
    theoretical values discussed in Sect.~\ref{subsec:slt_connection_QCDspectrum}.

  \item $\SLTthk$: $B=B_\text{th}$, $C=C_\text{th}$ and $Q=0$ are fixed
    parameters, while $\kappa$ is a free parameter.
 
 \end{itemize}

The main difference between our $\LT$ parametrization, the
highest-rank result by COMPETE, and the PDG parametrization is in the
energy scale appearing in the leading term $\ln^2 s$.
In the COMPETE analysis, the energy scale $s_0$ is a free parameter,
which does not depend on the scattering particles.
Our energy scale, on the other hand, is fixed and depends only on the
masses of the scattering particles, $s_0^{ab} =  m_am_b$.
In the PDG analysis this scale depends on the masses of
the colliding particles and on a universal mass scale also entering
their parametrization of the coefficient $B$, so it contains both a
fixed and a free part.

\section{Dataset and Methodology}
\label{sec:slt_dataset_methodology}

In this section, we present our dataset and the methodology used in our fits.

\subsection{Dataset}
\label{subsec:slt_dataset}

Our dataset comprises data from meson-baryon and baryon-baryon scattering,
namely $pp$, $\ppbar$, $pn$, $\bar{p}n$, $\pi^{\pm}p$, $K^{\pm}p$ and $K^{\pm}n$,
in the center of mass energy range 5 GeV $\leq \sqrt{s} \leq $ 8 TeV. 
Specifically, for $pp$ scattering in the LHC energies, we have included in the fits
the data\footnote{This analysis was developed before the publication of the ATLAS measurement at 8 TeV.}
T1-T5, T8, T9 and A1 of Table~\ref{tab:data_sigtot_LHC}. 
More details are given in Section~\ref{sec:data_fits-data}.

We considered in the fits only data from accelerator experiments. 
However, some cosmic-ray data~\cite{ARGO_YBJ:2009,PierreAuger:2012,TA:2015}
are shown in the figures just to illustrate the trend with the energy,
but they were not included in the fits. Since their uncertainties
are large, we do not expect great deviation from the results obtained without them. 

In all cases, we have treated the data points as independent, 
including those that have the same energy. For all data we have considered
statistic and systematic uncertainties added in quadrature.

We stress that we are not including data from reactions that involve photons
or deuterons and we do not constrain our fits using the data for the $\rho$
parameter, as done in the previous chapter and by COMPETE and PDG.

Finally, we mention that there are nine points available 
for $\Sigma^-p$ scattering in the energy region of
interest~\cite{PDG_data_website}.   
Including these points makes the fits more unstable (due to the
absence of data in the corresponding crossed channel).
On the other hand, these data present large errors and they do not affect the final result.
Therefore, in what follows we will consider only fits without this dataset.

\subsection{Methodology}
\label{subsec:slt_methodology}

As discussed in Sect.~\ref{sec:slt_nonperturbative_approach}, 
the leading and subleading terms were obtained for meson-meson scattering.
The authors in Ref.~\cite{Giordano_Meggiolaro:2014} argue that they expect 
that the leading term would be the same if one does the calculation for 
meson-baryon and baryon-baryon scattering. Here, we also assume that the
same is true for the subleading terms, at least in the phenomenological context.
This assumption has also a practical motivation: we expect that the subleading
term becomes important only at high energies and we have data for meson-baryon 
scattering only up to mid energies (around 25 GeV). Just for $pp$ and $\ppbar$ 
(therefore baryon-baryon scattering) we have data available at the TeV scale thanks
to Tevatron ($\ppbar$) and to LHC ($pp$).\footnote{Remember that we are not 
considering data from cosmic-rays in the data reductions.} 
For this reason, we consider first only 
fits using Eqs.~\eqref{eq:sigtot-par-notation}, \eqref{eq:sigtot-le}, \eqref{eq:sigtot-he}
to $pp$ and $\ppbar$ data and after that, we study fits to all reactions 
(Table~\ref{tab:data_sigtot_otherreac}).

\begin{figure}[htb!]
\centering
\includegraphics[scale=1]{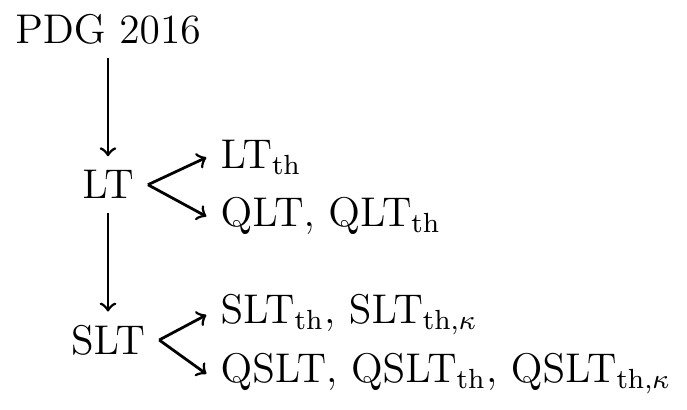}
\caption{\label{fig:IV_scheme} Initial-value scheme used in the fits.}
\end{figure}

In order to start from a solid and updated result, we decided to use
as initial values for the $\LT$ fit the results reported in the 2016 edition
of the Review of Particle Physics by PDG~\cite{PDG:2016},
and then use the results of $\LT$ fits as initial
values for the $\SLT$ fits. In this way, fitting first $\LT$
(that essentially corresponds to the PDG parametrization) 
we create a reference for discussing differences when we include the subleading
terms as well ($\SLT$, $\QLT$), instead of comparing directly with the PDG
result. However, this procedure presented some problems when considering
the fit to all hadronic data (see below in
Sect.~\ref{subsubsec:slt_res_allreactions}). In that case, we
decided to use as initial values for the parameters $B$ and $C$ in
the $\SLT$ fit to all data the results obtained in the $\SLT$ fit to
$\pp$/$\ppbar$ data. For the other parameters, we used the results
obtained in the $\LT$ fit to all data. The detailed 
scheme is shown in Fig.~\ref{fig:IV_scheme}, where $X \rightarrow
Y$ means that the results of variant $X$ were used as
initial values for the fit with variant~$Y$. 

As discussed in Chapter \ref{chapt:data_fits}, we use the reduced chi-squared as
a measure of goodness of fit~\cite{MINUIT_Manual:1994,Bevington_Robinson_book:1992}.

At last, we note that the variant $\LTthk$ is the same as $\LT$, except for a rescaling of the parameter $B \to \kappa B_\text{th}$.
Therefore, it will not bet considered in our fits and discussions.

\section{Fit Results}
\label{sec:slt_fit_results}

In this section, we present the results obtained in the fits, first to $pp$ and $\ppbar$
data only and after that to all reactions, considering all the variants. 
Finally, we compare and discuss the obtained results.

\subsection{Fits to $pp$ and $\ppbar$ Data}
\label{subsec:slt_results_fits_pp_ppbar}

\begin{figure}[h!]
\centering
\includegraphics[scale=0.5]{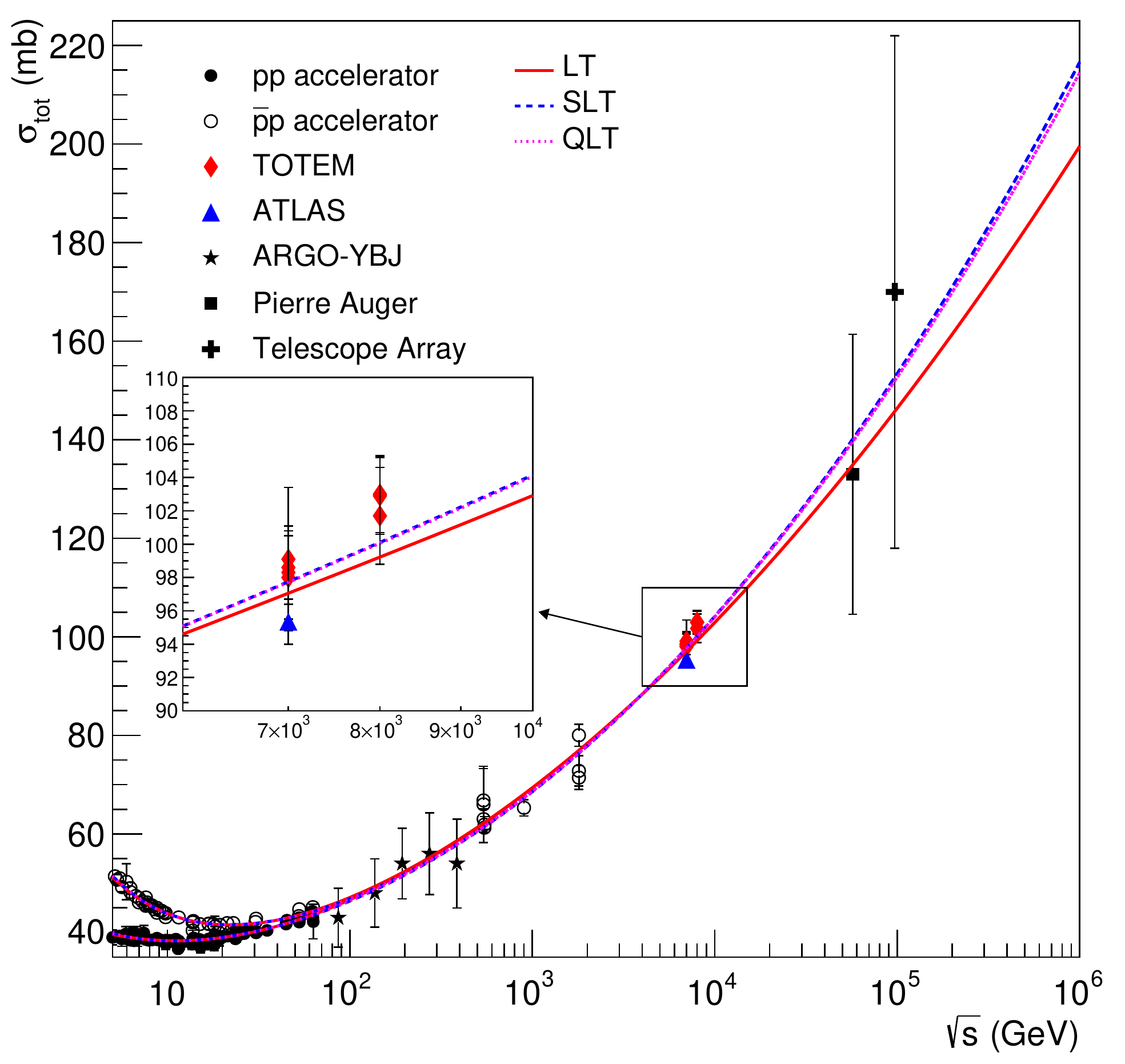}
\caption{\label{fig:res_pp_pbarp}Results of fits with $\LT$ ($\kappa=1$, $B$ free, $C=0$, $Q=0$),
$\SLT$ ($\kappa=1$, $B$ and $C$ free, $Q=0$) and $\QLT$ ($\kappa=1$, $B$ free, $C=0$, $Q$ free) to $pp$ and $\ppbar$ data.} 
\end{figure}

The parameters obtained in fits to $\pp$ and $\ppbar$ data with $\LT$,
$\SLT$ and $\QLT$ are shown in Table~\ref{tab:res_pp_pbarp}, those with $\LTth$,
$\SLTth$ and $\SLTthk$ in Table~\ref{tab:res_pp_pbarp_SLTkf} and
those with $\QLTth$, $\QSLTth$ and $\QSLTthk$ in Table~\ref{tab:res_pp_pbarp_QSLTkf}.
The curves calculated with the parameters of $\LT$, $\SLT$ and $\QLT$
are compared to the experimental data in Fig.~\ref{fig:res_pp_pbarp}, 
and those of $\SLTthk$ and $\QSLTthk$ in Fig.~\ref{fig:res_pp_pbarp_SLTkf_QSLTkf}.
Below we discuss the results that we have obtained using the different variants,
first without the inclusion of the $\ln s$ term, $Q=0$ (itens a, b and c) and
after that by including it (item d).

\subsubsection{a) Fits with $\LT$ ($\kappa=1$, $B$ free, $C=0$, $Q=0$) and $\SLT$ ($\kappa=1$, $B$ and $C$ free, $Q=0$)}
\label{subsubsec:slt_res_pp_pbarp}

The results obtained with variants $\LT$ and $\SLT$ present good description of data.
There is a small decrease in the value of $\chi^2/\nu$ going from $\LT$ to $\SLT$.
However, as discussed in Section~\ref{subsec:slt_methodology}, we cannot favour one variant 
with respect to the other on the basis of this value. Given that both are~$\lesssim~1$,
we can say that both variants result in good fits to data.

In $\SLT$, we have obtained $C<0$ as expected and $C\neq0$ within the uncertainty. 
The negative value of $C$ causes an increase in the $B$ parameter and also in $\AP$. 
The uncertainty in $\AP$ increases one order of magnitude, but the relative
uncertainty is still small ($\sim$ 5.6\%). 

Given the small difference between the values of $\chi^2/\nu$ of $\LT$ and $\SLT$
mentioned above, we cannot claim that the fit with the subleading term represents
an improvement with respect to the fit without it. We can only say that we are able
to estimate the value of $C$.

Regarding the Reggeon trajectories, $b_2$ is practically stable and we see an increase
of $b_1$ when $C$ is a free parameter. The values of $b_1$ and $b_2$ in $\SLT$ are compatible
within the uncertainties. Therefore, we may suppose that the presence of the subleading
term turns the intercept of the trajectories degenerate $\alpha_{\mathds{R}1}(0)\approx\alpha_{\mathds{R}2}(0)$,
at least within the errors. Observe that we also see an increase in $A_1$ while $A_2$ is stable.
A similar effect was observed by COMPETE in Ref.~\cite{COMPETE:2002a}
when discussing their highest-rank result (similar to $\LT$).
In that case, the degeneracy of the Reggeon intercepts was attributed to
a decreasing contribution of the log-squared term for $s<s_h$, where
$s_h$ is the energy scale determined in the fit. 

It is important to stress that the values of $b_1$ and $b_2$ obtained in $\LT$ are
not far from the values obtained by the PDG and also in other
analyses (for instance, the one by Menon and Silva
in Ref.~\cite{Menon_Silva:2013b} and those presented in Chapter~\ref{chapt:rise_sigmatot}).

\begin{table}[htb]
  \centering
  \caption{\label{tab:res_pp_pbarp}Results of fits with $\LT$ ($\kappa=1$, $B$ free, $C=0$, $Q=0$), 
  $\SLT$ ($\kappa=1$, $B$ and $C$ free, $Q=0$) and $\QLT$ ($\kappa=1$, $B$ free, $C=0$, $Q$ free) to $\sigmatot$ data of $\pp$ and $\ppbar$ scattering.
    Parameters $A_1$, $A_2$, $\AP$, $B$, $C$ and $Q$
    are in mb, while $b_1$, $b_2$ and $\kappa$ are dimensionless.} 
  \begin{tabular}{c|ccc}\hline\hline
               & \multicolumn{3}{c}{Fits to $\sigmatot$}\\\hline
               & $\LT$      & $\SLT$      & $\QLT$      \\\hline
  $B$          & 0.2269(38) &  0.349(29)  &  0.311(19)  \\ 
  $C$          & 0 (fixed)  & -0.95(21)   &  0 (fixed)  \\
  $\kappa$     & 1 (fixed)  &  1 (fixed)  &  1 (fixed)  \\
  $Q$          & 0 (fixed)  &  0 (fixed)  & -2.40(48)   \\
  $b_1$        & 0.342(15)  &  0.560(76)  &  0.586(89)  \\
  $b_2$        & 0.539(15)  &  0.541(16)  &  0.541(16)  \\
  $A_1$        & 56.8(1.7)  &  64.4(8.2)  &  60.6(8.7)  \\
  $A_2$        & 35.2(2.5)  &  35.6(2.5)  &  35.6(2.5)  \\
  $\AP$        & 24.77(60)  &  35.7(2.0)  &  41.7(3.0)  \\\hline
  $\chi^2/\nu$ & 0.972      &  0.933      &  0.934      \\
  $\nu$        & 165        &  164        &  164        \\\hline\hline
  \end{tabular}                                          
\end{table}

As already commented in Section~\ref{subsec:slt_otheranalysis}, the subleading term $\sim \ln s\cdot \ln\ln s$
also appears in other approaches. In Ref.~\cite{Diez_etal:2015}, this term arises from an AdS/CFT approach
together with a $\sim\ln s$ term [see Eq.~\eqref{eq:sigtot-Ads/CFT}].
The authors also perform fits to $\sigmatot$
data from $pp$ and $\ppbar$ scattering in order to determine what is the dominant
subleading contribution to $\sigmatot$: $\ln s$ or $\ln s \ln\ln s$. 
They also parametrize the low-energy contributions to
$\sigmatot$ with Reggeon terms, however with some fixed parameters in the values expected 
from Regge theory and others using sum rules in the resonance region (following the works
by Block and Halzen~\cite{Block_Halzen:2006}). Their results favor the $\ln s$ term over
$\ln s\ln\ln s$ contribution.

In any case, it is important to stress that their results are
different from ours in several aspects: (i) they fix low-energy parameters,
which we do not; (ii) as commented in Sect.~\ref{subsec:slt_otheranalysis}, their theoretical 
result does not address the dependence with the colliding particles, while in Ref.~\cite{Giordano_Meggiolaro:2014}
the $\ln s \ln\ln s$ is universal and the $\ln s$ term is reaction dependent; (iii) they use as cutoff 
energy to the fits $\sqrt{s_\text{min}}~=~6~\text{GeV}$, while here we have used $\sqrt{s_\text{min}}=5~\text{GeV}$.

\subsubsection{b) Fits with $\LTth$ ($\kappa=1$, $B=B_\text{th}$, $C=0$, $Q=0$) and $\SLTth$ ($\kappa=1$, $B=B_\text{th}$, $C=C_\text{th}$, $Q=0$)}
\label{subsubsec:slt_res_pp_pbarp_SLTth}

Now, we present the results from fits where $B$ and $C$ are fixed
to the theoretical values associated with the three states from QCD discussed in Section~\ref{subsec:slt_connection_QCDspectrum}, 
and $\kappa=1$ fixed. 
The selected results from fits without the logarithmic term are shown in Table~\ref{tab:res_pp_pbarp_SLTkf};
the corresponding curves ($\kappa=1$) are not displayed.
In what follows, we divide the results in the variants and in the values of $B$ and $C$ used as input.

\begin{itemize}
 \item \textbf{$B_\text{th}$ and $C_\text{th}$ for $\Omega^\pm$ baryon}
 
 The values of $B$ and $C$ calculated with the mass and spin of $\Omega^\pm$ 
 baryon are displayed in Eq.~\eqref{eq:B-Cth-Omega}, namely $B_\text{th} = 0.22$~mb
 and $C_\text{th} = -0.44$~mb. In $\LTth$, we have $B=B_\text{th}$ and $C=0$ fixed. 
 This result is close to the one obtained using $\LT$. This can be seen by the 
 value of $B$ in the second column in Table~\ref{tab:res_pp_pbarp}: 
 0.2269~mb~$\sim$~0.23~mb, very close to $B_\text{th}$. The other parameters 
 and the value of $\chi^2/\nu$ present small variations compared to $\LT$. 
 We also have a good description of the experimental data, not shown in figures.
 
 On the other hand, in $\SLTth$, where now $C = C_\text{th}$ is fixed,
 the result is not satisfactory as the previous variant. In fact, the 
 value of $\chi^2/\nu$ is considerably high, indicating a poor description
 of the data. 
 We would like to point out that we have practically no change in the parameters associated
 with the odd signature Reggeon contribution, while the intercept of the even
 trajectory increases going from $\LTth$ to $\SLTth$. The same happens to the $A_1$ parameter.
 
 \item \textbf{$B_\text{th}$ and $C_\text{th}$ for glueball $2^{++}$ state}
 
 In this case, we consider the (quenched) mass and spin of the glueball $2^{++}$ state,
 giving $B_\text{th}=-C_\text{th}=0.42$~mb [Eq.~\eqref{eq:B-Cth-g2pp}]. In both variants
 we get a poor description of the data with $\chi^2/\nu \sim 3$. The intercept of the even
 Reggeon trajectory is too small when compared with other cases and we have a negative 
 ``constant'' Pomeron contribution ($\AP<0$). Besides, the fits present a 
 non-positive definite error matrix. In this case, although the fit has converged,
 the error matrix may have some calculation problems and we cannot fully trust
 in the estimation of uncertainties of the free parameters \cite{MINUIT_Manual:1994}.
 For this reason, we decided not to show these results in Table~\ref{tab:res_pp_pbarp_SLTkf}.
 
 The description of data are similar for the two variants, with overestimation of 
 $\sigmatot$ at LHC energies.
 
 \item \textbf{$B_\text{th}$ and $C_\text{th}$ for glueball $3^{+-}$ state}
 
 Considering the glueball $3^{+-}$ state, we have (using again the quenched mass)
 $B_\text{th}=0.78$ mb and $C_\text{th}~=~-~0.39$~mb [Eq.~\eqref{eq:B-Cth-g3pm}]. 
 In these cases, the minimizer did not converge and, therefore, no solution was obtained.
 
\end{itemize}

\begin{table}[htb]
   \centering
   \caption{\label{tab:res_pp_pbarp_SLTkf}Results of fits with
     $\LTth$ ($\kappa=1$, $B=B_\text{th}$, $C=0$, $Q=0$),
     $\SLTth$ ($\kappa=1$, $B=B_\text{th}$, $C=C_\text{th}$, $Q=0$) and
     $\SLTthk$ ($\kappa$ free, $B=B_\text{th}$, $C=C_\text{th}$, $Q=0$) to
     $\sigmatot$ data of $\pp$ and $\ppbar$ scattering.
     The values of $B$ and $C$ are fixed to the
     theoretical values calculated with the masses and the spins of
     the $\Omega^{\pm}$ baryon, the $2^{++}$ glueball state and the
     $3^{+-}$ glueball state (quenched values), while the parameter $Q$
     is fixed to zero. For the units of measurement of the parameters,
     see Table~\ref{tab:res_pp_pbarp}.}
   \begin{tabular}{c|ccc|c|c}\hline\hline
                & \multicolumn{3}{c|}{$\Omega^\pm$ baryon} & {$2^{++}$ glueball} & {$3^{+-}$ glueball} \\\hline
                & $\LTth$ & $\SLTth$ & $\SLTthk$ & $\SLTthk$ & $\SLTthk$ \\\hline
  $B_\text{th}$ & 0.22 (fixed)&  0.22 (fixed) &  0.22 (fixed)  &  0.42 (fixed)  &  0.78 (fixed)      \\
  $C_\text{th}$ & 0 (fixed)   & -0.44 (fixed) & -0.44 (fixed)  & -0.42 (fixed)  & -0.39 (fixed)      \\
   $\kappa$     & 1 (fixed)   &  1 (fixed)    &  1.377(18)     &  0.6159(96)    &  0.3097(51)        \\
   $Q$          & 0 (fixed)   &  0 (fixed)    &  0 (fixed)     &  0 (fixed)     &  0 (fixed)         \\
   $b_1$        & 0.365(10)   &  0.743(20)    &  0.458(20)     &  0.385(17)     &  0.361(17)         \\
   $b_2$        & 0.539(15)   &  0.528(16)    &  0.540(15)     &  0.539(15)     &  0.539(15)         \\
   $A_1$        & 58.5(1.7)   &  115.3(8.5)   &  57.5(3.2)     &  56.0(2.2)     &  56.3(2.0)         \\
   $A_2$        & 35.3(2.5)   &  33.7(2.4)    &  35.4(2.5)     &  35.3(2.4)     &  35.2(2.4)         \\
   $\AP$        & 25.75(21)   &  35.862(74)   &  32.17(29)     &  28.13(46)     &  26.38(55)         \\\hline
   $\chi^2/\nu$ & 0.987       &  3.59         &  0.937         &  0.957         &  0.965             \\
   $\nu$        & 166         &  166          &  165           &  165           &  165               \\\hline\hline
   \end{tabular}
\end{table}

\subsubsection{c) Fits with $\SLTthk$ ($\kappa$ free, $B=B_\text{th}$, $C=C_\text{th}$, $Q=0$)}
\label{subsubsec:slt_res_pp_pbarp_SLTthk}

\begin{figure}[htb!]
\centering
\subfloat[$\SLTthk$]{\label{fig:res_pp_pbarp_SLTkf_panel_a}\includegraphics[scale=0.39]{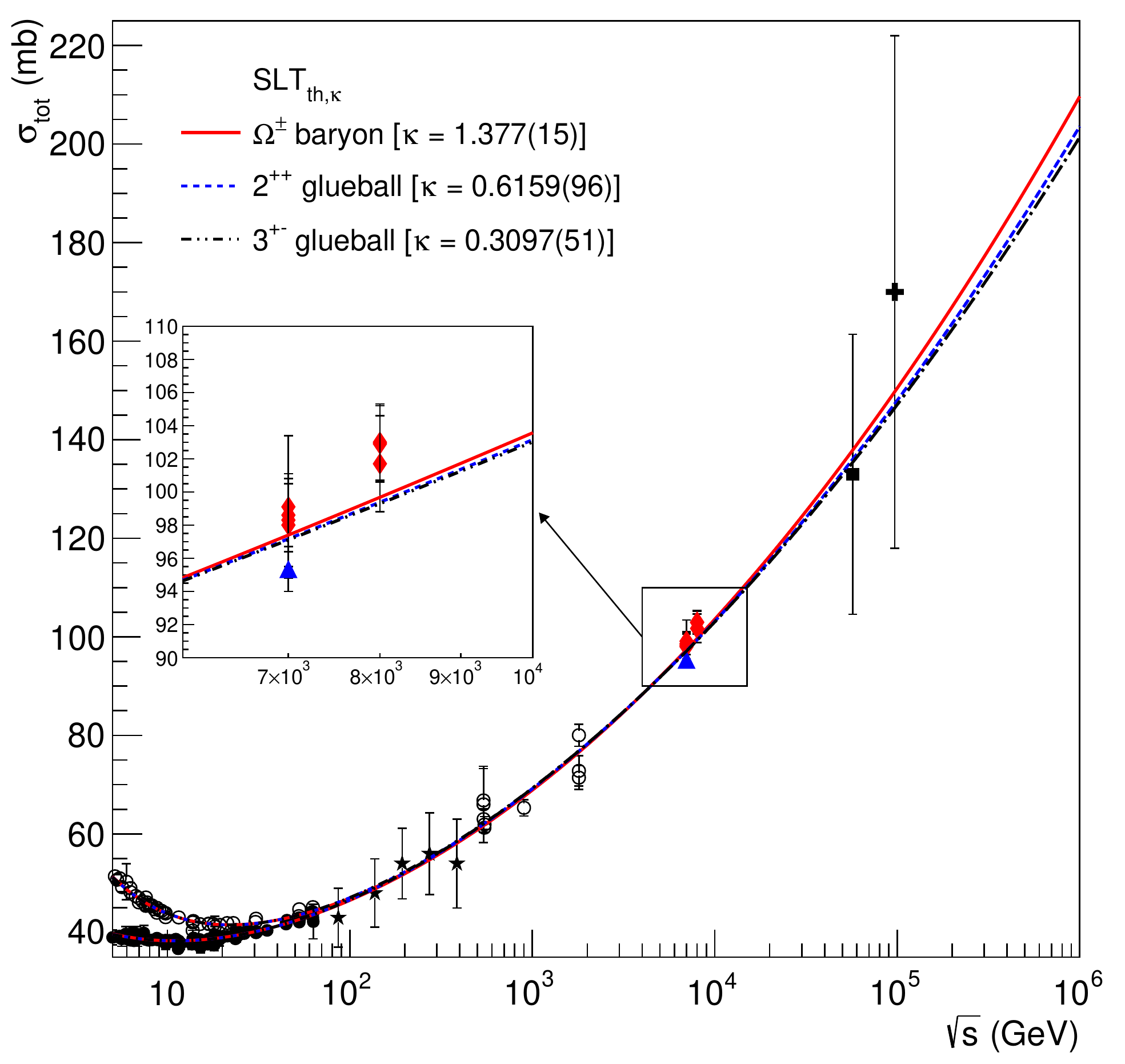}}\hfill
\subfloat[$\QSLTthk$]{\label{fig:res_pp_pbarp_QSLTkf_panel_b}\includegraphics[scale=0.39]{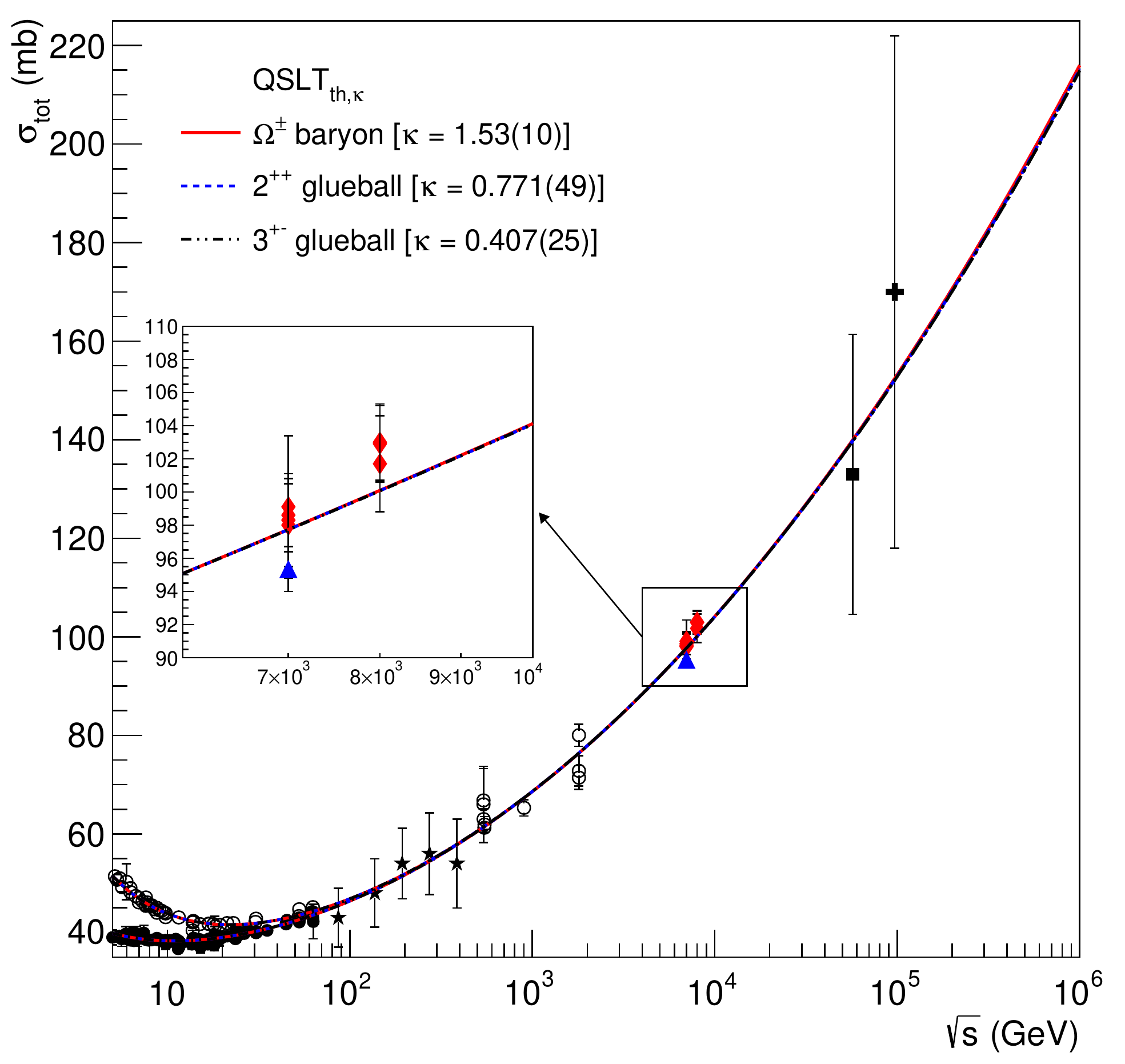}}
\caption{\label{fig:res_pp_pbarp_SLTkf_QSLTkf}Results of fits with 
         \subref{fig:res_pp_pbarp_SLTkf_panel_a}~$\SLTthk$~($\kappa$~free, $B=B_\text{th}$, $C=C_\text{th}$, $Q=0$) 
     and \subref{fig:res_pp_pbarp_QSLTkf_panel_b}~$\QSLTthk$~($\kappa$~free, $B=B_\text{th}$, $C=C_\text{th}$, $Q$ free) 
     to $pp$ and  $\ppbar$ data for $B_\text{th}$ and $C_\text{th}$ calculated from the  $\Omega^\pm$ baryon,
     and the $2^{++}$ and $3^{+-}$ glueball states. The legend of data is the same of Figure~\ref{fig:res_pp_pbarp}.} 
\end{figure}

In these variants, we consider both $B$ and $C$ fixed with $\kappa$ free. 
The parameters obtained in the fits are shown in Table~\ref{tab:res_pp_pbarp_SLTkf} and
the corresponding curves for the three states in Fig.~\ref{fig:res_pp_pbarp_SLTkf_QSLTkf}~\subref{fig:res_pp_pbarp_SLTkf_panel_a}.
With these variants, we may infer the asymptotic scenario for the colliding particle,
since according to Eq.~\eqref{eq:ratio-eltot-kappa}, the asymptotic ratio between 
$\sigmael$ and $\sigmatot$ is given by $\kappa/2$. 

In all the cases considered for $B_\text{th}$ and $C_\text{th}$ values,
the fits have good quality ($\chi^2/\nu \lesssim 1$) with small differences
in the $\chi^2$ value among them. We see small variations of some parameters,
for instance, $A_1$, $b_1$ and $\AP$. Apart from these differences, 
the description of data is the same in the fitted energy range for all cases.
As can be seen in Figure~\ref{fig:res_pp_pbarp_SLTkf_QSLTkf}~\subref{fig:res_pp_pbarp_SLTkf_panel_a},
the difference at the LHC energies is very small and the results start to be different only at 
cosmic-ray energies.

Regarding the value of $\kappa$, using the values of $B_\text{th}$ and $C_\text{th}$
from the $\Omega^\pm$ baryon, we get $\kappa > 1$, therefore an antishadowing scenario.
For the glueball cases, we get $\kappa < 1$, hence a grey-disk scenario, being the
value from the $2^{++}$ case larger than the value from the $3^{+-}$ case. The connection 
with the ratio $\sigmael/\sigmatot$ will be discussed in Section~\ref{subsec:slt_results_ratio_eltot_kappa}.

\subsubsection{d) Fits with the inclusion of the logarithmic term ($\QLT$, $\QSLTth$, $\QSLTth$)}
\label{subsubsec:slt_results_pp_ppbar_log}

In this section, we discuss the effects of including
the logarithmic term ($Q$ as a free parameter) in the fits discussed above.
The fit results are displayed in Table~\ref{tab:res_pp_pbarp_QSLTkf} and the 
curves for the three states and variant $\QSLTthk$ (discussed below) in Fig.~\ref{fig:res_pp_pbarp_SLTkf_QSLTkf}\subref{fig:res_pp_pbarp_QSLTkf_panel_b}.

Considering first the variants with $B$ and/or $C$ as free parameters, 
the logarithmic term (in variant $\QLT$) describes the data
in the same way as $\SLT$, i.e., with the same $\chi^2/\nu$,
as can be seen in Table~\ref{tab:res_pp_pbarp} and in
Fig.~\ref{fig:res_pp_pbarp}.
However, the case in which both subleading terms are present ($\QSLT$) did not converge
when considering only $pp$ and $\ppbar$ data. Given that the $\ln s$ term
and the $\ln s \cdot \ln \ln s$ term describe the data in similar ways,
we attribute the non-convergence to a competition between these two
contributions. 

Instead, regarding the fits with $B$ and/or $C$ fixed
(see results in Table~\ref{tab:res_pp_pbarp_QSLTkf} and in 
Fig.~\ref{fig:res_pp_pbarp_SLTkf_QSLTkf}\subref{fig:res_pp_pbarp_QSLTkf_panel_b}), 
we observe the same behavior seen in the fits without the logarithmic term.
It is worth mentioning an improvement in the $\chi^2/\nu$ value in $\QSLTth$,
when compared to $\SLTth$, in the $\Omega^\pm$ case.
We also see small improvements in the $\chi^2$ value for $\QSLTthk$ for the
glueball cases and, moreover, we notice that the inclusion of the logarithmic
term in $\QSLTthk$ leads to Reggeon trajectories which are degenerate within
the errors. In all cases, the central value of the $\kappa$ parameter
increases when the logarithmic contribution is present.
The description of the data is the same for all the theoretical values
considered, as can be seen in 
Figure~\ref{fig:res_pp_pbarp_SLTkf_QSLTkf}\subref{fig:res_pp_pbarp_QSLTkf_panel_b}.

\begin{table}[htb]
   \centering
   \caption{\label{tab:res_pp_pbarp_QSLTkf}Results of fits with
     $\QLTth$ ($\kappa=1$, $B=B_\text{th}$, $C=0$, $Q$ free),
     $\QSLTth$ ($\kappa=1$, $B=B_\text{th}$, $C=C_\text{th}$, $Q$ free) and
     $\QSLTthk$ ($\kappa$ free, $B=B_\text{th}$, $C=C_\text{th}$, $Q$ free) to
     $\sigmatot$ data of $\pp$ and $\ppbar$ scattering. 
     The values of $B$ and $C$ are fixed to the
     theoretical values calculated with the masses and the spins of
     the $\Omega^{\pm}$ baryon, the $2^{++}$ glueball state and the
     $3^{+-}$ glueball state (quenched values). For the units of 
     measurement of the parameters, see Table~\ref{tab:res_pp_pbarp}.}
    \begin{tabular}{c|ccc|c|c}\hline\hline
                & \multicolumn{3}{c|}{$\Omega^\pm$ baryon} & {$2^{++}$ glueball} & {$3^{+-}$ glueball} \\\hline
                & $\QLTth$   & $\QSLTth$    & $\QSLTthk$    & $\QSLTthk$    & $\QSLTthk$          \\\hline
  $B_\text{th}$ & 0.22 (fixed)&  0.22 (fixed) &  0.22 (fixed)  &  0.42 (fixed)  &  0.78 (fixed)        \\
  $C_\text{th}$ & 0 (fixed)   & -0.44 (fixed) & -0.44 (fixed)  & -0.42 (fixed)  & -0.39 (fixed)        \\
   $\kappa$     & 1 (fixed)   &  1 (fixed)    &  1.53(10)      &  0.771(49)     &  0.407(25)           \\
   $Q$          & 0.19(12)    &  1.86(13)     & -0.69(41)      & -1.58(45)      & -2.00(47)            \\
   $b_1$        & 0.335(21)   &  0.311(18)    &  0.566(82)     &  0.576(86)     &  0.581(88)           \\
   $b_2$        & 0.539(15)   &  0.538(14)    &  0.541(16)     &  0.541(16)     &  0.541(16)           \\
   $A_1$        & 57.9(1.5)   &  63.3(1.3)    &  63.2(8.0)     &  61.8(8.3)     &  61.2(8.5)           \\
   $A_2$        & 35.2(2.5)   &  35.1(2.3)    &  35.6(2.5)     &  35.6(2.5)     &  35.6(2.5)           \\
   $\AP$        & 23.4(1.6)   &  16.0(1.8)    &  37.4(3.0)     &  39.6(3.0)     &  40.7(3.0)           \\\hline
   $\chi^2/\nu$ & 0.977       &  0.985        &  0.933         &  0.934         &  0.934               \\
   $\nu$        & 165         &  165          &  164           &  164           &  164                 \\\hline\hline
   \end{tabular}
\end{table}

\subsection{Fits to All Reactions}
\label{subsec:slt_results_fits_allreactions}

\begin{figure}[htb!]
\centering
\includegraphics[scale=0.8]{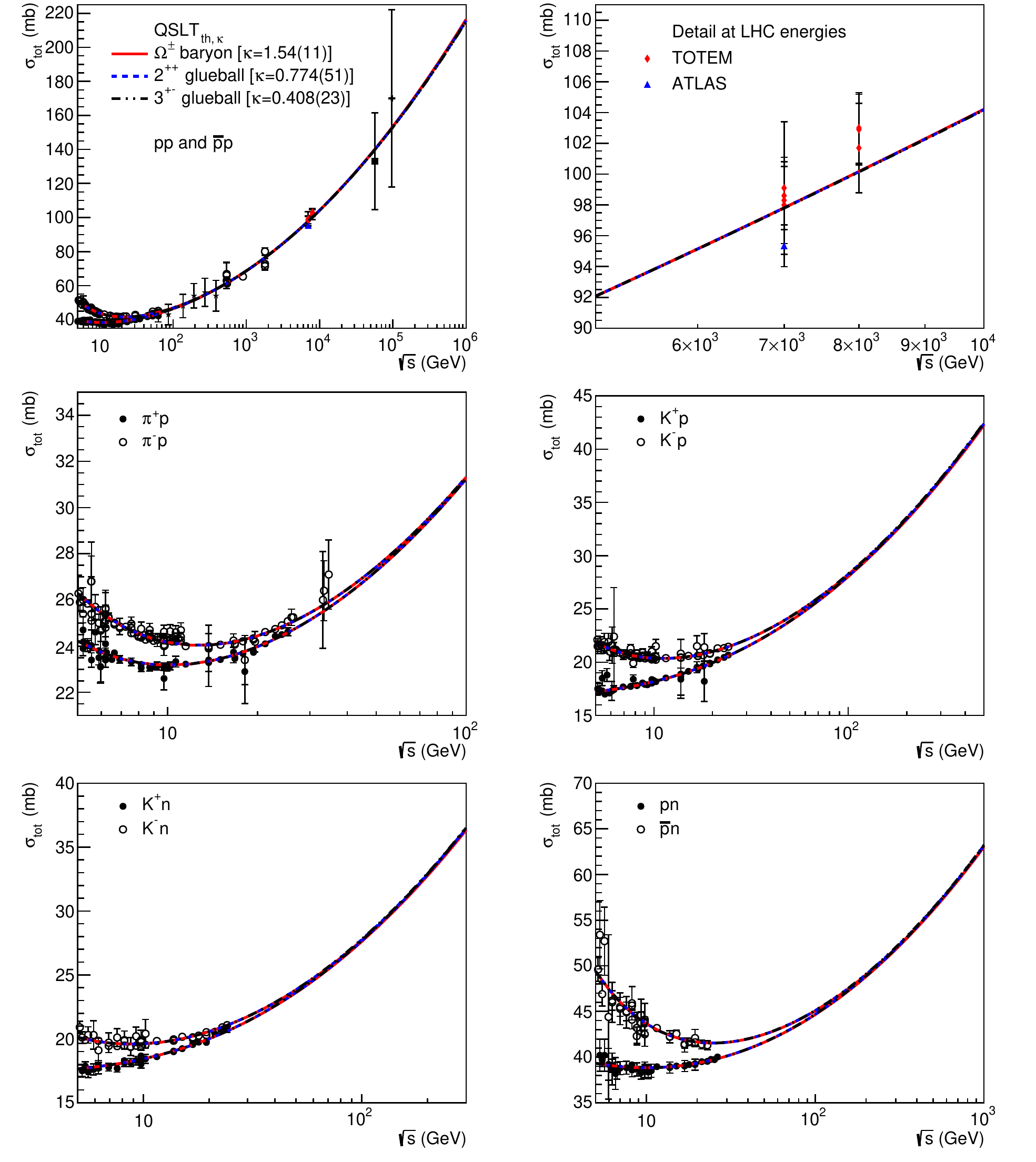}
\caption{\label{fig:res_allreactions_woSmp_QSLTkf}Results of fits with
 $\QSLTthk$ ($\kappa$ free, $B=B_\text{th}$, $C=C_\text{th}$, $Q$ free) to all
 data for $B_\text{th}$ and $C_\text{th}$ calculated from the
 $\Omega^\pm$ baryon, and the $2^{++}$ and $3^{+-}$ glueball states.
 The legend for the curves is shown in the top-left panel. For the
 legend of $pp$ and $\ppbar$ data see Figure~\ref{fig:res_pp_pbarp}.}
\end{figure}

In this section we present and discuss the results obtained from fits
to the full dataset (data from meson-baryon and baryon-baryon
scattering) with the variants $\LT$, $\SLT$, $\QLT$, $\QSLT$ 
(Table~\ref{tab:res_allreactions_woSmp}), 
with the variants $\LTth$, $\SLTth$ 
and $\SLTthk$ (Table~\ref{tab:res_allreactions_SLTkf_woSmp}),
and with the variants $\QLTth$, $\QSLTth$ 
and $\QSLTthk$ (Table~\ref{tab:res_allreactions_QSLTkf_woSmp}).
The comparison of the curves of $\SLTthk$ and $\QSLTthk$ with the experimental
data is shown in Figs.~\ref{fig:res_allreactions_woSmp_SLTkf} and
\ref{fig:res_allreactions_woSmp_QSLTkf}, respectively.

\subsubsection{a) Fits with $\LT$ ($\kappa=1$, $B$ free, $C=0$, $Q=0$) and $\SLT$ ($\kappa=1$, $B$ and $C$ free, $Q=0$)}
\label{subsubsec:slt_res_allreactions}

The results are presented in Table~\ref{tab:res_allreactions_woSmp}.
With respect to $\LT$, we have a good description of data with $\chi^2/\nu\sim 1$.
The same is true for $\SLT$, that we now discuss. When considering
this variant, more care was needed regarding the initial values. 
Following the same scheme for the
choice of initial values as in the analysis of $pp$ and $\ppbar$ data (see
Fig.~\ref{fig:IV_scheme}), the resulting fit has a
non-positive definite error matrix.
In order to obtain a more reliable result (with an accurate error matrix),
we changed the initial value for the parameters $B$ and $C$: we used the values
obtained in the fit to $pp$ and $\ppbar$ data with $\SLT$ (Table~\ref{tab:res_pp_pbarp})
in place of the values obtained in the fits to all reactions with $\LT$. Namely, 
$B=0.349$~mb and $C=-0.95$~mb instead of $B=0.2433$~mb and $C=0$. 
With this choice we obtained a more reliable result with an accurate error matrix.

The result presents $C<0$ as expected, but with smaller
magnitude and uncertainty than in the $\SLT$ fit to $pp$ and $\ppbar$
data, although the relative uncertainty is the same ($\sim$ 22\%).
We attribute this to the presence of more data
at low-energies. On the other hand, the $\chi^2/\nu$ is practically the same. 
It is important to mention that here we are increasing the effect of low-energy 
data in the estimation of $C$ compared to the $pp$/$\ppbar$
fits, since we have more low-energy than high-energy data points in the present dataset. 
In fact, we have nonzero correlation coefficients between low- and high-energy parameters,
indicating the influence of the low-energy data in the determination of
$C$ in the fit. In Eq.~\eqref{eq:corr-SLT-allreactions-woSmp} we show some correlation coefficients
that illustrate this point (see also table 6 in Ref.~\cite{Menon_Silva:2013a}):
\begin{equation}
 \corr(C,b_1) = 0.946, \quad \corr(C,A_1^{pp})=0.757, \quad \corr(C,A_1^{\pi p})=0.822.
 \label{eq:corr-SLT-allreactions-woSmp}
\end{equation}

Apart from these general aspects of the fits, there is still one point that demands some comments. 
This point concerns the negative value of the parameter $\AP^{\pi p}$ 
that appears in $\LT$, while it changes to a positive value in $\SLT$.
This is the only negative critical Pomeron contribution (although with large errors) in this set of fits. 
Taking into account the property of factorization of the residues of the Regge
poles~\cite{Barone_Predazzi_book:2002} (see also the comments in Ref.~\cite{COMPETE:2002a}), 
this parameter should be positive. However, factorization is only proven for simple poles and
is valid when only one trajectory dominates \cite{Barone_Predazzi_book:2002}.
Note that in this analysis the dominant term is the triple-pole Pomeron ($\ln^2 s$) and not the critical Pomeron.
In any case, the value of $\AP$ is affected by the
choice of the energy scale in the leading and subleading terms
in Eq.~\eqref{eq:sigtot-he}.
Therefore, we cannot exclude this result based on factorization. 
On the other hand, even with $\AP^{\pi p}<0$, we do not have a negative Pomeron 
contribution (from the combination of this constant term plus the $\ln^2 s$ term) in $\LT$.

\begin{table}[htb!]
  \centering
  \caption{\label{tab:res_allreactions_woSmp}Results of fits with
    $\LT$ ($\kappa=1$, $B$ free, $C=0$, $Q=0$), $\SLT$ ($\kappa=1$, $B$ and $C$ free, $Q=0$), 
    $\QLT$ ($\kappa=1$, $B$ free, $C=0$, $Q$ free) and $\QSLT$ ($\kappa=1$, $B$, $C$ and $Q$ free) to $\sigmatot$ (all data).
    For the units of measurement of the parameters, see Table~\ref{tab:res_pp_pbarp}.} 
 \begin{tabular}{c|ccccc}\hline\hline
                & $\LT$      &  $\SLT$     & $\QLT$      & $\QSLT$    \\\hline
  $B$           & 0.2433(46) &  0.2652(96) &  0.1646(73) &  0.363(28) \\ 
  $C$           & 0 (fixed)  & -0.200(44)  &  0 (fixed)  & -1.32(16)  \\ 
  $\kappa$      & 1 (fixed)  &  1 (fixed)  &  1 (fixed)  &  1 (fixed) \\
  $b_1$         & 0.222(11)  &  0.2420(85) &  0.2536(92) &  0.545(79) \\ 
  $b_2$         & 0.5128(99) &  0.513(11)  &  0.530(11)  &  0.532(11) \\\hline
  $A_1^{pp}$    & 47.86(62)  &  44.33(91)  &  69.8(1.7)  &  64.8(1.7) \\ 
  $A_2^{pp}$    & 30.8(1.4)  &  30.8(1.5)  &  33.73(1.7) &  34.2(1.7) \\
  $\AP^{pp}$    & 19.0(1.1)  &  22.61(22)  &  6.5(1.3)   &  33.2(2.6) \\
  $Q^{pp}$      & 0 (fixed)  &  0 (fixed)  &  2.10(15)   &  0.94(17)  \\\hline
  $A_1^{pn}$    & 47.2(1.1)  &  43.6(1.3)  &  43.7(5.8)  &  33.2(7.3) \\ 
  $A_2^{pn}$    & 27.4(1.5)  &  27.5(1.6)  &  29.6(1.8)  &  29.9(1.8) \\ 
  $\AP^{pn}$    & 19.2(1.1)  &  22.86(36)  &  22.1(3.8)  &  40.2(2.5) \\
  $Q^{pn}$      & 0 (fixed)  &  0 (fixed)  &  0.48(41)   &  0.025(0.24)\\\hline
  $A_1^{\pi p}$ & 70.37(99)  &  67.9(1.7)  &  63.3(2.3)  &  69(13)    \\ 
  $A_2^{\pi p}$ & 15.7(1.0)  &  15.8(1.1)  &  16.6(1.2)  &  16.9(1.2) \\ 
  $\AP^{\pi p}$ & -3.3(1.3)  &  0.80(31)   &  1.4(1.1)   &  24.3(2.1) \\
  $Q^{\pi p}$   & 0 (fixed)  & 0 (fixed)   &  0.49(12)   & -0.28(10)  \\\hline
  $A_1^{Kp}$    & 3.42(57)   &  30.31(73)  &  26.3(2.2)  &  16.8(2.4) \\ 
  $A_2^{Kp}$    & 17.54(91)  &  17.56(96)  &  18.9(1.1)  &  19.1(1.1) \\ 
  $\AP^{Kp}$    & 1.77(85)   &  5.09(11)   &  6.1(1.1)   &  17.13(92) \\
  $Q^{Kp}$      & 0 (fixed)  &  0 (fixed)  &  0.31(13)   &  0.496(79) \\\hline
  $A_1^{Kn}$    & 32.72(73)  &  28.76(77)  &  16.8(1.3)  &  7.4(4.4)  \\ 
  $A_2^{Kn}$    & 9.28(69)   &  9.30(71)   &  10.15(77)  &  10.28(79) \\ 
  $\AP^{Kn}$    & 1.93(84)   &  5.22(14)   &  10.81(70)  &  18.1(1.0) \\
  $Q^{Kn}$      & 0 (fixed)  &  0 (fixed)  & -0.159(33)  &  0.35(11)  \\\hline
  $\chi^2/\nu$  & 1.060      &  1.063      &  0.791      &  0.766     \\
  $\nu$         & 532        &  531        &  527        &  526       \\\hline\hline
  \end{tabular}
\end{table}

\subsubsection{b) Fits with $\LTth$ ($\kappa=1$, $B=B_\text{th}$, $C=0$) and $\SLTth$ ($\kappa=1$, $B=B_\text{th}$, $C=C_\text{th}$)}
\label{subsubsec:slt_res_allreactions_SLTth}
 
We now discuss the results obtained with $B$ and/or $C$ fixed at the theoretical values
discussed in Section~\ref{subsec:slt_connection_QCDspectrum} in the fits to all data.

 \begin{itemize}
 \item \textbf{$B_\text{th}$ and $C_\text{th}$ for $\Omega^\pm$ baryon}
 
 The results (2nd and 3rd columns of Table~\ref{tab:res_allreactions_SLTkf_woSmp})
 obtained in this case are satisfactory for $\LTth$. In the first case,
 we have a small increase of the $\chi^2/\nu$ value in comparison with $\LT$, although
 this variation is small. For $\SLTth$, we get $\chi^2/\nu\sim 2$ while in $\SLT$
 we have $\sim 1$. However, this increase is less than that observed in the fits
 to $pp$ and $\ppbar$ data only. Contrary to the fit $\LT$, here in $\LTth$ we have
 that all $\AP^i>0$.
 
 \item \textbf{$B_\text{th}$ and $C_\text{th}$ for glueball $2^{++}$ state}
 
 The results obtained here with $\LTth$ and $\SLTth$ have the same features
 of the fits to $pp$ and $\ppbar$ data, for example, the small $b_1$ parameter. 
 We also note that almost all $\AP^i <0$. The $\chi^2/\nu$ values are around 1.5,
 with similar description of data for both variants. 
 Regarding $pp$ and $\ppbar$, the fits overestimate the data at the LHC energies,
 reaching the upper error bar of the TOTEM data. 
 Additionally, both fits have a non-positive definite error matrix.
  
 \item \textbf{$B_\text{th}$ and $C_\text{th}$ for glueball $3^{+-}$ state}
 
 Again, using the mass and spin of the glueball $3^{+-}$ state, the fits did not converge. 
 
\end{itemize}
 
 \begin{table}[htb!]
  \centering
  \caption{\label{tab:res_allreactions_SLTkf_woSmp}Results of fits with
    $\LTth$ ($\kappa=1$, $B=B_\text{th}$, $C=0$, $Q=0$),
    $\SLTth$ ($\kappa=1$, $B=B_\text{th}$, $C=C_\text{th}$, $Q=0$) and
    $\SLTthk$ ($\kappa$ free, $B=B_\text{th}$, $C=C_\text{th}$, $Q=0$) to
    $\sigmatot$ (all data). The values of
    $B$ and $C$ are fixed to the theoretical values calculated with
    the masses and the spins of the $\Omega^{\pm}$ baryon, the
    $2^{++}$ glueball state and the $3^{+-}$ glueball state (quenched
    values), while the parameters $Q^{ab}$ are fixed to zero.
    For the units of measurement of the parameters, see Table~\ref{tab:res_pp_pbarp}.}
  \begin{tabular}{c|ccc|c|c}\hline\hline
                     & \multicolumn{3}{c|}{$\Omega^\pm$ baryon}& {$2^{++}$ glueball} & {$3^{+-}$ glueball}      \\\hline
                     & $\LTth$      & $\SLTth$      & $\SLTthk$     & $\SLTthk$     & $\SLTthk$       \\\hline
  $B_\text{th}$      & 0.22 (fixed) &  0.22 (fixed) &  0.22 (fixed) &  0.42 (fixed) &  0.78 (fixed)  \\ 
  $C_\text{th}$      & 0 (fixed)    & -0.44 (fixed) & -0.44 (fixed) & -0.42 (fixed) & -0.39 (fixed)  \\ 
  $\kappa$           & 1 (fixed)    &  1 (fixed)    &  1.439(23)    &  0.653(12)    &  0.3303(64)    \\
  $Q$                & 0 (fixed)    &  0 (fixed)    &  0 (fixed)    &  0 (fixed)
    &  0 (fixed)         \\
  $b_1$              & 0.2744(66)   &  0.554(13)    &  0.292(14)    &  0.249(13)    &  0.234(12)     \\ 
  $b_2$              & 0.5141(97)   &  0.515(11)    &  0.514(10)    &  0.513(11)    &  0.513(11)     \\\hline 
  $A_1^{pp}$         & 47.04(71)    &  59.0(2.7)    &  37.99(87)    &  43.12(57)    &  45.54(58)     \\ 
  $A_2^{pp}$         & 31.0(1.4)    &  31.4(1.6)    &  30.9(1.4)    &  30.8(1.5)    &  30.8(1.5)     \\ 
  $\AP^{pp}$         & 23.40(24)    &  35.159(77)   &  29.22(51)    &  23.76(82)    &  21.29(98)     \\\hline 
  $A_1^{pn}$         & 46.3(1.2)    &  57.8(3.2)    &  37.3(1.2)    &  42.4(1.1)    &  44.9(1.1)     \\ 
  $A_2^{pn}$         & 27.6(1.5)    &  27.9(1.7)    &  27.5(1.6)    &  27.5(1.6)    &  27.4(1.6)     \\ 
  $\AP^{pn}$         & 23.64(37)    &  35.24(15)    &  29.44(56)    &  24.01(87)    &  21.5(1.0)     \\\hline 
  $A_1^{\pi p}$      & 73.5(1.5)    &  136.9(8.2)   &  64.8(2.3)    &  67.0(1.5)    &  68.6(1.2)     \\ 
  $A_2^{\pi p}$      & 16.11(98)    &  16.7(1.1)    &  16.1(1.0)    &  15.9(1.1)    &  15.8(1.1)     \\ 
  $\AP^{\pi p}$      & 19.84(28)    &  15.921(83)   &  8.27(65)     &  20.8(1.0)    & -0.69(1.2)     \\\hline
  $A_1^{Kp}$         & 32.53(51)    &  28.1(1.4)    &  22.61(49)    &  28.95(47)    &  31.68(54)     \\ 
  $A_2^{Kp}$         & 17.67(88)    &  17.66(98)    &  17.57(92)    &  17.55(96)    &  17.54(97)     \\ 
  $\AP^{Kp}$         & 5.39(18)     &  15.450(48)   &  11.09(37)    &  6.15(64)     &  3.90(78)      \\\hline
  $A_1^{Kn}$         & 30.89(72)    &  23.8(1.8)    &  20.86(64)    &  27.38(71)    &  30.16(72)     \\
  $A_2^{Kn}$         & 9.35(69)     &  9.43(73)     &  9.33(70)     &  9.30(71)     &  9.29(72)      \\ 
  $\AP^{Kn}$         & 5.48(20)     &  15.388(71)   &  11.18(36)    &  6.27(64)     &  4.04(77)      \\\hline 
  $\chi^2/\nu$       & 1.108        &  1.966        &  1.071        &  1.062        &  1.061         \\
  $\nu$              & 533          &  533          &  532          &  532          &  532           \\\hline\hline
  \end{tabular}
\end{table}

\begin{figure}[htb!]
\centering
\includegraphics[scale=0.8]{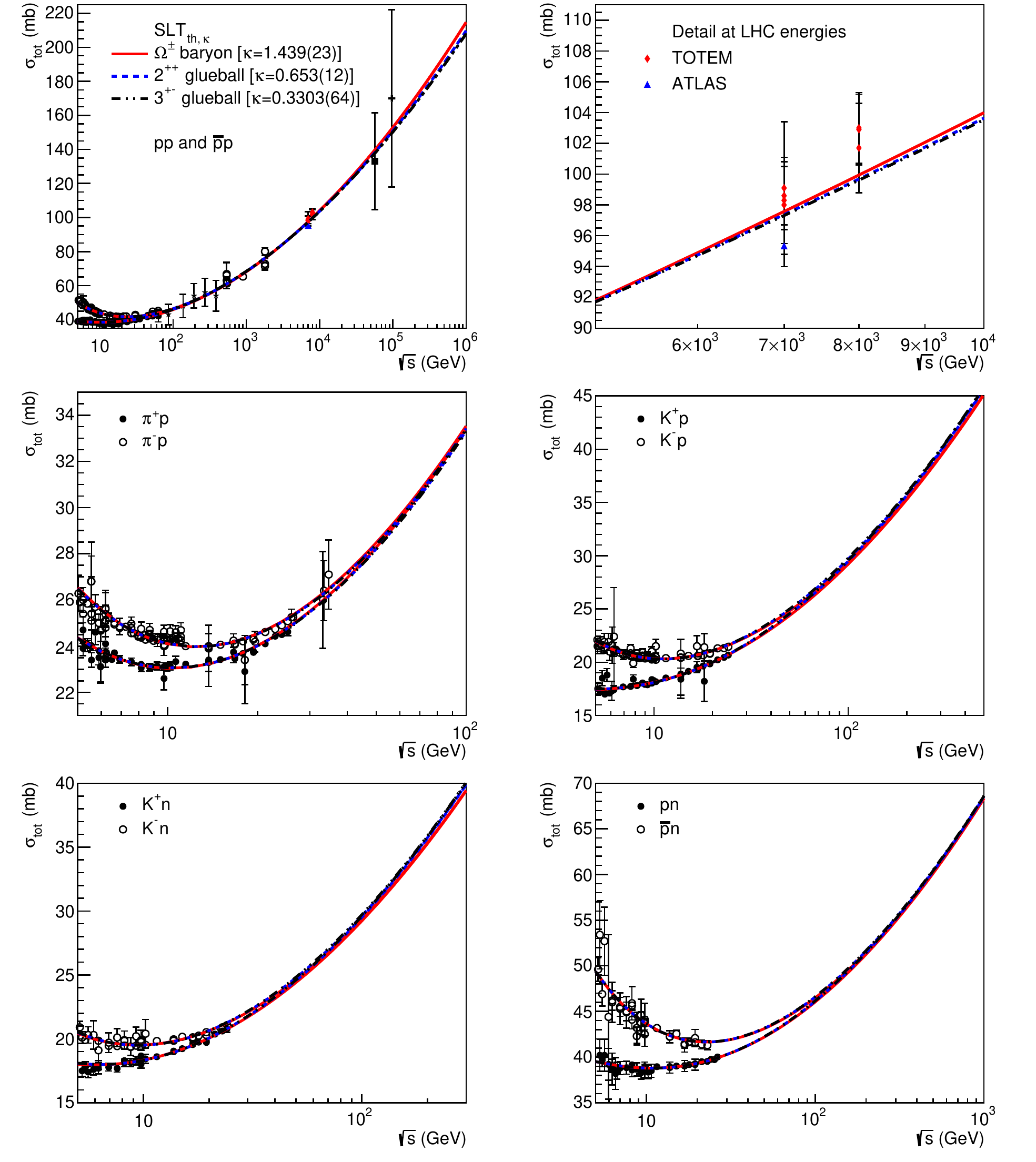}
\caption{\label{fig:res_allreactions_woSmp_SLTkf}Results of fits with
 $\SLTthk$ ($\kappa$ free, $B=B_\text{th}$, $C=C_\text{th}$, $Q=0$) to all
 data for $B_\text{th}$ and $C_\text{th}$ calculated from the
 $\Omega^\pm$ baryon, and the $2^{++}$ and $3^{+-}$ glueball states.
 The legend for the curves is shown in the top-left panel. For the
 legend of $pp$ and $\ppbar$ data see Figure~\ref{fig:res_pp_pbarp}.}
\end{figure}

\subsubsection{c) Fits with $\SLTthk$ ($\kappa$ free, $B=B_\text{th}$, $C=C_\text{th}$)}
\label{subsubsec:slt_res_allreactions_SLTthk}
 
Finally, we discuss the results of the fits with $\kappa$ as a free parameter.
We show in 4th, 5th and 6th columns of Table~\ref{tab:res_allreactions_SLTkf_woSmp} 
all the parameters determined in the fits with $\SLTthk$.
The $\chi^2/\nu$ values are close to 1, indicating that a minimum has been reached.
Furthermore, we get a good description of data in all cases. 
As in the fit with $\SLT$, the subleading term $\AP^{\pi p}$ is positive,
except when we consider the glueball $3^{+-}$ state, which has a negative 
central value and, considering its uncertainty, is compatible with zero.
 
The description of data (Fig.~\ref{fig:res_allreactions_woSmp_SLTkf}) is also similar
to the result of the fit to $pp$ and $\ppbar$ data: in the energy range of the fit, all
cases give the same description, presenting small differences in the extrapolation. 
We may say that all cases are compatible.
 
Concerning the $\kappa$ value, we have $\kappa > 1$ for the $\Omega^{\pm}$ baryon and $\kappa < 1$ 
for the glueball states $2^{++}$ and $3^{+-}$, being the latter smaller than the former, 
as obtained from fits to $pp$ and $\ppbar$ data only,
but with different central values. The connection with ratio $\sigmael/\sigmatot$ 
will be discussed in Section~\ref{subsec:slt_results_ratio_eltot_kappa}.

\begin{table}[htb!]
  \centering
  \caption{\label{tab:res_allreactions_QSLTkf_woSmp}Results of fits with
    $\QLTth$ ($\kappa=1$, $B=B_\text{th}$, $C=0$, $Q$ free),
    $\QSLTth$ ($\kappa=1$, $B=B_\text{th}$, $C=C_\text{th}$, $Q$ free) and
    $\QSLTthk$ ($\kappa$ free, $B=B_\text{th}$, $C=C_\text{th}$, $Q$ free) to
    $\sigmatot$ (all data). The values of
    $B$ and $C$ are fixed to the theoretical values calculated with
    the masses and the spins of the $\Omega^{\pm}$ baryon, the
    $2^{++}$ glueball state and the $3^{+-}$ glueball state (quenched
    values). For the units of measurement of the parameters, see Table~\ref{tab:res_pp_pbarp}.}
  \begin{tabular}{c|ccc|c|c}\hline\hline
                     & \multicolumn{3}{c|}{$\Omega^\pm$ baryon}& {$2^{++}$ glueball} & {$3^{+-}$ glueball}\\\hline
                     & $\QLTth$      & $\QSLTth$     & $\QSLTthk$    & $\QSLTthk$    & $\QSLTthk$    \\\hline
  $B_\text{th}$      &  0.22 (fixed) &  0.22 (fixed) &  0.22 (fixed) &  0.42 (fixed) &  0.78 (fixed) \\
  $C_\text{th}$      &  0 (fixed)    & -0.44 (fixed) & -0.44 (fixed) & -0.42 (fixed) & -0.39 (fixed) \\
  $\kappa$           &  1 (fixed)    &  1 (fixed)    &  1.54(11)     &  0.774(51)    &  0.408(23)    \\
  $b_1$              &  0.331(20)    &  0.307(18)    &  0.565(93)    &  0.576(92)    &  0.582(80)    \\
  $b_2$              &  0.531(11)    &  0.531(11)    &  0.532(11)    &  0.532(11)    &  0.532(11)    \\\hline
  $A_1^{pp}$         &  57.5(1.4)    &  63.0(1.2)    &  62.3(8.7)    &  61.0(8.8)    &  60.4(7.8)    \\ 
  $A_2^{pp}$         &  33.9(1.7)    &  33.9(1.7)    &  34.2(1.8)    &  34.2(1.8)    &  34.2(1.8)    \\ 
  $\AP^{pp}$         &  23.2(1.5)    &  15.8(1.8)    &  37.5(3.3)    &  39.8(3.1)    &  40.9(2.7)    \\
  $Q^{pp}$           &  0.20(12)     &  1.88(13)     & -0.71(46)     & -1.61(47)     & -2.03(42)     \\\hline 
  $A_1^{pn}$         &  32.9(8.8)    &  38.3(9.0)    &  30.4(9.8)    &  29(10)       &  28(10)       \\ 
  $A_2^{pn}$         &  29.7(1.8)    &  29.7(1.8)    &  29.9(1.8)    &  29.9(1.8)    &  29.9(1.8)    \\ 
  $\AP^{pn}$         &  34.7(5.4)    &  28.3(5.0)    &  44.0(3.2)    &  46.1(3.0)    &  47.1(2.7)    \\
  $Q^{pn}$           & -1.11(53)     &  0.49(57)     & -1.57(47)     & -2.45(47)     & -2.86(44)     \\\hline 
  $A_1^{\pi p}$      &  52.7(3.1)    &  57.7(2.6)    &  66.7(1.5)    &  65.6(1.5)    &  65(13)       \\ 
  $A_2^{\pi p}$      &  16.7(1.2)    &  16.7(1.2)    &  16.9(1.2)    &  16.8(1.2)    &  16.8(1.2)    \\ 
  $\AP^{\pi p}$      &  16.6(1.7)    &  9.6(2.0)     &  28.6(2.8)    &  30.8(2.7)    &  31.9(2.3)    \\
  $Q^{\pi p}$        & -1.28(15)     &  0.36(16)     & -1.94(40)     & -2.84(42)     & -3.26(38)     \\\hline
  $A_1^{Kp}$         &  17.0(2.7)    &  21.9(1.2)    &  12.7(3.5)    &  10.4(3.6)    &  9.3(3.6)     \\ 
  $A_2^{Kp}$         &  19.0(1.1)    &  19.0(1.1)    &  19.1(1.1)    &  19.1(1.1)    &  19.1(1.1)    \\ 
  $\AP^{Kp}$         &  15.6(1.4)    &  10.10(99)    &  21.0(1.6)    &  23.0(1.5)    &  24.0(1.3)    \\
  $Q^{Kp}$           & -1.04(15)     &  0.481(91)    & -1.11(27)     & -1.98(30)     & -2.39(27)     \\\hline
  $A_1^{Kn}$         &  8.7(5.1)     &  13.5(5.2)    &  25.6(6.7)    &  0.036(7.1)   & -1.2(6.9)     \\
  $A_2^{Kn}$         &  10.21(81)    &  10.20(80)    &  10.28(81)    &  10.27(81)    &  10.27(81)    \\ 
  $\AP^{Kn}$         &  18.5(2.3)    &  13.4(2.6)    &  22.0(1.6)    &  24.0(1.5)    &  24.9(1.4)    \\
  $Q^{Kn}$           & -1.36(27)     &  0.13(29)     & -1.25(27)     & -2.12(29)     & -2.53(27)     \\\hline 
  $\chi^2/\nu$       &  0.778        &  0.781        &  0.764        &  0.764        &  0.764        \\
  $\nu$              &  528          &  528          &  527          &  527          &  527          \\\hline\hline
  \end{tabular}
\end{table}

\subsubsection{d) Fits with inclusion of the logarithmic term ($\QLT$, $\QSLT$, $\QSLTth$, $\QSLTthk$)}
\label{subsubsec:slt_results_allreactions_log}

Overall, the inclusion of the logarithmic term in the fits to all reactions
decreases the value of $\chi^2/\nu$ from about 1 (for the $\SLT$ fit) to about
0.8 (for the $\QLT$ and $\QSLT$ fits), as can be seen in Table~\ref{tab:res_allreactions_woSmp}.
Of course, some care is needed in judging 
the implications of this change in the $\chi^2/\nu$. Following the approach of
Ref.~\cite{Giordano_Meggiolaro:2014}, the coefficient of the logarithmic term,
differently from $B$ and $C$, is expected to be reaction dependent.
This means the inclusion of only one more free parameter when analysing the
$\pp$/$\ppbar$ case, but when we consider all other reactions we have to
include one free parameter for each pair of crossed channels, therefore
five new free parameters.
Analysing the values of $Q^{ab}$ 
(see Table~\ref{tab:res_allreactions_woSmp}), we see that some of them change
their sign when going from $\QLT$ to $\QSLT$. We understand that
the presence of the subleading term of type $\ln s \cdot \ln\ln s$
creates some competition between the two subleading contributions. 
We recall that the fit $\QSLT$ to only $pp$ and $\ppbar$ data did not converge.

The fit results obtained with $B$ and/or $C$ fixed are shown in Table~\ref{tab:res_allreactions_QSLTkf_woSmp}.
We see that these cases present the same features of the fits to only $pp$ and $\ppbar$ data,
but with a decrease of the $\chi^2/\nu$ values.
When $\kappa$ is considered as a free parameter, we obtain the same scenarios that have been obtained
without the logarithmic term, but now with a slightly larger central value for the
parameter $\kappa$.

\subsection{Asymptotic Results for the Ratio $\sigmael/\sigmatot$ from the Parameter $\kappa$} 
\label{subsec:slt_results_ratio_eltot_kappa}

In Section~\ref{subsec:slt_connection_QCDspectrum}, we showed that the asymptotic ratio 
between the integrated elastic cross section and the total cross section is related
to the parameter $\kappa$ through a simple relation
\begin{equation}
 \frac{\sigmael}{\sigmatot} \sim \frac{\kappa}{2} \quad (s \to \infty).
\end{equation}

Therefore, it is straightforward to obtain these asymptotic values from
the fits discussed in the previous sections. 

In Table~\ref{tab:ratio_eltot_pp_ppbar_allreactions} we summarize all the
resulting values for the ratio $\sigmael/\sigmatot = \kappa/2$ (together with the corresponding
uncertainties, calculated via standard error propagation).
From our discussion on the asymptotic scenarios (Section~\ref{sec:ratioX_asymp_values}),
the fits with the $\Omega^\pm$ baryon values indicate an asymptotic scenario 
in the anti-shadowing regime, since $\sigmael/\sigmatot > 0.5$. 
On the other hand, the values inferred from the fits with $B$ and $C$
fixed to the values obtained from the $2^{++}$ glueball state indicate a
\textit{grey-disk} scenario, with $\sigmael/\sigmatot \sim 0.31 \div 0.38$.  
This value is in agreement with the asymptotic ratio obtained in other studies,
as discussed in Section~\ref{sec:ratioX_comments_semitransparent}.

Using the mass and spin of the $3^{+-}$ glueball state, we also get a
\textit{grey-disk} scenario: however, the resulting asymptotic value is smaller
than the experimental data available so far. 
From Table~\ref{tab:ratio_eltot_pp_ppbar_allreactions}, we see that the value for
this ratio is around $0.16 \div 0.20$, while the experimental value at
the highest energy reached in accelerator experiments (namely 8 TeV) is approximately 0.27.
Since till now the data indicate a rising trend with energy
(see for instance Fig.~\ref{fig:data_X}),
this means that if this scenario is reliable, then the data should
present a local maximum and then decrease as the energy increases
until reaching the asymptotic value. Although there seems to be
no theoretical reason to exclude this type of behavior, it seems
quite unlikely to happen, and we would rather expect a smooth
rise with energy until the asymptotic value
is reached.

\begin{table}[htb!]
 \centering
 \caption{\label{tab:ratio_eltot_pp_ppbar_allreactions}Ratio
  $\sigmael/\sigmatot = \kappa/2$ with $\kappa$ determined
  from the fits $\SLTthk$ and $\QSLTthk$ to $pp$ and $\bar{p}p$ data only,
  and also from fits to data from all reactions.
  Uncertainties are calculated with standard error propagation.}
 \begin{tabular}{c|cc|cc}\hline\hline
 $\sigmael/\sigmatot = \kappa/2$ &\multicolumn{2}{c|}{Fits to $pp/\bar{p}p$ data only} & \multicolumn{2}{c}{Fits to all reactions}\\\hline
 & $\SLTthk$ & $\QSLTthk$ & $\SLTthk$ & $\QSLTthk$ \\\hline
 $\Omega^\pm$ baryon & 0.6885(91)   & 0.765(50)   & 0.720(12)    & 0.770(55)     \\
 $2^{++}$ glueball   & 0.3080(48)   & 0.385(25)   & 0.3265(60)   & 0.387(26)     \\
 $3^{+-}$ glueball   & 0.1548(26)   & 0.203(13)   & 0.1652(32)   & 0.204(12)     \\\hline\hline
\end{tabular}
\end{table}

We can also evaluate the value of the $B$ parameter using the values of $\kappa$ 
obtained from the fits with $B_\text{th}$ and the relation $B=\kappa B_\text{th}$.
These values are shown in Table~\ref{tab:B_SLTthk_QSLTthk_fit_pp_ppbar_allreactions}.
We notice that the inclusion of the logarithmic term ($Q\neq0$) leads to $B$ values
slightly larger than in the case of $Q=0$.
Moreover, the $B$ values associated with the $2^{++}$ glueball and variant $\SLTthk$, namely $\sim 0.25 - 0.27$~mb
are consistent with the PDG 2016 result, $B\sim 0.27$~mb (Eq.~\eqref{eq:result-pdg2016}) and the fit results discussed in Chapter~\ref{chapt:rise_sigmatot}
to ensemble T+A through the FMS-L2 and AU-L$\gamma=2$ models, $\sim 0.25$~mb (Tables~\ref{tab:res_FMS_L2_Lgamma_T_A_TA} and \ref{tab:res_AU_Lgamma2_Lgamma_T_A_TA}).

 \begin{table}[htb!]
 \centering
 \caption{\label{tab:B_SLTthk_QSLTthk_fit_pp_ppbar_allreactions}Value of the coefficient
  $B = \kappa B_\text{th}$ in mb with $\kappa$ determined
  from the fits $\SLTthk$ and $\QSLTthk$ to $pp$ and $\bar{p}p$ data only,
  and also from fits to data for all reactions.
  Uncertainties are calculated with standard error propagation.}
 \begin{tabular}{c|cc|cc}\hline\hline
 $B = \kappa B_\text{th}$ (mb) &\multicolumn{2}{c|}{Fits to $pp/\bar{p}p$ data only} & \multicolumn{2}{c}{Fits to all reactions}\\\hline
 & $\SLTthk$ & $\QSLTthk$ & $\SLTthk$ & $\QSLTthk$ \\\hline
 $\Omega^\pm$ baryon & 0.303(4)     & 0.337(22)   & 0.317(5)     & 0.339(24)     \\
 $2^{++}$ glueball   & 0.259(4)     & 0.324(21)   & 0.274(5)     & 0.325(21)     \\
 $3^{+-}$ glueball   & 0.242(4)     & 0.317(20)   & 0.258(5)     & 0.318(18)     \\\hline\hline
 \end{tabular}
\end{table}

\section{Summary and Conclusions}
\label{sec:slt_conclusions}

In this chapter, we performed a phenomenological analysis of total cross section data from
hadronic scattering in order to estimate the contribution of a subleading term to $\sigmatot$ of type 
$\ln s \cdot \ln\ln s$, recently obtained in a nonperturbative QCD approach \cite{Giordano_Meggiolaro:2014}
(see also Refs.~\cite{Martin_Roy:2014,Nastase_Sonnenschein:2015,Diez_etal:2015} for similar results).
An important feature of the results of Ref.~\cite{Giordano_Meggiolaro:2014} is that the coefficient of this subleading term
together with the coefficient of the leading term ($\sim \ln^2 s$) are \textit{universal} (i.e. independent of colliding particles)
and are connected with the stable spectrum of QCD. Besides this, another subleading term is present in the form $\sim \ln s$.
Contrary to the previous one, this term is reaction-dependent. Precisely, we have considered as parametrization to the total cross section
in the high-energy domain
\begin{equation}
 \sigmatot^{ab}(s)\! \underset{s\to\infty}{\sim}\!
 B \ln^2\left(\frac{s}{s_0^{ab}}\right) +
 C \ln\left(\frac{s}{s_0^{ab}}\right)
 \ln\left[\ln\left(\frac{s}{s_0^{ab}}\right)\right] +
 Q^{ab} \ln\left(\frac{s}{s_0^{ab}}\right) \,,
 \label{eq:sigtot-asymp-concl}
\end{equation}

\noindent where $s_0^{ab}=m_am_b$. As usual, we used Reggeon exchange contributions
to $\sigmatot$ to describe its behavior at low energies. We performed fits to $pp$ and $\ppbar$
scattering with $\sqrt{s}>5$ GeV including the data obtained at the LHC by the TOTEM 
and the ATLAS Collaborations. 
In order to test the universality of both leading and subleading terms, we have also included 
data from other meson-baryon and baryon-baryon scattering.

In general, we have obtained a good description of the data with
parametrizations that generalise the highest-rank result of the
COMPETE Collaboration, by including a $\ln s$ and/or a $\ln s\cdot \ln\ln s$
term besides the $\ln^2 s$ term.
However, with the dataset presently available and as discussed in Section~\ref{sec:slt_fit_results} (see, in particular, the 
results of the fits $\LT$, $\SLT$, $\QLT$ and $\QSLT$
in Tables~\ref{tab:res_pp_pbarp} and \ref{tab:res_allreactions_woSmp} and in
Figure~\ref{fig:res_pp_pbarp}), this type of analysis, in which $B$, $C$ and $Q$ are treated as
free parameters, was not conclusive, because of the ``competition'' between
the two subleading terms $\ln s \cdot \ln \ln s$ and $\ln s$ in the range of energy
considered. The same conclusion was obtained in Ref.~\cite{Diez_etal:2015}, in which
an analysis similar to ours was performed.
In particular, the quality of the fits to $\pp$ and $\ppbar$ data (i.e., the
dataset more sensitive to the high-energy behavior) with only one of
the two subleading terms or without any of them are comparable, so
that we cannot decide which one is the best.
When fitting our parametrizations to all reactions, the value of $\chi^2/\nu$
decreases from about 1, for the $\SLT$ fit, to about 0.8, for the $\QLT$ and $\QSLT$
fits, but, again, no conclusive statement can be made.
In fact, this decrease in the $\chi^2/\nu$ can be attributed to the addition 
of five new free parameters $Q^{ab}$. We expect that further studies,
including more recent results at 2.76 TeV (TOTEM) and 8 TeV (ATLAS),
as well as new measurements at 13 TeV may provide new insights to this topic.

Nonetheless, we took advantage of the fact that the universal coefficients are connected with the QCD spectrum to write
$B=\kappa B_\text{th}$ and $C=\kappa C_\text{th}$, where $B_\text{th}$ and $C_\text{th}$ are the theoretical values
related to the spectrum
and $\kappa$ is connected with the asymptotic value of the ratio $\sigmael/\sigmatot$ (Section~\ref{sec:slt_nonperturbative_approach}).
We then performed fits where $B_\text{th}$ and $C_\text{th}$ are fixed at the values discussed in
Section~\ref{subsec:slt_connection_QCDspectrum}
and $\kappa$ is either fixed at 1 (corresponding to a black disk) or considered a free parameter.

As showed in Section~\ref{subsec:slt_results_ratio_eltot_kappa},
the results obtained with the $\Omega^\pm$ baryon values
indicate an asymptotic scenario in the anti-shadowing regime, 
i.e. $\sigmael/\sigmatot>0.5$, while with the glueball states $2^{++}$ and $3^{+-}$,
the grey-disk scenario was obtained. In the case of the $2^{++}$ glueball state, 
we have $\sigmael/\sigmatot \sim 0.31 \div 0.38$, in agreement with
the results of Chapter~\ref{chapt:ratio_eltot} and other studies~\cite{Fagundes_Menon_Silva:2013a,Menon_Silva:2013a,Menon_Silva:2013b,
Kohara_etal:2014b,Dremin:2014,Dremin:2015a,Dremin:2015b,Desgrolard:1999,Desgrolard:2000}.
On the other hand, with the $3^{+-}$ state the asymptotic result indicates $\sigmael/\sigmatot \sim 0.16 \div 0.20$, 
which is below the experimental values measured at the LHC ($\sim 0.27$ at 8 TeV).

Of course, we cannot claim that this semi-transparent result confirms the results of the other approaches. 
However, it is interesting to note that a similar value for this ratio is obtained from an analysis of only the total
cross section, provided that some theoretical input on $B_\text{th}$ (and $C_\text{th}$) is considered.
Moreover, we recall that the $2^{++}$ glueball state is also a 
``historical'' candidate for the soft (standard) Pomeron trajectory (simple pole),
as commented in Section~\ref{subsec:rise_Regge_simple_pole}.

%% file: ch_conclusions_f.tex
%
%
%
%
\cleardoublepage



\chapter{Conclusions}\label{chapt:conclusions}

In this thesis, 
we were interested in the elastic scattering of hadrons at high-energies,
especially in proton-proton and antiproton-proton scattering. Given the nonperturbative character
of the elastic scattering (small transferred momenta) and the still absence of a full nonperturbative QCD
description, elastic scattering is studied by means of empirical and phenomenological
analyses, some of them presenting a basis on QCD.

The measurements performed at the LHC by the TOTEM and ATLAS Collaborations at the energies of 7 and 8~TeV
for $pp$ scattering were very important to test the models constructed in the last decades, as well as
to refine these models and their predictions for higher energies.

Here we have presented studies on physical quantities related to the elastic scattering
with focus on cross-sections, namely the ratio $\sigmael/\sigmatot$,
and the energy dependence of $\sigmatot$, which is related to the elastic scattering amplitude
by the optical theorem and the $\rho$ parameter.
Along the text, we have discussed an empirical analysis of the ratio 
$\sigmael/\sigmatot$~\cite{Fagundes_Menon_Silva:2016a}, a phenomenological study of the
rise of the total cross section based on the Regge-Gribov 
formalism~\cite{Fagundes_Menon_Silva:2017a,Fagundes_Menon_Silva:2017b}, and finally, the study of the 
subleading contributions to $\sigmatot$
based on recent asymptotic results obtained in a nonperturbative QCD approach \cite{Giordano_Meggiolaro_Silva:2017}. 
In what follows, we highlight the main results of each chapter.

The empirical analysis of the ratio $X=\sigmael/\sigmatot$ was presented in Chapter~\ref{chapt:ratio_eltot}.
We have considered four empirical
parametrizations for fits to the available experimental data on that ratio, including those obtained at the LHC. 
The asymptotic value of the ratio $X$ is represented by the $A$ parameter. 
In what concerns this parameter, two types of fits were performed.
In the \textit{constrained} fit, $A$ was fixed at some chosen values inspired by empirical,
phenomenological, and theoretical results. 
From these fits, all statistically equivalent, we were able to conclude that the black-disk scenario 
($A=0.5$) is not the only possible solution considering the available data. In the other type of fit, named \textit{unconstrained}, 
we let $A$ be a free parameter. The results are compatible with a semi-transparent scenario,
with average value $A_g=0.30\pm0.12$. In this case, the proton may behave asymptotically as a grey disk.

We stress that this novel result (first obtained in Ref.~\cite{Fagundes_Menon_Silva:2013a}) contrasts with
the usual phenomenological picture predicting an asymptotic black disk and is in agreement with the results obtained
in other subsequent analyses.

In Chapter~\ref{chapt:rise_sigmatot}, we presented a phenomenological study on the energy dependence of the total cross section
and the $\rho$ parameter. The parametrizations considered
are based on the Regge-Gribov formalism~\cite{Collins_book:1977,Barone_Predazzi_book:2002,Donnachie_etal_book:2002}.
Two leading terms were considered: the $\ln^2 s$ term (that corresponds to a triple-pole Pomeron contribution and was denoted by L2 model)
and the $\ln^\gamma s$ term (introduced by Amaldi \etal \cite{Amaldi:1977}, denoted L$\gamma$ model), where $\gamma$ is a real free parameter.
Given the tension between the 8 TeV $\sigmatot$ measurements by TOTEM and ATLAS, we have considered
three ensembles of data (besides the data above 5~GeV and below the LHC energies):
one in which only TOTEM data were included, another one in which only ATLAS data were included and a third one that comprises
all data available (TOTEM + ATLAS). We performed fits to $\sigmatot$ and $\rho$ data and we explored two analytical approaches to connect the real and imaginary parts of the elastic scattering amplitude: 
Derivative Dispersion Relations (DDR) with the introduction of the
effective subtraction constant (called FMS approach) and Asymptotic Uniqueness (AU approach).

We noticed that the analytic result obtained for L$\gamma$ models in the FMS approach can be
understood as a generalization of the L2 model for a real exponent of the $\ln s$. 
The same is not true for the AU approach, in which oscillatory energy-dependent 
factors appear in front of the $\ln^\gamma$ term. The presence of such factors is difficult to interpret within Regge-Gribov formalisms. 
Furthermore, the approach which was used seems to affect the value obtained for the $\gamma$ parameter. 
In the FMS approach, we have $\gamma \gtrsim 2 $ while AU leads to values of $\gamma \lesssim 2$. 
Taking the two possible scenarios for the rise of $\sigmatot$ showed by TOTEM and ATLAS data, we can infer a range of possible values of $\gamma$ taking into account 
the estimated parameter uncertainties: $1.9\lesssim \gamma \lesssim 2.4$ in FMS approach and $1.7\lesssim \gamma \lesssim 2.2$ in AU approach.
At last, we recall that when considering TOTEM and ATLAS data separately, TOTEM data indicates a faster rise than ATLAS data.
We stress that in all known analysis with $\gamma$ as a free fit parameter and data reduction to both $\sigmatot$
and $\rho$ data the results do not favor $\gamma=2$, but real values (above or below 2). If that is really the case,
the associated singularity cannot be a triple pole. In a mathematical context, 
we have shown that a branch point can be associated with a real (not-integer) $\gamma$ value.

Finally, in Chapter~\ref{chapt:SLT_sigmatot}, we studied the subleading contributions to $\sigmatot$. 
The study is based on the recent results by Giordano and Meggiolaro~\cite{Giordano_Meggiolaro:2014}, 
in which a nonperturbative approach for the elastic scattering is employed. 
The asymptotic result predicts that $\sigmatot$ has a leading term $\ln^2 s$ with universal coefficient $B$
and a subleading term $\ln s \ln\ln s$, also with a universal coefficient $C$. 
In turn, these two coefficients are related to the spectrum of QCD. Another subleading term 
$\ln s$ is also infered, however, its coefficient $Q$ is reaction dependent. 
We developed several fits to experimental data on $pp$ and $\ppbar$ scattering and also to data
from meson-baryon and other baryon-baryon scattering, in order to estimate the contribution of these two subleading terms. 
We identified a competition between the two subleading terms that does not allow us
to state which is dominant one at the present energies.
We also explored the presence of an extra parameter $\kappa$ to write $B =\kappa B_\text{th}$
and $C=\kappa C_\text{th}$, where $B_\text{th}$ and $C_\text{th}$ are the values given by the QCD spectrum.
In this case, we were able to infer the asymptotic scenario for the colliding particles, 
since $\kappa/2$ corresponds to the ratio $X=\sigmael/\sigmatot$.
We tested three possible theoretical inputs for $B_\text{th}$ and $C_\text{th}$. In one of them,
related to the $2^{++}$ glueball state, we get a grey disk scenario with $X\sim 0.3$, 
in accordance with the results presented in Chapter~\ref{chapt:ratio_eltot} and quoted analyses.
We recall that the $2^{++}$ glueball is a historical candidate for the soft
(simple-pole) Pomeron trajectory (see Fig.~\ref{fig:pomeron_trajectory}).

We understand that the results obtained here provided new insights on the study of the hadronic cross sections at high and asymptotic
energies. Despite the present tension between the TOTEM and ATLAS results, the expected new reanalysis of data from Run 1 and
new data from Run 2 at 13~TeV can shed some light on the subject, allowing further developments along the lines presented and discussed in this study.
Moreover, new experimental data from other soft scattering states (single and double dissociation),
and the study of these processes
can certainly contribute to a better understanding of the QCD soft sector.

%% file: app_ddr_f.tex
%
\cleardoublepage



\chapter{From Integral to Derivative Dispersion Relations}\label{app:DDR}

In this appendix, we show how the Derivative Dispersion Relations (DDR) used in Chapter~\ref{chapt:rise_sigmatot}
can be obtained from the Integral Dispersion Relations (IDR).

Following Ref.~\cite{Avila_Menon:2004}, let us consider the IDR for the even amplitude, given by Eq.~\eqref{eq:IDR-even},
\begin{equation}
 \Real F_{+}(s) = K + \frac{2s^2}{\pi}\,P\int_{s_\text{th}}^{\infty} ds' \left[\frac{1}{s'^2-s^2}\right] \frac{\Imag F_{+}(s')}{s'},
\end{equation}

\noindent where $K$ is the subtraction constant and $P$ denotes the Cauchy principal value.

Making the change of variable $s' = e^{\xi'}$, $s = e^{\xi}$ and defining $g(\xi')\equiv \Imag F_+(e^{\xi'})/\xi'$,
the above equation reads
\begin{align}
 \Real F_+ (e^{\xi}) - K  & = \frac{2e^{2\xi}}{\pi} P\int_{\ln s_0}^{\infty} \frac{g(e^{\xi'})e^{\xi'}}{e^{2\xi'}-e^{2\xi}}d\xi'\nonumber\\
  & = \frac{e^{\xi}}{\pi} P\int_{\ln s_0}^{\infty} \frac{g(e^{\xi'})}{\sinh(\xi'-\xi)}d\xi'.\label{eq:IDR-even-xi}
\end{align}

If $g$ is an analytical function of its argument, we can write $g(\xi')$ as a power series of $\xi'$
\begin{equation}
 g(\xi') = \sum_{n=0}^\infty \frac{g^{(n)}(\xi)}{n!}(\xi'-\xi)^n
\end{equation}

\noindent and calculate the integral in Eq.~\eqref{eq:IDR-even-xi} term by term.
Moreover, we consider the high-energy limit (see Sect.~\ref{subsec:rise_IDR}),
which consists in taking the limit $s_0\to 0$, i.e. $\ln s_0 \to -\infty$. Therefore,
\begin{equation}
 \Real F_+ (e^{\xi}) - K = \frac{e^\xi}{\pi}\sum_{n=0}^\infty \frac{g^{(n)}(\xi)}{n!} P\int_{-\infty}^\infty \frac{(\xi'-\xi)^n}{\sinh(\xi'-\xi)}d\xi'.
\end{equation}

Considering a new change of variable $x=\xi'-\xi$, we can rewrite the above equation
\begin{equation}
 \Real F_+ (e^{\xi}) - K = e^\xi\sum_{n=0}^\infty \frac{g^{(n)}(\xi)}{n!} I_n,
\end{equation}

\noindent where
\begin{equation}
 I_n = \frac{1}{\pi}\int_{-\infty}^{\infty} \frac{x^n}{\sinh x} dx.
\end{equation}

For $n$ even, we have $I_n=0$. For $n$ odd, we calculate this integral with the auxiliary integral
\begin{equation}
 J(a) = \frac{1}{\pi}P\int_{-\infty}^{\infty} \frac{e^{ax}}{\sinh x}dx = \tan\left(\frac{a\pi}{2}\right),
\end{equation}

\noindent such that
\begin{equation}
 I_n = \frac{d^n}{da^n} J(a)\bigg\rvert_{a=0}.
\end{equation}

Therefore,
\begin{align}
 \Real F_+ (e^{\xi}) - K & = e^\xi\sum_{n=0}^\infty \frac{g^{(n)}(\xi)}{n!} \frac{d^n}{da^n}\tan\left(\frac{\pi a}{2}\right)\bigg\rvert_{a=0}\nonumber\\
 & = e^{\xi}\tan\left(\frac{\pi}{2}\frac{d}{d\xi}\right)g(\xi), 
\end{align}

\noindent where the series expansion is implicit in the tangent operator. 
Finally, we can rewrite this equation in the variable $s$ as
\begin{equation}
 \frac{\Real F_+(s)}{s} = \frac{K}{s} + \tan\left(\frac{\pi}{2}\frac{d}{d\ln s}\right)\frac{\Imag F_+(s)}{s}.
\end{equation}

A similar procedure can be done for the odd amplitude, Eq.~\eqref{eq:IDR-odd}, and we obtain
\begin{equation}
 \frac{\Real F_-(s)}{s} = \tan\left[\frac{\pi}{2}\left(1+\frac{d}{d\ln s}\right)\right] \frac{\Imag F_-(s)}{s}.
\end{equation}

For practical use, it is convenient to consider the series expansion around the origin of the trigonometric operators. 
Moreover, for the odd case, we have to consider the substitution
\begin{equation}
 \tan\left[\frac{\pi}{2}\left(1+\frac{d}{d\ln s}\right)\right] \to -\cot\left[\frac{\pi}{2}\frac{d}{d\ln s}\right].\nonumber
\end{equation}

In this way, the series expansions read
\begin{align}
\frac{\Real F_{+}(s)}{s} & = \frac{K}{s} + \left[ \frac{\pi}{2} \frac{d}{d\ln s} + \frac{1}{3} \left(\frac{\pi}{2}\frac{d}{d \ln s}\right)^3 +
\frac{2}{15} \left(\frac{\pi}{2}\frac{d}{d \ln s}\right)^5 + \dots \right] \frac{\Imag F_{+}(s)}{s}, \nonumber\\[10pt]
\frac{\Real F_{-}(s)}{s}  &= - \int \left\{\frac{d}{d\ln s}\left[\cot \left( \frac{\pi}{2}\frac{d}{d\ln s} \right)\right]\frac{\Imag F_{-}(s)}{s} \right\} d\ln s \nonumber \\
                          &= - \frac{2}{\pi}\int \left\{ \left[ 1 - \frac{1}{3} \left(\frac{\pi}{2}\frac{d}{d \ln s}\right)^2 \frac{1}{45} \left(\frac{\pi}{2}\frac{d}{d \ln s}\right)^4
                             - \dots \right] \frac{\Imag F_{-}(s)}{s} \right\} \, d \ln s. \nonumber
\end{align}

\noindent which correspond to relations introduced by Kang and Nicolescu \cite{Kang_Nicolescu:1975},
Eqs.~\eqref{eq:DDR-KN-even} and \eqref{eq:DDR-KN-odd}.

%% file: app_mellin_f.tex
%
\cleardoublepage



\chapter{The Mellin Transform: Singularities and Asymptotics}\label{app:mellin}

In this appendix, we discuss the singularity that may be associated
with the L$\gamma$ law by means of the Mellin transform (see, for example, Ref.~\cite{Forshaw:1997}, Sect. 2.8).
This subject is treated in more detail in Ref.~\cite{CapelasdeOliveira:2017}.

The Mellin transform connects a function $f(x)$ defined on the positive real axis
$0<x<\infty$ to a function $F(z)$ defined on the complex plane
throught the relation \cite{Bertrand:2000,Oosthuizen:2013}
\begin{equation}
 F(z) = {\mathcal{M}}[f(x);z] = \int_0^\infty f(x)x^{z-1}dx.
 \label{eq:Mellin-def}
\end{equation}

We call $F(z)$ the Mellin transform of $f(x)$.
The above integral exists in the so-called \textit{strip of definition} $a_1 < \Real[z] < a_2$, 
where the real numbers $a_1$ and $a_2$ depend on the function $f(x)$.

The inverse is given by
\begin{equation}
 f(x) = \frac{1}{2\pi i}\int_{a-i\infty}^{a+i\infty} F(z) x^{-z} dz,
 \label{eq:Mellin-inverse}
\end{equation}

\noindent where $a$ is some $\Real[z]$ inside strip of definition.

An interesting result concerning the Mellin transform, 
which is our interest here, is the relation between the asymptotic behavior of $f(x)$ and
the associated singularity in the complex plane (see, for example, Ref.~\cite{Oosthuizen:2013}). 
This can be attained calculating the Mellin transform of $f$.

Before applying it to our case, let us first recall a property of the Mellin transform.
If we make $z \to -z$ in Eq.~\eqref{eq:Mellin-def}, we get
\begin{equation}
 F(-z) = \mathcal{M}[f(1/x);z].
\end{equation}

Defining $F(-z) \equiv G(\ell)$, where $\ell$ is the complex angular momentum, and by changing the variable $x = 1/s$, we get
\begin{equation}
 G(\ell) = \int_0^\infty s^{-\ell -1}f(1/s)ds.
\end{equation}

Following Ref.~\cite{Forshaw:1997}, we are interested in the singularity 
associated with the $(s/s_0)^\alpha \ln^\gamma(s/s_0)$ contribution to the total cross section, with $s_0=4m_p^2$
and $\gamma$ a real (not integer) number. As discussed in Section~\ref{subsec:rise_further_comments_Lgamma}, 
this function is defined for $s\geq s_0$. Denoting $\tilde{s}=s/s_0$, we can write
\begin{equation}
 f(1/\tilde{s}) = \theta(\tilde{s}-1)\tilde{s}^{\alpha}\ln^\gamma \tilde{s},
 \label{eq:Mellin-f-deff}
\end{equation}

\noindent there $\theta$ is the Heavside function.

The Mellin transform of this function is given by
\begin{align}
 G(\ell) & = \int_0^\infty \tilde{s}^{-\ell -1}\theta(\tilde{s}-1)\tilde{s}^{\alpha}\ln^\gamma \tilde{s}\,d\tilde{s}\nonumber\\
         & = \int_1^\infty \tilde{s}^{\alpha -\ell -1}\ln^\gamma \tilde{s}\,d\tilde{s}.\nonumber
\end{align}

With the change of variable $\tilde{s} = 1/\xi$, we obtain
\begin{equation}
 G(\ell) = \int_0^1 \xi^{(\ell-\alpha)  -1}\ln^\gamma \left(\frac{1}{\xi}\right)\,d\xi.\nonumber
\end{equation}

From Ref.~\cite{Gradshteyn_etal_book:1980}, page 550, formula 4.272.6, we obtain
\begin{equation}
 G(\ell) = \frac{\Gamma(\gamma+1)}{(\ell - \alpha)^{\gamma+1}},
 \label{eq:Mellin-Lgamma}
\end{equation}

\noindent for $\gamma>-1$, $\Real[\ell-\alpha]>0$ and where $\Gamma$ is the Euler gamma function.

Therefore, according to Eq.~\eqref{eq:Mellin-Lgamma}, for $\gamma$ real (not integer), the function defined in Eq.~\eqref{eq:Mellin-f-deff} is associated 
to a branch point at $\ell=\alpha$. When $\gamma = n$, with $n>0 $ an integer, we have a pole of order $n+1$.

%% file: app_publications_f.tex
%
%
%
\cleardoublepage



\chapter{Related Publications}\label{app:publications}

In this appendix, we list the publications related to the development of this research program.
The publications are organised in those directly related to the thesis, those
produced in the period, but do not correspond to the main part of the thesis, and the
proceedings publications.

\section{Main publications}

\begin{itemize}
  
  \item ``\textbf{Exploring central opacity and asymptotic scenarios in elastic hadron scattering}''\\
   D.A. Fagundes, M.J Menon and P.V.R.G. Silva\\
   Nucl. Phys. A \textbf{946} (2016) 194-226; arXiv:1509.04108 [hep-ph]\\
   Related to Chapter~\ref{chapt:ratio_eltot}.

  \item ``\textbf{Bounds on the rise of total cross section from LHC7 and LHC8 data}''\\
   D.A. Fagundes, M.J. Menon and P.V.R.G. Silva\\
   Nucl. Phys. A \textbf{966} (2017) 185-196; arXiv:1703.07486 [hep-ph]\\
   Related to Chapter~\ref{chapt:rise_sigmatot}.

  \item ``\textbf{Leading components in forward elastic hadron scattering: Derivative dispersion relations and asymptotic uniqueness}''\\
   D.A. Fagundes, M.J. Menon and P.V.R.G. Silva\\
   Int. J. Mod. Phys. A \textbf{32} (2017) no.32, 1750184;  arXiv:1705.01504 [hep-ph]\\
   Related to Chapter~\ref{chapt:rise_sigmatot}.
 
  \item ``\textbf{Investigation of the leading and subleading high-energy behavior of hadron-hadron total 
   cross sections using a best-fit analysis of hadronic scattering data}''\\
   M. Giordano, E. Meggiolaro and P.V.R.G. Silva\\
   Phys. Rev. D \textbf{96} (2017) no.3, 034015; arXiv:1703.00244 [hep-ph]\\
   Related to Chapter~\ref{chapt:SLT_sigmatot}.

\end{itemize}

\section{Other Publications}

\begin{itemize}
 
 \item ``\textbf{Fine structure of the diffraction cone: from ISR to the LHC}''\\
   D.A. Fagundes, L. Jenkovszky, E.Q. Miranda, G. Pancheri, P.V.R.G. Silva\\
   Int. J. Mod. Phys. A \textbf{31} (2016) no.28\&29, 1645022\\
   Also appearing in: Gribov-85 Memorial Volume: Exploring Quantum Field Theory: pp. 180-194 (World Scientific, 2016); arXiv:1509.02197 [hep-ph]\\
   Proceedings of Gribov-85 Memorial Workshop on Theoretical Physics of XXI Century: Chernogolovka, Russia, June 7-20, 2015.
   
 \item ``\textbf{Analytic components for the hadronic total cross-section: Fractional calculus and Mellin transform}''\\
   E. Capelas de Oliveira, M.J. Menon and P.V.R.G. Silva\\
   arXiv:1708.01255 [hep-ph] 
   
\end{itemize}

\section{Proceedings}

\begin{itemize}
 
 \item ``\textbf{Asymptotic scenarios for the proton’s central opacity: An empirical study}''\\
   D.A. Fagundes, M.J Menon and P.V.R.G. Silva\\
   AIP Conf. Proc. \textbf{1654} (2015) 050005; arXiv:1410.4423 [hep-ph]\\
   Proceedings of 8th International Workshop on Diffraction in High Energy Physics (Diffraction 2014): Primo\v{s}ten, Croatia, September 10-16, 2014.
  
 \item ``\textbf{Updating an empirical analysis on the proton’s central opacity and asymptotia}''\\
   D.A. Fagundes, M.J Menon and P.V.R.G. Silva\\
   J. Phys. Conf. Ser. \textbf{706} (2016) no.5, 052027; arXiv:1505.01233 [hep-ph]\\
   Proceedings of 13th International Workshop on Hadron Physics: Angra dos Reis, Rio de Janeiro, Brazil, March 22-27, 2015.
   
 \end{itemize}

%% file: PhD_PVRGSilva_f.bbl
\begin{thebibliography}{100}
\expandafter\ifx\csname url\endcsname\relax
  \def\url#1{\texttt{#1}}\fi
\expandafter\ifx\csname urlprefix\endcsname\relax\def\urlprefix{URL }\fi
\providecommand{\bibinfo}[2]{#2}
\providecommand{\eprint}[2][]{\url{#2}}

\bibitem{Halzen_Martin_book:1984}
\bibinfo{author}{Halzen, F.} \& \bibinfo{author}{Martin, A.~D.}
\newblock \emph{\bibinfo{title}{{Quarks and Leptons: An Introductory Course in
  Modern Particle Physics}}} (\bibinfo{publisher}{Wiley}, \bibinfo{address}{New
  York, USA}, \bibinfo{year}{1984}).

\bibitem{Giordano_Meggiolaro:2014}
\bibinfo{author}{Giordano, M.} \& \bibinfo{author}{Meggiolaro, E.}
\newblock \bibinfo{title}{{Hadronic total cross sections at high energy and the
  QCD spectrum}}.
\newblock \emph{\bibinfo{journal}{JHEP}} \textbf{\bibinfo{volume}{03}},
  \bibinfo{pages}{002} (\bibinfo{year}{2014}).
\newblock \eprint{1311.3133}.

\bibitem{Donnachie:1979}
\bibinfo{author}{Donnachie, A.} \& \bibinfo{author}{Landshoff, P.~V.}
\newblock \bibinfo{title}{{Elastic Scattering at Large $t$}}.
\newblock \emph{\bibinfo{journal}{Z. Phys.}} \textbf{\bibinfo{volume}{C2}},
  \bibinfo{pages}{55} (\bibinfo{year}{1979}).
\newblock \bibinfo{note}{[Erratum: Z. Phys.C2,372(1979)]}.

\bibitem{TOTEM:2011a}
\bibinfo{author}{Antchev, G.} \emph{et~al.}
\newblock \bibinfo{title}{{Proton-proton elastic scattering at the LHC energy
  of $\sqrt{s} = 7$~TeV}}.
\newblock \emph{\bibinfo{journal}{Europhys. Lett.}}
  \textbf{\bibinfo{volume}{95}}, \bibinfo{pages}{41001} (\bibinfo{year}{2011}).
\newblock \eprint{1110.1385}.

\bibitem{Block:2011b}
\bibinfo{author}{Block, M.~M.} \& \bibinfo{author}{Halzen, F.}
\newblock \bibinfo{title}{{Forward hadronic scattering at 7 TeV: Predictions
  for the LHC: An Update}}.
\newblock \emph{\bibinfo{journal}{Phys. Rev.}} \textbf{\bibinfo{volume}{D83}},
  \bibinfo{pages}{077901} (\bibinfo{year}{2011}).
\newblock \eprint{1102.3163}.

\bibitem{Bourrely:2003}
\bibinfo{author}{Bourrely, C.}, \bibinfo{author}{Soffer, J.} \&
  \bibinfo{author}{Wu, T.~T.}
\newblock \bibinfo{title}{{Impact picture phenomenology for $\pi^\pm p$, $K^\pm
  p$ and $p p$, $\bar{p} p$ elastic scattering at high-energies}}.
\newblock \emph{\bibinfo{journal}{Eur. Phys. J.}}
  \textbf{\bibinfo{volume}{C28}}, \bibinfo{pages}{97--105}
  (\bibinfo{year}{2003}).
\newblock \eprint{hep-ph/0210264}.

\bibitem{Islam:2009}
\bibinfo{author}{Islam, M.~M.}, \bibinfo{author}{Kaspar, J.} \&
  \bibinfo{author}{Luddy, R.~J.}
\newblock \bibinfo{title}{{Deep-Elastic $pp$ Scattering at LHC from Low-$x$
  Gluons}}.
\newblock \emph{\bibinfo{journal}{Mod. Phys. Lett.}}
  \textbf{\bibinfo{volume}{A24}}, \bibinfo{pages}{485--496}
  (\bibinfo{year}{2009}).
\newblock \eprint{0804.0455}.

\bibitem{Jenkovszky:2011}
\bibinfo{author}{Jenkovszky, L.}, \bibinfo{author}{Kuprash, O.},
  \bibinfo{author}{Lamsa, J.} \& \bibinfo{author}{Orava, R.}
\newblock \bibinfo{title}{{Low-Mass Diffraction at the LHC}}.
\newblock \emph{\bibinfo{journal}{Mod. Phys. Lett.}}
  \textbf{\bibinfo{volume}{A26}}, \bibinfo{pages}{2029--2037}
  (\bibinfo{year}{2011}).
\newblock \eprint{1106.3299}.

\bibitem{Petrov:2003}
\bibinfo{author}{Petrov, V.~A.}, \bibinfo{author}{Predazzi, E.} \&
  \bibinfo{author}{Prokudin, A.}
\newblock \bibinfo{title}{{Coulomb interference in high-energy $pp$ and
  $\bar{p}p$ scattering}}.
\newblock \emph{\bibinfo{journal}{Eur. Phys. J.}}
  \textbf{\bibinfo{volume}{C28}}, \bibinfo{pages}{525--533}
  (\bibinfo{year}{2003}).
\newblock \eprint{hep-ph/0206012}.

\bibitem{TOTEM:2011b}
\bibinfo{author}{Antchev, G.} \emph{et~al.}
\newblock \bibinfo{title}{{First measurement of the total proton-proton cross
  section at the LHC energy of $\sqrt{s}$ = 7~TeV}}.
\newblock \emph{\bibinfo{journal}{Europhys. Lett.}}
  \textbf{\bibinfo{volume}{96}}, \bibinfo{pages}{21002} (\bibinfo{year}{2011}).
\newblock \eprint{1110.1395}.

\bibitem{TOTEM:2013a}
\bibinfo{author}{Antchev, G.} \emph{et~al.}
\newblock \bibinfo{title}{{Measurement of proton-proton elastic scattering and
  total cross-section at $\sqrt{s}$ = 7~TeV}}.
\newblock \emph{\bibinfo{journal}{Europhys. Lett.}}
  \textbf{\bibinfo{volume}{101}}, \bibinfo{pages}{21002}
  (\bibinfo{year}{2013}).

\bibitem{TOTEM:2013c}
\bibinfo{author}{Antchev, G.} \emph{et~al.}
\newblock \bibinfo{title}{{Luminosity-independent measurements of total,
  elastic and inelastic cross-sections at $\sqrt{s} = 7$~TeV}}.
\newblock \emph{\bibinfo{journal}{Europhys. Lett.}}
  \textbf{\bibinfo{volume}{101}}, \bibinfo{pages}{21004}
  (\bibinfo{year}{2013}).

\bibitem{TOTEM:2013d}
\bibinfo{author}{Antchev, G.} \emph{et~al.}
\newblock \bibinfo{title}{{Luminosity-Independent Measurement of the
  Proton-Proton Total Cross Section at $\sqrt{s}=8$ TeV}}.
\newblock \emph{\bibinfo{journal}{Phys. Rev. Lett.}}
  \textbf{\bibinfo{volume}{111}}, \bibinfo{pages}{012001}
  (\bibinfo{year}{2013}).

\bibitem{TOTEM:2016}
\bibinfo{author}{Antchev, G.} \emph{et~al.}
\newblock \bibinfo{title}{{Measurement of elastic $pp$ scattering at
  $\sqrt{\hbox {$s$}} = \hbox {8}$ TeV in the Coulomb-nuclear interference
  region: determination of the $\mathbf {\rho }$-parameter and the total
  cross-section}}.
\newblock \emph{\bibinfo{journal}{Eur. Phys. J.}}
  \textbf{\bibinfo{volume}{C76}}, \bibinfo{pages}{661} (\bibinfo{year}{2016}).
\newblock \eprint{1610.00603}.

\bibitem{TOTEM_Kaspar:2017}
\bibinfo{author}{Ka\v{s}par, J.}
\newblock \bibinfo{title}{{Soft diffraction at the LHC}}.
\newblock \emph{\bibinfo{journal}{Presented at XLVII International Symposium on
  Multiparticle Dynamics (ISMD 2017)}}  (\bibinfo{year}{2017}).
\newblock
  \urlprefix\url{http://indico.nucleares.unam.mx/event/1180/session/20/contribution/50}.

\bibitem{ATLAS:2014}
\bibinfo{author}{Aad, G.} \emph{et~al.}
\newblock \bibinfo{title}{{Measurement of the total cross section from elastic
  scattering in $pp$ collisions at $\sqrt{s}=7$ TeV with the ATLAS detector}}.
\newblock \emph{\bibinfo{journal}{Nucl. Phys. B}}
  \textbf{\bibinfo{volume}{889}}, \bibinfo{pages}{486--548}
  (\bibinfo{year}{2014}).
\newblock \eprint{1408.5778}.

\bibitem{ATLAS:2016}
\bibinfo{author}{Aaboud, M.} \emph{et~al.}
\newblock \bibinfo{title}{{Measurement of the total cross section from elastic
  scattering in $pp$ collisions at $\sqrt{s}=8$ TeV with the ATLAS detector}}.
\newblock \emph{\bibinfo{journal}{Phys. Lett. B}}
  \textbf{\bibinfo{volume}{761}}, \bibinfo{pages}{158--178}
  (\bibinfo{year}{2016}).
\newblock \eprint{1607.06605}.

\bibitem{Barone_Predazzi_book:2002}
\bibinfo{author}{Barone, V.} \& \bibinfo{author}{Predazzi, E.}
\newblock \emph{\bibinfo{title}{{High-Energy Particle Diffraction}}}, vol.
  \bibinfo{volume}{565} of \emph{\bibinfo{series}{Texts and Monographs in
  Physics}} (\bibinfo{publisher}{Springer-Verlag}, \bibinfo{address}{Berlin
  Heidelberg}, \bibinfo{year}{2002}).
\newblock
  \urlprefix\url{http://www-spires.fnal.gov/spires/find/books/www?cl=QC794.6.C6B37::2002}.

\bibitem{Dremin:2014}
\bibinfo{author}{Dremin, I.~M.}
\newblock \bibinfo{title}{{Critical regime of proton elastic scattering at the
  LHC}}.
\newblock \emph{\bibinfo{journal}{JETP Lett.}} \textbf{\bibinfo{volume}{99}},
  \bibinfo{pages}{243--245} (\bibinfo{year}{2014}).
\newblock \eprint{1401.3106}.

\bibitem{Dremin:2015a}
\bibinfo{author}{Dremin, I.~M.}
\newblock \bibinfo{title}{{Torus or black disk?}}
\newblock \emph{\bibinfo{journal}{Bull. Lebedev Phys. Inst.}}
  \textbf{\bibinfo{volume}{42}}, \bibinfo{pages}{21--25}
  (\bibinfo{year}{2015}).
\newblock \bibinfo{note}{[Kratk. Soobshch. Fiz.42,no.1,8(2015)]},
  \eprint{1404.4142}.

\bibitem{Dremin:2015b}
\bibinfo{author}{Dremin, I.~M.}
\newblock \bibinfo{title}{{Interaction region of high energy protons}}.
\newblock \emph{\bibinfo{journal}{Phys. Usp.}} \textbf{\bibinfo{volume}{58}},
  \bibinfo{pages}{61--70} (\bibinfo{year}{2015}).
\newblock \eprint{1406.2153}.

\bibitem{Froissart:1961}
\bibinfo{author}{Froissart, M.}
\newblock \bibinfo{title}{{Asymptotic behavior and subtractions in the
  Mandelstam representation}}.
\newblock \emph{\bibinfo{journal}{Phys. Rev.}} \textbf{\bibinfo{volume}{123}},
  \bibinfo{pages}{1053--1057} (\bibinfo{year}{1961}).

\bibitem{Martin:1963}
\bibinfo{author}{Martin, A.}
\newblock \bibinfo{title}{{Unitarity and high-energy behavior of scattering
  amplitudes}}.
\newblock \emph{\bibinfo{journal}{Phys. Rev.}} \textbf{\bibinfo{volume}{129}},
  \bibinfo{pages}{1432--1436} (\bibinfo{year}{1963}).

\bibitem{Martin:1965}
\bibinfo{author}{Martin, A.}
\newblock \bibinfo{title}{{Extension of the axiomatic analyticity domain of
  scattering amplitudes by unitarity. 1.}}
\newblock \emph{\bibinfo{journal}{Nuovo Cim. A}} \textbf{\bibinfo{volume}{42}},
  \bibinfo{pages}{930--953} (\bibinfo{year}{1965}).

\bibitem{Lukaszuk_Martin:1967}
\bibinfo{author}{Lukaszuk, L.} \& \bibinfo{author}{Martin, A.}
\newblock \bibinfo{title}{{Absolute upper bounds for $\pi \pi$ scattering}}.
\newblock \emph{\bibinfo{journal}{Nuovo Cim. A}} \textbf{\bibinfo{volume}{52}},
  \bibinfo{pages}{122--145} (\bibinfo{year}{1967}).

\bibitem{Eden:1971}
\bibinfo{author}{Eden, R.~J.}
\newblock \bibinfo{title}{{Theorems on high energy collisions of elementary
  particles}}.
\newblock \emph{\bibinfo{journal}{Rev. Mod. Phys.}}
  \textbf{\bibinfo{volume}{43}}, \bibinfo{pages}{15--35}
  (\bibinfo{year}{1971}).

\bibitem{Azimov:2011}
\bibinfo{author}{Azimov, {\relax Ya}.}
\newblock \bibinfo{title}{{How Robust is the Froissart Bound?}}
\newblock \emph{\bibinfo{journal}{Phys. Rev.}} \textbf{\bibinfo{volume}{D84}},
  \bibinfo{pages}{056012} (\bibinfo{year}{2011}).
\newblock \eprint{1104.5314}.

\bibitem{Azimov:2012a}
\bibinfo{author}{Azimov, {\relax Ya}.~I.}
\newblock \bibinfo{title}{{Froissart Bounds for Amplitudes and Cross Sections
  at High Energies}}  (\bibinfo{year}{2012}).
\newblock \eprint{1204.0984}.

\bibitem{Azimov:2012b}
\bibinfo{author}{Azimov, {\relax Ya}.}
\newblock \bibinfo{title}{{What is the real meaning of the Froissart theorem?}}
\newblock In \emph{\bibinfo{booktitle}{{Published in the Proceedings of the
  International Workshop HSQCD 2012, eds. V.T.Kim and L.N.Lipatov (PNPI,
  Gatchina, 2012) p.22}}} (\bibinfo{year}{2012}).
\newblock
  \urlprefix\url{http://inspirehep.net/record/1128463/files/arXiv:1208.4304.pdf}.
\newblock \eprint{1208.4304}.

\bibitem{Martin:2009}
\bibinfo{author}{Martin, A.}
\newblock \bibinfo{title}{{The Froissart bound for inelastic cross-sections}}.
\newblock \emph{\bibinfo{journal}{Phys. Rev. D}} \textbf{\bibinfo{volume}{80}},
  \bibinfo{pages}{065013} (\bibinfo{year}{2009}).
\newblock \eprint{0904.3724}.

\bibitem{PDG_data_website}
\bibinfo{howpublished}{\url{http://pdg.lbl.gov/2015/hadronic-xsections/hadron.html}}.

\bibitem{COMPETE:2002b}
\bibinfo{author}{Cudell, J.~R.} \emph{et~al.}
\newblock \bibinfo{title}{{Benchmarks for the forward observables at RHIC, the
  Tevatron Run II and the LHC}}.
\newblock \emph{\bibinfo{journal}{Phys. Rev. Lett.}}
  \textbf{\bibinfo{volume}{89}}, \bibinfo{pages}{201801}
  (\bibinfo{year}{2002}).
\newblock \eprint{hep-ph/0206172}.

\bibitem{PDG:2016}
\bibinfo{author}{Patrignani, C.} \emph{et~al.}
\newblock \bibinfo{title}{{Review of Particle Physics}}.
\newblock \emph{\bibinfo{journal}{Chin. Phys. C}}
  \textbf{\bibinfo{volume}{40}}, \bibinfo{pages}{100001}
  (\bibinfo{year}{2016}).

\bibitem{TOTEM:2015}
\bibinfo{author}{Antchev, G.} \emph{et~al.}
\newblock \bibinfo{title}{{Evidence for non-exponential elastic proton-proton
  differential cross-section at low $|t|$ and $\sqrt{s}$ = 8~TeV by TOTEM}}.
\newblock \emph{\bibinfo{journal}{Nucl. Phys. B}}
  \textbf{\bibinfo{volume}{899}}, \bibinfo{pages}{527--546}
  (\bibinfo{year}{2015}).
\newblock \eprint{1503.08111}.

\bibitem{Fagundes_etal:2015c}
\bibinfo{author}{Fagundes, D.~A.}, \bibinfo{author}{Jenkovszky, L.},
  \bibinfo{author}{Miranda, E.~Q.}, \bibinfo{author}{Pancheri, G.} \&
  \bibinfo{author}{Silva, P. V. R.~G.}
\newblock \bibinfo{title}{{Fine structure of the diffraction cone: from ISR to
  the LHC}}.
\newblock In \emph{\bibinfo{booktitle}{{Gribov-85 Memorial Workshop on
  Theoretical Physics of XXI Century Chernogolovka, Russia, June 17-20, 2015}}}
  (\bibinfo{year}{2015}).
\newblock
  \urlprefix\url{http://inspirehep.net/record/1392451/files/arXiv:1509.02197.pdf}.
\newblock \eprint{1509.02197}.

\bibitem{PDG:2014}
\bibinfo{author}{Olive, K.~A.} \emph{et~al.}
\newblock \bibinfo{title}{{Review of Particle Physics}}.
\newblock \emph{\bibinfo{journal}{Chin. Phys. C}}
  \textbf{\bibinfo{volume}{38}}, \bibinfo{pages}{090001}
  (\bibinfo{year}{2014}).

\bibitem{ARGO_YBJ:2009}
\bibinfo{author}{Aielli, G.} \emph{et~al.}
\newblock \bibinfo{title}{{Proton-air cross section measurement with the
  ARGO-YBJ cosmic ray experiment}}.
\newblock \emph{\bibinfo{journal}{Phys. Rev. D}} \textbf{\bibinfo{volume}{80}},
  \bibinfo{pages}{092004} (\bibinfo{year}{2009}).
\newblock \eprint{0904.4198}.

\bibitem{PierreAuger:2012}
\bibinfo{author}{Abreu, P.} \emph{et~al.}
\newblock \bibinfo{title}{{Measurement of the proton-air cross-section at
  $\sqrt{s}=57$ TeV with the Pierre Auger Observatory}}.
\newblock \emph{\bibinfo{journal}{Phys. Rev. Lett.}}
  \textbf{\bibinfo{volume}{109}}, \bibinfo{pages}{062002}
  (\bibinfo{year}{2012}).
\newblock \eprint{1208.1520}.

\bibitem{TA:2015}
\bibinfo{author}{Abbasi, R.~U.} \emph{et~al.}
\newblock \bibinfo{title}{{Measurement of the proton-air cross section with
  Telescope Array's Middle Drum detector and surface array in hybrid mode}}.
\newblock \emph{\bibinfo{journal}{Phys. Rev.}} \textbf{\bibinfo{volume}{D92}},
  \bibinfo{pages}{032007} (\bibinfo{year}{2015}).
\newblock \eprint{1505.01860}.

\bibitem{ROOT_website}
\bibinfo{howpublished}{\url{http://root.cern.ch/}}.

\bibitem{Bevington_Robinson_book:1992}
\bibinfo{author}{Bevington, P.} \& \bibinfo{author}{Robinson, D.}
\newblock \emph{\bibinfo{title}{{Data Reduction and Error Analysis for the
  Physical Sciences}}} (\bibinfo{publisher}{McGraw-Hill},
  \bibinfo{address}{Massachusetts}, \bibinfo{year}{1992}).

\bibitem{MINUIT_Manual:1994}
\bibinfo{author}{James, F.}
\newblock \emph{\bibinfo{title}{{MINUIT Function Minimization and Error
  Analysis: Reference Manual Version 94.1}}} (\bibinfo{year}{1994}).

\bibitem{James:1975}
\bibinfo{author}{James, F.} \& \bibinfo{author}{Roos, M.}
\newblock \bibinfo{title}{{Minuit: A System for Function Minimization and
  Analysis of the Parameter Errors and Correlations}}.
\newblock \emph{\bibinfo{journal}{Comput. Phys. Commun.}}
  \textbf{\bibinfo{volume}{10}}, \bibinfo{pages}{343--367}
  (\bibinfo{year}{1975}).

\bibitem{Fagundes_Menon_Silva:2013a}
\bibinfo{author}{Fagundes, D.~A.}, \bibinfo{author}{Menon, M.~J.} \&
  \bibinfo{author}{Silva, P. V. R.~G.}
\newblock \bibinfo{title}{{On the rise of the proton-proton cross-sections at
  high energies}}.
\newblock \emph{\bibinfo{journal}{J. Phys. G}} \textbf{\bibinfo{volume}{40}},
  \bibinfo{pages}{065005} (\bibinfo{year}{2013}).
\newblock \eprint{1208.3456}.

\bibitem{Menon_Silva:2013a}
\bibinfo{author}{Menon, M.~J.} \& \bibinfo{author}{Silva, P. V. R.~G.}
\newblock \bibinfo{title}{{An updated analysis on the rise of the hadronic
  total cross-section at the LHC energy region}}.
\newblock \emph{\bibinfo{journal}{Int. J. Mod. Phys. A}}
  \textbf{\bibinfo{volume}{28}}, \bibinfo{pages}{1350099}
  (\bibinfo{year}{2013}).
\newblock \eprint{1212.5096}.

\bibitem{Menon_Silva:2013b}
\bibinfo{author}{Menon, M.~J.} \& \bibinfo{author}{Silva, P. V. R.~G.}
\newblock \bibinfo{title}{{A study on analytic parametrizations for
  proton–proton cross-sections and asymptotia}}.
\newblock \emph{\bibinfo{journal}{J. Phys. G}} \textbf{\bibinfo{volume}{40}},
  \bibinfo{pages}{125001} (\bibinfo{year}{2013}).
\newblock \bibinfo{note}{[Erratum: J. Phys. G {\bf 41}, 019501(2014)]},
  \eprint{1305.2947}.

\bibitem{Fagundes_Menon_Silva:2017b}
\bibinfo{author}{Fagundes, D.~A.}, \bibinfo{author}{Menon, M.~J.} \&
  \bibinfo{author}{Silva, P. V. R.~G.}
\newblock \bibinfo{title}{{Leading components in forward elastic hadron
  scattering: Derivative dispersion relations and asymptotic uniqueness}}.
\newblock \emph{\bibinfo{journal}{Int. J. Mod. Phys.}}
  \textbf{\bibinfo{volume}{A32}}, \bibinfo{pages}{1750184}
  (\bibinfo{year}{2017}).
\newblock \eprint{1705.01504}.

\bibitem{Fagundes_Menon_Silva:2015}
\bibinfo{author}{Fagundes, D.~A.}, \bibinfo{author}{Menon, M.~J.} \&
  \bibinfo{author}{Silva, P. V. R.~G.}
\newblock \bibinfo{title}{{Asymptotic scenarios for the proton’s central
  opacity: An empirical study}}.
\newblock \emph{\bibinfo{journal}{AIP Conf. Proc.}}
  \textbf{\bibinfo{volume}{1654}}, \bibinfo{pages}{050005}
  (\bibinfo{year}{2015}).
\newblock \eprint{1410.4423}.

\bibitem{Fagundes_Menon_Silva:2016a}
\bibinfo{author}{Fagundes, D.~A.}, \bibinfo{author}{Menon, M.~J.} \&
  \bibinfo{author}{Silva, P. V. R.~G.}
\newblock \bibinfo{title}{{Exploring central opacity and asymptotic scenarios
  in elastic hadron scattering}}.
\newblock \emph{\bibinfo{journal}{Nucl. Phys. A}}
  \textbf{\bibinfo{volume}{946}}, \bibinfo{pages}{194--226}
  (\bibinfo{year}{2016}).
\newblock \eprint{1509.04108}.

\bibitem{Fagundes_Menon_Silva:2016b}
\bibinfo{author}{Fagundes, D.~A.}, \bibinfo{author}{Menon, M.~J.} \&
  \bibinfo{author}{Silva, P. V. R.~G.}
\newblock \bibinfo{title}{{Updating an empirical analysis on the proton’s
  central opacity and asymptotia}}.
\newblock \emph{\bibinfo{journal}{J. Phys. Conf. Ser.}}
  \textbf{\bibinfo{volume}{706}}, \bibinfo{pages}{052027}
  (\bibinfo{year}{2016}).
\newblock \eprint{1505.01233}.

\bibitem{Bellandi:1991}
\bibinfo{author}{Bellandi, J.} \emph{et~al.}
\newblock \bibinfo{title}{{On the behavior of inelasticity at very
  high-energy}}.
\newblock \emph{\bibinfo{journal}{Phys. Lett.}}
  \textbf{\bibinfo{volume}{B262}}, \bibinfo{pages}{102--104}
  (\bibinfo{year}{1991}).

\bibitem{ForwardLHC:2016}
\bibinfo{author}{Akiba, K.} \emph{et~al.}
\newblock \bibinfo{title}{{LHC Forward Physics}}.
\newblock \emph{\bibinfo{journal}{J. Phys.}} \textbf{\bibinfo{volume}{G43}},
  \bibinfo{pages}{110201} (\bibinfo{year}{2016}).
\newblock \eprint{1611.05079}.

\bibitem{Ulrich_etal:2009}
\bibinfo{author}{Ulrich, R.}, \bibinfo{author}{Engel, R.},
  \bibinfo{author}{Muller, S.}, \bibinfo{author}{Schussler, F.} \&
  \bibinfo{author}{Unger, M.}
\newblock \bibinfo{title}{{Proton-Air Cross Section and Extensive Air
  Showers}}.
\newblock \emph{\bibinfo{journal}{Nucl. Phys. Proc. Suppl.}}
  \textbf{\bibinfo{volume}{196}}, \bibinfo{pages}{335--340}
  (\bibinfo{year}{2009}).
\newblock \eprint{0906.3075}.

\bibitem{Engel:1998}
\bibinfo{author}{Engel, R.}, \bibinfo{author}{Gaisser, T.~K.},
  \bibinfo{author}{Lipari, P.} \& \bibinfo{author}{Stanev, T.}
\newblock \bibinfo{title}{{Proton proton cross-section at $\sqrt{s}$ similar to
  30 TeV}}.
\newblock \emph{\bibinfo{journal}{Phys. Rev.}} \textbf{\bibinfo{volume}{D58}},
  \bibinfo{pages}{014019} (\bibinfo{year}{1998}).
\newblock \eprint{hep-ph/9802384}.

\bibitem{Engel:2000}
\bibinfo{author}{Engel, R.}
\newblock \bibinfo{title}{{Total cross-sections and diffraction}}.
\newblock \emph{\bibinfo{journal}{Nucl. Phys. Proc. Suppl.}}
  \textbf{\bibinfo{volume}{82}}, \bibinfo{pages}{221--231}
  (\bibinfo{year}{2000}).

\bibitem{Chou:1969}
\bibinfo{author}{Chou, T.~T.} \& \bibinfo{author}{Yang, C.-N.}
\newblock \bibinfo{title}{{Model of Elastic High-Energy Scattering}}.
\newblock \emph{\bibinfo{journal}{Phys. Rev.}} \textbf{\bibinfo{volume}{170}},
  \bibinfo{pages}{1591--1596} (\bibinfo{year}{1968}).

\bibitem{Bourrely:1984}
\bibinfo{author}{Bourrely, C.}, \bibinfo{author}{Soffer, J.} \&
  \bibinfo{author}{Wu, T.~T.}
\newblock \bibinfo{title}{{Impact Picture Expectations for Very High-Energy
  Elastic $p p$ and $p \bar{p}$ Scattering}}.
\newblock \emph{\bibinfo{journal}{Nucl. Phys.}}
  \textbf{\bibinfo{volume}{B247}}, \bibinfo{pages}{15--28}
  (\bibinfo{year}{1984}).

\bibitem{Bourrely:1988}
\bibinfo{author}{Bourrely, C.}, \bibinfo{author}{Soffer, J.} \&
  \bibinfo{author}{Wu, T.~T.}
\newblock \bibinfo{title}{{Impact Picture Predictions for $\bar{p}p$ and $pp$
  Elastic Scattering at {CERN} Collider, Fnal Collider, {LHC} and {SSC}}}.
\newblock \emph{\bibinfo{journal}{Z. Phys.}} \textbf{\bibinfo{volume}{C37}},
  \bibinfo{pages}{369--375} (\bibinfo{year}{1988}).

\bibitem{Bourrely:2011}
\bibinfo{author}{Bourrely, C.}, \bibinfo{author}{Soffer, J.} \&
  \bibinfo{author}{Wu, T.~T.}
\newblock \bibinfo{title}{{Determination of the forward slope in $p~p$ and
  $\bar p~p$ elastic scattering up to LHC energy}}.
\newblock \emph{\bibinfo{journal}{Eur. Phys. J.}}
  \textbf{\bibinfo{volume}{C71}}, \bibinfo{pages}{1601} (\bibinfo{year}{2011}).
\newblock \eprint{1011.1756}.

\bibitem{Block:2012}
\bibinfo{author}{Block, M.~M.} \& \bibinfo{author}{Halzen, F.}
\newblock \bibinfo{title}{{New experimental evidence that the proton develops
  asymptotically into a black disk}}.
\newblock \emph{\bibinfo{journal}{Phys. Rev.}} \textbf{\bibinfo{volume}{D86}},
  \bibinfo{pages}{051504} (\bibinfo{year}{2012}).
\newblock \eprint{1208.4086}.

\bibitem{Khoze:2015a}
\bibinfo{author}{Khoze, V.~A.}, \bibinfo{author}{Martin, A.~D.} \&
  \bibinfo{author}{Ryskin, M.~G.}
\newblock \bibinfo{title}{{Elastic scattering and Diffractive dissociation in
  the light of LHC data}}.
\newblock \emph{\bibinfo{journal}{Int. J. Mod. Phys.}}
  \textbf{\bibinfo{volume}{A30}}, \bibinfo{pages}{1542004}
  (\bibinfo{year}{2015}).
\newblock \eprint{1402.2778}.

\bibitem{Gotsman:2015a}
\bibinfo{author}{Gotsman, E.}, \bibinfo{author}{Levin, E.} \&
  \bibinfo{author}{Maor, U.}
\newblock \bibinfo{title}{{A comprehensive model of soft interactions in the
  LHC era}}.
\newblock \emph{\bibinfo{journal}{Int. J. Mod. Phys.}}
  \textbf{\bibinfo{volume}{A30}}, \bibinfo{pages}{1542005}
  (\bibinfo{year}{2015}).
\newblock \eprint{1403.4531}.

\bibitem{Gotsman:2015b}
\bibinfo{author}{Gotsman, E.}, \bibinfo{author}{Levin, E.} \&
  \bibinfo{author}{Maor, U.}
\newblock \bibinfo{title}{{CGC/saturation approach for soft interactions at
  high energy: a two channel model}}.
\newblock \emph{\bibinfo{journal}{Eur. Phys. J.}}
  \textbf{\bibinfo{volume}{C75}}, \bibinfo{pages}{179} (\bibinfo{year}{2015}).
\newblock \eprint{1502.05202}.

\bibitem{Block:2015}
\bibinfo{author}{Block, M.~M.}, \bibinfo{author}{Durand, L.},
  \bibinfo{author}{Ha, P.} \& \bibinfo{author}{Halzen, F.}
\newblock \bibinfo{title}{{Eikonal fit to $pp$ and $\bar{p}p$ scattering and
  the edge in the scattering amplitude}}.
\newblock \emph{\bibinfo{journal}{Phys. Rev.}} \textbf{\bibinfo{volume}{D92}},
  \bibinfo{pages}{014030} (\bibinfo{year}{2015}).
\newblock \eprint{1505.04842}.

\bibitem{Nemes:2015}
\bibinfo{author}{Nemes, F.}, \bibinfo{author}{Csörgő, T.} \&
  \bibinfo{author}{Csanád, M.}
\newblock \bibinfo{title}{{Excitation function of elastic $pp$ scattering from
  a unitarily extended Bialas–Bzdak model}}.
\newblock \emph{\bibinfo{journal}{Int. J. Mod. Phys.}}
  \textbf{\bibinfo{volume}{A30}}, \bibinfo{pages}{1550076}
  (\bibinfo{year}{2015}).
\newblock \eprint{1505.01415}.

\bibitem{Selyugin:2014}
\bibinfo{author}{Selyugin, O.~V.}
\newblock \bibinfo{title}{{Total cross sections and $\rho$ at high energy}}.
\newblock \emph{\bibinfo{journal}{Nucl. Phys.}}
  \textbf{\bibinfo{volume}{A922}}, \bibinfo{pages}{180--190}
  (\bibinfo{year}{2014}).
\newblock \eprint{1312.1271}.

\bibitem{Grau_etal:2012}
\bibinfo{author}{Grau, A.}, \bibinfo{author}{Pacetti, S.},
  \bibinfo{author}{Pancheri, G.} \& \bibinfo{author}{Srivastava, Y.~N.}
\newblock \bibinfo{title}{{Checks of Asymptotia in $pp$ Elastic Scattering at
  LHC}}.
\newblock \emph{\bibinfo{journal}{Phys. Lett.}}
  \textbf{\bibinfo{volume}{B714}}, \bibinfo{pages}{70--75}
  (\bibinfo{year}{2012}).
\newblock \eprint{1206.1076}.

\bibitem{Wibig:2011}
\bibinfo{author}{Wibig, T.}
\newblock \bibinfo{title}{{Elastic scattering at 7 TeV and high energy cross
  section for cosmic ray studies}}.
\newblock \emph{\bibinfo{journal}{J. Phys.}} \textbf{\bibinfo{volume}{G39}},
  \bibinfo{pages}{085003} (\bibinfo{year}{2012}).
\newblock \eprint{1111.0441}.

\bibitem{Fagundes:2011}
\bibinfo{author}{Fagundes, D.~A.}, \bibinfo{author}{Luna, E. G.~S.},
  \bibinfo{author}{Menon, M.~J.} \& \bibinfo{author}{Natale, A.~A.}
\newblock \bibinfo{title}{{Aspects of a Dynamical Gluon Mass Approach to
  Elastic Hadron Scattering at LHC}}.
\newblock \emph{\bibinfo{journal}{Nucl. Phys.}}
  \textbf{\bibinfo{volume}{A886}}, \bibinfo{pages}{48--70}
  (\bibinfo{year}{2012}).
\newblock \eprint{1112.4680}.

\bibitem{Wu:2011}
\bibinfo{author}{Wu, T.~T.}, \bibinfo{author}{Martin, A.},
  \bibinfo{author}{Roy, S.~M.} \& \bibinfo{author}{Singh, V.}
\newblock \bibinfo{title}{{An upper bound on the total inelastic cross-section
  as a function of the total cross-section}}.
\newblock \emph{\bibinfo{journal}{Phys. Rev.}} \textbf{\bibinfo{volume}{D84}},
  \bibinfo{pages}{025012} (\bibinfo{year}{2011}).
\newblock \eprint{1011.1349}.

\bibitem{Dremin:2013}
\bibinfo{author}{Dremin, I.~M.}
\newblock \bibinfo{title}{{Elastic scattering of hadrons}}.
\newblock \emph{\bibinfo{journal}{Phys. Usp.}} \textbf{\bibinfo{volume}{56}},
  \bibinfo{pages}{3--28} (\bibinfo{year}{2013}).
\newblock \bibinfo{note}{[Usp. Fiz. Nauk183,3(2013)]}, \eprint{1206.5474}.

\bibitem{Cartiglia:2013}
\bibinfo{author}{Cartiglia, N.}
\newblock \bibinfo{title}{{Measurement of the proton-proton total, elastic,
  inelastic and diffractive cross sections at 2, 7, 8 and 57 TeV}}
  (\bibinfo{year}{2013}).
\newblock \eprint{1305.6131}.

\bibitem{Block:2011}
\bibinfo{author}{Block, M.~M.} \& \bibinfo{author}{Halzen, F.}
\newblock \bibinfo{title}{{Experimental Confirmation that the Proton is
  Asymptotically a Black Disk}}.
\newblock \emph{\bibinfo{journal}{Phys. Rev. Lett.}}
  \textbf{\bibinfo{volume}{107}}, \bibinfo{pages}{212002}
  (\bibinfo{year}{2011}).
\newblock \eprint{1109.2041}.

\bibitem{Troshin:2012}
\bibinfo{author}{Troshin, S.~M.} \& \bibinfo{author}{Tyurin, N.~E.}
\newblock \bibinfo{title}{{Deep--elastic scattering and asymptotics}}.
\newblock \emph{\bibinfo{journal}{Phys. Lett.}}
  \textbf{\bibinfo{volume}{B707}}, \bibinfo{pages}{558--561}
  (\bibinfo{year}{2012}).
\newblock \eprint{1111.4454}.

\bibitem{Troshin_Tyurin:1993}
\bibinfo{author}{Troshin, S.~M.} \& \bibinfo{author}{Tyurin, N.~E.}
\newblock \bibinfo{title}{{Beyond the black disk limit}}.
\newblock \emph{\bibinfo{journal}{Phys. Lett. B}}
  \textbf{\bibinfo{volume}{316}}, \bibinfo{pages}{175--177}
  (\bibinfo{year}{1993}).
\newblock \eprint{hep-ph/9307250}.

\bibitem{Troshin_Tyurin:2007}
\bibinfo{author}{Troshin, S.~M.} \& \bibinfo{author}{Tyurin, N.~E.}
\newblock \bibinfo{title}{{Reflective scattering from unitarity saturation}}.
\newblock \emph{\bibinfo{journal}{Int. J. Mod. Phys. A}}
  \textbf{\bibinfo{volume}{22}}, \bibinfo{pages}{4437--4449}
  (\bibinfo{year}{2007}).
\newblock \eprint{hep-ph/0701241}.

\bibitem{Alkin:2014}
\bibinfo{author}{Alkin, A.}, \bibinfo{author}{Martynov, E.},
  \bibinfo{author}{Kovalenko, O.} \& \bibinfo{author}{Troshin, S.~M.}
\newblock \bibinfo{title}{{Impact-parameter analysis of TOTEM data at the LHC:
  Black disk limit exceeded}}.
\newblock \emph{\bibinfo{journal}{Phys. Rev.}} \textbf{\bibinfo{volume}{D89}},
  \bibinfo{pages}{091501} (\bibinfo{year}{2014}).
\newblock \eprint{1403.8036}.

\bibitem{Menon_etal_sig:1994}
\bibinfo{author}{Menon, A.}, \bibinfo{author}{Mehrotra, K.},
  \bibinfo{author}{Mohan, C.} \& \bibinfo{author}{Ranka, S.}
\newblock \bibinfo{title}{{Characterization of a Class of Sigmoid Functions
  with Applications to Neural Networks}}.
\newblock \emph{\bibinfo{journal}{Electrical Engineering and Computer Science
  Technical Reports.}} \bibinfo{pages}{Paper 152} (\bibinfo{year}{1994}).
\newblock \urlprefix\url{http://surface.syr.edu/eecs_techreports/152/}.

\bibitem{Kucharavy_deGuio:2011}
\bibinfo{author}{Kucharavy, D.} \& \bibinfo{author}{Guio, R.~D.}
\newblock \bibinfo{title}{Application of s-shaped curves}.
\newblock \emph{\bibinfo{journal}{Procedia Engineering}}
  \textbf{\bibinfo{volume}{9}}, \bibinfo{pages}{559 -- 572}
  (\bibinfo{year}{2011}).
\newblock
  \urlprefix\url{http://www.sciencedirect.com/science/article/pii/S1877705811001597}.

\bibitem{Fagundes:2012}
\bibinfo{author}{Fagundes, D.~A.} \& \bibinfo{author}{Menon, M.~J.}
\newblock \bibinfo{title}{{Total Hadronic Cross Section and the Elastic Slope:
  An Almost Model-Independent Connection}}.
\newblock \emph{\bibinfo{journal}{Nucl. Phys.}}
  \textbf{\bibinfo{volume}{A880}}, \bibinfo{pages}{1--11}
  (\bibinfo{year}{2012}).
\newblock \eprint{1112.5115}.

\bibitem{Fagundes:2013}
\bibinfo{author}{Fagundes, D.~A.} \& \bibinfo{author}{Menon, M.~J.}
\newblock \bibinfo{title}{{Hadronic Cross Sections, Elastic Slope and Physical
  Bounds}}.
\newblock \emph{\bibinfo{journal}{AIP Conf. Proc.}}
  \textbf{\bibinfo{volume}{1520}}, \bibinfo{pages}{297--299}
  (\bibinfo{year}{2013}).
\newblock \eprint{1208.0510}.

\bibitem{Pumplin:1973}
\bibinfo{author}{Pumplin, J.}
\newblock \bibinfo{title}{{Eikonal models for diffraction dissociation on
  nuclei}}.
\newblock \emph{\bibinfo{journal}{Phys. Rev.}} \textbf{\bibinfo{volume}{D8}},
  \bibinfo{pages}{2899--2903} (\bibinfo{year}{1973}).

\bibitem{Pumplin:1982}
\bibinfo{author}{Pumplin, J.}
\newblock \bibinfo{title}{{Diffractive Processes}}.
\newblock \emph{\bibinfo{journal}{Phys. Scripta}}
  \textbf{\bibinfo{volume}{25}}, \bibinfo{pages}{191--197}
  (\bibinfo{year}{1982}).

\bibitem{Lipari:2013}
\bibinfo{author}{Lipari, P.} \& \bibinfo{author}{Lusignoli, M.}
\newblock \bibinfo{title}{{Interpretation of the measurements of total, elastic
  and diffractive cross sections at LHC}}.
\newblock \emph{\bibinfo{journal}{Eur. Phys. J.}}
  \textbf{\bibinfo{volume}{C73}}, \bibinfo{pages}{2630} (\bibinfo{year}{2013}).
\newblock \eprint{1305.7216}.

\bibitem{ALICE:2013}
\bibinfo{author}{Abelev, B.} \emph{et~al.}
\newblock \bibinfo{title}{{Measurement of inelastic, single- and
  double-diffraction cross sections in proton--proton collisions at the LHC
  with ALICE}}.
\newblock \emph{\bibinfo{journal}{Eur. Phys. J.}}
  \textbf{\bibinfo{volume}{C73}}, \bibinfo{pages}{2456} (\bibinfo{year}{2013}).
\newblock \eprint{1208.4968}.

\bibitem{CMS:2013}
\bibinfo{author}{Chatrchyan, S.} \emph{et~al.}
\newblock \bibinfo{title}{{Measurement of the inelastic proton-proton cross
  section at $\sqrt{s}=7$ TeV}}.
\newblock \emph{\bibinfo{journal}{Phys. Lett.}}
  \textbf{\bibinfo{volume}{B722}}, \bibinfo{pages}{5--27}
  (\bibinfo{year}{2013}).
\newblock \eprint{1210.6718}.

\bibitem{CMS:2015}
\bibinfo{author}{Khachatryan, V.} \emph{et~al.}
\newblock \bibinfo{title}{{Measurement of diffraction dissociation cross
  sections in $pp$ collisions at $\sqrt{s}$ = 7 TeV}}.
\newblock \emph{\bibinfo{journal}{Phys. Rev.}} \textbf{\bibinfo{volume}{D92}},
  \bibinfo{pages}{012003} (\bibinfo{year}{2015}).
\newblock \eprint{1503.08689}.

\bibitem{UA5:1986a}
\bibinfo{author}{Alner, G.~J.} \emph{et~al.}
\newblock \bibinfo{title}{{Antiproton-proton cross sections at 200 and 900 GeV
  c.m. energy}}.
\newblock \emph{\bibinfo{journal}{Z. Phys.}} \textbf{\bibinfo{volume}{C32}},
  \bibinfo{pages}{153--161} (\bibinfo{year}{1986}).

\bibitem{UA5:1986b}
\bibinfo{author}{Ansorge, R.~E.} \emph{et~al.}
\newblock \bibinfo{title}{{Diffraction Dissociation at the {CERN} Pulsed
  Collider at {CM} Energies of 900-{GeV} and 200-{GeV}}}.
\newblock \emph{\bibinfo{journal}{Z. Phys.}} \textbf{\bibinfo{volume}{C33}},
  \bibinfo{pages}{175} (\bibinfo{year}{1986}).

\bibitem{CDF:2001}
\bibinfo{author}{Affolder, T.} \emph{et~al.}
\newblock \bibinfo{title}{{Double diffraction dissociation at the Fermilab
  Tevatron collider}}.
\newblock \emph{\bibinfo{journal}{Phys. Rev. Lett.}}
  \textbf{\bibinfo{volume}{87}}, \bibinfo{pages}{141802}
  (\bibinfo{year}{2001}).
\newblock \eprint{hep-ex/0107070}.

\bibitem{CDF:1994b}
\bibinfo{author}{Abe, F.} \emph{et~al.}
\newblock \bibinfo{title}{{Measurement of $\bar{p}p$ single diffraction
  dissociation at $\sqrt{s} = 546$ GeV and 1800 GeV}}.
\newblock \emph{\bibinfo{journal}{Phys. Rev.}} \textbf{\bibinfo{volume}{D50}},
  \bibinfo{pages}{5535--5549} (\bibinfo{year}{1994}).

\bibitem{CDF:1994a}
\bibinfo{author}{Abe, F.} \emph{et~al.}
\newblock \bibinfo{title}{{Measurement of the $\bar{p}p$ total cross-section at
  $\sqrt{s} = 546$ GeV and 1800 GeV}}.
\newblock \emph{\bibinfo{journal}{Phys. Rev. D}} \textbf{\bibinfo{volume}{50}},
  \bibinfo{pages}{5550--5561} (\bibinfo{year}{1994}).

\bibitem{Sunday_etal:2012}
\bibinfo{author}{Sunday, J.}, \bibinfo{author}{James, A.~A.},
  \bibinfo{author}{Ibijola, E.~A.}, \bibinfo{author}{Ogunrinde, R.~B.} \&
  \bibinfo{author}{Ogunyebi, S.~N.}
\newblock \bibinfo{title}{{A Computational Approach to Verhulst-Pearl Model}}.
\newblock \emph{\bibinfo{journal}{IOSR Journal of Mathematics (IOSR-JM)}}
  \textbf{\bibinfo{volume}{4}}, \bibinfo{pages}{6--13} (\bibinfo{year}{2012}).

\bibitem{Tsoularis:2001}
\bibinfo{author}{Tsoularis, A.}
\newblock \bibinfo{title}{{Analysis of Logistic Growth Models}}.
\newblock \emph{\bibinfo{journal}{Res. Lett. Inf. Math. Sci.}}
  \textbf{\bibinfo{volume}{2}}, \bibinfo{pages}{23--46} (\bibinfo{year}{2001}).

\bibitem{Lopez_etal:2010}
\bibinfo{author}{Lopez, R.}, \bibinfo{author}{Morin, B.} \&
  \bibinfo{author}{Suslov, S.}
\newblock \bibinfo{title}{{Logistic Models with Time-Dependent Coefficients and
  Some of Their Applications}}  (\bibinfo{year}{2010}).
\newblock \eprint{1008.2534}.

\bibitem{Selyugin:2008}
\bibinfo{author}{Selyugin, O.~V.}, \bibinfo{author}{Cudell, J.~R.} \&
  \bibinfo{author}{Predazzi, E.}
\newblock \bibinfo{title}{{Analytic properties of different unitarization
  schemes}}.
\newblock \emph{\bibinfo{journal}{Eur. Phys. J. ST}}
  \textbf{\bibinfo{volume}{162}}, \bibinfo{pages}{37--42}
  (\bibinfo{year}{2008}).
\newblock \eprint{0712.0621}.

\bibitem{Cudell:2009}
\bibinfo{author}{Cudell, J.~R.}, \bibinfo{author}{Predazzi, E.} \&
  \bibinfo{author}{Selyugin, O.~V.}
\newblock \bibinfo{title}{{New analytic unitarisation schemes}}.
\newblock \emph{\bibinfo{journal}{Phys. Rev.}} \textbf{\bibinfo{volume}{D79}},
  \bibinfo{pages}{034033} (\bibinfo{year}{2009}).
\newblock \eprint{0812.0735}.

\bibitem{Bourrely:2015}
\bibinfo{author}{Bourrely, C.}
\newblock \bibinfo{title}{{Parton Distribution Functions properties of the
  statistical model}}  (\bibinfo{year}{2015}).
\newblock \eprint{1507.03752}.

\bibitem{Cline:1973}
\bibinfo{author}{Cline, D.}, \bibinfo{author}{Halzen, F.} \&
  \bibinfo{author}{Luthe, J.}
\newblock \bibinfo{title}{{High transverse momentum secondaries and rising
  total cross-sections in cosmic ray interactions}}.
\newblock \emph{\bibinfo{journal}{Phys. Rev. Lett.}}
  \textbf{\bibinfo{volume}{31}}, \bibinfo{pages}{491--494}
  (\bibinfo{year}{1973}).

\bibitem{Afek:1980}
\bibinfo{author}{Afek, Y.}, \bibinfo{author}{Leroy, C.},
  \bibinfo{author}{Margolis, B.} \& \bibinfo{author}{Valin, P.}
\newblock \bibinfo{title}{{Differential and Total Proton Cross-sections,
  Particle Production and the Parton Model}}.
\newblock \emph{\bibinfo{journal}{Phys. Rev. Lett.}}
  \textbf{\bibinfo{volume}{45}}, \bibinfo{pages}{85} (\bibinfo{year}{1980}).

\bibitem{Gaisser:1985}
\bibinfo{author}{Gaisser, T.~K.} \& \bibinfo{author}{Halzen, F.}
\newblock \bibinfo{title}{{Soft Hard Scattering in the TeV Range}}.
\newblock \emph{\bibinfo{journal}{Phys. Rev. Lett.}}
  \textbf{\bibinfo{volume}{54}}, \bibinfo{pages}{1754} (\bibinfo{year}{1985}).

\bibitem{LHeureux:1985}
\bibinfo{author}{L'Heureux, P.}, \bibinfo{author}{Margolis, B.} \&
  \bibinfo{author}{Valin, P.}
\newblock \bibinfo{title}{{Quark - Gluon Model for Diffraction at
  High-Energies}}.
\newblock \emph{\bibinfo{journal}{Phys. Rev.}} \textbf{\bibinfo{volume}{D32}},
  \bibinfo{pages}{1681--1691} (\bibinfo{year}{1985}).

\bibitem{Pancheri:1985}
\bibinfo{author}{Pancheri, G.} \& \bibinfo{author}{Srivastava, Y.}
\newblock \bibinfo{title}{{Jets in Minimum Bias Physics}}.
\newblock \emph{\bibinfo{journal}{Conf. Proc.}}
  \textbf{\bibinfo{volume}{C850313}}, \bibinfo{pages}{28}
  (\bibinfo{year}{1985}).
\newblock \bibinfo{note}{[Phys. Lett.B159,69(1985)]}.

\bibitem{Durand:1987}
\bibinfo{author}{Durand, L.} \& \bibinfo{author}{Hong, P.}
\newblock \bibinfo{title}{{QCD and Rising Total Cross-Sections}}.
\newblock \emph{\bibinfo{journal}{Phys. Rev. Lett.}}
  \textbf{\bibinfo{volume}{58}}, \bibinfo{pages}{303--306}
  (\bibinfo{year}{1987}).

\bibitem{Capella:1986}
\bibinfo{author}{Capella, A.}, \bibinfo{author}{Tran Thanh~Van, J.} \&
  \bibinfo{author}{Kwiecinski, J.}
\newblock \bibinfo{title}{{Minijets, QCD and Unitarity}}.
\newblock \emph{\bibinfo{journal}{Phys. Rev. Lett.}}
  \textbf{\bibinfo{volume}{58}}, \bibinfo{pages}{2015} (\bibinfo{year}{1987}).

\bibitem{DiasdeDeus:1987}
\bibinfo{author}{Dias~de Deus, J.} \& \bibinfo{author}{Kwiecinski, J.}
\newblock \bibinfo{title}{{Semi-hard {QCD}: Minijets and Elastic Scattering}}.
\newblock \emph{\bibinfo{journal}{Phys. Lett.}}
  \textbf{\bibinfo{volume}{B196}}, \bibinfo{pages}{537--542}
  (\bibinfo{year}{1987}).

\bibitem{Block_Gregores:1999}
\bibinfo{author}{Block, M.~M.}, \bibinfo{author}{Gregores, E.~M.},
  \bibinfo{author}{Halzen, F.} \& \bibinfo{author}{Pancheri, G.}
\newblock \bibinfo{title}{{Photon-proton and photon-photon scattering from
  nucleon-nucleon forward amplitudes}}.
\newblock \emph{\bibinfo{journal}{Phys. Rev.}} \textbf{\bibinfo{volume}{D60}},
  \bibinfo{pages}{054024} (\bibinfo{year}{1999}).
\newblock \eprint{hep-ph/9809403}.

\bibitem{Fagundes:2015}
\bibinfo{author}{Fagundes, D.~A.}, \bibinfo{author}{Grau, A.},
  \bibinfo{author}{Pancheri, G.}, \bibinfo{author}{Srivastava, Y.~N.} \&
  \bibinfo{author}{Shekhovtsova, O.}
\newblock \bibinfo{title}{{Soft edge of hadron scattering and minijet models
  for the total and inelastic $pp$ cross sections at LHC and beyond}}.
\newblock \emph{\bibinfo{journal}{Phys. Rev.}} \textbf{\bibinfo{volume}{D91}},
  \bibinfo{pages}{114011} (\bibinfo{year}{2015}).
\newblock \eprint{1504.04890}.

\bibitem{UA1_Albajar:1988}
\bibinfo{author}{Albajar, C.} \emph{et~al.}
\newblock \bibinfo{title}{{Production of Low Transverse Energy Clusters in
  $\bar{p}p$ Collisions at $\sqrt{s}$ = 0.2-0.9 TeV and their Interpretation in
  Terms of QCD Jets}}.
\newblock \emph{\bibinfo{journal}{Nucl. Phys.}}
  \textbf{\bibinfo{volume}{B309}}, \bibinfo{pages}{405--425}
  (\bibinfo{year}{1988}).

\bibitem{Fialkowski:1975}
\bibinfo{author}{Fialkowski, K.} \& \bibinfo{author}{Miettinen, H.~I.}
\newblock \bibinfo{title}{{Semitransparent Hadrons from Multichannel Absorption
  Effects}}.
\newblock \emph{\bibinfo{journal}{Nucl. Phys.}}
  \textbf{\bibinfo{volume}{B103}}, \bibinfo{pages}{247--257}
  (\bibinfo{year}{1976}).

\bibitem{Sukhatme_Henyey:1976}
\bibinfo{author}{Sukhatme, U.~P.} \& \bibinfo{author}{Henyey, F.~S.}
\newblock \bibinfo{title}{{Unitarity Bounds on Diffraction Dissociation}}.
\newblock \emph{\bibinfo{journal}{Nucl. Phys.}}
  \textbf{\bibinfo{volume}{B108}}, \bibinfo{pages}{317--326}
  (\bibinfo{year}{1976}).

\bibitem{Lipari:2009}
\bibinfo{author}{Lipari, P.} \& \bibinfo{author}{Lusignoli, M.}
\newblock \bibinfo{title}{{Multiple Parton Interactions in Hadron Collisions
  and Diffraction}}.
\newblock \emph{\bibinfo{journal}{Phys. Rev.}} \textbf{\bibinfo{volume}{D80}},
  \bibinfo{pages}{074014} (\bibinfo{year}{2009}).
\newblock \eprint{0908.0495}.

\bibitem{Achilli:2011}
\bibinfo{author}{Achilli, A.} \emph{et~al.}
\newblock \bibinfo{title}{{Total and inelastic cross-sections at LHC at
  $\sqrt{s} = 7$ TeV and beyond}}.
\newblock \emph{\bibinfo{journal}{Phys. Rev.}} \textbf{\bibinfo{volume}{D84}},
  \bibinfo{pages}{094009} (\bibinfo{year}{2011}).
\newblock \eprint{1102.1949}.

\bibitem{Kohara_etal:2014a}
\bibinfo{author}{Kohara, A.~K.}, \bibinfo{author}{Ferreira, E.} \&
  \bibinfo{author}{Kodama, T.}
\newblock \bibinfo{title}{{pp Interaction at Very High Energies in Cosmic Ray
  Experiments}}.
\newblock \emph{\bibinfo{journal}{J. Phys.}} \textbf{\bibinfo{volume}{G41}},
  \bibinfo{pages}{115003} (\bibinfo{year}{2014}).
\newblock \eprint{1406.5773}.

\bibitem{Kohara_etal:2014b}
\bibinfo{author}{Kohara, A.~K.}, \bibinfo{author}{Ferreira, E.} \&
  \bibinfo{author}{Kodama, T.}
\newblock \bibinfo{title}{{pp Elastic Scattering at LHC Energies}}.
\newblock \emph{\bibinfo{journal}{Eur. Phys. J.}}
  \textbf{\bibinfo{volume}{C74}}, \bibinfo{pages}{3175} (\bibinfo{year}{2014}).
\newblock \eprint{1408.1599}.

\bibitem{Kohara_etal:2015a}
\bibinfo{author}{Kohara, A.~K.}, \bibinfo{author}{Ferreira, E.} \&
  \bibinfo{author}{Kodama, T.}
\newblock \bibinfo{title}{{pp interactions in extended air showers}}.
\newblock \emph{\bibinfo{journal}{EPJ Web Conf.}}
  \textbf{\bibinfo{volume}{99}}, \bibinfo{pages}{10002} (\bibinfo{year}{2015}).
\newblock \eprint{1410.8467}.

\bibitem{Kohara_etal:2015b}
\bibinfo{author}{Kohara, A.~K.}, \bibinfo{author}{Ferreira, E.} \&
  \bibinfo{author}{Kodama, T.}
\newblock \bibinfo{title}{{Elastic Amplitudes and Observables in pp
  Scattering}}.
\newblock \emph{\bibinfo{journal}{AIP Conf. Proc.}}
  \textbf{\bibinfo{volume}{1654}}, \bibinfo{pages}{050003}
  (\bibinfo{year}{2015}).
\newblock \eprint{1411.3518}.

\bibitem{Roy:2017}
\bibinfo{author}{Roy, S.~M.}
\newblock \bibinfo{title}{{Pomeron pole plus grey disk model: real parts,
  inelastic cross sections and LHC data}}.
\newblock \emph{\bibinfo{journal}{Phys. Lett.}}
  \textbf{\bibinfo{volume}{B764}}, \bibinfo{pages}{180--185}
  (\bibinfo{year}{2017}).
\newblock \eprint{1602.03627}.

\bibitem{Dremin:2017}
\bibinfo{author}{Dremin, I.~M.}
\newblock \bibinfo{title}{{Unexpected behaviour of cross sections of high
  energy protons}}.
\newblock \emph{\bibinfo{journal}{EPJ Web Conf.}}
  \textbf{\bibinfo{volume}{145}}, \bibinfo{pages}{10003}
  (\bibinfo{year}{2017}).

\bibitem{Kohara:2017}
\bibinfo{author}{Kohara, A.~K.}
\newblock \bibinfo{title}{{Analyticity and scaling property of pp and p\=p
  forward scattering amplitudes}}  (\bibinfo{year}{2017}).
\newblock \eprint{1710.06961}.

\bibitem{Fagundes_Menon_Silva:2017a}
\bibinfo{author}{Fagundes, D.~A.}, \bibinfo{author}{Menon, M.~J.} \&
  \bibinfo{author}{Silva, P. V. R.~G.}
\newblock \bibinfo{title}{{Bounds on the rise of total cross section from LHC7
  and LHC8 data}}.
\newblock \emph{\bibinfo{journal}{Nucl. Phys.}}
  \textbf{\bibinfo{volume}{A966}}, \bibinfo{pages}{185--196}
  (\bibinfo{year}{2017}).
\newblock \eprint{1703.07486}.

\bibitem{Pancheri_Srivastava:2017}
\bibinfo{author}{Pancheri, G.} \& \bibinfo{author}{Srivastava, Y.~N.}
\newblock \bibinfo{title}{{Introduction to the physics of the total
  cross-section at LHC}}.
\newblock \emph{\bibinfo{journal}{Eur. Phys. J.}}
  \textbf{\bibinfo{volume}{C77}}, \bibinfo{pages}{150} (\bibinfo{year}{2017}).
\newblock \eprint{1610.10038}.

\bibitem{Kaspar:2011}
\bibinfo{author}{Kaspar, J.}, \bibinfo{author}{Kundrat, V.},
  \bibinfo{author}{Lokajicek, M.} \& \bibinfo{author}{Prochazka, J.}
\newblock \bibinfo{title}{{Phenomenological models of elastic nucleon
  scattering and predictions for LHC}}.
\newblock \emph{\bibinfo{journal}{Nucl. Phys.}}
  \textbf{\bibinfo{volume}{B843}}, \bibinfo{pages}{84--106}
  (\bibinfo{year}{2011}).

\bibitem{Fiore:2009}
\bibinfo{author}{Fiore, R.} \emph{et~al.}
\newblock \bibinfo{title}{{Forward Physics at the LHC: Elastic Scattering}}.
\newblock \emph{\bibinfo{journal}{Int. J. Mod. Phys.}}
  \textbf{\bibinfo{volume}{A24}}, \bibinfo{pages}{2551--2599}
  (\bibinfo{year}{2009}).
\newblock \eprint{0810.2902}.

\bibitem{Collins_book:1977}
\bibinfo{author}{Collins, P. D.~B.}
\newblock \emph{\bibinfo{title}{{An Introduction to Regge Theory and
  High-Energy Physics}}}.
\newblock Cambridge Monographs on Mathematical Physics
  (\bibinfo{publisher}{Cambridge Univ. Press}, \bibinfo{address}{Cambridge,
  UK}, \bibinfo{year}{2009}).
\newblock
  \urlprefix\url{http://www-spires.fnal.gov/spires/find/books/www?cl=QC793.3.R4C695}.

\bibitem{Donnachie_etal_book:2002}
\bibinfo{author}{Donnachie, S.}, \bibinfo{author}{Dosch, H.~G.},
  \bibinfo{author}{Nachtmann, O.} \& \bibinfo{author}{Landshoff, P.}
\newblock \bibinfo{title}{{Pomeron physics and QCD}}.
\newblock \emph{\bibinfo{journal}{Camb. Monogr. Part. Phys. Nucl. Phys.
  Cosmol.}} \textbf{\bibinfo{volume}{19}}, \bibinfo{pages}{1--347}
  (\bibinfo{year}{2002}).

\bibitem{COMPETE:2002a}
\bibinfo{author}{Cudell, J.~R.} \emph{et~al.}
\newblock \bibinfo{title}{{Hadronic scattering amplitudes: Medium-energy
  constraints on asymptotic behavior}}.
\newblock \emph{\bibinfo{journal}{Phys. Rev. D}} \textbf{\bibinfo{volume}{65}},
  \bibinfo{pages}{074024} (\bibinfo{year}{2002}).
\newblock \eprint{hep-ph/0107219}.

\bibitem{PDG:2010}
\bibinfo{author}{Nakamura, K.} \emph{et~al.}
\newblock \bibinfo{title}{{Review of particle physics}}.
\newblock \emph{\bibinfo{journal}{J. Phys.}} \textbf{\bibinfo{volume}{G37}},
  \bibinfo{pages}{075021} (\bibinfo{year}{2010}).

\bibitem{PDG:2012}
\bibinfo{author}{Beringer, J.} \emph{et~al.}
\newblock \bibinfo{title}{{Review of Particle Physics (RPP)}}.
\newblock \emph{\bibinfo{journal}{Phys. Rev.}} \textbf{\bibinfo{volume}{D86}},
  \bibinfo{pages}{010001} (\bibinfo{year}{2012}).

\bibitem{Amaldi:1977}
\bibinfo{author}{Amaldi, U.} \emph{et~al.}
\newblock \bibinfo{title}{{The Real Part of the Forward Proton Proton
  Scattering Amplitude Measured at the CERN Intersecting Storage Rings}}.
\newblock \emph{\bibinfo{journal}{Phys. Lett.}} \textbf{\bibinfo{volume}{66B}},
  \bibinfo{pages}{390--394} (\bibinfo{year}{1977}).

\bibitem{Augier:1993}
\bibinfo{author}{Augier, C.} \emph{et~al.}
\newblock \bibinfo{title}{{Predictions on the total cross-section and real part
  at LHC and SSC}}.
\newblock \emph{\bibinfo{journal}{Phys. Lett.}}
  \textbf{\bibinfo{volume}{B315}}, \bibinfo{pages}{503--506}
  (\bibinfo{year}{1993}).

\bibitem{Bueno_Velasco:1996}
\bibinfo{author}{Bueno, A.} \& \bibinfo{author}{Velasco, J.}
\newblock \bibinfo{title}{{A Comparative study on two characteristic
  parametrizations for high-energy $p p$ and $\bar{p} p$ total
  cross-sections}}.
\newblock \emph{\bibinfo{journal}{Phys. Lett.}}
  \textbf{\bibinfo{volume}{B380}}, \bibinfo{pages}{184--188}
  (\bibinfo{year}{1996}).
\newblock \eprint{hep-ph/9605321}.

\bibitem{Cudell:1997}
\bibinfo{author}{Cudell, J.~R.}, \bibinfo{author}{Kang, K.} \&
  \bibinfo{author}{Kim, S.~K.}
\newblock \bibinfo{title}{{Bounds on the soft pomeron intercept}}.
\newblock \emph{\bibinfo{journal}{Phys. Lett.}}
  \textbf{\bibinfo{volume}{B395}}, \bibinfo{pages}{311--331}
  (\bibinfo{year}{1997}).
\newblock \eprint{hep-ph/9601336}.

\bibitem{Luna_Menon:2003}
\bibinfo{author}{Luna, E. G.~S.} \& \bibinfo{author}{Menon, M.~J.}
\newblock \bibinfo{title}{{Extrema bounds for the soft pomeron intercept}}.
\newblock \emph{\bibinfo{journal}{Phys. Lett.}}
  \textbf{\bibinfo{volume}{B565}}, \bibinfo{pages}{123--130}
  (\bibinfo{year}{2003}).
\newblock \eprint{hep-ph/0305280}.

\bibitem{Luna_Menon_Montanha2004}
\bibinfo{author}{Luna, E. G.~S.}, \bibinfo{author}{Menon, M.~J.} \&
  \bibinfo{author}{Montanha, J.}
\newblock \bibinfo{title}{{An Analysis on extrema and constrained bounds for
  the soft Pomeron intercept}}.
\newblock \emph{\bibinfo{journal}{Nucl. Phys. A}}
  \textbf{\bibinfo{volume}{745}}, \bibinfo{pages}{104--120}
  (\bibinfo{year}{2004}).
\newblock \eprint{hep-ph/0408211}.

\bibitem{Igi:2002}
\bibinfo{author}{Igi, K.} \& \bibinfo{author}{Ishida, M.}
\newblock \bibinfo{title}{{Investigations of the $\pi N$ total cross-sections
  at high-energies using new FESR: $\log\nu$ or $(\log \nu)^2$}}.
\newblock \emph{\bibinfo{journal}{Phys. Rev.}} \textbf{\bibinfo{volume}{D66}},
  \bibinfo{pages}{034023} (\bibinfo{year}{2002}).
\newblock \eprint{hep-ph/0202163}.

\bibitem{Igi:2005}
\bibinfo{author}{Igi, K.} \& \bibinfo{author}{Ishida, M.}
\newblock \bibinfo{title}{{Predictions of $p p$, $\bar{p} p$ total cross
  section and $\rho$ ratio at LHC and cosmic-ray energies}}.
\newblock \emph{\bibinfo{journal}{Phys. Lett.}}
  \textbf{\bibinfo{volume}{B622}}, \bibinfo{pages}{286--294}
  (\bibinfo{year}{2005}).
\newblock \eprint{hep-ph/0505058}.

\bibitem{Block_Halzen:2004}
\bibinfo{author}{Block, M.~M.} \& \bibinfo{author}{Halzen, F.}
\newblock \bibinfo{title}{{Evidence for the saturation of the Froissart
  bound}}.
\newblock \emph{\bibinfo{journal}{Phys. Rev.}} \textbf{\bibinfo{volume}{D70}},
  \bibinfo{pages}{091901} (\bibinfo{year}{2004}).
\newblock \eprint{hep-ph/0405174}.

\bibitem{Block_Halzen:2005}
\bibinfo{author}{Block, M.~M.} \& \bibinfo{author}{Halzen, F.}
\newblock \bibinfo{title}{{New evidence for the saturation of the Froissart
  bound}}.
\newblock \emph{\bibinfo{journal}{Phys. Rev.}} \textbf{\bibinfo{volume}{D72}},
  \bibinfo{pages}{036006} (\bibinfo{year}{2005}).
\newblock \bibinfo{note}{[Erratum: Phys. Rev.D72,039902(2005)]},
  \eprint{hep-ph/0506031}.

\bibitem{Fagundes_Menon_Silva:2012a}
\bibinfo{author}{Fagundes, D.~A.}, \bibinfo{author}{Menon, M.~J.} \&
  \bibinfo{author}{Silva, P. V. R.~G.}
\newblock \bibinfo{title}{{Total Hadronic Cross Section Data and the
  Froissart-Martin Bound}}.
\newblock \emph{\bibinfo{journal}{Braz. J. Phys.}}
  \textbf{\bibinfo{volume}{42}}, \bibinfo{pages}{452--464}
  (\bibinfo{year}{2012}).
\newblock \eprint{1112.4704}.

\bibitem{Svensson:1967}
\bibinfo{author}{Svensson, B. E.~Y.}
\newblock \bibinfo{title}{{High-energy phenomenology and regge poles}}
  (\bibinfo{year}{1967}).
\newblock \urlprefix\url{http://cds.cern.ch/record/868665}.

\bibitem{Chew:1961}
\bibinfo{author}{Chew, G.~F.} \& \bibinfo{author}{Frautschi, S.~C.}
\newblock \bibinfo{title}{{Principle of Equivalence for All Strongly
  Interacting Particles Within the S Matrix Framework}}.
\newblock \emph{\bibinfo{journal}{Phys. Rev. Lett.}}
  \textbf{\bibinfo{volume}{7}}, \bibinfo{pages}{394--397}
  (\bibinfo{year}{1961}).

\bibitem{Chew:1962}
\bibinfo{author}{Chew, G.~F.} \& \bibinfo{author}{Frautschi, S.~C.}
\newblock \bibinfo{title}{{Regge Trajectories and the Principle of Maximum
  Strength for Strong Interactions}}.
\newblock \emph{\bibinfo{journal}{Phys. Rev. Lett.}}
  \textbf{\bibinfo{volume}{8}}, \bibinfo{pages}{41--44} (\bibinfo{year}{1962}).

\bibitem{Eden_book:1967}
\bibinfo{author}{Eden, R.~J.}
\newblock \emph{\bibinfo{title}{{High Energy Collisions of Elementary
  Particles}}} (\bibinfo{publisher}{Cambridge University Press},
  \bibinfo{address}{Cambridge}, \bibinfo{year}{1967}).

\bibitem{Barger:1968}
\bibinfo{author}{Barger, V.}, \bibinfo{author}{Olsson, M.} \&
  \bibinfo{author}{Reeder, D.~D.}
\newblock \bibinfo{title}{{Asymptotic projections of scattering models}}.
\newblock \emph{\bibinfo{journal}{Nucl. Phys.}} \textbf{\bibinfo{volume}{B5}},
  \bibinfo{pages}{411--430} (\bibinfo{year}{1968}).

\bibitem{Abatzis:1994}
\bibinfo{author}{Abatzis, S.} \emph{et~al.}
\newblock \bibinfo{title}{{Observation of a narrow scalar meson at 1450 MeV in
  the reaction $pp\to p_{f}(\pi^+\pi^-\pi^+\pi^-)p_{s}$ at 450 GeV/c using the
  CERN Omega spectrometer}}.
\newblock \emph{\bibinfo{journal}{Physics Letters B}}
  \textbf{\bibinfo{volume}{324}}, \bibinfo{pages}{509 -- 514}
  (\bibinfo{year}{1994}).
\newblock
  \urlprefix\url{http://www.sciencedirect.com/science/article/pii/0370269394902313}.

\bibitem{Donnachie:2013}
\bibinfo{author}{Donnachie, A.} \& \bibinfo{author}{Landshoff, P.~V.}
\newblock \bibinfo{title}{{$pp$ and $\bar pp$ total cross sections and elastic
  scattering}}.
\newblock \emph{\bibinfo{journal}{Phys. Lett.}}
  \textbf{\bibinfo{volume}{B727}}, \bibinfo{pages}{500--505}
  (\bibinfo{year}{2013}).
\newblock \bibinfo{note}{[Erratum: Phys. Lett.B750,669(2015)]},
  \eprint{1309.1292}.

\bibitem{Low:1975}
\bibinfo{author}{Low, F.~E.}
\newblock \bibinfo{title}{{A Model of the Bare Pomeron}}.
\newblock \emph{\bibinfo{journal}{Phys. Rev.}} \textbf{\bibinfo{volume}{D12}},
  \bibinfo{pages}{163--173} (\bibinfo{year}{1975}).

\bibitem{Nussinov:1975}
\bibinfo{author}{Nussinov, S.}
\newblock \bibinfo{title}{{Colored Quark Version of Some Hadronic Puzzles}}.
\newblock \emph{\bibinfo{journal}{Phys. Rev. Lett.}}
  \textbf{\bibinfo{volume}{34}}, \bibinfo{pages}{1286--1289}
  (\bibinfo{year}{1975}).

\bibitem{Cudell:1991}
\bibinfo{author}{Cudell, J.~R.} \& \bibinfo{author}{Ross, D.~A.}
\newblock \bibinfo{title}{{Theory and phenomenology of the gluon propagator
  from the Dyson-Schwinger equation in QCD}}.
\newblock \emph{\bibinfo{journal}{Nucl. Phys.}}
  \textbf{\bibinfo{volume}{B359}}, \bibinfo{pages}{247--261}
  (\bibinfo{year}{1991}).

\bibitem{Halzen:1992}
\bibinfo{author}{Halzen, F.}, \bibinfo{author}{Krein, G.~I.} \&
  \bibinfo{author}{Natale, A.~A.}
\newblock \bibinfo{title}{{Relating the QCD pomeron to an effective gluon
  mass}}.
\newblock \emph{\bibinfo{journal}{Phys. Rev.}} \textbf{\bibinfo{volume}{D47}},
  \bibinfo{pages}{295--298} (\bibinfo{year}{1993}).

\bibitem{Canfora:2017}
\bibinfo{author}{Canfora, F.~E.} \emph{et~al.}
\newblock \bibinfo{title}{{Double nonperturbative gluon exchange: An update on
  the soft-Pomeron contribution to pp scattering}}.
\newblock \emph{\bibinfo{journal}{Phys. Rev.}} \textbf{\bibinfo{volume}{C96}},
  \bibinfo{pages}{025202} (\bibinfo{year}{2017}).
\newblock \eprint{1707.02109}.

\bibitem{Eden_Landshoff:1964}
\bibinfo{author}{Eden, R.~J.} \& \bibinfo{author}{Landshoff, P.~V.}
\newblock \bibinfo{title}{{Higher Order Poles in the $S$ Matrix}}.
\newblock \emph{\bibinfo{journal}{Phys. Rev.}} \textbf{\bibinfo{volume}{136}},
  \bibinfo{pages}{B1817--B1820} (\bibinfo{year}{1964}).
\newblock \urlprefix\url{https://link.aps.org/doi/10.1103/PhysRev.136.B1817}.

\bibitem{Gribov_Migdal:1968a}
\bibinfo{author}{Gribov, V.~N.} \& \bibinfo{author}{Migdal, A.~A.}
\newblock \bibinfo{title}{{Properties of the Pomeranchuk pole and the branch
  cuts related to it at low momentum transfer}}.
\newblock \emph{\bibinfo{journal}{Sov. J. Nucl. Phys.}}
  \textbf{\bibinfo{volume}{8}}, \bibinfo{pages}{583--590}
  (\bibinfo{year}{1969}).
\newblock \bibinfo{note}{[Yad. Fiz.8,1002(1968)]}.

\bibitem{Gribov_Migdal:1968b}
\bibinfo{author}{Gribov, V.~N.} \& \bibinfo{author}{Migdal, A.~A.}
\newblock \bibinfo{title}{{The Pomeranchuk quasi-stable pole and diffraction
  scattering at ultrahigh-energy}}.
\newblock \emph{\bibinfo{journal}{Sov. J. Nucl. Phys.}}
  \textbf{\bibinfo{volume}{8}}, \bibinfo{pages}{703} (\bibinfo{year}{1969}).
\newblock \bibinfo{note}{[Yad. Fiz.8,1213(1968)]}.

\bibitem{Bronzan_Kane_Sukhatme:1974}
\bibinfo{author}{Bronzan, J.~B.}, \bibinfo{author}{Kane, G.~L.} \&
  \bibinfo{author}{Sukhatme, U.~P.}
\newblock \bibinfo{title}{{Obtaining Real Parts of Scattering Amplitudes
  Directly from Cross-Section Data Using Derivative Analyticity Relations}}.
\newblock \emph{\bibinfo{journal}{Phys. Lett. B}}
  \textbf{\bibinfo{volume}{49}}, \bibinfo{pages}{272--276}
  (\bibinfo{year}{1974}).

\bibitem{Avila_Menon:2004}
\bibinfo{author}{\'Avila, R.~F.} \& \bibinfo{author}{Menon, M.~J.}
\newblock \bibinfo{title}{{Critical analysis of derivative dispersion relations
  at high-energies}}.
\newblock \emph{\bibinfo{journal}{Nucl. Phys. A}}
  \textbf{\bibinfo{volume}{744}}, \bibinfo{pages}{249--272}
  (\bibinfo{year}{2004}).
\newblock \eprint{hep-ph/0309028}.

\bibitem{Donnachie:1992}
\bibinfo{author}{Donnachie, A.} \& \bibinfo{author}{Landshoff, P.~V.}
\newblock \bibinfo{title}{{Total cross-sections}}.
\newblock \emph{\bibinfo{journal}{Phys. Lett.}}
  \textbf{\bibinfo{volume}{B296}}, \bibinfo{pages}{227--232}
  (\bibinfo{year}{1992}).
\newblock \eprint{hep-ph/9209205}.

\bibitem{Byron_Fuller_book:1970}
\bibinfo{author}{Byron, F. W.~J.} \& \bibinfo{author}{Fuller, R.~W.}
\newblock \emph{\bibinfo{title}{{Mathematics of Classical and Quantum
  Physics}}} (\bibinfo{publisher}{Dover Publications}, \bibinfo{address}{New
  York}, \bibinfo{year}{1992}).

\bibitem{Goldberger:1957}
\bibinfo{author}{Goldberger, M.~L.}, \bibinfo{author}{Nambu, Y.} \&
  \bibinfo{author}{Oehme, R.}
\newblock \bibinfo{title}{Dispersion relations for nucleon-nucleon scattering}.
\newblock \emph{\bibinfo{journal}{Annals of Physics}}
  \textbf{\bibinfo{volume}{2}}, \bibinfo{pages}{226 -- 282}
  (\bibinfo{year}{1957}).
\newblock
  \urlprefix\url{http://www.sciencedirect.com/science/article/pii/0003491657900301}.

\bibitem{Soding:1964}
\bibinfo{author}{Söding, P.}
\newblock \bibinfo{title}{Real part of the proton-proton and proton-antiproton
  forward scattering amplitude at high energies}.
\newblock \emph{\bibinfo{journal}{Physics Letters}}
  \textbf{\bibinfo{volume}{8}}, \bibinfo{pages}{285 -- 287}
  (\bibinfo{year}{1964}).
\newblock
  \urlprefix\url{http://www.sciencedirect.com/science/article/pii/S0031916364918979}.

\bibitem{Bertini:1996}
\bibinfo{author}{Bertini, M.}, \bibinfo{author}{Giffon, M.},
  \bibinfo{author}{Jenkovszky, L.~L.} \& \bibinfo{author}{Paccanoni, F.}
\newblock \bibinfo{title}{{Real part of the scattering amplitude and model
  analysis}}.
\newblock \emph{\bibinfo{journal}{Nuovo Cim.}} \textbf{\bibinfo{volume}{A109}},
  \bibinfo{pages}{257--270} (\bibinfo{year}{1996}).

\bibitem{Gradshteyn_etal_book:1980}
\bibinfo{author}{Gradshteyn, I.} \& \bibinfo{author}{Ryzhik, I.}
\newblock \emph{\bibinfo{title}{{Table of Integrals, Series, and Products}}}
  (\bibinfo{publisher}{Academic Press}, \bibinfo{address}{San Diego, CA},
  \bibinfo{year}{1980}).

\bibitem{Abramowitz_Stegun_book:1964}
\bibinfo{author}{Abramowitz, M.} \& \bibinfo{author}{Stegun, I.~A.}
\newblock \emph{\bibinfo{title}{{Handbook of Mathematical Function}}}
  (\bibinfo{publisher}{Dover}, \bibinfo{address}{New York},
  \bibinfo{year}{1964}).

\bibitem{Avila_Menon:2006}
\bibinfo{author}{\'Avila, R.~F.} \& \bibinfo{author}{Menon, M.~J.}
\newblock \bibinfo{title}{{Extended derivative dispersion relations}}.
\newblock In \emph{\bibinfo{booktitle}{{Sense of Beauty in Physics:
  Miniconference in Honor of Adriano Di Giacomo on his 70th Birthday Pisa,
  Italy, January 26-27, 2006}}} (\bibinfo{year}{2006}).
\newblock \eprint{hep-ph/0601194}.

\bibitem{Avila_Menon:2007a}
\bibinfo{author}{\'Avila, R.~F.} \& \bibinfo{author}{Menon, M.~J.}
\newblock \bibinfo{title}{{Derivative dispersion relations above the physical
  threshold}}.
\newblock \emph{\bibinfo{journal}{Braz. J. Phys.}}
  \textbf{\bibinfo{volume}{37}}, \bibinfo{pages}{358--367}
  (\bibinfo{year}{2007}).
\newblock \eprint{hep-ph/0512166}.

\bibitem{Ferreira_Sesma:2008}
\bibinfo{author}{Ferreira, E.} \& \bibinfo{author}{Sesma, J.}
\newblock \bibinfo{title}{{Representation of Integral Dispersion Relations by
  Local Forms}}.
\newblock \emph{\bibinfo{journal}{J. Math. Phys.}}
  \textbf{\bibinfo{volume}{49}}, \bibinfo{pages}{033504}
  (\bibinfo{year}{2008}).
\newblock \eprint{0707.4266}.

\bibitem{Ferreira_Sesma:2013}
\bibinfo{author}{Ferreira, E.} \& \bibinfo{author}{Sesma, J.}
\newblock \bibinfo{title}{{Two-point derivative dispersion relations}}.
\newblock \emph{\bibinfo{journal}{Journal of Mathematical Physics}}
  \textbf{\bibinfo{volume}{54}}, \bibinfo{pages}{033507}
  (\bibinfo{year}{2013}).
\newblock \urlprefix\url{http://dx.doi.org/10.1063/1.4795116}.
\newblock \eprint{http://dx.doi.org/10.1063/1.4795116}.

\bibitem{Avila_phdthesis:2009}
\bibinfo{author}{\'Avila, R.~F.}
\newblock \emph{\bibinfo{title}{{Analitical, Empirical and Phenomenological
  Aspects of the Elastic Hadron Scattering at High Energies}}}.
\newblock Ph.D. thesis, \bibinfo{school}{IMECC-UNICAMP},
  \bibinfo{address}{Campinas, SP} (\bibinfo{year}{2009}).
\newblock
  \urlprefix\url{http://www.bibliotecadigital.unicamp.br/document/?code=000436923\&opt=1}.
\newblock \bibinfo{note}{(Available in Portuguese.)}.

\bibitem{Avila_Menon:2007b}
\bibinfo{author}{\'Avila, R.~F.} \& \bibinfo{author}{Menon, M.~J.}
\newblock \bibinfo{title}{{From integral to derivative dispersion relations}}.
\newblock \emph{\bibinfo{journal}{Braz. J. Phys.}}
  \textbf{\bibinfo{volume}{37}}, \bibinfo{pages}{661--664}
  (\bibinfo{year}{2007}).

\bibitem{Avila_Menon:2004b}
\bibinfo{author}{\'Avila, R.~F.} \& \bibinfo{author}{Menon, M.~J.}
\newblock \bibinfo{title}{{Derivative dispersion relations}}.
\newblock \emph{\bibinfo{journal}{PoS}} \textbf{\bibinfo{volume}{WC2004}},
  \bibinfo{pages}{043} (\bibinfo{year}{2004}).
\newblock \eprint{hep-ph/0411401}.

\bibitem{Martynov:2004}
\bibinfo{author}{Martynov, E.}, \bibinfo{author}{Cudell, J.~R.} \&
  \bibinfo{author}{Selyugin, O.~V.}
\newblock \bibinfo{title}{{Integral and derivative dispersion relations,
  analysis of the forward scattering data}}.
\newblock \emph{\bibinfo{journal}{Eur. Phys. J.}}
  \textbf{\bibinfo{volume}{C33}}, \bibinfo{pages}{S533} (\bibinfo{year}{2004}).
\newblock \eprint{hep-ph/0311019}.

\bibitem{Cudell:2003}
\bibinfo{author}{Cudell, J.~R.}, \bibinfo{author}{Martynov, E.} \&
  \bibinfo{author}{Selyugin, O.~V.}
\newblock \bibinfo{title}{{Integral and derivative dispersion relations in the
  analysis of the data on $pp$ and $\bar{p}p$ forward scattering}}.
\newblock In \emph{\bibinfo{booktitle}{{International Conference on New Trends
  in High-Energy Physics (Experiment, Phenomenology, Theory) Alushta, Crimea,
  Ukraine, May 24-31, 2003}}} (\bibinfo{year}{2003}).
\newblock \eprint{hep-ph/0307254}.

\bibitem{Ferreira:2017a}
\bibinfo{author}{Ferreira, E.}, \bibinfo{author}{Kohara.A.K.} \&
  \bibinfo{author}{Sesma, J.}
\newblock \bibinfo{title}{{New properties of the Lerch's transcendent}}.
\newblock \emph{\bibinfo{journal}{Journal of Number Theory}}
  \textbf{\bibinfo{volume}{172}}, \bibinfo{pages}{21 -- 31}
  (\bibinfo{year}{2017}).
\newblock
  \urlprefix\url{http://www.sciencedirect.com/science/article/pii/S0022314X16302219}.

\bibitem{Ferreira:2017b}
\bibinfo{author}{Ferreira, E.}, \bibinfo{author}{Kohara, A.~K.} \&
  \bibinfo{author}{Sesma, J.}
\newblock \bibinfo{title}{{Exact treatment of dispersion relations in pp and
  p\=p elastic scattering}}  (\bibinfo{year}{2017}).
\newblock \eprint{1704.08866}.

\bibitem{Kang_Nicolescu:1975}
\bibinfo{author}{Kang, K.} \& \bibinfo{author}{Nicolescu, B.}
\newblock \bibinfo{title}{{Models for Hadron - Hadron Scattering at
  High-Energies and Rising Total Cross-Sections}}.
\newblock \emph{\bibinfo{journal}{Phys. Rev.}} \textbf{\bibinfo{volume}{D11}},
  \bibinfo{pages}{2461} (\bibinfo{year}{1975}).

\bibitem{Block:2006}
\bibinfo{author}{Block, M.~M.}
\newblock \bibinfo{title}{{Hadronic forward scattering: Predictions for the
  Large Hadron Collider and cosmic rays}}.
\newblock \emph{\bibinfo{journal}{Phys. Rept.}} \textbf{\bibinfo{volume}{436}},
  \bibinfo{pages}{71--215} (\bibinfo{year}{2006}).
\newblock \eprint{hep-ph/0606215}.

\bibitem{Block_Cahn:1985}
\bibinfo{author}{Block, M.~M.} \& \bibinfo{author}{Cahn, R.~N.}
\newblock \bibinfo{title}{{High-Energy $p \bar{p}$ and $p p$ Forward Elastic
  Scattering and Total Cross-Sections}}.
\newblock \emph{\bibinfo{journal}{Rev. Mod. Phys.}}
  \textbf{\bibinfo{volume}{57}}, \bibinfo{pages}{563} (\bibinfo{year}{1985}).

\bibitem{UA4:1984}
\bibinfo{author}{Bozzo, M.} \emph{et~al.}
\newblock \bibinfo{title}{{Measurement of the Proton-antiproton Total and
  Elastic Cross-Sections at the CERN SPS Collider}}.
\newblock \emph{\bibinfo{journal}{Phys. Lett.}}
  \textbf{\bibinfo{volume}{147B}}, \bibinfo{pages}{392--398}
  (\bibinfo{year}{1984}).

\bibitem{E710:1992}
\bibinfo{author}{Amos, N.~A.} \emph{et~al.}
\newblock \bibinfo{title}{{Measurement of $\rho$, the ratio of the real to
  imaginary part of the $\bar{p} p$ forward elastic scattering amplitude, at
  $\sqrt{s}$ = 1.8 TeV}}.
\newblock \emph{\bibinfo{journal}{Phys. Rev. Lett.}}
  \textbf{\bibinfo{volume}{68}}, \bibinfo{pages}{2433--2436}
  (\bibinfo{year}{1992}).

\bibitem{E811:1999}
\bibinfo{author}{Avila, C.} \emph{et~al.}
\newblock \bibinfo{title}{{A Measurement of the proton-antiproton total
  cross-section at $\sqrt{s}$ = 1.8 TeV}}.
\newblock \emph{\bibinfo{journal}{Phys. Lett. B}}
  \textbf{\bibinfo{volume}{445}}, \bibinfo{pages}{419--422}
  (\bibinfo{year}{1999}).

\bibitem{Cheng_Wu:1970}
\bibinfo{author}{Cheng, H.} \& \bibinfo{author}{Wu, T.~T.}
\newblock \bibinfo{title}{{Limit of Cross-Sections at Infinite Energy}}.
\newblock \emph{\bibinfo{journal}{Phys. Rev. Lett.}}
  \textbf{\bibinfo{volume}{24}}, \bibinfo{pages}{1456--1460}
  (\bibinfo{year}{1970}).

\bibitem{Giordano_Meggiolaro:2015}
\bibinfo{author}{Giordano, M.} \& \bibinfo{author}{Meggiolaro, E.}
\newblock \bibinfo{title}{{Comments on high-energy total cross sections in
  QCD}}.
\newblock \emph{\bibinfo{journal}{Phys. Lett.}}
  \textbf{\bibinfo{volume}{B744}}, \bibinfo{pages}{263--267}
  (\bibinfo{year}{2015}).
\newblock \eprint{1411.0553}.

\bibitem{Bertrand:2000}
\bibinfo{author}{Bertrand, J.}, \bibinfo{author}{Bertrand, P.} \&
  \bibinfo{author}{Ovarlez, J.}
\newblock \bibinfo{title}{{The Mellin Transform}}.
\newblock In \bibinfo{editor}{Poularikas, A.~D.} (ed.)
  \emph{\bibinfo{booktitle}{{The Transforms and Applications Handbook: Second
  Edition.}}} (\bibinfo{publisher}{CRC Press LLC}, \bibinfo{address}{Boca
  Raton}, \bibinfo{year}{2000}).

\bibitem{CapelasdeOliveira:2017}
\bibinfo{author}{Capelas~de Oliveira, E.}, \bibinfo{author}{Menon, M.~J.} \&
  \bibinfo{author}{Silva, P. V. R.~G.}
\newblock \bibinfo{title}{{Analytic components for the hadronic total
  cross-section: Fractional calculus and Mellin transform}}
  (\bibinfo{year}{2017}).
\newblock \eprint{1708.01255}.

\bibitem{Giordano_Meggiolaro_Silva:2017}
\bibinfo{author}{Giordano, M.}, \bibinfo{author}{Meggiolaro, E.} \&
  \bibinfo{author}{Silva, P. V. R.~G.}
\newblock \bibinfo{title}{{Investigation of the leading and subleading
  high-energy behavior of hadron-hadron total cross sections using a best-fit
  analysis of hadronic scattering data}}.
\newblock \emph{\bibinfo{journal}{Phys. Rev.}} \textbf{\bibinfo{volume}{D96}},
  \bibinfo{pages}{034015} (\bibinfo{year}{2017}).
\newblock \eprint{1703.00244}.

\bibitem{Nachtmann:1991}
\bibinfo{author}{Nachtmann, O.}
\newblock \bibinfo{title}{{Considerations concerning diffraction scattering in
  quantum chromodynamics}}.
\newblock \emph{\bibinfo{journal}{Annals Phys.}}
  \textbf{\bibinfo{volume}{209}}, \bibinfo{pages}{436--478}
  (\bibinfo{year}{1991}).

\bibitem{Dosch_etal:1994}
\bibinfo{author}{Dosch, H.~G.}, \bibinfo{author}{Ferreira, E.} \&
  \bibinfo{author}{Kr{\"a}mer, A.}
\newblock \bibinfo{title}{{Nonperturbative QCD treatment of high-energy hadron
  hadron scattering}}.
\newblock \emph{\bibinfo{journal}{Phys. Rev. D}} \textbf{\bibinfo{volume}{50}},
  \bibinfo{pages}{1992--2015} (\bibinfo{year}{1994}).
\newblock \eprint{hep-ph/9405237}.

\bibitem{Nachtmann:1997_1}
\bibinfo{author}{Nachtmann, O.}
\newblock \bibinfo{title}{{Perturbative and nonperturbative aspects of quantum
  field theory. Proceedings, 35. Internationale Universit{\"a}tswochen f{\"u}r
  Kern- und Teilchenphysik}}.
\newblock \emph{\bibinfo{journal}{Lect. Notes Phys.}}
  \textbf{\bibinfo{volume}{479}}, \bibinfo{pages}{pp.1--86}
  (\bibinfo{year}{1997}).

\bibitem{Berger_Nachtmann:1999}
\bibinfo{author}{Berger, E.~R.} \& \bibinfo{author}{Nachtmann, O.}
\newblock \bibinfo{title}{{Differential cross-sections for high-energy elastic
  hadron hadron scattering in nonperturbative QCD}}.
\newblock \emph{\bibinfo{journal}{Eur. Phys. J. C}}
  \textbf{\bibinfo{volume}{7}}, \bibinfo{pages}{459--473}
  (\bibinfo{year}{1999}).
\newblock \eprint{hep-ph/9808320}.

\bibitem{Shoshi_etal:2002}
\bibinfo{author}{Shoshi, A.~I.}, \bibinfo{author}{Steffen, F.~D.} \&
  \bibinfo{author}{Pirner, H.~J.}
\newblock \bibinfo{title}{{$S$ matrix unitarity, impact parameter profiles,
  gluon saturation and high-energy scattering}}.
\newblock \emph{\bibinfo{journal}{Nucl. Phys. A}}
  \textbf{\bibinfo{volume}{709}}, \bibinfo{pages}{131--183}
  (\bibinfo{year}{2002}).
\newblock \eprint{hep-ph/0202012}.

\bibitem{Verlinde_Verlinde:1993}
\bibinfo{author}{Verlinde, H.~L.} \& \bibinfo{author}{Verlinde, E.~P.}
\newblock \bibinfo{title}{{QCD at high-energies and two-dimensional field
  theory}}  (\bibinfo{year}{1993}).
\newblock \eprint{hep-th/9302104}.

\bibitem{Korchemsky:1994}
\bibinfo{author}{Korchemsky, G.~P.}
\newblock \bibinfo{title}{{On Near forward high-energy scattering in QCD}}.
\newblock \emph{\bibinfo{journal}{Phys. Lett. B}}
  \textbf{\bibinfo{volume}{325}}, \bibinfo{pages}{459--466}
  (\bibinfo{year}{1994}).
\newblock \eprint{hep-ph/9311294}.

\bibitem{Korchemskaya_Korchemsky:1995}
\bibinfo{author}{Korchemskaya, I.~A.} \& \bibinfo{author}{Korchemsky, G.~P.}
\newblock \bibinfo{title}{{High-energy scattering in QCD and cross
  singularities of Wilson loops}}.
\newblock \emph{\bibinfo{journal}{Nucl. Phys. B}}
  \textbf{\bibinfo{volume}{437}}, \bibinfo{pages}{127--162}
  (\bibinfo{year}{1995}).
\newblock \eprint{hep-ph/9409446}.

\bibitem{Meggiolaro:1996}
\bibinfo{author}{Meggiolaro, E.}
\newblock \bibinfo{title}{{A Remark on the high-energy quark quark scattering
  and the eikonal approximation}}.
\newblock \emph{\bibinfo{journal}{Phys. Rev. D}} \textbf{\bibinfo{volume}{53}},
  \bibinfo{pages}{3835--3845} (\bibinfo{year}{1996}).
\newblock \eprint{hep-th/9506043}.

\bibitem{Meggiolaro:2001}
\bibinfo{author}{Meggiolaro, E.}
\newblock \bibinfo{title}{{Eikonal propagators and high-energy parton parton
  scattering in gauge theories}}.
\newblock \emph{\bibinfo{journal}{Nucl. Phys. B}}
  \textbf{\bibinfo{volume}{602}}, \bibinfo{pages}{261--288}
  (\bibinfo{year}{2001}).
\newblock \eprint{hep-ph/0009261}.

\bibitem{Meggiolaro:1997}
\bibinfo{author}{Meggiolaro, E.}
\newblock \bibinfo{title}{{The High-energy quark quark scattering: From
  Minkowskian to Euclidean theory}}.
\newblock \emph{\bibinfo{journal}{Z. Phys. C}} \textbf{\bibinfo{volume}{76}},
  \bibinfo{pages}{523--535} (\bibinfo{year}{1997}).
\newblock \eprint{hep-th/9602104}.

\bibitem{Meggiolaro:1998}
\bibinfo{author}{Meggiolaro, E.}
\newblock \bibinfo{title}{{The Analytic continuation of the high-energy quark
  quark scattering amplitude}}.
\newblock \emph{\bibinfo{journal}{Eur. Phys. J. C}}
  \textbf{\bibinfo{volume}{4}}, \bibinfo{pages}{101--106}
  (\bibinfo{year}{1998}).
\newblock \eprint{hep-th/9702186}.

\bibitem{Meggiolaro:2002}
\bibinfo{author}{Meggiolaro, E.}
\newblock \bibinfo{title}{{The Analytic continuation of the high-energy
  parton-parton scattering amplitude with an IR cutoff}}.
\newblock \emph{\bibinfo{journal}{Nucl. Phys. B}}
  \textbf{\bibinfo{volume}{625}}, \bibinfo{pages}{312--326}
  (\bibinfo{year}{2002}).
\newblock \eprint{hep-ph/0110069}.

\bibitem{Meggiolaro:2005}
\bibinfo{author}{Meggiolaro, E.}
\newblock \bibinfo{title}{{On the loop-loop scattering amplitudes in Abelian
  and non-Abelian gauge theories}}.
\newblock \emph{\bibinfo{journal}{Nucl. Phys. B}}
  \textbf{\bibinfo{volume}{707}}, \bibinfo{pages}{199--214}
  (\bibinfo{year}{2005}).
\newblock \eprint{hep-ph/0407084}.

\bibitem{Giordano_Meggiolaro:2009}
\bibinfo{author}{Giordano, M.} \& \bibinfo{author}{Meggiolaro, E.}
\newblock \bibinfo{title}{{A Nonperturbative foundation of the
  Euclidean-Minkowskian duality of Wilson-loop correlation functions}}.
\newblock \emph{\bibinfo{journal}{Phys. Lett. B}}
  \textbf{\bibinfo{volume}{675}}, \bibinfo{pages}{123--132}
  (\bibinfo{year}{2009}).
\newblock \eprint{0902.4145}.

\bibitem{LambertW}
\bibinfo{author}{Corless, R.~M.}, \bibinfo{author}{Gonnet, G.~H.},
  \bibinfo{author}{Hare, D.~E.}, \bibinfo{author}{Jeffrey, D.~J.} \&
  \bibinfo{author}{Knuth, D.~E.}
\newblock \bibinfo{title}{{On the Lambert W function}}.
\newblock \emph{\bibinfo{journal}{Adv. Comput. Math.}}
  \textbf{\bibinfo{volume}{5}}, \bibinfo{pages}{329--359}
  (\bibinfo{year}{1996}).

\bibitem{Martin_Roy:2014}
\bibinfo{author}{Martin, A.} \& \bibinfo{author}{Roy, S.~M.}
\newblock \bibinfo{title}{{Froissart Bound on Total Cross-section without
  Unknown Constants}}.
\newblock \emph{\bibinfo{journal}{Phys. Rev. D}} \textbf{\bibinfo{volume}{89}},
  \bibinfo{pages}{045015} (\bibinfo{year}{2014}).
\newblock \eprint{1306.5210}.

\bibitem{Diez_etal:2015}
\bibinfo{author}{Errasti~D\'iez, V.}, \bibinfo{author}{Godbole, R.~M.} \&
  \bibinfo{author}{Sinha, A.}
\newblock \bibinfo{title}{{Improvements to the Froissart bound from AdS/CFT}}.
\newblock \emph{\bibinfo{journal}{Phys. Lett. B}}
  \textbf{\bibinfo{volume}{746}}, \bibinfo{pages}{285--292}
  (\bibinfo{year}{2015}).
\newblock \eprint{1504.05754}.

\bibitem{Nastase_Sonnenschein:2015}
\bibinfo{author}{Nastase, H.} \& \bibinfo{author}{Sonnenschein, J.}
\newblock \bibinfo{title}{{More on Heisenberg’s model for high energy
  nucleon-nucleon scattering}}.
\newblock \emph{\bibinfo{journal}{Phys. Rev. D}} \textbf{\bibinfo{volume}{92}},
  \bibinfo{pages}{105028} (\bibinfo{year}{2015}).
\newblock \eprint{1504.01328}.

\bibitem{Heisenberg:1952}
\bibinfo{author}{Heisenberg, W.}
\newblock \bibinfo{title}{{Mesonenerzeugung als Stosswellenproblem}}.
\newblock \emph{\bibinfo{journal}{Z. Phys.}} \textbf{\bibinfo{volume}{133}},
  \bibinfo{pages}{65} (\bibinfo{year}{1952}).

\bibitem{Block_Halzen:2006}
\bibinfo{author}{Block, M.~M.} \& \bibinfo{author}{Halzen, F.}
\newblock \bibinfo{title}{{Duality as a robust constraint on the LHC cross
  section}}.
\newblock \emph{\bibinfo{journal}{Phys. Rev. D}} \textbf{\bibinfo{volume}{73}},
  \bibinfo{pages}{054022} (\bibinfo{year}{2006}).
\newblock \eprint{hep-ph/0510238}.

\bibitem{Desgrolard:1999}
\bibinfo{author}{Desgrolard, P.}, \bibinfo{author}{Jenkovszky, L.~L.} \&
  \bibinfo{author}{Struminsky, B.}
\newblock \bibinfo{title}{{Where is the black disc limit?}}
\newblock \emph{\bibinfo{journal}{Eur. Phys. J.}}
  \textbf{\bibinfo{volume}{C11}}, \bibinfo{pages}{145--151}
  (\bibinfo{year}{1999}).
\newblock \eprint{hep-ph/9901437}.

\bibitem{Desgrolard:2000}
\bibinfo{author}{Desgrolard, P.}, \bibinfo{author}{Jenkovszky, L.~L.} \&
  \bibinfo{author}{Struminsky, B.~V.}
\newblock \bibinfo{title}{{Unitarity, (anti)shadowing, and black-disk limit}}.
\newblock \emph{\bibinfo{journal}{Phys. Atom. Nucl.}}
  \textbf{\bibinfo{volume}{63}}, \bibinfo{pages}{891--896}
  (\bibinfo{year}{2000}).
\newblock \bibinfo{note}{[Yad. Fiz.63,962(2000)]}.

\bibitem{Forshaw:1997}
\bibinfo{author}{Forshaw, J.~R.} \& \bibinfo{author}{Ross, D.~A.}
\newblock \bibinfo{title}{{Quantum Chromodynamics and the Pomeron}}.
\newblock \emph{\bibinfo{journal}{Cambridge Lect. Notes Phys.}}
  \textbf{\bibinfo{volume}{9}}, \bibinfo{pages}{1--248} (\bibinfo{year}{1997}).

\bibitem{Oosthuizen:2013}
\bibinfo{author}{Oosthuizen, J.}
\newblock \bibinfo{title}{{The Mellin Transform}}.
\newblock
  \urlprefix\url{http://math.sun.ac.za/wp-content/uploads/2013/02/Hons-Projek.pdf}.

\end{thebibliography}
